\documentclass[aps,prd,nofootinbib]{revtex4-2}
\usepackage{amsmath}
\usepackage{amsfonts}
\usepackage{amssymb}
\usepackage{makeidx}
\usepackage{graphicx,epsfig}
\usepackage{color}
\usepackage{subfigure}
\usepackage{ulem}
\usepackage{multirow}
\usepackage{epstopdf}
\usepackage{dcolumn}
\usepackage{comment}
\usepackage{caption}
\usepackage{subcaption}
\usepackage{epstopdf}
\usepackage{dcolumn}
\usepackage{slashed}
\usepackage{makecell}
\usepackage{url}
\usepackage{bm}
\usepackage[section]{placeins}
\usepackage[colorlinks=true, citecolor=blue, urlcolor = blue, linkcolor= red, bookmarks=true]{hyperref}
\newcommand{\beq}{\begin{equation}}
	\newcommand{\eeq}{\end{equation}}
\newcommand{\bea}{\begin{eqnarray}}
	\newcommand{\eea}{\end{eqnarray}}

\begin{document}
	
	\title {Double heavy quarkonia production with color-octet channels at Z factory and at the CEPC/FCC-ee }

    \author{ Xiao-Peng Wang $^{(a)}$}
    \author{Guang-Zhi Xu $^{(a)}$}
    \email{ xuguangzhi@lnu.edu.cn}
    \author{Kui-Yong Liu $^{(b)}$}
    \email{liukuiyong@lnu.edu.cn}
    \affiliation{ {\footnotesize (a)~School of Physics, Liaoning University, Shenyang 110036, China}\\
    {\footnotesize (b)~School of Physics, Liaoning Normal University, Dalian 116029, China}}
	
	\date{\today}
	
	\begin{abstract}
	   Within the NRQCD framework, we calculate the exclusive production of double heavy quarkonium(double charmonium and double bottomonium)  at future super $Z$ factory and at the CEPC/FCC-ee. The color-octet(CO) channels in the $\gamma^*/Z^*$-propagated process are considered along with the color-singlet(CS) channels. We found that the contributions of CO states to the total cross section are significant or dominant for many processes within energy region at $Z$ factory and at the CEPC/FCC-ee. The experimental measurements will help us to verify the CO mechanism. Among these CO channels, the gluon fragmentation into $^3S_1^{8}$ states is most important. Thus, the comparison between the theoretical results and future data will give a strong constraint to the matrix elements $\langle\mathcal{O}\left(^3S_1^{[8]}\right)\rangle$. Additionally, we consider the relativistic corrections to both the CS and CO channels which decrease the cross sections significantly. Specially, the $K$ factors are about $0.5$ for most charmonium channels. We get estimates of the events for double heavy quarkonium production. The final events of  $J/\psi+\eta_c$, $J/\psi+J/\psi$, $\Upsilon+\eta_b$, $\Upsilon+\Upsilon$ production would be (22, 570, 71, 61) and (206, 5343, 665, 576) at the CEPC (2-year) and at the FCC-ee (4-year) for the $Z$ factory mode, respectively.
		
	\end{abstract}
	
	\maketitle
	
	\section{Introduction
} \label{introduce}

Heavy quarkonium is a bound state of a heavy quark ($Q$) and its antiquark ($\bar{Q}$) with a nonrelativistic nature. Nonrelativistic QCD (NRQCD) \cite{NRQCD} has been widely accepted to study heavy quarkonium production and decay. Under this framework, the heavy quarkonium production can be factorized into short-distance coefficients (SDCs) and long-distance matrix elements (LDMEs). The SDCs, which represent the production of intermediate $Q\bar{Q}$ pairs, are perturbatively calculable. The LDMEs, which are non-perturbative universal parameters, describe the hadronization of the $Q\bar{Q}$ pair into a physical quarkonium state. These LDMEs can be obtained from potential models, lattice QCD calculations, or extracted from experimental data. 

The color-octet mechanism (COM) is a key component of the NRQCD approach, introduced by systematically accounting for the higher-Fock components of a quarkonium state. Unlike the color-singlet model (CSM), the intermediate $Q\bar{Q}$ pair in COM can be in a color-octet (CO) configuration with different quantum numbers from the final quarkonium state. With the COM, infrared divergences encountered in the CSM can be eliminated\cite{PwaveIR}, and the $\psi'$ anomaly observed at hadron colliders can be explained\cite{psianomalye1,psianomalye2,psianomalyt}, which is considered a significant achievement of NRQCD. However, there are still some challenges that question the NRQCD approach. One of these is the universality problem of the CO LDMEs. LDMEs extracted from hadronic scattering measurements by different theoretical groups have shown inconsistencies\cite{universalityt1,universalityt2,universalityt3,universalityt4}. Additionally, for inclusive and exclusive charmonium production at B factories, higher-order corrections to the color-singlet (CS) channel often saturate the data, leaving little room for the COM
\cite{Bfactorye1,Bfactorye2,KYLiu1,RCbraaten,LOHag,Bfactorynlo1,Bfactorynlo2,RC1,RC2,RC3,RCbodwin,Bfactorynnlo,huangxd}.We refer the reader to the following reviews \cite{review1,review2,review3,review4,review5} for more information on the current status of heavy quarkonia physics. 

Several future $e^+e^-$ collider projects are proposed, including the Circular Electron Positron Collider (CEPC) \cite{CEPC}, the Future Circular Collider (FCC) \cite{FCC,FCC1}, and the GigaZ program of the International Linear Collider (ILC) \cite{ILC1,ILC2,ILC3,ILC4}. Additionally, a super Z factory with a high luminosity of $\mathcal{L} \simeq 10^{34-36} \text{cm}^{-2}\text{s}^{-1}$ \cite{Zfact} is also proposed. These facilities provide excellent opportunities to revisit both exclusive and inclusive heavy quarkonium production, thereby allowing for a more thorough verification of the COM.
Within the CSM, the exclusive production of double heavy quarkonia in $e^+e^-$ annihilation at Z factories has been extensively studied at leading order (LO) \cite{LOHag,LOccchengu,LOccLikhode}. Next-to-leading order (NLO) QCD corrections have also been considered \cite{Berezhnoy,LuoxuanNLO,Belov}. Specifically, NLO QCD corrections for double S-wave $B_c$ meson production \cite{Berezhn-nlo-doubleBc} and associated S-wave charmonium-bottomonium production \cite{Belov-lo-charm+bottom} have been obtained.
Recently, Liao et al. \cite{Liao_2023,liaoqili2,liaoqili} have investigated the production of double charmonia, double bottomonia, and double $B_c$ mesons at LO. Pure QED processes also contribute observably to these production mechanisms \cite{Belov}.
On the other hand, it has been found that the contributions of the COM are significant in semi-exclusive processes \cite{sunzhan}. Therefore, it is also necessary to study the effects of COM in both exclusive and inclusive heavy quarkonia production processes at future $e^+e^-$ colliders.

In the present paper, we will study the production of double heavy quarkonia (double charmonia or double bottomonia) 
within the COM framework. The final states for these processes include:
$J/\psi+\eta_c, J/\psi+J/\psi,  J/\psi+\chi_{cJ}, J/\psi+h_{c}, \eta_c+\eta_c, \eta_c+h_{c}, \eta_c+\chi_{cJ}, \Upsilon+\eta_{b}, \Upsilon+\Upsilon, \Upsilon+\chi_{bJ}, \Upsilon+h_b,  \eta_b+\eta_b, \eta_b+h_{b}, \eta_b+\chi_{bJ}$.

The rest of this paper is organized as follows. In Section \ref{solution}, we present the formula and method used in our study. In Section \ref{inputldmes}, we provide the input parameters and LDMEs utilized in this work. In Section \ref{results}, we analyze all possible production channels and present the results of the cross sections and differential cross sections. We also discuss the generation of events and the associated uncertainties. Finally, a summary is given in Section \ref{summary}.
	
	\section{Framework} \label{solution}
		
	\subsection{Diagrams and amplitudes}

		Within the NRQCD framework, the production cross sections can be divided into two parts: 
            SDCs and LDMEs:\cite{NRQCD}
		\begin{equation}
			\label{eq:factorization} \hat{\sigma}(e^++e^-{\rightarrow}H_1+H_2)=\sum_{mn}\frac{F_{mn}}{m_{Q1}^{d_m-4}m_{Q2}^{d_n-4}}\langle0|\mathcal{O}_m^{H_1}|0\rangle\langle0|\mathcal{O}_n^{H_2}|0\rangle.
		\end{equation}
			SDCs $F_{mn}$ are labeled by the subscripts $m$ and $n$ for different production channels. 
        The factor $m_Q^{d_{m,n}-4}$ is introduced to make the SDCs $F_{mn}$ dimensionless.
The  Fock state expansion formalism for heavy quarkonium up-to the order of $\mathcal{O}(v^2)$ can be expressed as bellow, where $v$ is the relative velocity between quark and anti-quark in heavy quarkonium rest frame.
		\begin{widetext}	
		\begin{figure}
			\begin{tabular}{c c }
				\includegraphics[width=1.0\textwidth]{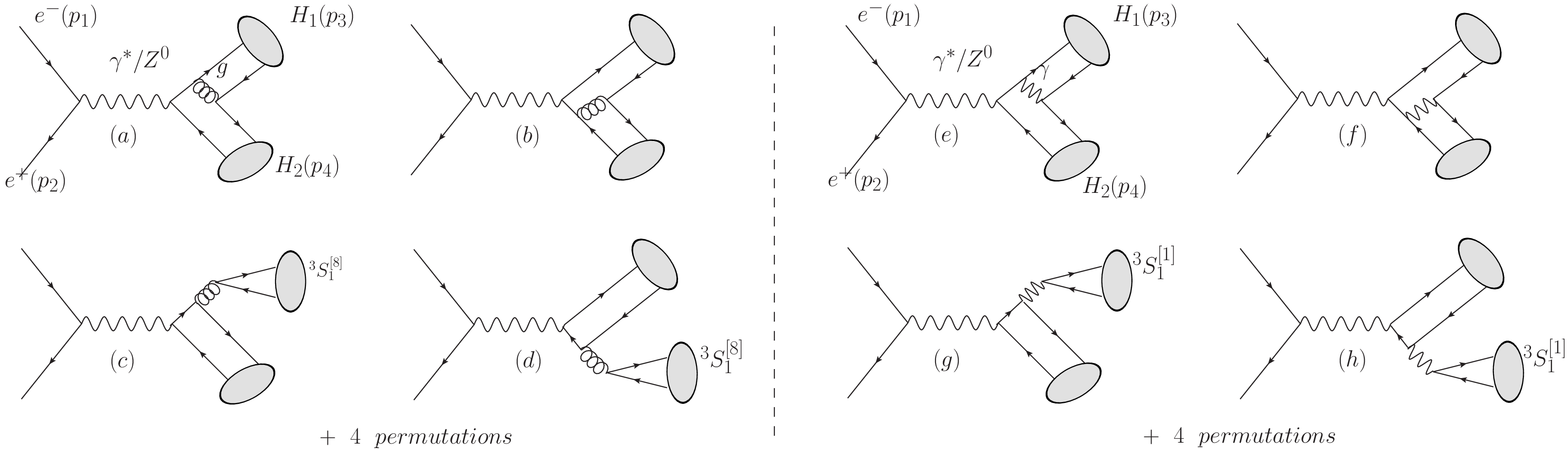}
			\end{tabular}
			\caption{CS/CO Feynman diagrams for $e^-(p_1)+e^+(p_2)\rightarrow \gamma^*/Z^0\rightarrow H_1(p_3)+H_2(p_4)$ at the tree level. The permutation diagrams can be obtained by reversing the quark line. The QCD diagrams are shown on  the left side (a, b, c, d), and the EW diagrams are shown on the right side (e, f, g, h). (c) and (d) pertain exclusively to the CO channels. (g) and (h) pertain exclusively to the CS channels.}
			\label{feynmandia}
		\end{figure}
		\FloatBarrier	
	\end{widetext}
	\bea
	|H\rangle=\mathcal{O}(1)|Q\bar{Q}(^{2S+1}L_J^{[1]})\rangle+\mathcal{O}(v)|Q\bar{Q}(^{2S+1}(L\pm1)_{J}^{[8]})g\rangle\cr
+\mathcal{O}(v^2)|Q\bar{Q}(^{2(S\pm1)+1}L_{J}^{[8]})g\rangle
	+\mathcal{O}(v^2)|Q\bar{Q}(^{2S+1}L_{J}^{[1,8]})gg\rangle+...
	\eea
    For specific heavy quarkonium mesons, the formula are written as, 
	\bea
	|J/\psi\rangle&=&\mathcal{O}(1)|c\bar{c}(^3S_1^1)\rangle+\mathcal{O}(v)|c\bar{c}(^3P_J^8)g\rangle+\mathcal{O}(v^2)|c\bar{c}(^1S_0^8)g\rangle+\mathcal{O}(v^2)|c\bar{c}(^3S_1^{8})gg\rangle+...\\
	|\eta_c\rangle&=&\mathcal{O}(1)|c\bar{c}(^1S_0^1)\rangle+\mathcal{O}(v)|c\bar{c}(^1P_1^8)g\rangle+\mathcal{O}(v^2)|c\bar{c}(^3S_1^8)g\rangle+\mathcal{O}(v^2)|c\bar{c}(^1S_0^{8})gg\rangle+...\\
	|\chi_{cJ}\rangle&=&\mathcal{O}(1)|c\bar{c}(^3P_J^1)\rangle+\mathcal{O}(v)|c\bar{c}(^3S_1^8)g\rangle+...\quad\quad\quad\quad\quad\\
	|h_{c}\rangle&=&\mathcal{O}(1)|c\bar{c}(^1P_1^1)\rangle+\mathcal{O}(v)|c\bar{c}(^1S_0^8)g\rangle+...\\
	|\Upsilon\rangle&=&\mathcal{O}(1)|b\bar{b}(^3S_1^1)\rangle+\mathcal{O}(v)|b\bar{b}(^3P_J^8)g\rangle+\mathcal{O}(v^2)|b\bar{b}(^1S_0^8)g\rangle+\mathcal{O}(v^2)|b\bar{b}(^3S_1^{8})gg\rangle+...\\
	|\eta_b\rangle&=&\mathcal{O}(1)|b\bar{b}(^1S_0^1)\rangle+\mathcal{O}(v)|b\bar{b}(^1P_1^8)g\rangle+\mathcal{O}(v^2)|b\bar{b}(^3S_1^8)g\rangle+\mathcal{O}(v^2)|b\bar{b}(^1S_0^{8})gg\rangle+...\\
	|\chi_{bJ}\rangle&=&\mathcal{O}(1)|b\bar{b}(^3P_J^1)\rangle+\mathcal{O}(v)|b\bar{b}(^3S_1^8)g\rangle+...\quad\quad\quad\quad\quad\\
	|h_{b}\rangle&=&\mathcal{O}(1)|b\bar{b}(^1P_1^1)\rangle+\mathcal{O}(v)|b\bar{b}(^1S_0^8)g\rangle+...
	\eea
    We will consider both the CS and CO intermediate states in the following calculations. 
	 
     Our calculations include the QCD contributions (at $\mathcal{O}(\alpha_s^2\alpha^2)$) and the electric-weak (EW) contributions (at $\mathcal{O}(\alpha^4)$) at tree level. We employed the FeynArts\cite{FA} package to generate the Feynman diagrams and amplitudes, and utilized the FeynCalc\cite{FC} package to manipulate the amplitudes. The typical diagrams for CS and CO channels are presented in Fig. \ref{feynmandia} through the single boson exchange process $e^+(p_1)+e^-(p_2)\to\gamma^*/Z^*(p)\to H_1(p_3) + H_2(p_4)$. 
     The t-channel EW diagrams (two bosons exchange process) are not presented which give small contributions except for $J/\psi$ pair and $\Upsilon$ pair production.
     The corresponding amplitudes in pertubative QCD/EW are expressed as,
	 \bea
    i\mathcal{M}=\sum_{n}\bar{v}(p_2)\mathcal{L}_{\mu}u(p_1)\mathcal{D}^{\mu\nu}\mathcal{A}^{(n)}_{\nu},
	 \eea
	 where the vertex $\mathcal{L}^{\mu}$ 
and propagator $D_{\mu\nu}$ are defined as, 
	 \bea
	  \mathcal{L}^\mu&=&-ie\gamma^\mu, ~~\frac{-ie}{4\cos\theta_W \sin\theta_W}\gamma^\mu(1-4\sin^2\theta_W-\gamma^5)\cr
	 \mathcal{D}_{\mu\nu}&=&\frac{-ig_{\mu\nu}}{p^2}, ~~\frac{-ig_{\mu\nu}}{p^2-m_Z^2+im_Z\Gamma_Z}
	 \eea
In the above definitions, the first (second) term corresponds to the process propagated by $\gamma^*$ ($Z^0$). 
$e = \sqrt{4\pi\alpha}$, 
$\theta_W$ stands for the Weinberg angle, and $m_Z$ and $\Gamma_Z$ denote the mass and width of the $Z^0$-boson, respectively. 
	In one certain channel, the hadron part of the amplitude is computed through the summation of all the diagrams, with each diagram being marked by a script, namely $\mathcal{A}^{(n)}$. For the $P$-wave state, the LO amplitude is expressed as the contraction of the orbital polarization vector $\varepsilon(L_z)$ and the derivative of the original amplitude with respect to the relative momentum (denoted by $q$) between the $Q\bar{Q}$ pair,   
    \bea
	 \mathcal{A}&=&\varepsilon^{\rho}(L_z)\frac{\partial}{\partial q^{\rho}}\mathcal{\tilde{A}}|_{q\rightarrow0}
	 \eea   

    Through the projection operator method, $Q\bar{Q}$ is projected onto specific spin and color quantum numbers. For instance, the original amplitudes of the diagrams labeled (a) and (c) in Fig. \ref{feynmandia} are written as below,

	 \bea
	 \mathcal{\tilde{A}}_\nu^{(a)}&=&g^2\textrm{Tr}\bigg[\Pi_1\gamma_{\rho}\frac{(\slashed{p}_{3Q}+\slashed{p}_{3\bar{Q}}+\slashed{p}_{4Q})+m_{Q}}{[(p_{3Q}+p_{3\bar{Q}}+p_{4Q})^2-m_{Q}^2](p_{3\bar{Q}}+p_{4Q})^2}\mathcal{V}_{\nu}\Pi_2\gamma^{\rho}\bigg]	\cr
\mathcal{\tilde{A}}_\nu^{(c)}&=&-g^2\textrm{Tr}\bigg[\gamma^{\rho}\Pi_1\bigg]~\textrm{Tr}\bigg[\Pi_2\gamma_{\rho}\frac{(\slashed{p}_{3Q}+\slashed{p}_{3\bar{Q}}+\slashed{p}_{4Q})+m_{Q}}{[(p_{3Q}+p_{3\bar{Q}}+p_{4Q})^2-m_{Q}^2](p_{3Q}+p_{3\bar{Q}})^2}\mathcal{V}_{\nu}\bigg]	 
	 \eea
    The phase difference between the two amplitudes with a sign change arises because one diagram can be obtained by exchanging the identical final-state fermions in the other. The vertex $\mathcal{V}_{\nu}$  for $\gamma/Z$ propagation are defined as follows, 
    \bea
    \mathcal{V}_{\nu}&=&iee_Q\gamma_{\nu}, ~~\frac{ie}{4\cos\theta_W \sin\theta_W}\gamma_{\nu}(1-4|e_Q|\sin^2\theta_W-\gamma^5)
    \eea
    where $e_Q$ is the charge of heavy quark.
The projection operator $\Pi_i$, with the subscript $i$ indicating the specific states, is constituted as the multiplicative combination of the spin projection and the color projection.
    The expressions of spin-singlet and spin-triplet projection operators are, 
	\bea
	\Pi(00)&\equiv&\sum_{s_Q s_{\bar{Q}}}\langle s_Q s_{\bar{Q}};00\rangle v(p_{\bar{Q}})\bar{u}(p_{Q})=\frac{1}{2\sqrt{2}(E_q+m_Q)}(-\slashed{p}_{\bar{Q}}+m_Q)\gamma_5\frac{\slashed{p}_Q+\slashed{p}_{\bar{Q}}+2E_q}{2E_q}(\slashed{p}_Q+m_Q)\\
	\Pi(1s_z)&\equiv&\sum_{s_Q s_{\bar{Q}}}\langle s_Q s_{\bar{Q}};1s_z\rangle v(p_{\bar{Q}})\bar{u}(p_{Q})=\frac{1}{2\sqrt{2}(E_q+m_Q)}(-\slashed{p}_{\bar{Q}}+m_Q)\slashed{\varepsilon}(s_z) \frac{\slashed{p}_Q+\slashed{p}_{\bar{Q}}+2E_q}{2E_q}(\slashed{p}_Q+m_Q)
		\label{projector}
	\eea
    where $\varepsilon(L_z)$ is the spin polarization vector, $E_q$ is the total energy of quark or antiquark in the rest frame of the meson.
    Color projection operators assume the form of $\sum_{ij}\frac{\delta_{ij}}{\sqrt{N_c}}$ in the case of CS, and $\sqrt{2}T^a_{ij},~a=1,2,\dots,8$ for CO.

    The requirement for the overall colorless nature of the final state dictates that both particles must be either in the CS state or in the CO state.
     For the CS channel, the color factors of diagram (a) and (b) are \(\frac{N_c^2 - 1}{2N_c}\), while for the CO channel, they are \(-\frac{\delta_{ab}}{2N_c}\) where $a,b$ are the color indexes of final states. In contrast, diagrams (c) and (d) can only be in the CO state, and the gluon fragment into the $^3S_1^{[8]}$ state. The total color factors are \(\frac{\delta_{ab}}{2}\). For the CS channel, the color factors of diagram (e) and (f) are $1$, while for the CO channel, they are $\delta_{ab}$. Diagrams (c) and (d) can only be in the CS state, and the final $\gamma^*/Z^0$ fragment into the $^3S_1^{[1]}$ state. The total color factors are $N_c$.

	 \subsection{Kinematics and relativistic corrections (RCs)}

     We compute the cross sections of each channel up to the NLO of $\mathcal{O}(v^2)$ and the conventional approach is employed to derive the RCs, as referenced in \cite{RC1,RC2,RC3,RCbraaten,YJLi,Li:2013nna}. For the $S$-wave state, the amplitudes are expanded to the second order of $|\vec{q}|$, whereas for the $P$-wave state, the expansion is carried out to the third order. Here,  $|\vec{q}|$ which is a Lorentz scalar represents the magnitude of the three-momentum of either $Q$ or $\bar{Q}$ within the rest frame of the meson. 
    Consequently, we have $E_q=\sqrt{m_Q^2+\vec{q}^2}$ and the meson mass can be approximated as $2E_{q}$. The momentum of $Q$($\bar{Q}$) in arbitrary frame is expressed as $p_Q=\frac{1}{2}p+q$ ($p_{\bar{Q}}=\frac{1}{2}p-q$) where $q$ is obtained via Lorentz boost from $(0,\vec{q})$. Therefore, after integrating out the angles of $q$, the corresponding relationship between the polynomial term of $q$ and $|\vec{q}|^2$ can be ascertained. $q^{\alpha}q^{\beta}\rightarrow\frac{|\vec{q}|^2}{3}\pi^{\alpha\beta}$ for $S$-wave state, $q^{\alpha}q^{\beta}q^{\sigma}\rightarrow\frac{|\vec{q}|^3}{5}(\pi^{\alpha\beta}\varepsilon^{\sigma}_{L_z}+\pi^{\alpha\sigma}\varepsilon^{\beta}_{L_z}+\pi^{\beta\sigma}\varepsilon^{\alpha}_{L_z})$ for $P$-wave state, where $\pi^{\alpha\beta}=-g^{\alpha\beta}+\frac{p^{\alpha}p^{\beta}}{p^2}$.  
    Furthermore, the normalization factor of $\sqrt{\frac{m_Q}{E_Q}}$ is incorporated into the amplitudes.

        The kinematics variables should be also expanded. For two-body final states, the Lorentz invariant Mandelstam variables are defined as,
	\bea
	s=(p_1+p_2)^2=(p_3+p_4)^2,
	t=(p_3-p_1)^2=(p_4-p_2)^2,
	u=(p_4-p_1)^2=(p_3-p_2)^2.
	\label{mandel}
	\eea
    with the relationship $s+t+u=M_{H_1}^2+M_{H_2}^2$.
    The variable $s$ is independent of $|\vec{q}|^2$, while $t$ and $u$ are dependent on $|\vec{q}|^2$. We define $t_0$ as the value of $t$ when $|\vec{q_1}| = |\vec{q_2}| = 0$, and $u_0$ as the value of $u$ when $|\vec{q_1}| = |\vec{q_2}| = 0$. Consequently, the following expansions are derived, as presented in \cite{YJLi}.
    \bea
	t&=&t_0-\frac{4[t_0+4m_Q^2]}{s-16m_Q^2}(|\vec{q_1}|^2+|\vec{q_2}|^2)+\mathcal{O}(v^4)\\
	u&=&u_0-\frac{4[u_0+4m_Q^2]}{s-16m_Q^2}(|\vec{q_1}|^2+|\vec{q_2}|^2)+\mathcal{O}(v^4)
	\eea

    The matrix elements of $\langle v^2\rangle$ are defined by the ratios of the NLO LDMEs  in $v^2$ to the LO LDMEs, 
    \bea
		\langle v^2\rangle\equiv\frac{\langle0|\mathcal{P}^H(^{2s+1}L_J)|0\rangle}{m_Q^2\langle0|\mathcal{O}^H(^{2s+1}L_J)|0\rangle}
	\eea
	which adhere to the velocity power scaling rules and are of the order of $v^2$. Moreover,
	\bea
		v^2=\langle v^2\rangle[1+\mathcal{O}(v^4)]
	\eea
    The four-Fermion operators $\mathcal{O}^{H}(^{2s + 1}L_J)$, which are of dimension-6 for the LO, and the dimension-8 four-Fermion operators $\mathcal{P}^{H}(^{2s + 1}L_J)$ utilized for the RCs, are defined in accordance with Ref. \cite{NRQCD}.

	\section{Parameters}\label{inputldmes}

	 We will take the following input parameters\cite{PDG} and LDMEs\cite{LDMEs} in the present work,
	\bea
	\label{input}
	&\alpha_s(2m_c)=0.26,~\alpha_s(2m_b)=0.18,~\alpha=1/137,\cr 
    &m_c=1.5~GeV,~m_b=4.7~GeV, m_Z=91.1876~ GeV,\cr
	&\Gamma_Z=2.4952~GeV,~\sin^2\theta_w=0.2312,~v^2_{c\bar{c}}=0.23,~v^2_{b\bar{b}}=0.1,
	\eea

    Here, the strong running coupling constant $\alpha_s$ is calculated by the two-loop formula,
	\bea
	\frac{\alpha_s(\mu)}{4\pi}=\frac{1}{\beta_0L}-\frac{\beta_1ln L}{\beta_0^3L^2}
	\eea
	where $L=ln(\mu^2/\Lambda^2_{QCD})$ with $\Lambda_{QCD}\simeq338MeV$, $\beta_0=(11/3)C_A-(4/3)T_fn_f$ and $\beta_1=(34/3)C_A^2-4C_FT_fn_f-(20/3)C_AT_f$ are the one-loop and two-loop coefficients of the QCD beta function, respectively. $n_f$ is the active quark flavors which is  set to 3 for heavy quarkonia.
	The values of $v^2$ for CS and CO are estimated via the Gremm-Kapustin relation\cite{GKrela1,GKrela2},
	\bea
	v^2=v_1^2=v_8^2=\frac{M_{Q\bar{Q}}-2m_Q^{pole}}{m_Q^{QCD}}
	\eea
	where $m_Q^{pole}$ is quark pole mass, $m_Q^{QCD}$ is the quark  mass in NRQCD, $M_{Q\bar{Q}}$ is  heavy quarkonium mass.

	The LDMEs of charmonium\cite{LDMEs1cc,LDMEs2,LDMEs3,LDMEs4,LDMEs5,LDMEs6,LDMEs7,LDMEs8,LDMEs9cc} are,
	\bea
	\langle\mathcal{O}^{J/\psi}[^3S_1^{[1]}]\rangle&=&1.2~GeV^3\cr
		\langle\mathcal{O}^{J/\psi}[^1S_0^{[8]}]\rangle&=&0.0180\pm0.0087~GeV^3\cr
		\langle\mathcal{O}^{J/\psi}[^3S_1^{[8]}]\rangle&=&0.0013\pm0.0013~GeV^3\cr
		\langle\mathcal{O}^{J/\psi}[^3P_0^{[8]}]\rangle&=&(0.0180\pm0.0087)m_c^2~GeV^3
	\eea
	\bea
		\langle\mathcal{O}^{\eta_c}[^1S_0^{[1]}]\rangle&=&\frac{1}{3}\times1.2~GeV^3\cr
		\langle\mathcal{O}^{\eta_c}[^1S_0^{[8]}]\rangle&=&\frac{1}{3}\times(0.0013\pm0.0013)~GeV^3\cr
		\langle\mathcal{O}^{\eta_c}[^3S_1^{[8]}]\rangle&=&0.0180\pm0.0087~GeV^3\cr
		\langle\mathcal{O}^{\eta_c}[^1P_1^{[8]}]\rangle&=&	3\times(0.0180\pm0.0087)m_c^2~GeV^3\nonumber
	\eea
	\bea
		\langle\mathcal{O}^{h_c}[^1P_1^{[1]}]\rangle&=&	3\times0.054m_c^2~GeV^3\cr
		\langle\mathcal{O}^{h_c}[^1S_0^{[8]}]\rangle&=&	3\times(0.00187\pm0.00025)~GeV^3\nonumber
	\eea
	\bea
		\langle\mathcal{O}^{\chi_{c0}}[^3P_0^{[1]}]\rangle&=&0.054m_c^2~GeV^3\cr
		\langle\mathcal{O}^{\chi_{c0}}[^3S_1^{[8]}]\rangle&=&0.00187\pm0.00025~GeV^3\cr
		\langle\mathcal{O}^{\chi_{c1}}[^3P_1^{[1]}]\rangle&=&3\times0.054m_c^2~GeV^3\cr
		\langle\mathcal{O}^{\chi_{c1}}[^3S_1^{[8]}]\rangle&=&3\times(0.00187\pm0.00025)~GeV^3\cr
		\langle\mathcal{O}^{\chi_{c2}}[^3P_2^{[1]}]\rangle&=&5\times0.054m_c^2~GeV^3\cr
		\langle\mathcal{O}^{\chi_{c2}}[^3S_1^{[8]}]\rangle&=&5\times(0.00187\pm0.00025)~GeV^3\nonumber
	\label{ldmescc}
	\eea
	and that of bottomonium\cite{LDMEs1cc,LDMEs2,LDMEs8,LDMEs9cc,LDMEs10bb} are:
		\bea
		\langle\mathcal{O}^{\Upsilon}[^3S_1^{[1]}]\rangle&=&10.9~GeV^3\cr
		\langle\mathcal{O}^{\Upsilon}[^1S_0^{[8]}]\rangle&=&(0.0121\pm0.0400)~GeV^3\cr
			\langle\mathcal{O}^{\Upsilon}[^3S_1^{[8]}]\rangle&=&(0.0477\pm0.0334)~GeV^3\cr
		\langle\mathcal{O}^{\Upsilon}[^3P_0^{[8]}]\rangle&=&5\times(0.0121\pm0.0400)m_b^2~GeV^3
\nonumber
	\eea
			\bea
		\langle\mathcal{O}^{\chi_{b0}}[^3P_0^{[1]}]\rangle&=&0.1m_b^2~GeV^3\cr
		\langle\mathcal{O}^{\chi_{b0}}[^3S_1^{[8]}]\rangle&=&0.1008~GeV^3\cr
		\langle\mathcal{O}^{\chi_{b1}}[^3P_1^{[1]}]\rangle&=&3\times0.1m_b^2~GeV^3\cr
		\langle\mathcal{O}^{\chi_{b1}}[^3S_1^{[8]}]\rangle&=&3\times0.1008~GeV^3\cr
			\langle\mathcal{O}^{\chi_{b2}}[^3P_2^{[1]}]\rangle&=&5\times0.1m_b^2~GeV^3\cr
		\langle\mathcal{O}^{\chi_{b2}}[^3S_1^{[8]}]\rangle&=&5\times0.1008~GeV^3\nonumber
	\eea
			\bea
		\langle\mathcal{O}^{\eta_{b}}[^1S_0^{[1]}]\rangle&=&\frac{1}{3}\langle\mathcal{O}^{\Upsilon}[^3S_1^{[1]}]\rangle=3.633~GeV^3\cr
		\langle\mathcal{O}^{\eta_{b}}[^1S_0^{[8]}]\rangle&=&\frac{1}{3}\langle\mathcal{O}^{\Upsilon}[^3S_1^{[8]}]\rangle=(0.0159\pm0.0111)~GeV^3\cr
		\langle\mathcal{O}^{\eta_{b}}[^3S_1^{[8]}]\rangle&=&\langle\mathcal{O}^{\Upsilon}[^1S_0^{[8]}]\rangle=(0.0121\pm0.0400)~GeV^3\cr
		\langle\mathcal{O}^{\eta_{b}}[^1P_1^{[8]}]\rangle&=&3\times\langle\mathcal{O}^{\Upsilon}[^3P_0^{[8]}]\rangle=3\times5\times(0.0121\pm0.0400)m_b^2~GeV^3\nonumber
	\eea
	\bea
			\langle\mathcal{O}^{h_{b}}[^1P_1^{[1]}]\rangle&=&3\times\langle\mathcal{O}^{\chi_{b0}}[^3P_0^{[1]}]\rangle=3\times0.1m_b^2 ~GeV^3\cr
				\langle\mathcal{O}^{h_{b}}[^1S_0^{[8]}]\rangle&=&3\times\langle\mathcal{O}^{\chi_{b0}}[^3S_1^{[8]}]\rangle=3\times0.1008~ GeV^3\nonumber
	\eea

The heavy quark spin symmetry (HQSS) is exhibited by them \cite{NRQCD}:

\bea
		\langle\mathcal{O}^{\chi_{cJ}}[^3P_J^{[1]}]\rangle&=&	(2J+1)\langle\mathcal{O}^{\chi_{c0}}[^3P_0^{[1]}]\rangle\cr
			\langle\mathcal{O}^{\chi_{cJ}}[^3S_1^{[8]}]\rangle&=&	(2J+1)\langle\mathcal{O}^{\chi_{c0}}[^3S_1^{[8]}]\rangle\cr
		\langle\mathcal{O}^{\eta_{c}}[^1S_0^{[1]}/^1S_0^{[8]}]\rangle&=&\frac{1}{3}		\langle\mathcal{O}^{J/\psi}[^3S_1^{[1]}/^3S_1^{[8]}]\rangle	,
		\langle\mathcal{O}^{\eta_{c}}[^3S_1^{[8]}]\rangle=	\langle\mathcal{O}^{J/\psi}[^1S_0^{[8]}]\rangle\cr
		\langle\mathcal{O}^{\eta_{c}}[^1P_1^{[8]}]\rangle&=&	3\langle\mathcal{O}^{J/\psi}[^3P_0^{[8]}]\rangle,
		\langle\mathcal{O}^{h_{c}}[^1P_1^{[1]}/^1S_0^{[8]}]\rangle=	3\langle\mathcal{O}^{\chi_{c0}}[^3P_0^{[1]}/^3S_1^{[8]}]\rangle
\eea
we also see that these parameters approximately satisfy the velocity scaling rule (VSR) of NRQCD\cite{NRQCD},
\bea
	\langle\mathcal{O}^{J/\psi}[^3S_1^{[1]}]\rangle\sim m_c^3v_c^3,		\langle\mathcal{O}^{\chi_{cJ}}[^3P_J^{[1]}]\rangle\sim m_c^5v_c^5,\cr
	\langle\mathcal{O}^{J/\psi}[^3S_1^{[8]}]\rangle\sim m_c^3v_c^7,
	\langle\mathcal{O}^{\chi_{cJ}}[^3S_1^{[8]}]\rangle\sim m_c^3v_c^5,\cr
\ldots\ldots\quad\quad\quad\quad\quad\quad\quad\quad\quad\quad\quad\quad
\eea

	 the CS LDMEs satisfy the quark potential model\cite{NRQCD},
	\bea
	\frac{		\langle\mathcal{O}^{J/\psi}[^3S_1^{[1]}]\rangle}{2N_c\times3}\simeq\frac{|R_{S}(0)|^2}{4\pi},
		\frac{		\langle\mathcal{O}^{\eta_c}[^1S_0^{[1]}]\rangle}{2N_c}\simeq\frac{|R_{S}(0)|^2}{4\pi}\cr
	\frac{		\langle\mathcal{O}^{\chi_{cJ}}[^3P_J^{[1]}]\rangle}{2N_c(2J+1)}(J=0,1,2)\simeq\frac{3|R'_{P}(0)|^2}{4\pi},	\frac{		\langle\mathcal{O}^{h_c}[^1P_1^{[1]}]\rangle}{2N_c\times3}\simeq\frac{3|R'_{P}(0)|^2}{4\pi}
	\eea

The CS LDMEs of charmonia used in this work, which correspond to the values of $|R_s(0)|^2=0.838, |R_p'(0)|^2=0.085$, are closely approximate the value ($|R_s(0)|^2=0.810, |R_p'(0)|^2=0.075$) in Ref. \cite{LDMEs11}, compared to maximal    ~($|R_s(0)|^2=2.458, |R_p'(0)|^2=0.322$\cite{LDBT}) and minimal ($|R_s(0)|^2=0.565, |R_p'(0)|^2=0.053$\cite{LDIO}) set in Refs. \cite{Liao_2023,LDLiao}, and that of bottomonia ~($|R_s(0)|^2=7.61, |R_p'(0)|^2=1.54$) are closely approximate the value ($|R_s(0)|^2=6.477, |R_p'(0)|^2=1.417$) in Ref. \cite{LDMEs11}, compared to maximal ($|R_s(0)|^2=16.12, |R_p'(0)|^2=5.874$\cite{LDBT}) and minimal ~($|R_s(0)|^2=5.298, |R_p'(0)|^2=1.111$\cite{LDCK}) set in Refs. \cite{Liao_2023,LDLiao}.
	
Analogous correlations among the LDMEs are also valid for bottomonia.

	\section{Results and discussion}\label{results}

		\subsection{Cross sections and CO contributions}\label{channels}

In the energy region of Z-factory, the pure EW contributions are also significant\cite{Belov}.
For the $J/\psi$ pair and $\Upsilon$ pair production, we consider the t-channel EW contributions which are significant either at B-factory or at the Z-factory\cite{Zhang:2010uia,rcresum,Bhatnagar:2024ykb}. For other processes, the t-channel EW contributions could be neglected compared with the s-channel contributions.
The $\gamma^*$-propagated processes contribute predominantly at the B-factory, but are negligible compared with the $Z^0$-propagated processes in the energy region of the Z-factory.		
And for the $\gamma^*$-propagated processes, due to the additional C-parity conservation constraint imposed by the $\gamma^*$ propagator, these processes exhibit fewer production channels. 
Therefore, in our calculations, the total cross sections will include the complete contributions from QCD and EW processes, as well as $\gamma^*/Z^0$-propagated processes. The interferences were also taken into account. 

In Table \ref{channelsanalysis}, we present the contributions stemming from all combinations of the CS and CO states that comply with the conservation laws of color, C-parity, and combined CP-parity.
Upon examining the table, it becomes evident that the CO channels contribute substantially more than the CS channels in several processes, particularly for the combinations involving the $^3S_1^{[8]}$ state. This indicates that the processes of gluon fragmentation into the intermediate $^3S_1^{[8]}$ state might play a crucial role. The significance of the $^3S_1^{[8]}$ state has also been emphasized in Refs. \cite{sunzhan,zhxc}, where the semi-exclusive heavy quarkonia production and the inclusive $\eta_Q$ production at the Z-factory were investigated.

  The cross-sections vary with centre-of-mass (c.m.) energy $\sqrt{s}$ are presented in Figs. \ref{z0cc} and \ref{z0bb}. 
  In the energy region of the B factory, $\gamma^*$ propagation process dominates. Near the energy of the $Z^0$ pole, $Z^0$ propagation process dominates due to the resonance effects. Differing from the case in B factory where the production of double charmonium is mainly due to the color singlet mechanism with the color octet contribution being negligible, at the Z factories, the color octet contribution prevails in a number of production processes. In Figs. \ref{z0ccco}, \ref{z0bbco}, the ratios $\sigma_{CO}/\sigma_{Total}$ versus $\sqrt{s}$ have been illustrated, enabling a more direct and intuitive perception of the color octet contributions to the total cross sections. It should be noted that in the production of double $\eta_c$ or double $\eta_b$, the contributions stemming from the CS channels vanish when the $t$-channel is left out of consideration.

 Relativistic corrections (RCs) exert a substantial impact on double charmonia production. At the \(Z^0\) pole and for higher collision energies, the $K$ factors associated with the NLO($v^2$) cross sections falls within the range of $0.3\sim0.5$. 
 And for bottomonia production, the influence of RCs is relatively minor. Here, the $K$ factor is in the range of $0.7\sim 0.8$. 
In Appendix \ref{appdA}, we depict the ratios of $\sigma_{NLO(v^2)}/\sigma_{LO}$ (i.e., the $K$ factors) as a function of $\sqrt{s}$ in Figs. \ref{z0cck} and \ref{z0bbk}. 
Correspondingly, in Appendix \ref{appdB}, we elucidate the ratios of the SDCs corresponding to the RCs and those at LO in the large energy limit. 
Notably, the NLO QCD corrections can also be remarkably significant, with the $K$ factor reaching as high as $2\sim5$, as reported in \cite{Berezhnoy,Belov}.

\begin{table}
	\caption{Production channels and corresponding cross sections (units:$\times10^{-4} fb$) at $Z^0$ pole in $\mathcal{O}(v^0)$. The percentage inside the brackets is the proportion of CO. The negligible results($<0.1\times 10^{-4} fb$) are not shown.
	}
	\begin{tabular}{|c| |cc| |cccc||cc||cc|}
		\hline
		$H_1$&\multicolumn{10}{c|}{ $H_2$}\\
		\hline
		\hline
		\multicolumn{11}{|c|}{ charmonia}\\
		\hline
		\hline
		$\eta_c$& $J/\psi$ &\makecell{32.1\\(66.1\%)}& $\chi_{cJ}$&\makecell{11.9$(J=0)$\\(42.7\%)}& \makecell{16.1$(J=1)$\\(95.0\%)}& \makecell{39.5$(J=2)$\\(64.3\%)}&$h_c$&\makecell{4.4\\(31.0\%)} &$\eta_c$& \makecell{39.2\\(100.0\%)}\\
		\hline
		\hline
		$^1S_0^{[1]}$     &$^3S_1^{[1]}$&10.9 &$^3P_J^{[1]}$ & 6.8&0.8&14.1&$^1P_1^{[1]}$ &3.1 & $^1S_0^{[1]}$& \\	
		\hline	
			$^3S_1^{[8]}$  &$^1S_0^{[8]}$&4.4  &$^3S_1^{[8]}$&2.0&6.1 &10.2&$^1S_0^{[8]}$&1.4&$^3S_1^{[8]}$  &9.8\\		
		~ &$^3S_1^{[8]}$ ~&~1.4     &~ &~&~ &~& & & &\\	
		~ &~$^3P_J^{[8]}$&~ 13.2    &~ &~&~ &~& & & & \\	
		\hline
		$^1S_0^{[8]}$     & $^3S_1^{[8]}$            &              &$^3S_1^{[8]}$ & & &0.1 & & &$^3S_1^{[8]}$  & 0.1\\	
		&    $^3P_J^{[8]}$         &              & & & & & & & $^1P_1^{[8]}$ & \\		
		\hline	
		$^1P_1^{[8]}$     &$^1S_0^{[8]}$&       &$^3S_1^{[8]}$ &3.0&9.1 &15.2&$^1S_0^{[8]}$ & &$^3S_1^{[8]}$&29.3\\	
		~ &$^3S_1^{[8]}$ ~&~  2.1  &~ &~&~ &~& && $^1P_1^{[8]}$& \\	
		~ &~$^3P_J^{[8]}$&~    &~ &~&~ &~&  & & & \\	
		\hline
		\hline
		$J/\psi$&$J/\psi$ & \makecell{2737.4\\(0.1\%)}& $\chi_{cJ}$&\makecell{4.1$(J=0)$\\(48.4\%)}& \makecell{8.2$(J=1)$\\(72.5\%)}& \makecell{13.1$(J=2)$\\(75.7\%)}&$h_c$&\makecell{33.3\\(0.3\%)} & & \\
		\hline
		\hline
		$^3S_1^{[1]}$&$^3S_1^{[1]}$ &2736.0  & $^3P_J^{[1]}$&2.1&2.3 &3.2 & $^1P_1^{[1]}$&33.2& & \\	
		\hline
			$^3S_{1}^{[8]}$&$^3S_1^{[8]}$ &0.1   &$^3S_1^{[8]}$ &0.1& 0.4& 0.7&$^1S_0^{[8]}$& 0.1& &\\	
		\hline
		$^1S_0^{[8]}$&$^3P_J^{[8]}$ &  &$^3S_1^{[8]}$ &0.5 &1.4 &2.3 & & & &\\	
		~        &$^3S_1^{[8]}$ &0.3  &~ &~ &~ &~ & & & & \\
		\hline
		$^3P_J^{[8]}$&$^3P_{J}^{[8]}$  &  &$^3S_1^{[8]}$ &1.4&4.1&6.9&$^1S_0^{[8]}$ & & & \\	
		~&$^3S_1^{[8]}$ &0.9  & & & & & &&& \\	
		\hline
		\hline
		\multicolumn{11}{|c|}{ bottomonia}\\
		\hline
		\hline
		$\eta_b$&$\Upsilon$ &\makecell{37.6\\(5.2\%)} & $\chi_{bJ}$&\makecell{8.1$(J=0)$\\(37.7\%)}& \makecell{15.0$(J=1)$\\(60.7\%)}& \makecell{28.5$(J=2)$\\(53.3\%)}&$h_b$&\makecell{7.8\\(9.6\%)}  &  $\eta_b$&\makecell{0.3\\(100.0\%)}\\
		\hline
		\hline
		$^1S_0^{[1]}$     &$^3S_1^{[1]}$& 35.7  &$^3P_J^{[1]}$ & 5.0&5.9&13.3&$^1P_1^{[1]}$ &7.1&$^1S_0^{[1]}$ & \\	
		\hline		
		$^3S_1^{[8]}$	&$^1S_0^{[8]}$& &$^3S_1^{[8]}$&0.3&1.0 &1.6&$^1S_0^{[8]}$&0.7 & $^3S_1^{[8]}$& \\	
		&$^3S_1^{[8]}$ &0.2      &~ &~&~ &~& & & & \\		
		&$^3P_J^{[8]}$ & 0.4     &~ &~&~ &~& & & & \\	
		\hline
		$^1S_0^{[8]}$     &   $^3S_1^{[8]}$          &   0.2           &$^3S_1^{[8]}$ & 0.3&1.0&1.6 &  && $^3S_1^{[8]}$& \\		
		~     &      $^3P_J^{[8]}$        &              &~& ~&~ &~ &  &&$^1P_1^{[8]}$&\\	
		\hline	
		$^1P_1^{[8]}$     &$^1S_0^{[8]}$&   &$^3S_1^{[8]}$ &2.4&7.2 &12.0&$^1S_0^{[8]}$ & &  $^3S_1^{[8]}$ &0.3 \\	
		&$^3S_1^{[8]}$& 1.1    &~ &~&~ &~& & &$^1P_1^{[8]}$  & \\	
		&$^3P_J^{[8]}$~&~      &~ &~&~ &~&  & & & \\	
		\hline
		\hline
		$\Upsilon$& $\Upsilon$&\makecell{36.7\\(6.1\%)} & $\chi_{bJ}$&\makecell{13.3$(J=0)$\\(39.6\%)}& \makecell{17.1$(J=1)$\\(92.4\%)}& \makecell{32.9$(J=2)$\\(80.0\%)}&$h_b$&\makecell{25.2\\(11.7\%)} && \\
		\hline
		\hline
		$^3S_1^{[1]}$&$^3S_1^{[1]}$ &  34.4 & $^3P_J^{[1]}$&8.0 &1.3 &6.6 & $^1P_1^{[1]}$&22.2& & \\	
		\hline
			$^3S_{1}^{[8]}$&$^3S_1^{[8]}$ & 0.3 &$^3S_1^{[8]}$& 1.3 &3.9 & 6.5&$^1S_0^{[8]}$&2.9 & &\\	
		\hline
		$^1S_0^{[8]}$& $^3P_J^{[8]}$& &$^3S_1^{[8]}$ & 0.2&0.7 &1.2 & & & & \\	
		~        & $^3S_1^{[8]}$& 0.1 &~ &~ &~ &~ & & & & \\
		\hline
		$^3P_J^{[8]}$&$^3P_{J}^{[8]}$ &0.1 &$^3S_1^{[8]}$ &3.7&11.2&18.7&$^1S_0^{[8]}$ &0.1 & &\\	
		~&$^3S_1^{[8]}$ & 1.8  & & & & & &&&\\	
		\hline
	
	\end{tabular}
	\label{channelsanalysis}
\end{table}
\FloatBarrier

  The numerical results of the total cross section at $\sqrt{s}=m_Z$ are illustrated in Table \ref{TCS1}. For comparison, we present the total cross sections of CS as well as CS plus CO, at the order of $v^0$ and $v^2$, respectively.
 Our LO CS results are in agreement with Refs. \cite{Belov,Liao_2023,LOccchengu,KYLiu1,KYLiu2} when using the same input parameters. \footnote{We find that a few processes are in contradiction with  Ref. \cite{Liao_2023},  our results of $\sigma(J/\psi+J/\psi)_{Z^0}$($\sigma(\Upsilon+\Upsilon)_{Z^0}$) are half of those in Ref. \cite{Liao_2023}, and the difference might be due to the particle identities. Moreover, in our results, $\sigma(J/\psi+h_{c})_{\gamma^*}$ and $\sigma(\Upsilon+h_{b})_{\gamma^*}$ are zero because of the C-parity conservation, which are also different from the values in Ref. \cite{Liao_2023}.} Our NLO($v^2$) CS result of $\gamma^*   \rightarrow J/\psi+\eta_{c}$ are consistent with Ref. \cite{RC1}.

 		\begin{table}
		\caption{ Production  cross sections (units:$\times 10^{-4}$ fb) of  the double heavy quarkonia at $\sqrt{s}=$91.1876 GeV. LO and NLO($v^2$) mean leading order and next-to-leading order results in the $v^2$ expansions, respectively. CS means the cross section of color-singlet channel, and CO means cross sections of all the color-octet channels. }
		\begin{tabular}{|c|c|c|c|c| |c|c|c|c|c|}
			\hline
			\multicolumn{5}{|c| |}{ charmonia}&\multicolumn{5}{c|}{ bottomonia}\\
			\hline
			~	&\multicolumn{2}{c|}{CS} &  \multicolumn{2}{c| |}{CS+CO}	&~	&\multicolumn{2}{c|}{CS}  & \multicolumn{2}{c|}{CS+CO}\\
			\hline
			~&~LO~&NLO($v^2$) &~LO~&NLO($v^2$)&	~&~LO~&NLO($v^2$) &~LO~&NLO($v^2$)\\
			\hline
			\hline
			$J/\psi+J/\psi$&2736.0&355.7&2737.4 & 356.2&	$\Upsilon+\Upsilon$&34.4&~36.8~&36.7 & 38.4\\
			\hline
			$J/\psi+\eta_c$& 10.9&6.2&32.1 & ~13.7~&	$\Upsilon+\eta_b$& 35.7&43.0&37.6 & ~44.3~\\
			\hline
			$J/\psi+h_{c}$& 33.2&~20.2~&33.3 & ~20.3~&	$\Upsilon+h_{b}$& 22.2&~20.4~&25.2 & 22.5\\
			\hline
			$J/\psi+\chi_{c0}$& 2.1&1.4&4.1 & 2.1&	$\Upsilon+\chi_{b0}$& 8.0&7.9&13.3 & 11.7\\
			\hline
			$J/\psi+\chi_{c1}$& 2.3&0.9&8.2 & 3.0&	$\Upsilon+\chi_{b1}$& 1.3&1.5&17.1 & 12.9\\
			\hline
			$J/\psi+\chi_{c2}$& 3.2&2.3&13.1 & ~5.9~&$\Upsilon+\chi_{b2}$& 6.6&6.4&32.9 & 25.4\\
			\hline
			$\eta_c+\eta_c$&0.0&0.0&39.2 & 13.3&	$\eta_b+\eta_b$&0.0&0.0&0.3& 0.2\\
			\hline
			$\eta_c+h_c$& 3.1&1.5&4.4 & 2.1	&$\eta_b+h_b$& 7.1&5.5&7.8 & 6.0\\
			\hline
			$\eta_c+\chi_{c0}$& 6.8&3.4&11.9 & ~5.4~&$\eta_b+\chi_{b0}$& 5.0&3.8&8.1 & 5.9\\
			\hline
			$\eta_c+\chi_{c1}$&0.8&~1.1~&16.1 & ~6.8~&	$\eta_b+\chi_{b1}$&5.9&6.7&15.0 & 12.8\\
			\hline
			$\eta_c+\chi_{c2}$& 14.1&~9.4~&39.5 & 19.0&	$\eta_b+\chi_{b2}$& 13.3&12.1&28.5 & 22.4\\
			\hline
		\end{tabular}
		\label{TCS1}
	\end{table}
	\FloatBarrier

In Figs. \ref{z0cccos} and \ref{z0bbcos}, we depict the differential cross section $d\sigma/d\cos\theta$, where $\theta$ denotes the scattering angle of $H_1$. It is evident that within the CS channel, a concave distribution pattern emerges for $J/\psi+\eta_c/\chi_{c1}/J/\psi, \eta_c+\chi_{c1}$ and $\Upsilon+\eta_b/\chi_{b1}/\Upsilon, \eta_b+\chi_{b1}$. In contrast, the remaining processes within this channel exhibit a bulge distribution. As for the total CO channel, a concave distribution is characteristic of all its processes.

 The differential cross section $\frac{d\sigma}{dp_t}$ can be written as:
 \bea
 \frac{d\sigma}{dp_t}=|\frac{d\cos\theta}{dp_t}|(\frac{d\sigma}{d\cos\theta})=\frac{p_t}{|\vec{p_3}|\sqrt{\vec{p_3}^2-p_t^2}}(\frac{d\sigma}{d\cos\theta})
 \label{pt1}
 \eea
 and the momentum of the $H_1$ heavy quarkonia:
 \bea
 \vec{p_3}=\frac{\sqrt{\lambda[s,(2E_{q3})^2,(2E_{q4})^2]}}{2\sqrt{s}}
 \label{pt2}
 \eea
 Combining the Eqs. (\ref{pt1}) and (\ref{pt2}), we can also get a $\mathcal{O}(v^2)$  expression:
 \bea
 \frac{d\sigma}{dp_t}= \frac{4p_ts^{3/2}[s(s-4p_t^2)-16m_Q^2(2p_t^2-s)(v^2-2)-256m_Q^4(v^2-1)]}{[s(s-16m_Q^2)]^{3/2}(s-16m_Q^2-4p_t^2)^{3/2}}(\frac{d\sigma}{d\cos\theta})+\mathcal{O}(v^4)
 \eea
  The results are shown in Figs. \ref{ccpt} and \ref{bbpt}, it's seen that the $p_t$ distribution change is more stable for the CO channel  than  for the CS channel.

\begin{widetext}
	\begin{figure*}[htbp]
		\begin{tabular}{c c c}
			\includegraphics[width=0.333\textwidth]{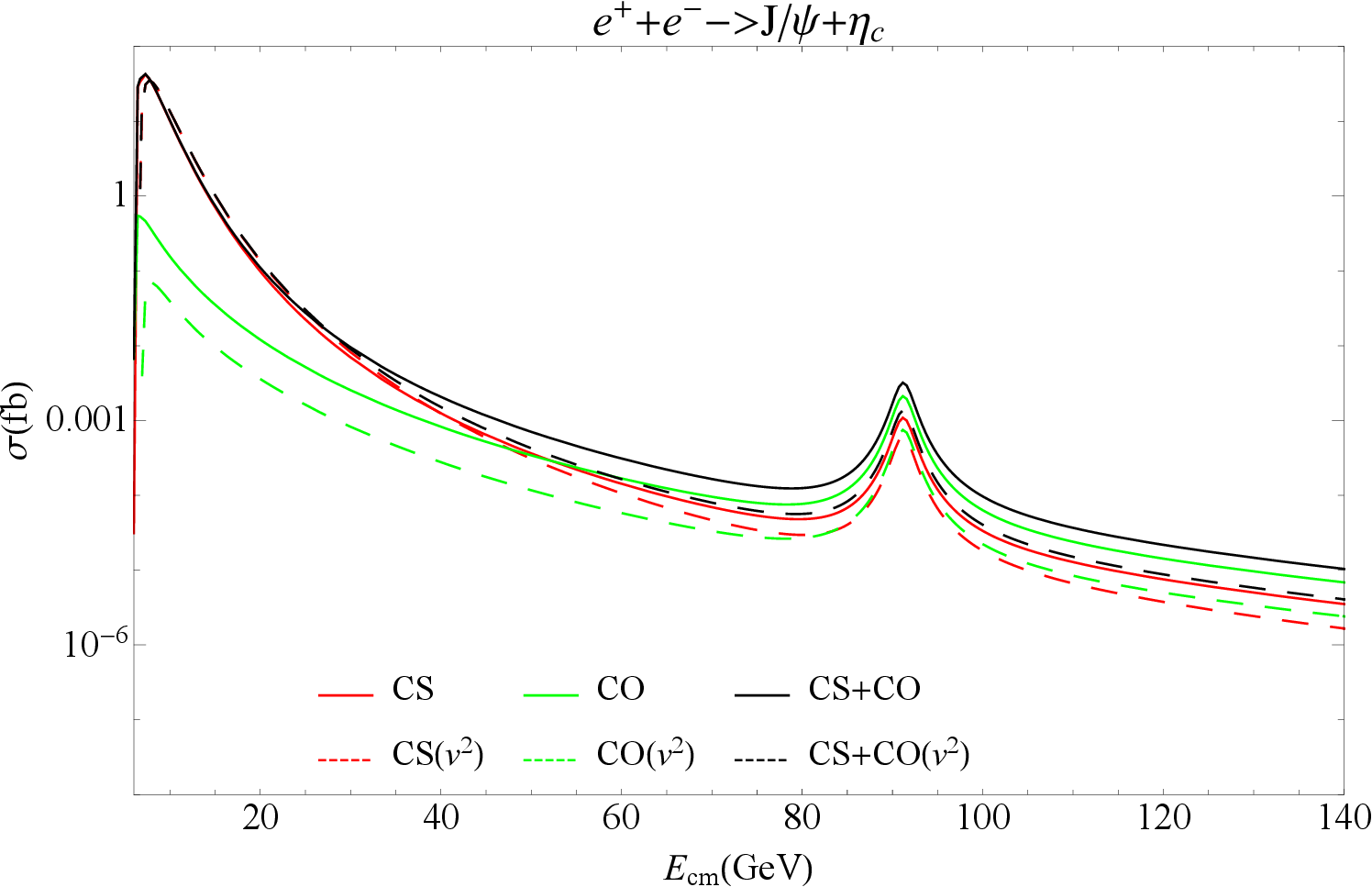}
			\includegraphics[width=0.333\textwidth]{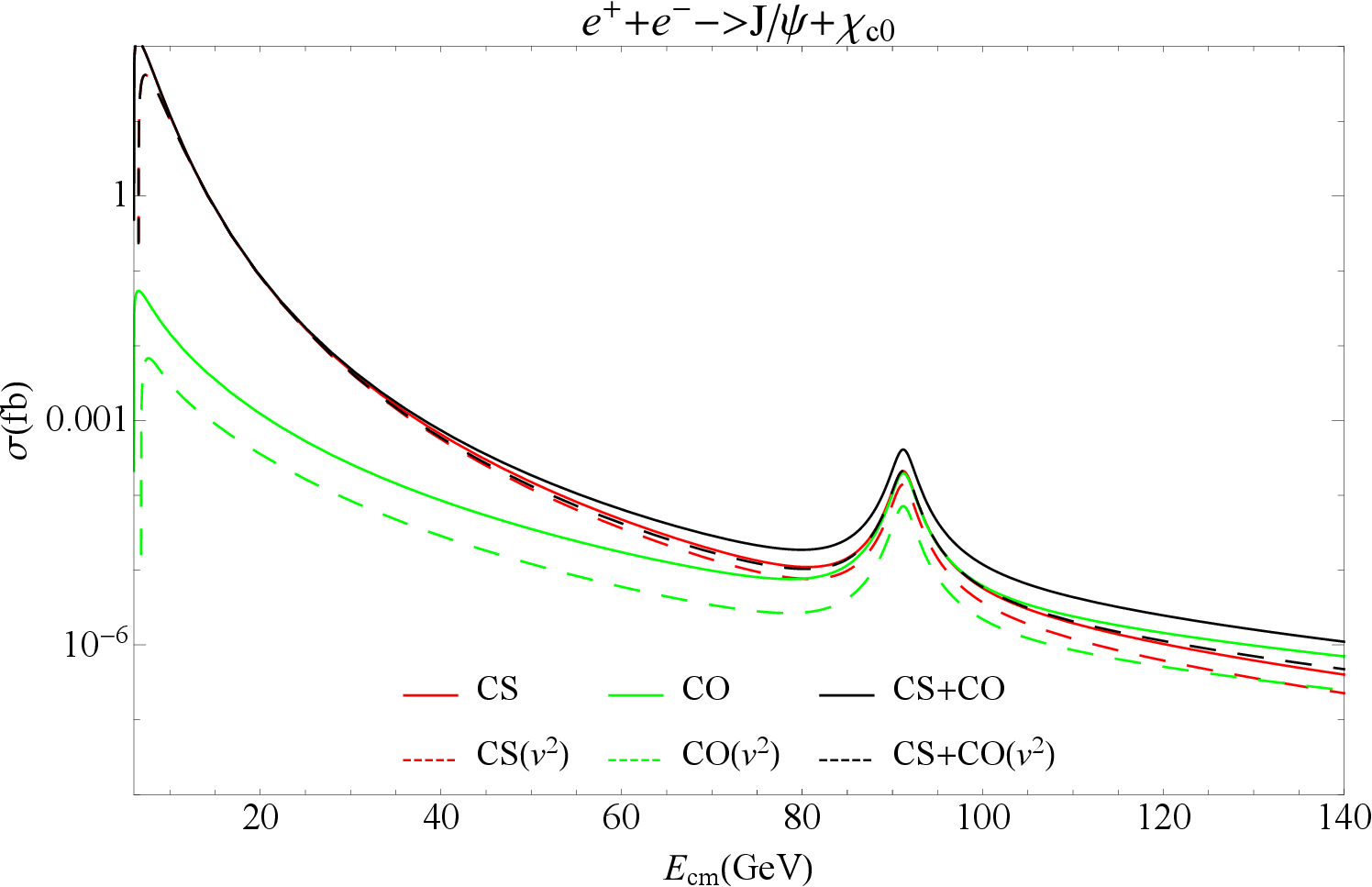}
				\includegraphics[width=0.333\textwidth]{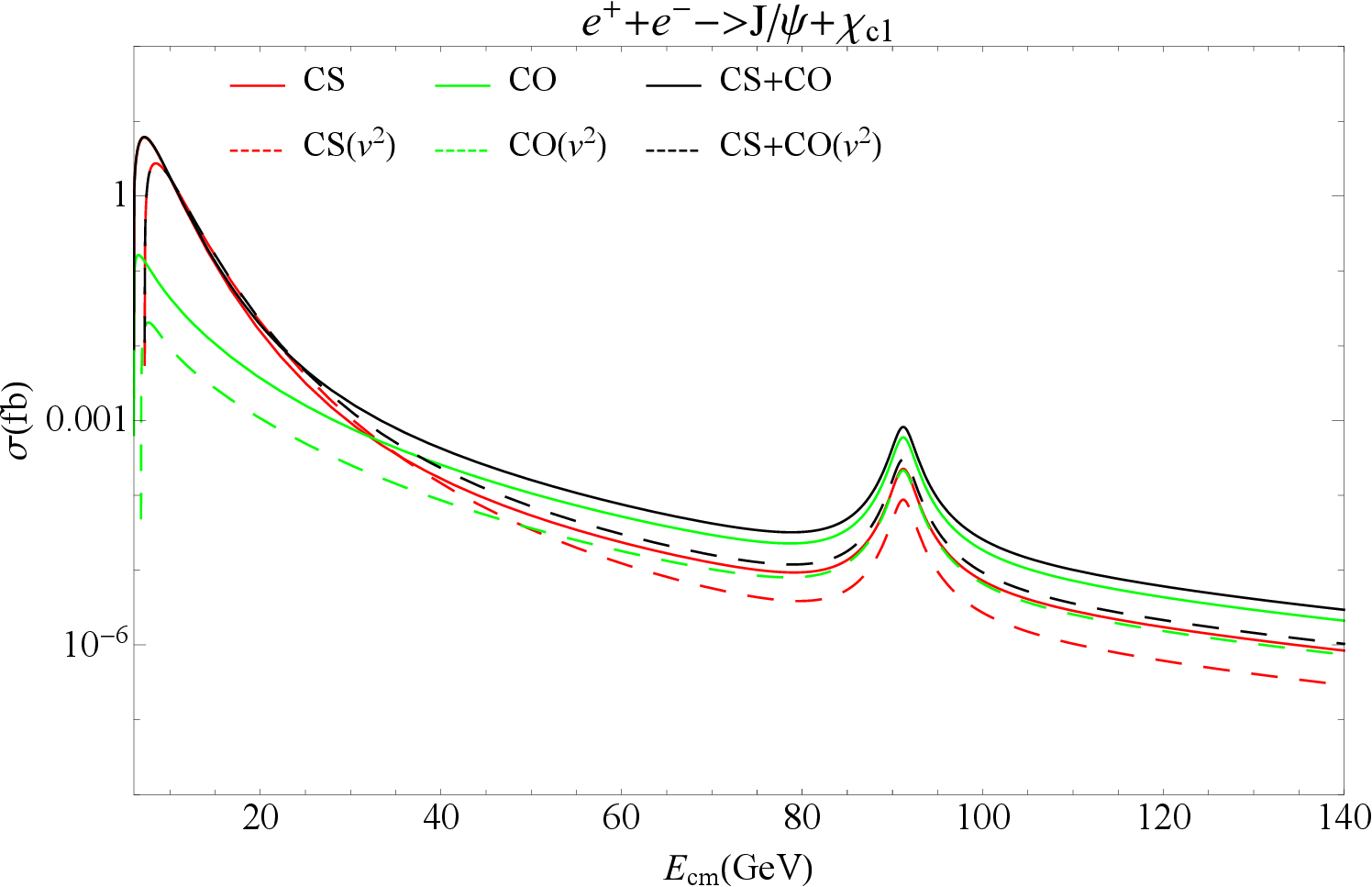}
		\end{tabular}
		\begin{tabular}{c c c}	
			\includegraphics[width=0.333\textwidth]{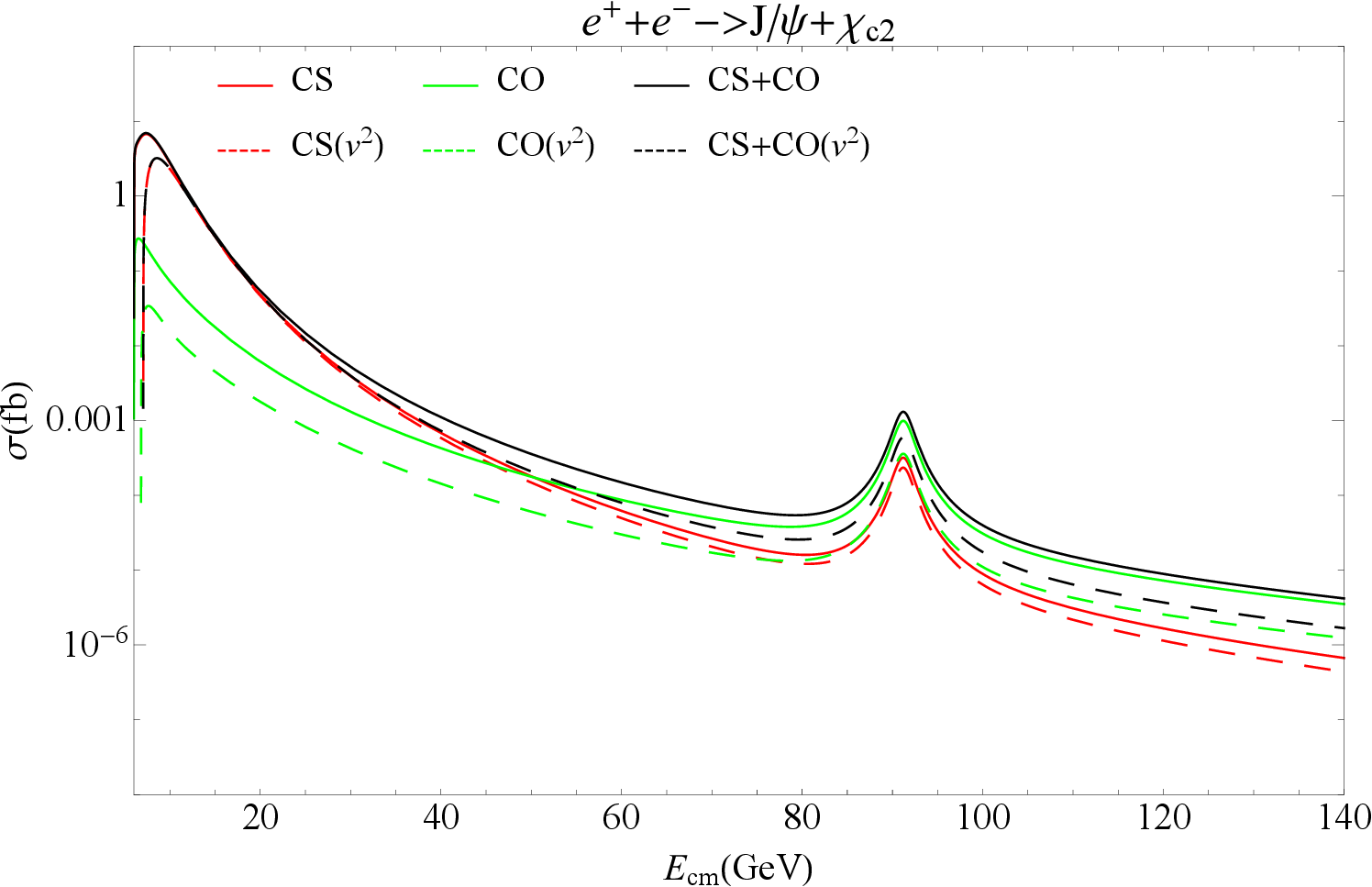}
				\includegraphics[width=0.333\textwidth]{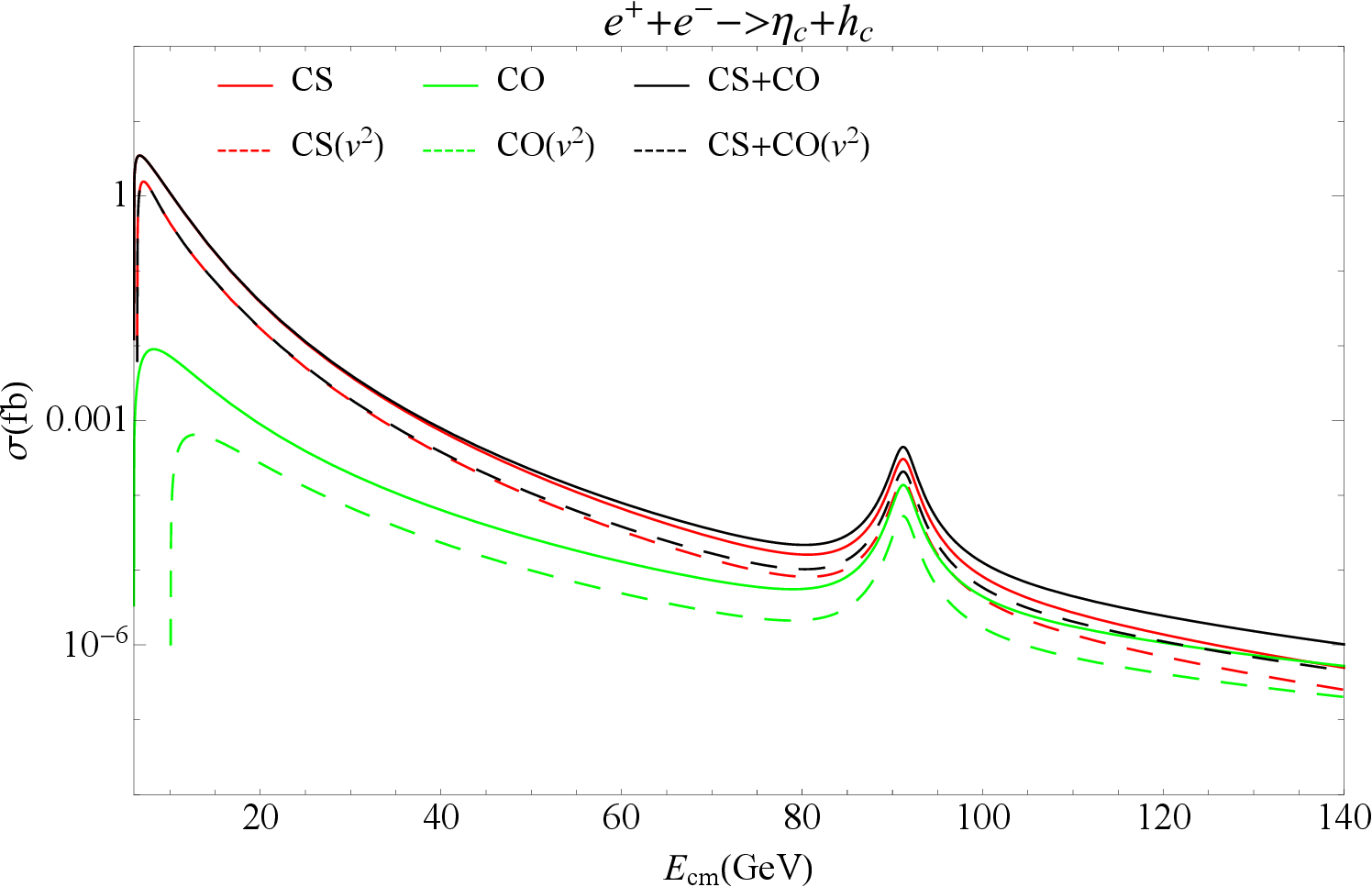}
			\includegraphics[width=0.333\textwidth]{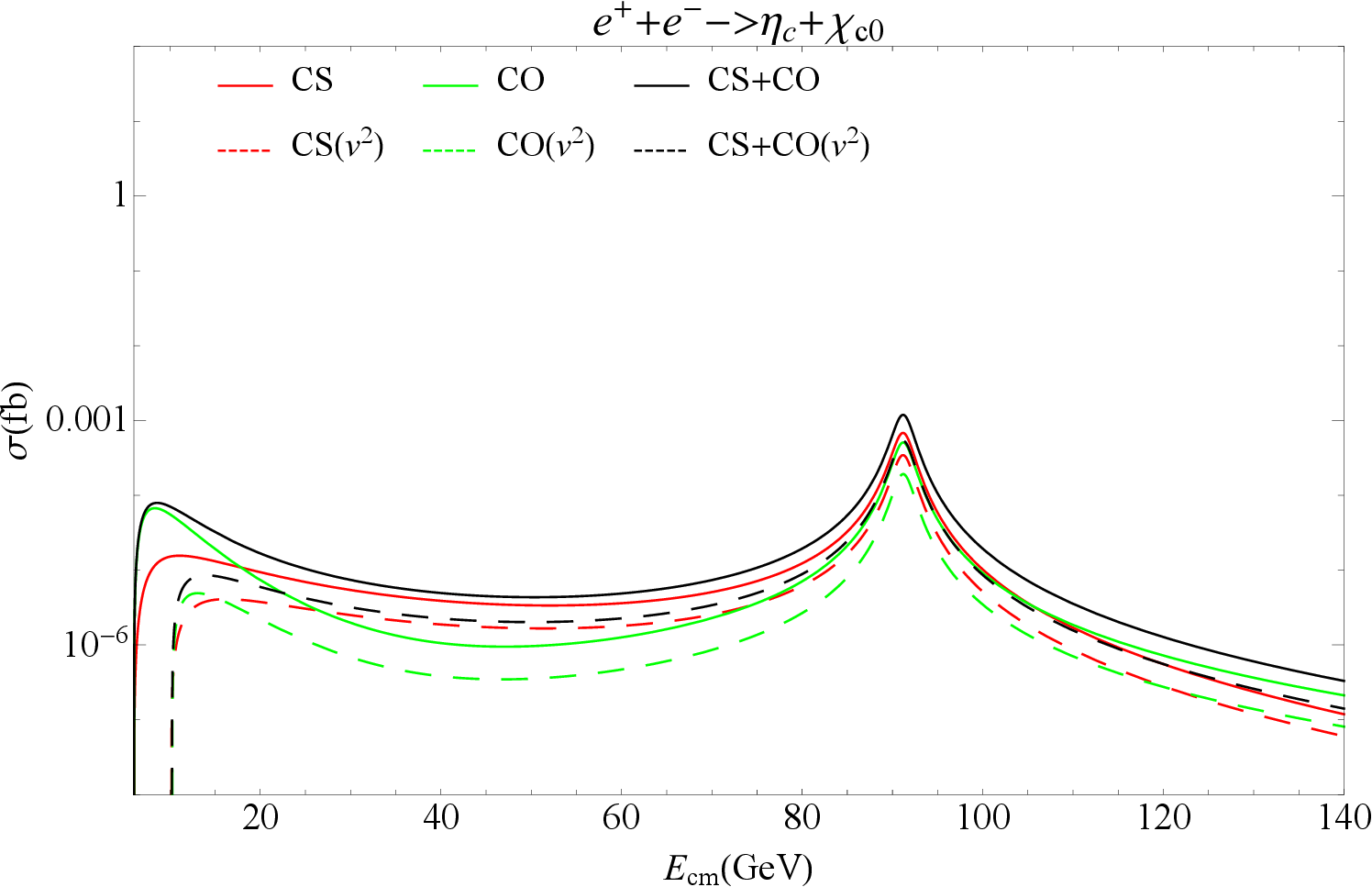}
		\end{tabular}
		\begin{tabular}{c c c}
			\includegraphics[width=0.333\textwidth]{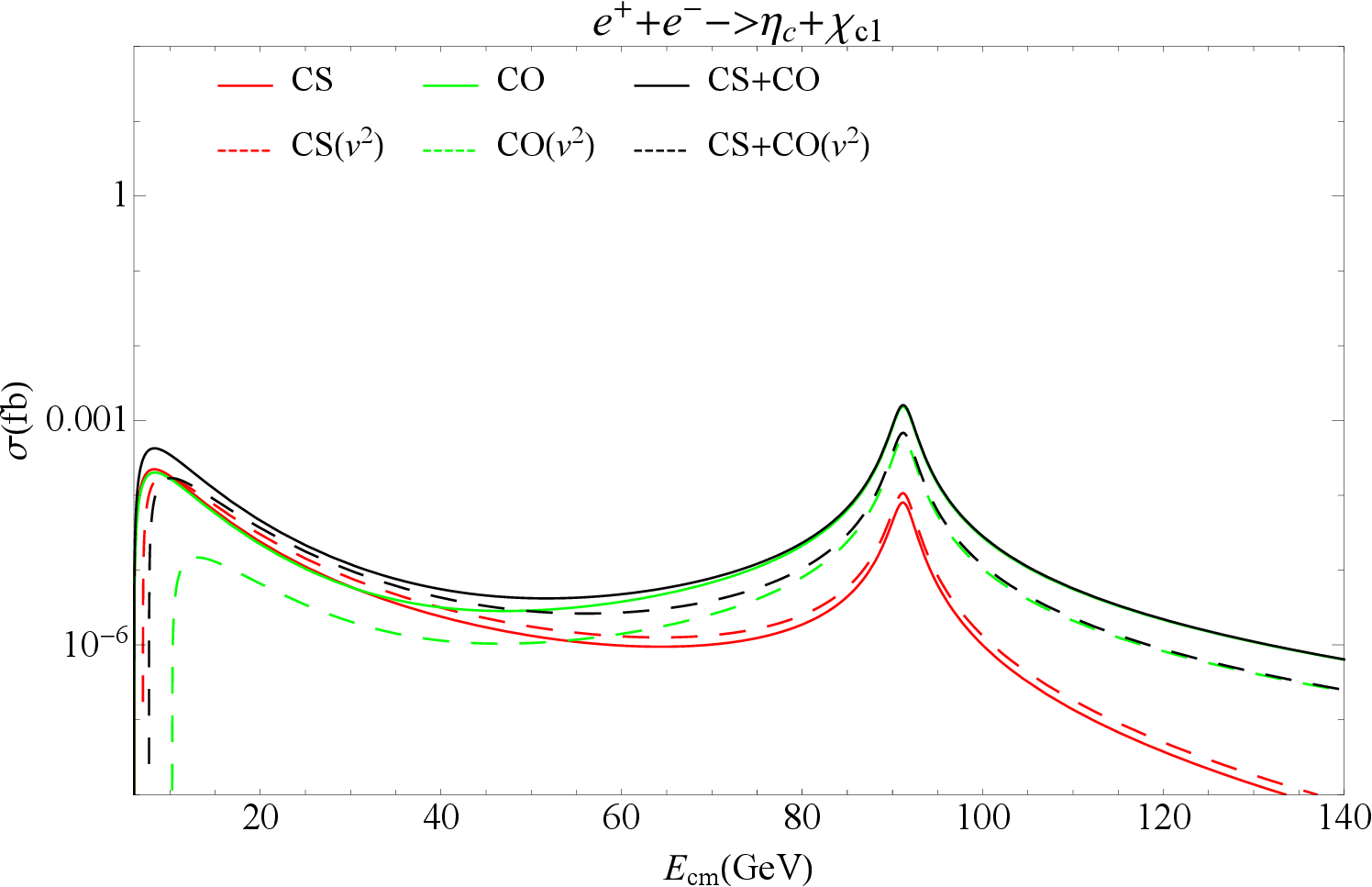}
		\includegraphics[width=0.333\textwidth]{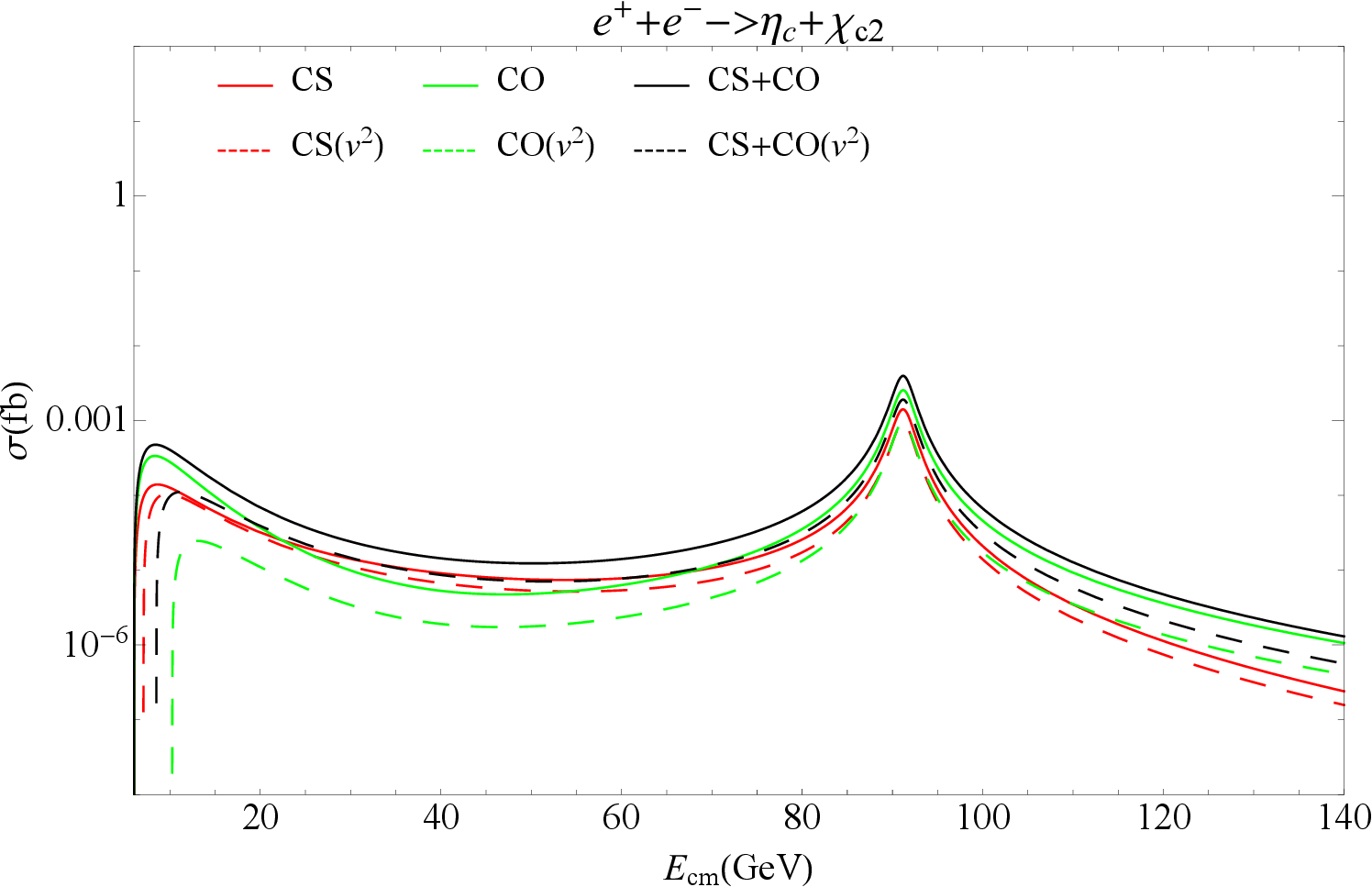}
		\includegraphics[width=0.333\textwidth]{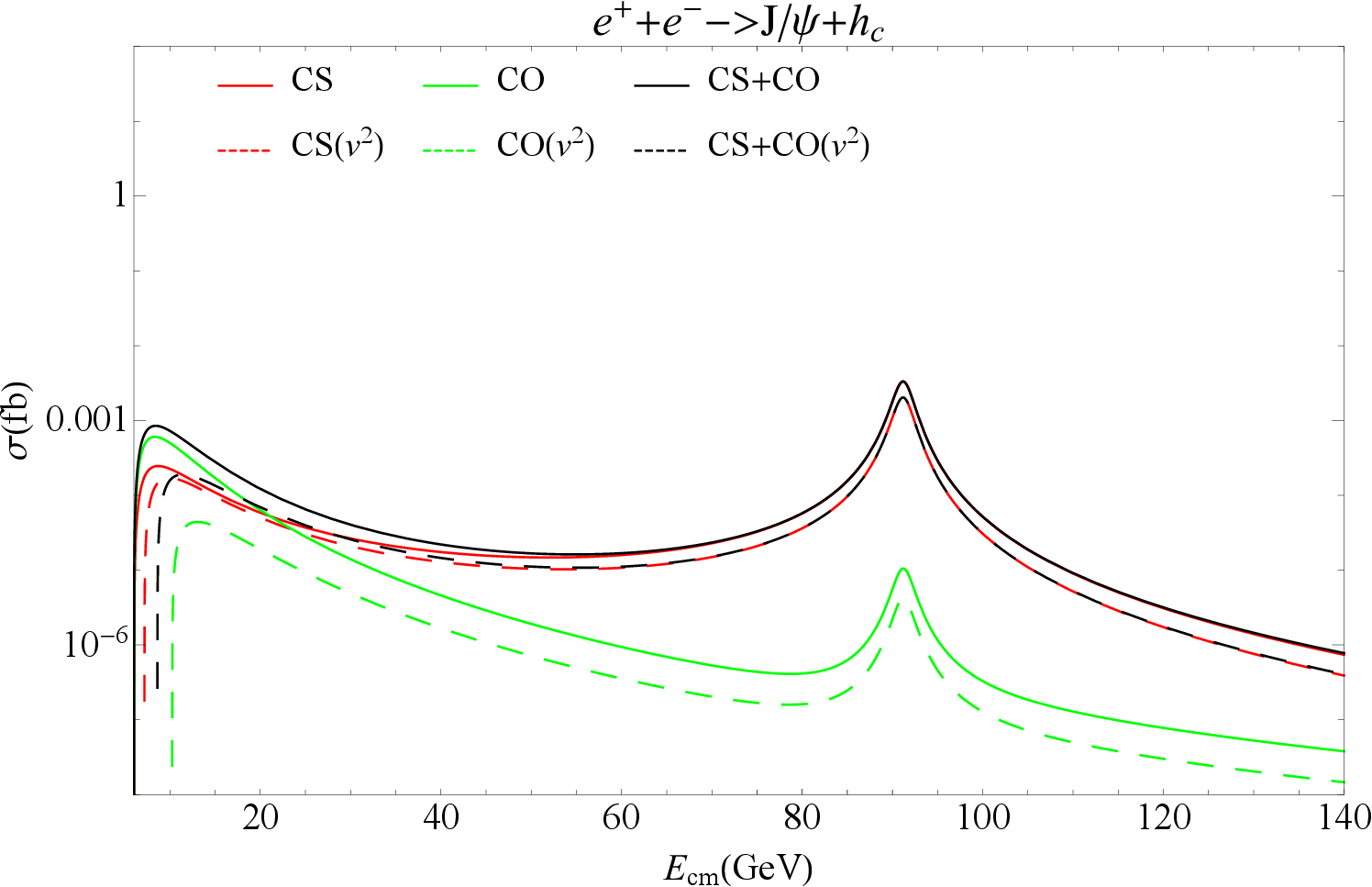}
		\end{tabular}
		\begin{tabular}{c c c }
			
			\includegraphics[width=0.333\textwidth]{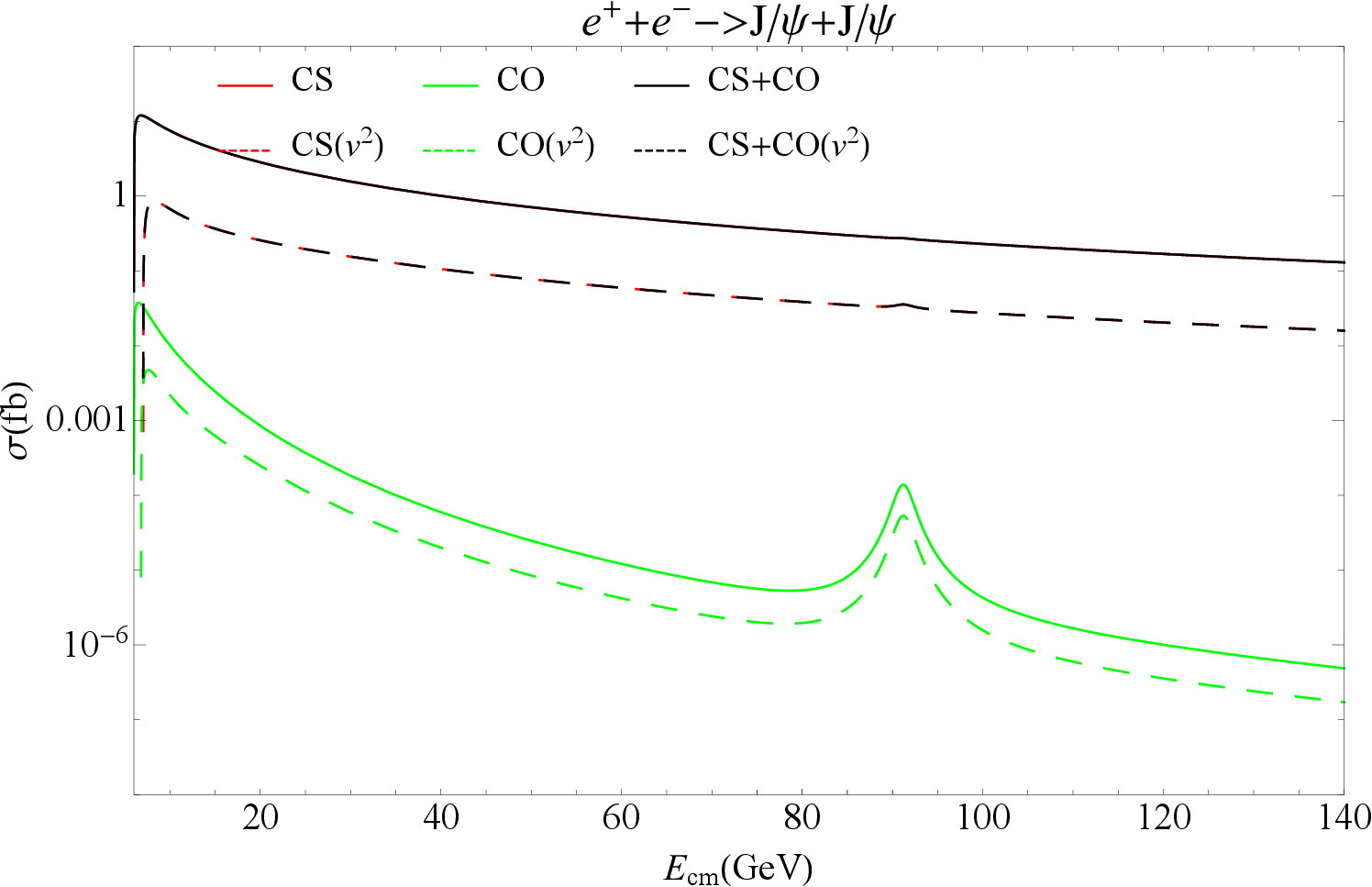}
				\includegraphics[width=0.333\textwidth]{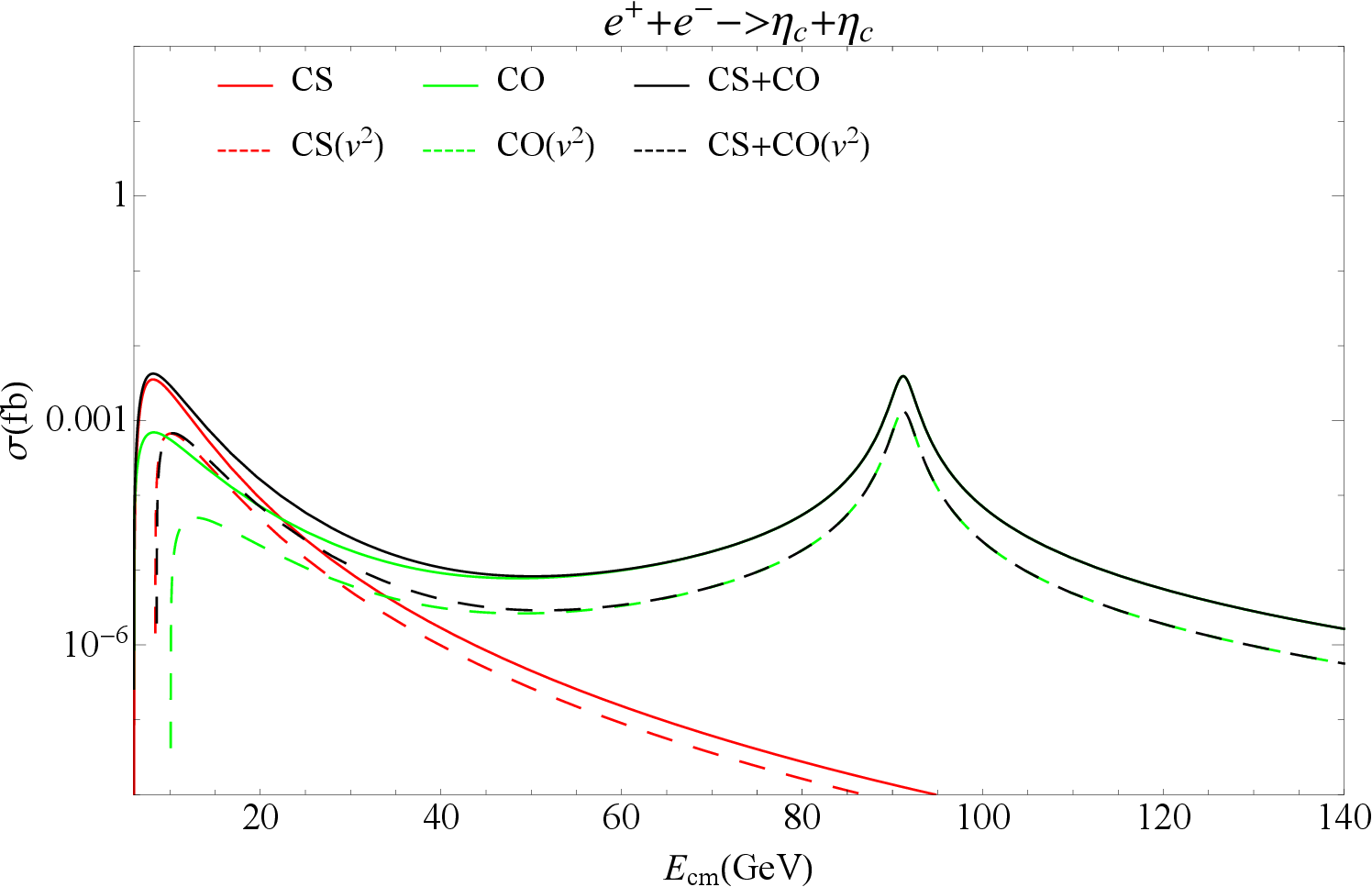}
			
		\end{tabular}
		\caption{ (Color online) Cross sections ($~\sigma~$) versus c.m. energy~($E_{cm}=\sqrt{s})$) ~for double  charmonium production. The solid line represents leading order (LO)  and the dashed line represents next-to-leading order in $v^2$ (NLO) results. The red line represents the CS channel, the green line represents the total CO channels and the black line represents the sum of  CS and CO. }
		\label{z0cc}
	\end{figure*}
		\FloatBarrier
\end{widetext}

	\begin{widetext}
	\begin{figure*}[htbp]
		\begin{tabular}{c c c}
			\includegraphics[width=0.333\textwidth]{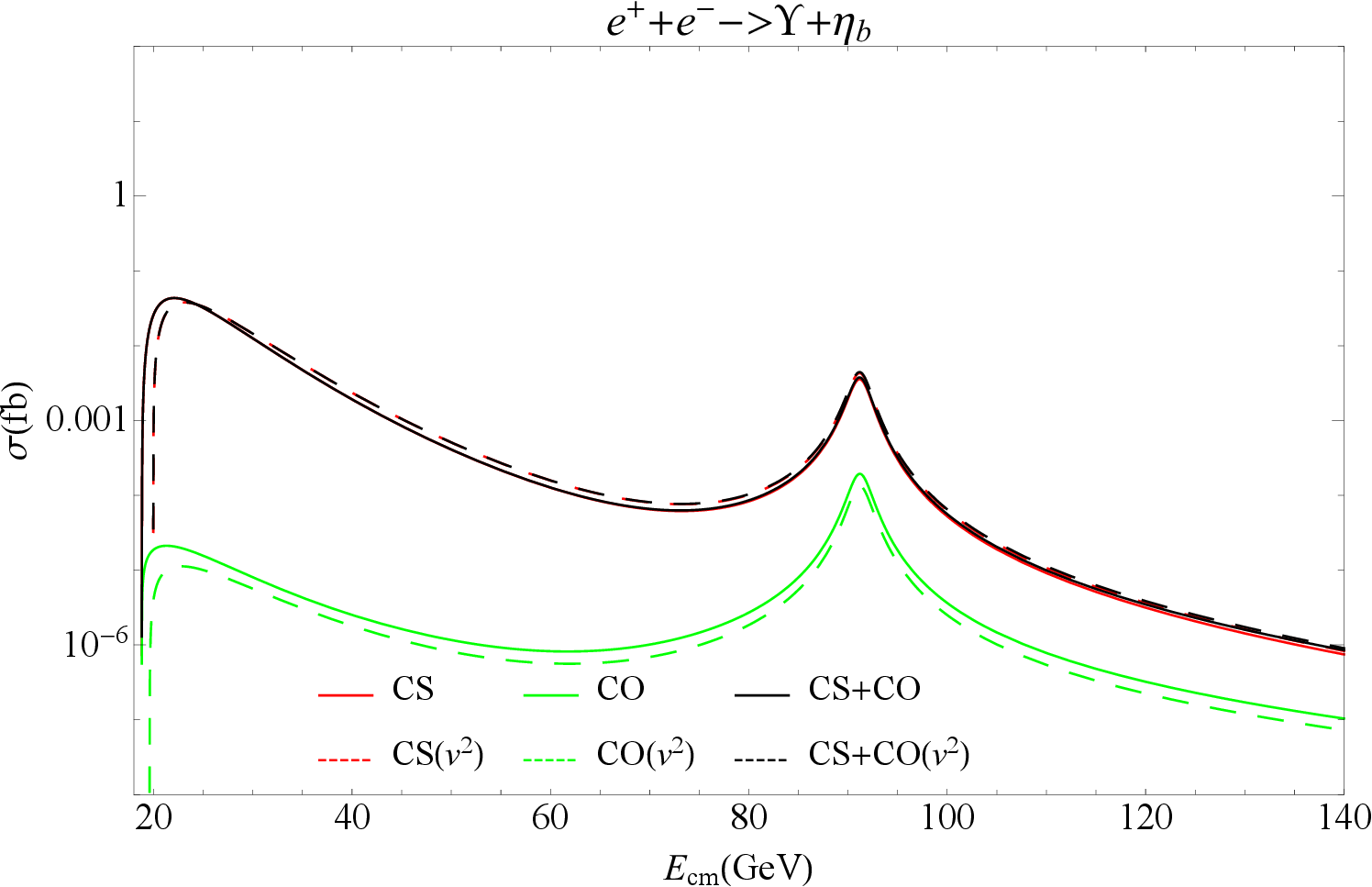}
			\includegraphics[width=0.333\textwidth]{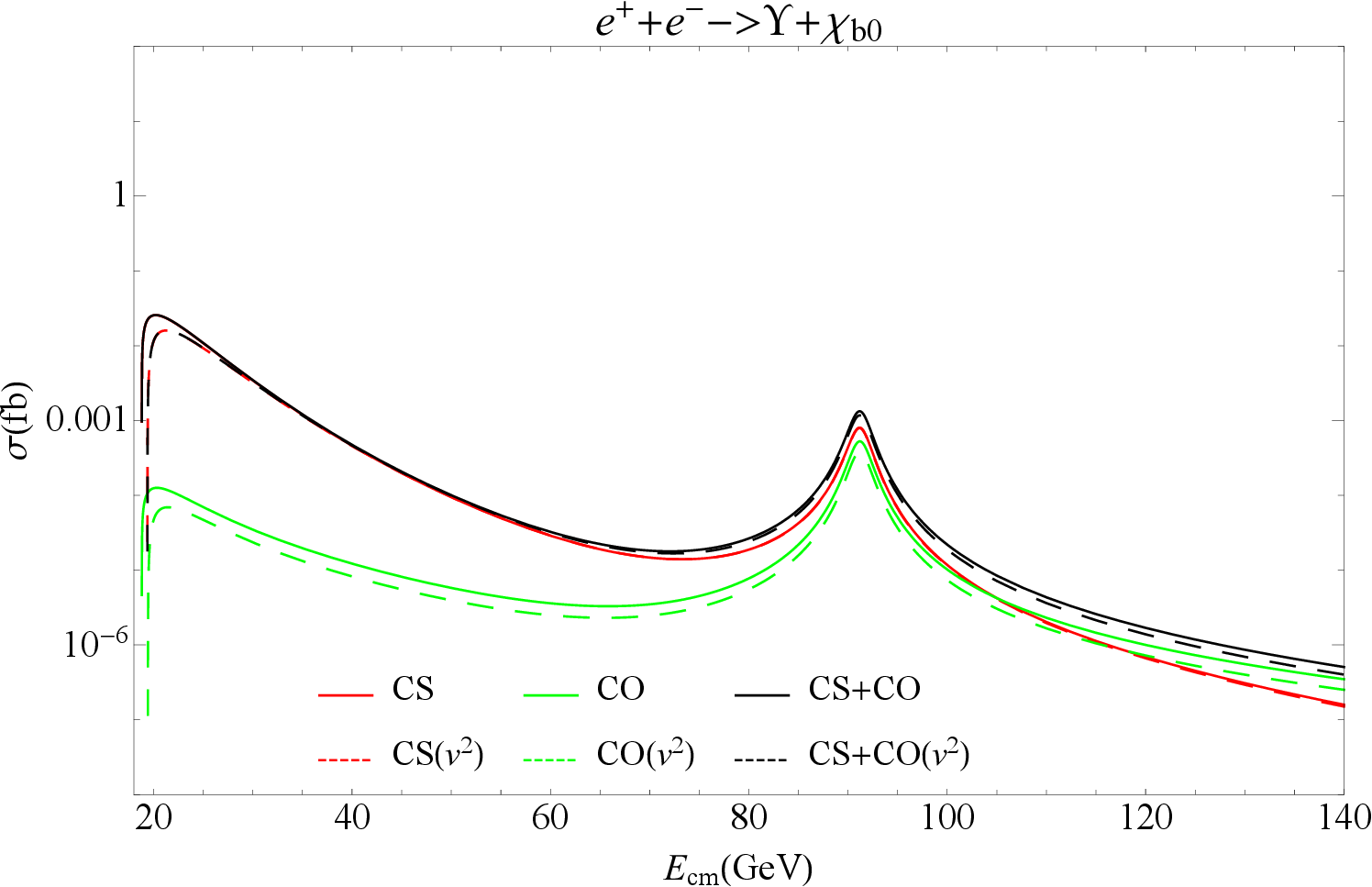}
			\includegraphics[width=0.333\textwidth]{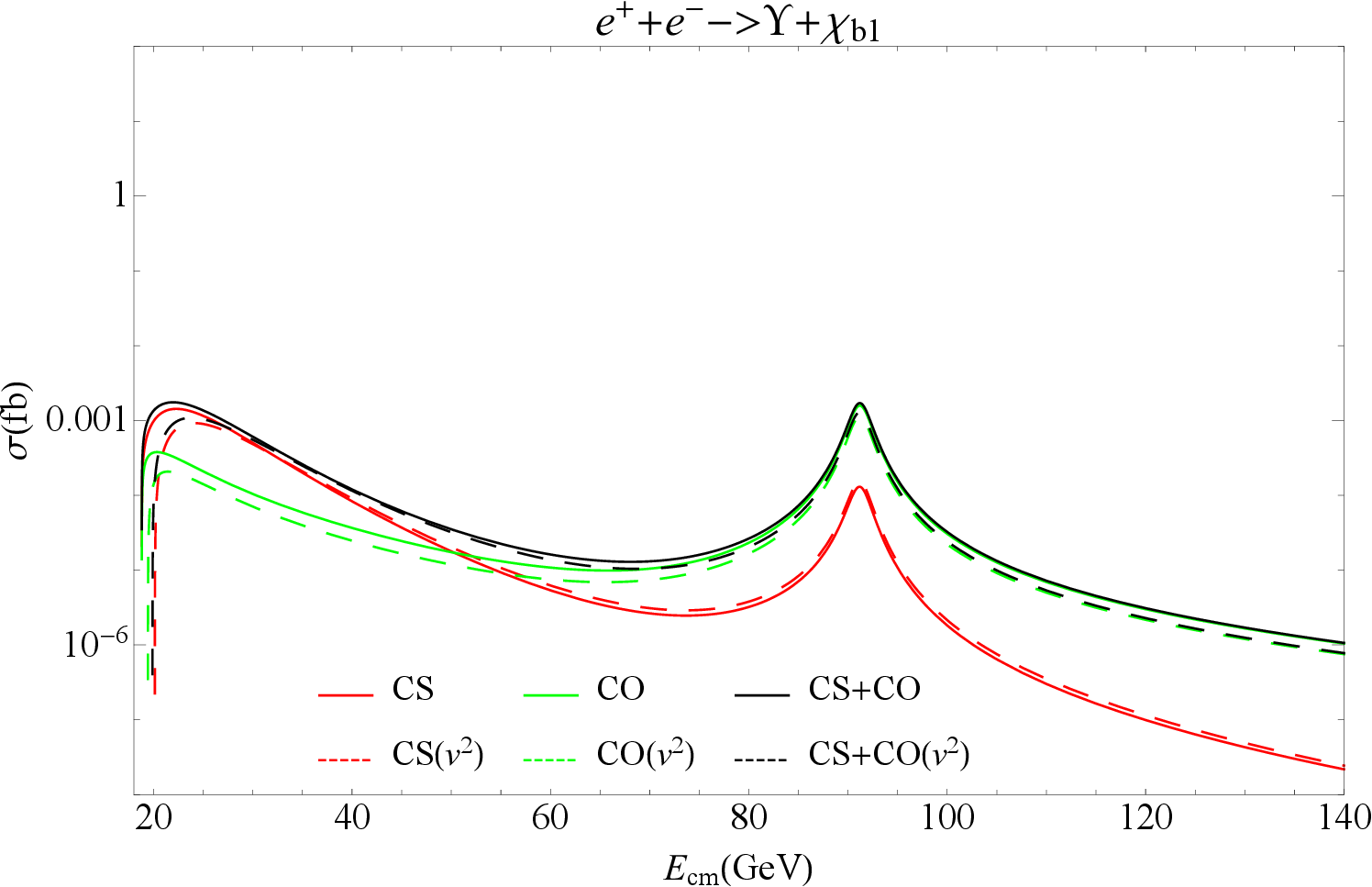}
		\end{tabular}
		\begin{tabular}{c c c}
			\includegraphics[width=0.333\textwidth]{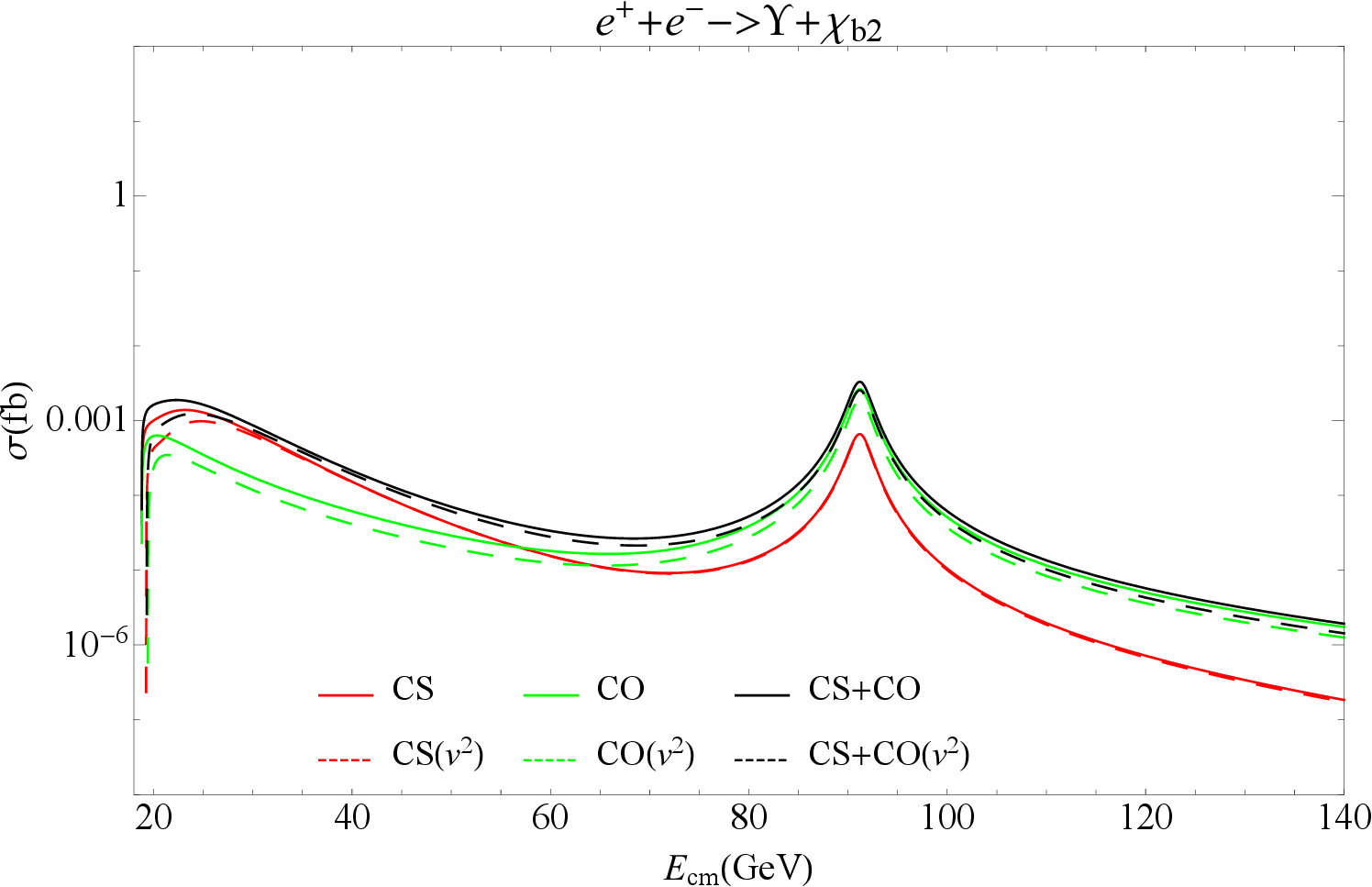}
			\includegraphics[width=0.333\textwidth]{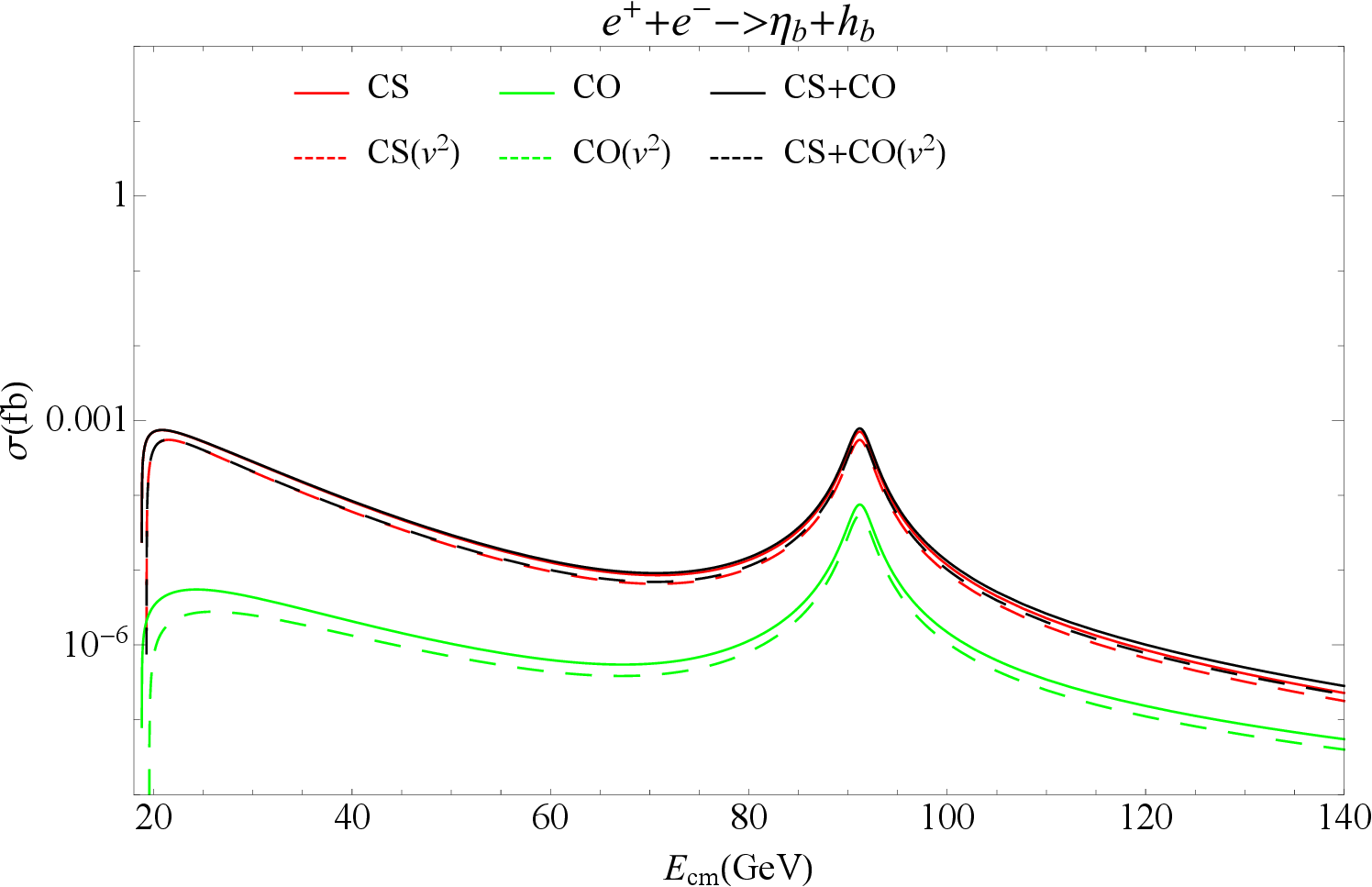}
			\includegraphics[width=0.333\textwidth]{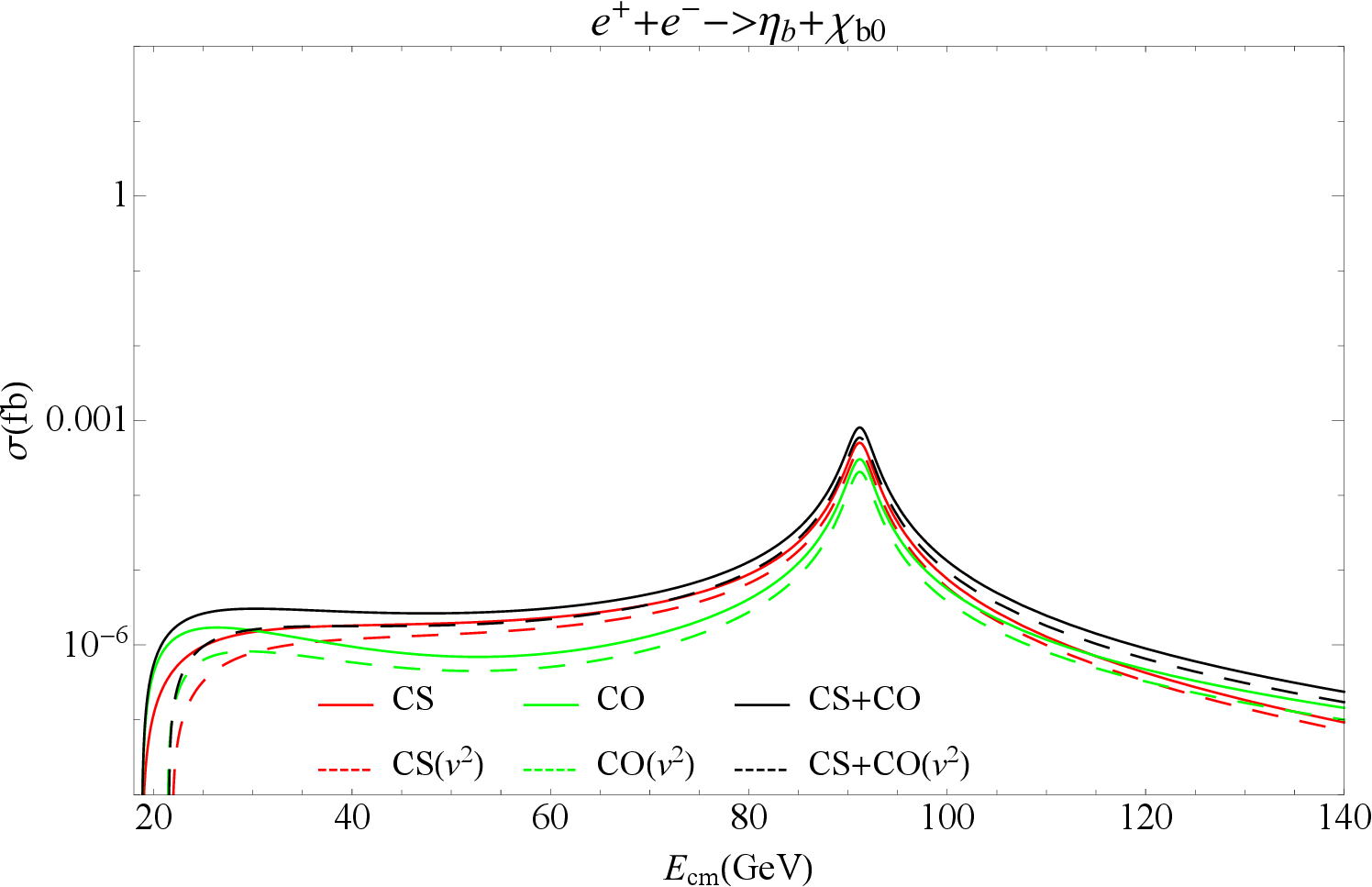}
		\end{tabular}
		\begin{tabular}{c c c}
			\includegraphics[width=0.333\textwidth]{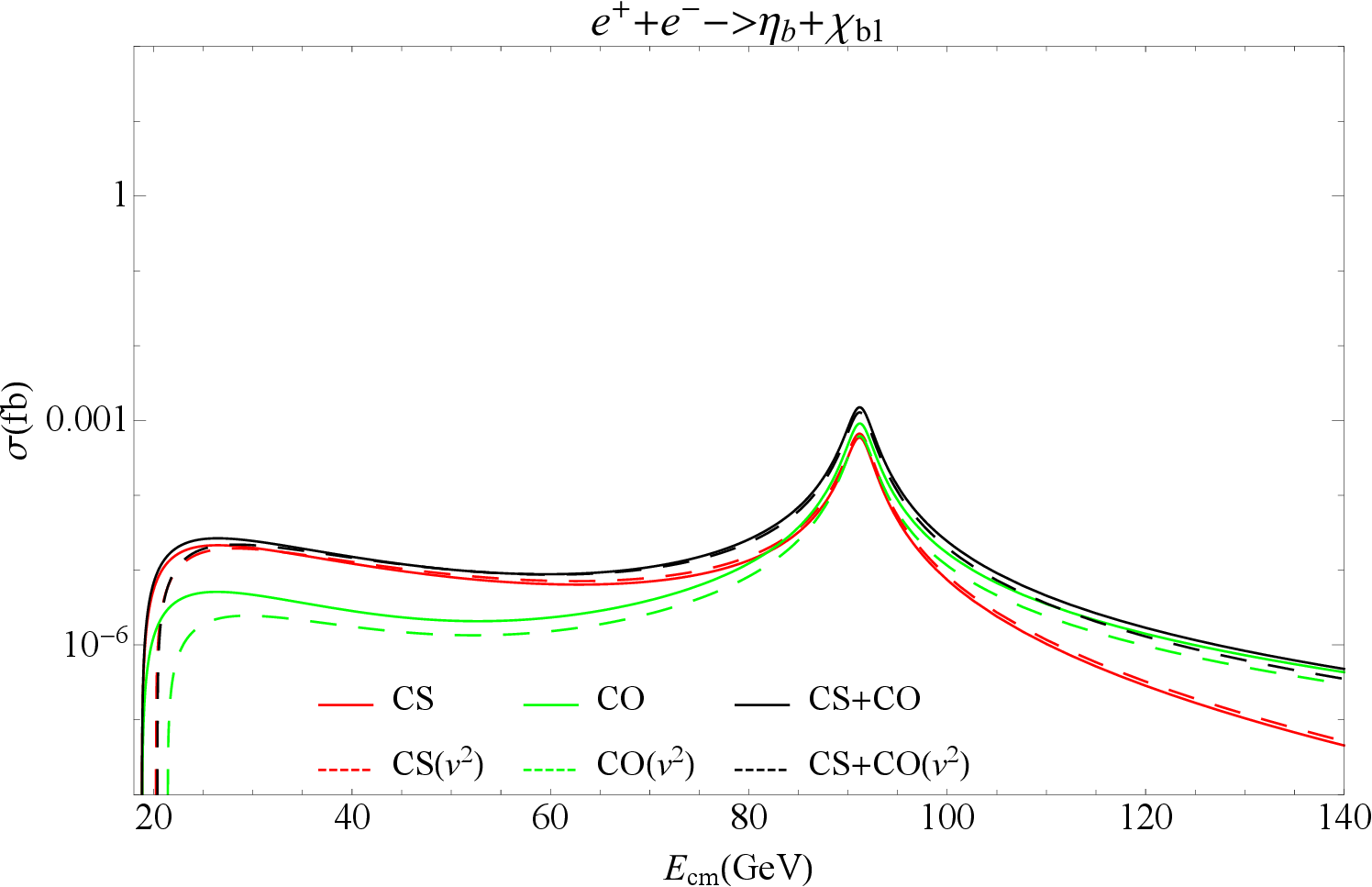}
			\includegraphics[width=0.333\textwidth]{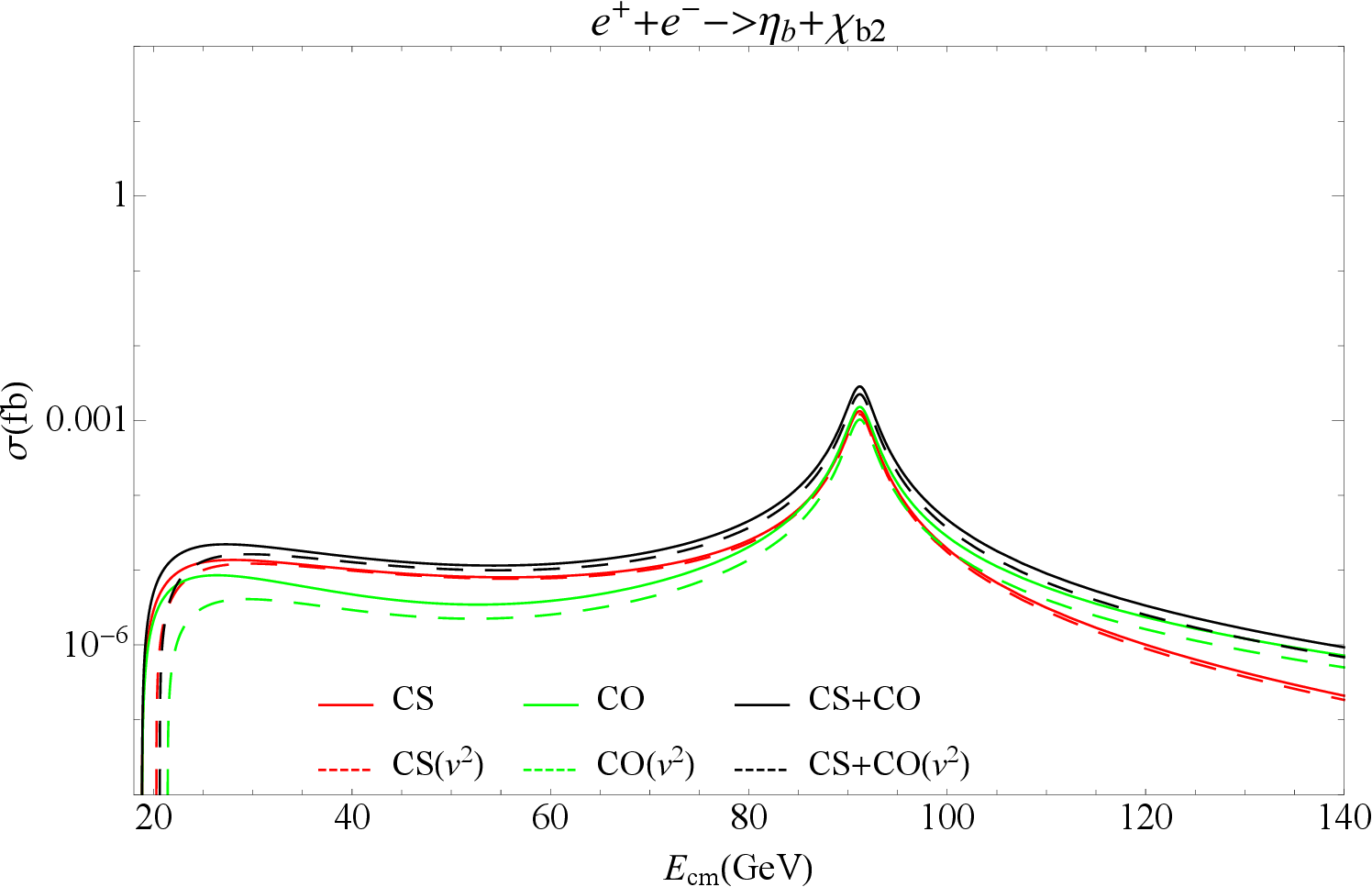}
			\includegraphics[width=0.333\textwidth]{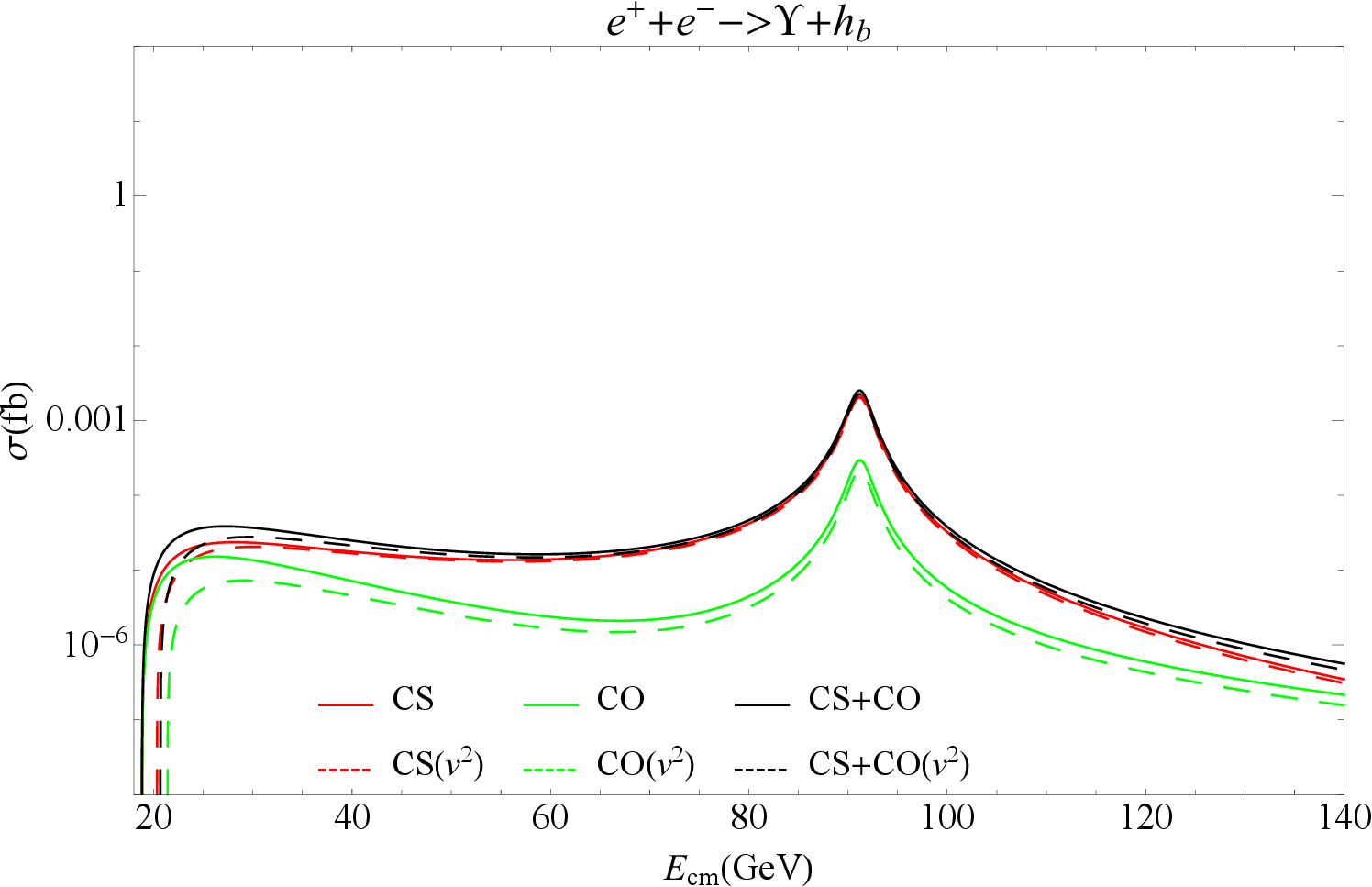}
		\end{tabular}
		\begin{tabular}{c c c }
			
			\includegraphics[width=0.333\textwidth]{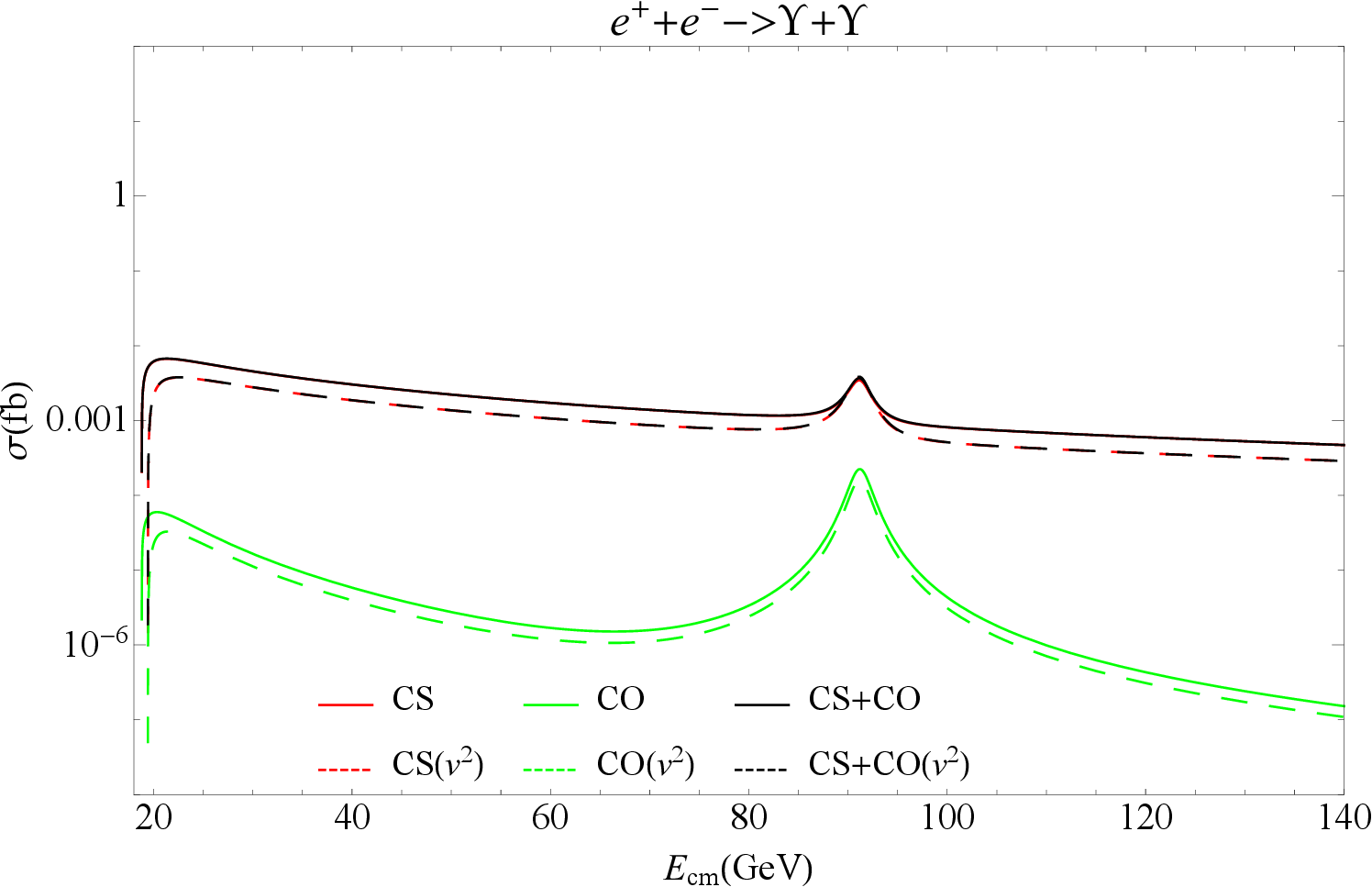}
				\includegraphics[width=0.333\textwidth]{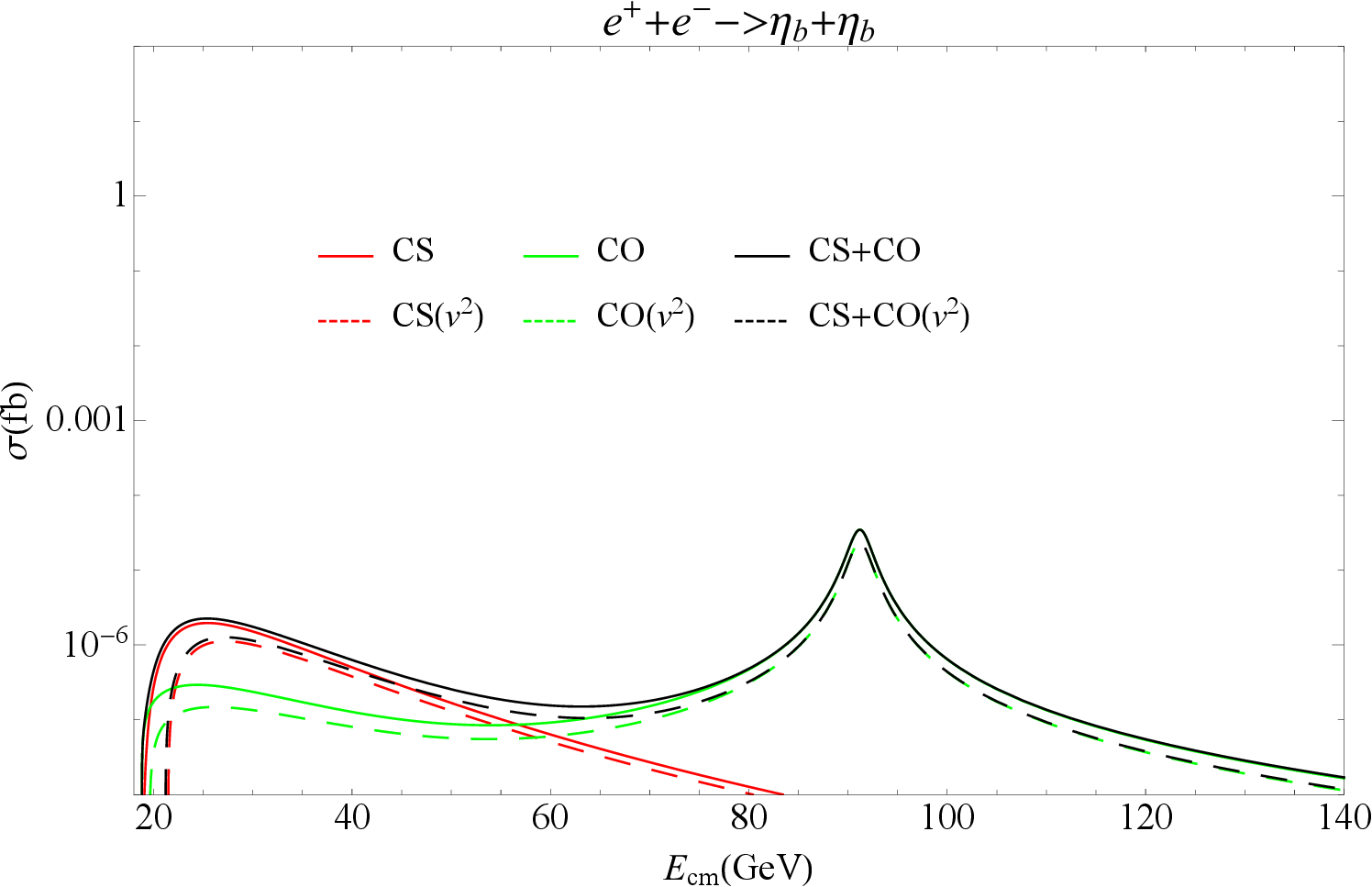}
			
		\end{tabular}
		\caption{ (Color online) Cross sections ($~\sigma~$) versus c.m. energy~($E_{cm}=\sqrt{s})$) ~for double  bottomonium production. The solid line represents leading order (LO)  and the dashed line represents next-to-leading order in $v^2$ (NLO) results. The red line represents the CS channel, the green line represents the total CO channels and the black line represents the sum of  CS and CO. }
		\label{z0bb}
	\end{figure*}
	\FloatBarrier
\end{widetext}
\begin{widetext}
	\begin{figure*}[htbp]
		\begin{tabular}{c c c}
			\includegraphics[width=0.333\textwidth]{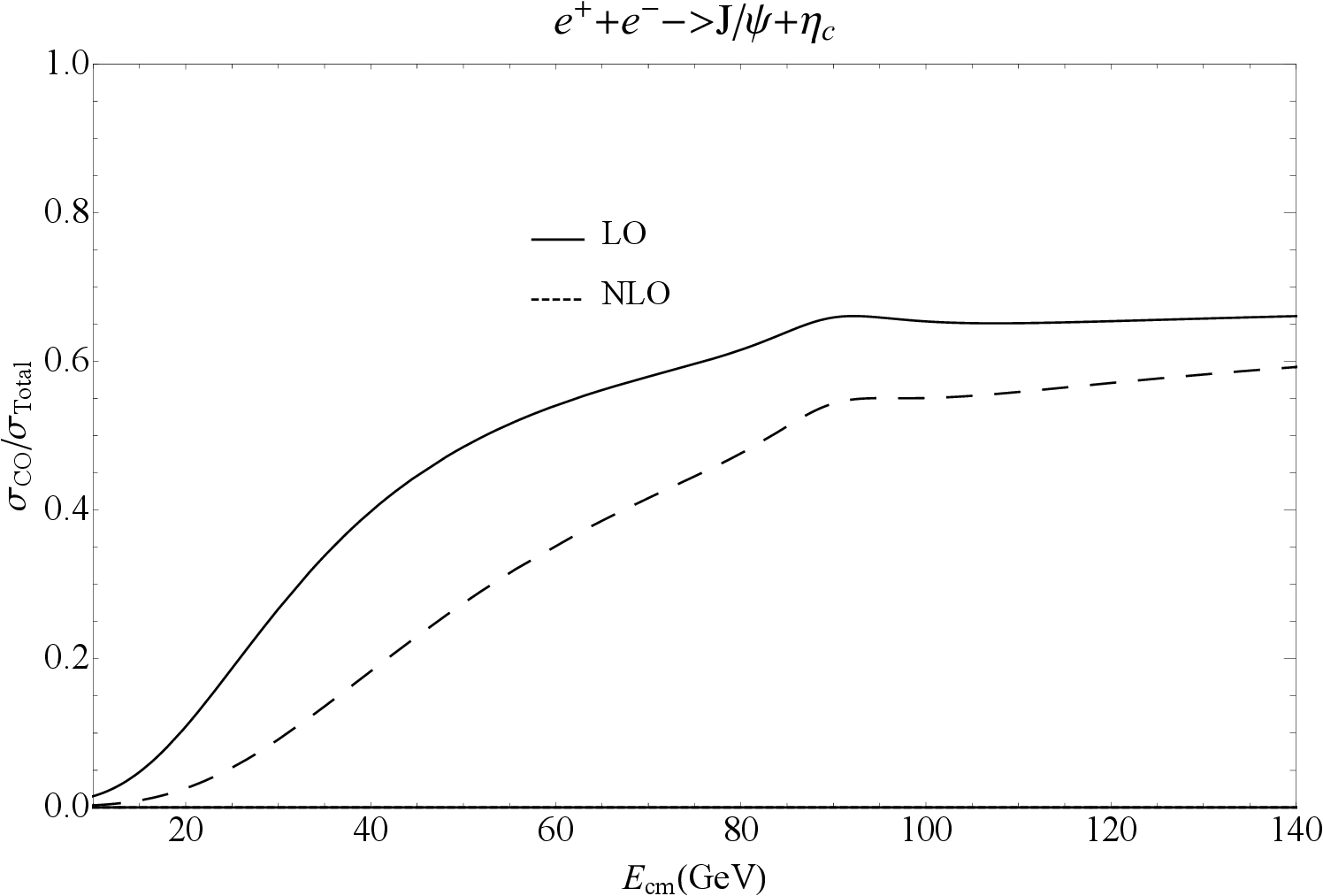}
			\includegraphics[width=0.333\textwidth]{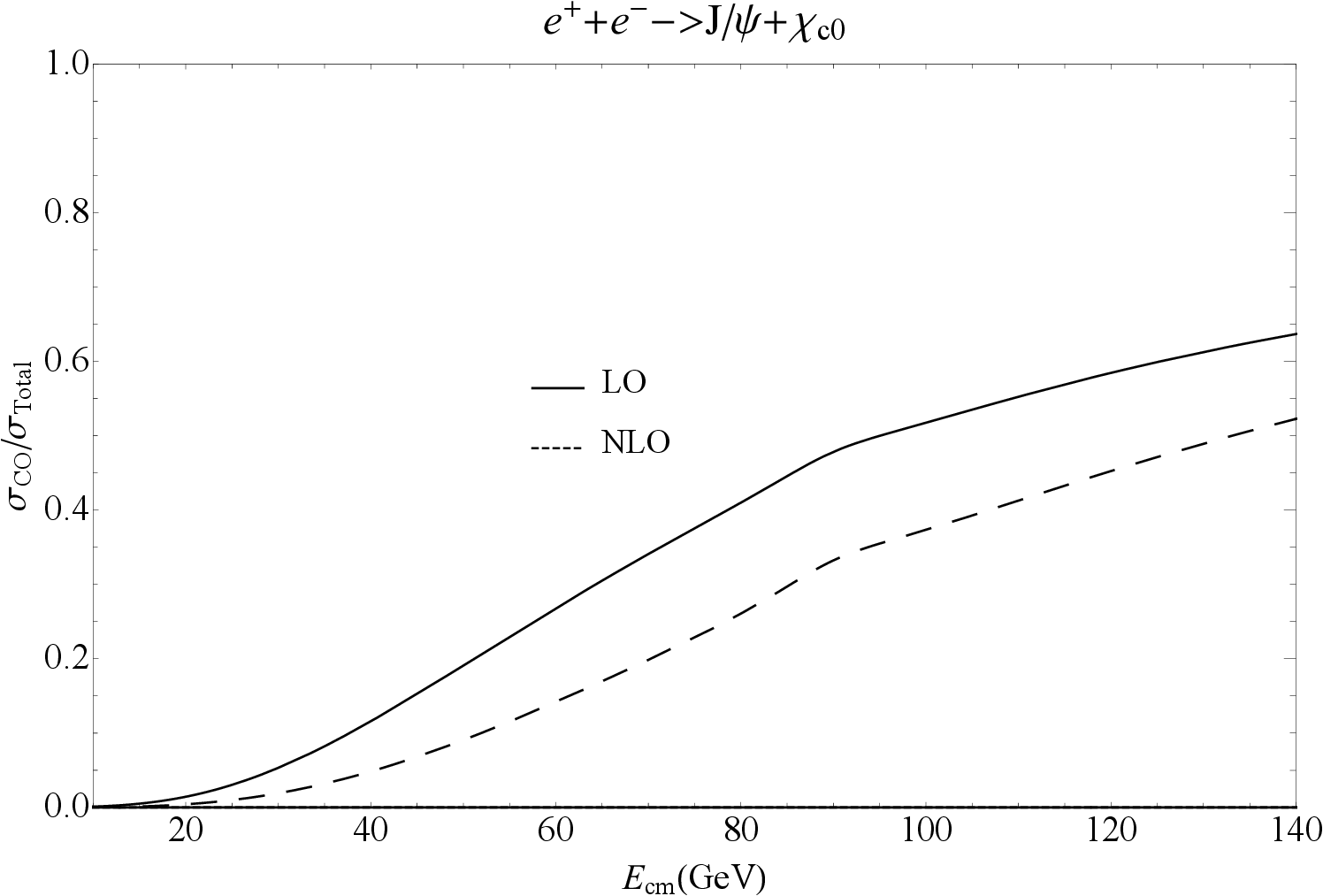}
			\includegraphics[width=0.333\textwidth]{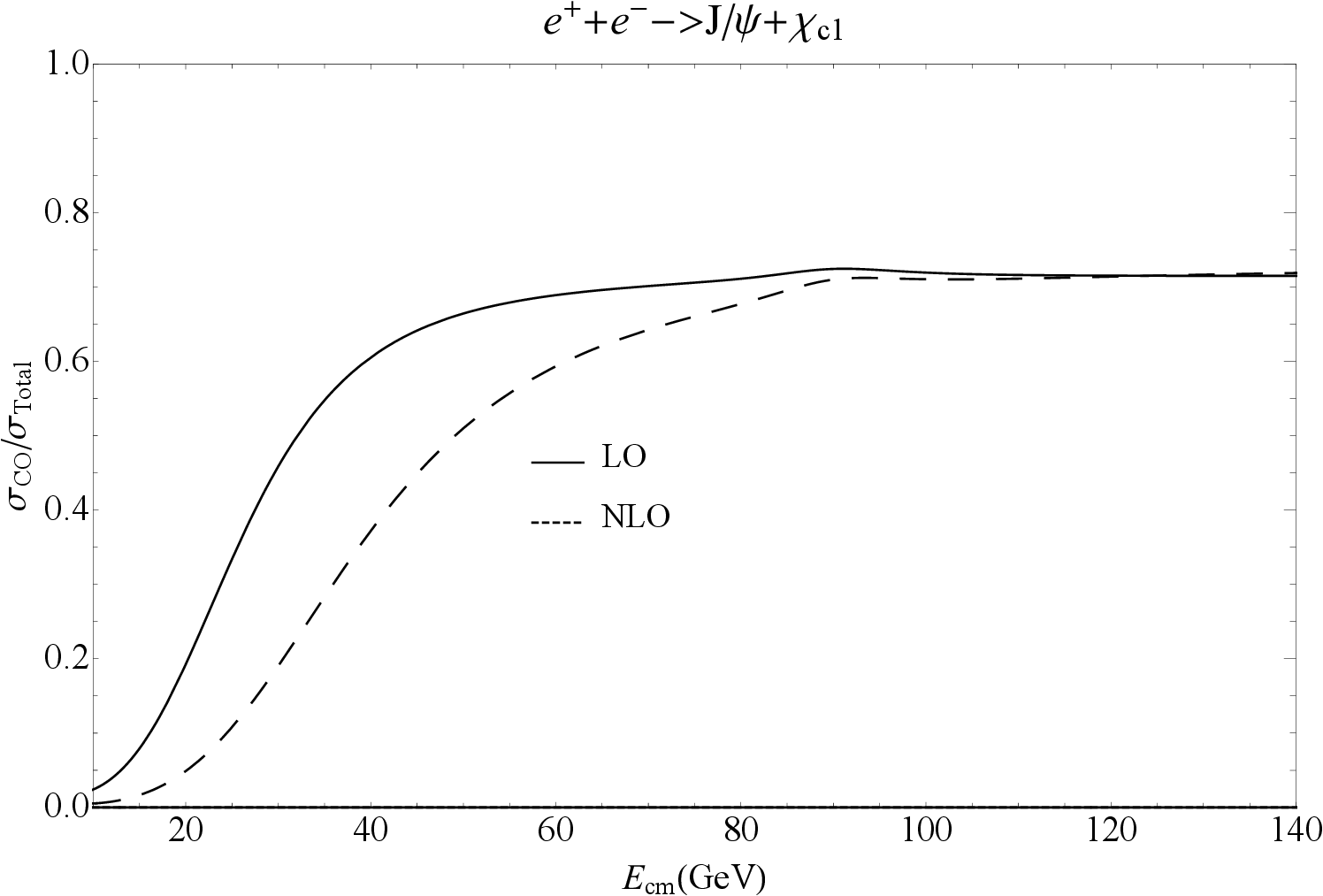}
		\end{tabular}
		\begin{tabular}{c c c}	
			\includegraphics[width=0.333\textwidth]{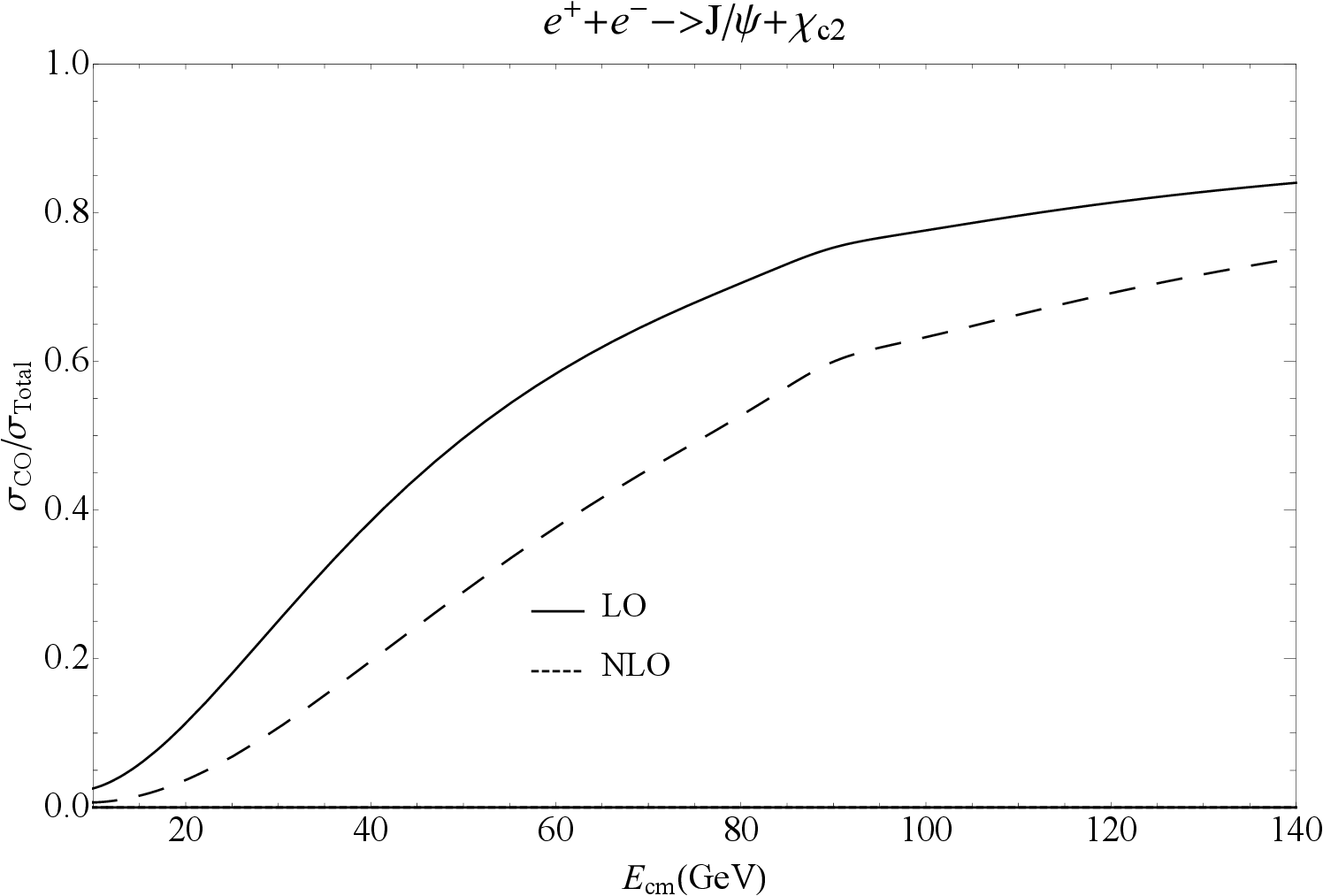}
			\includegraphics[width=0.333\textwidth]{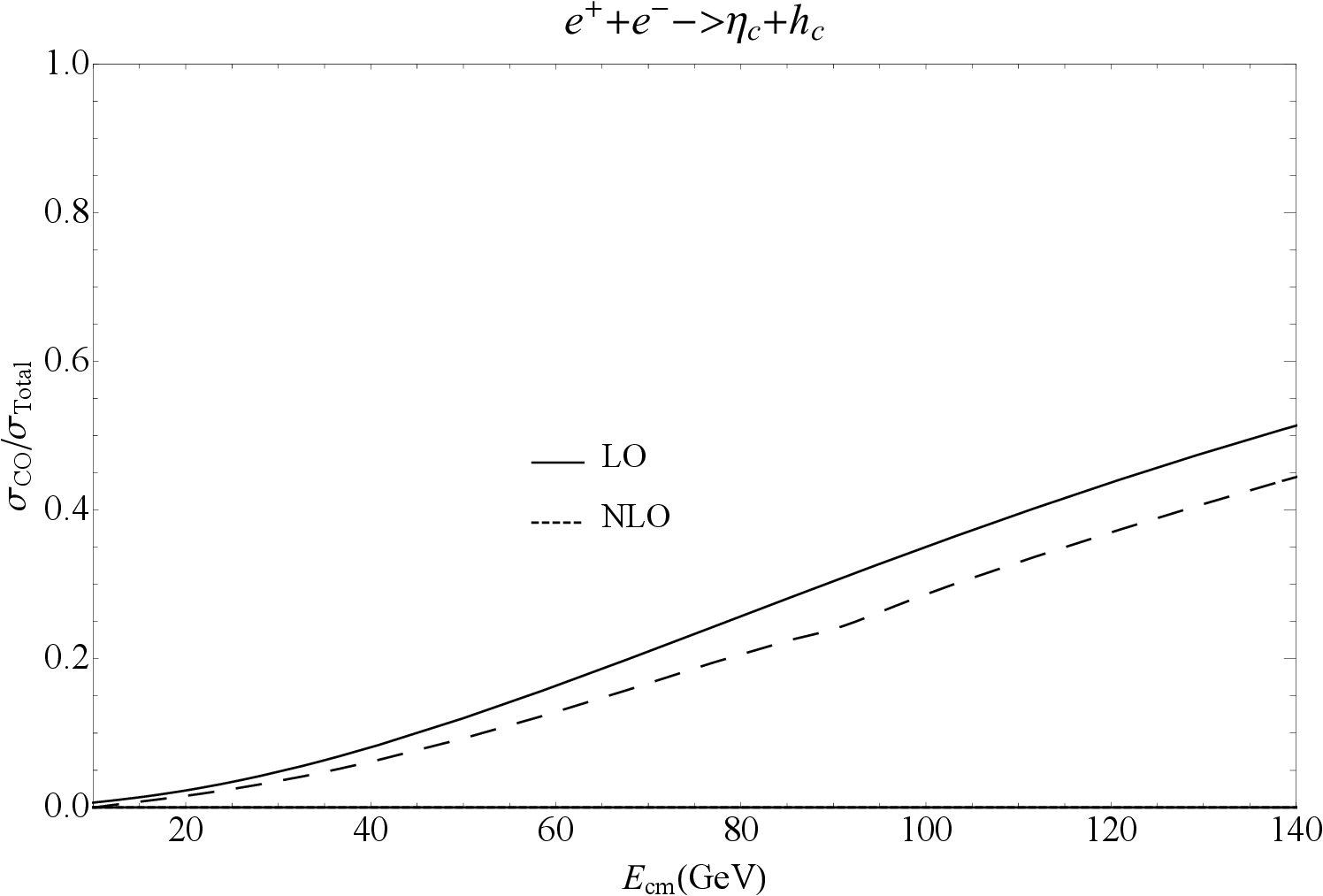}
			\includegraphics[width=0.333\textwidth]{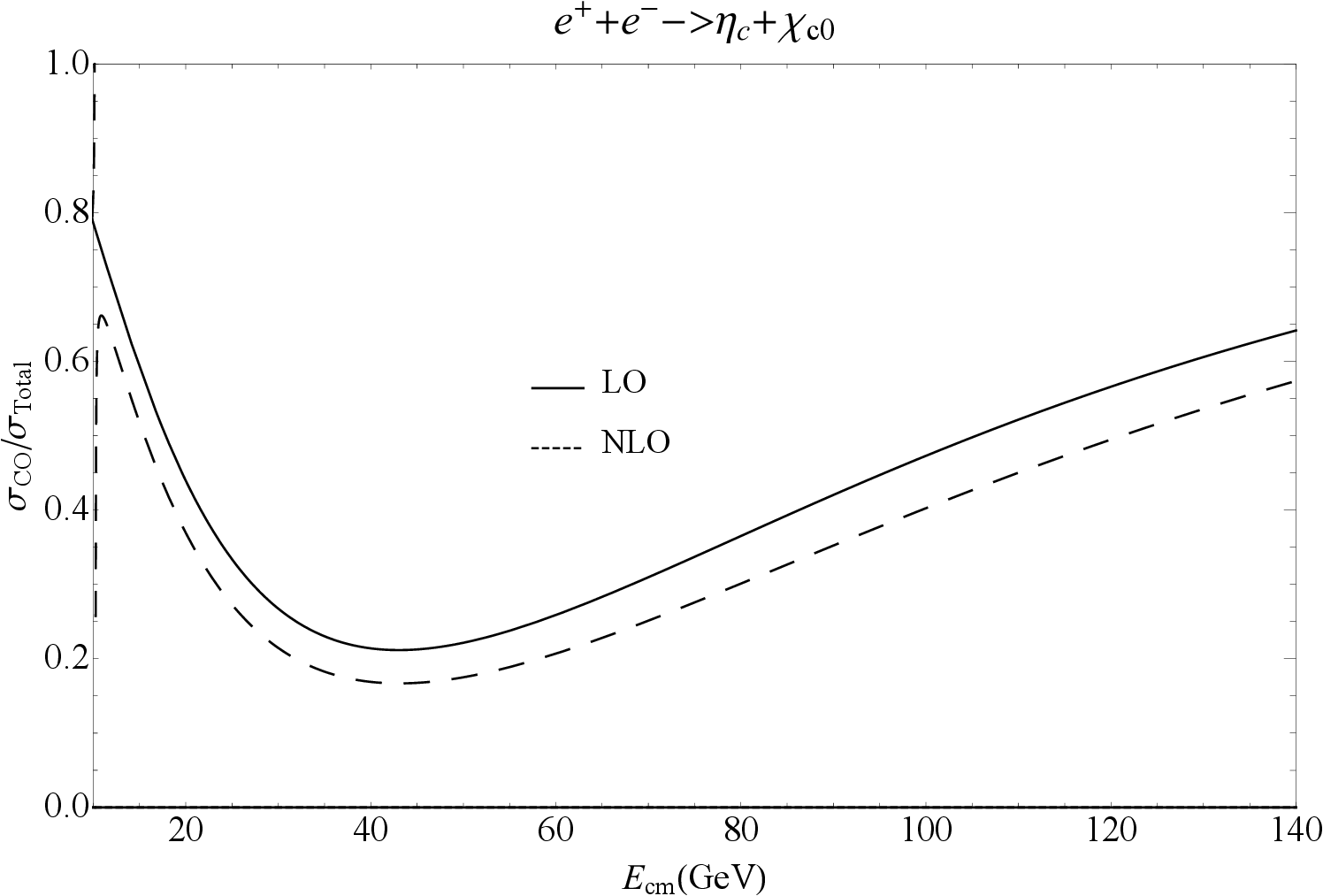}
		\end{tabular}
		\begin{tabular}{c c c}
			\includegraphics[width=0.333\textwidth]{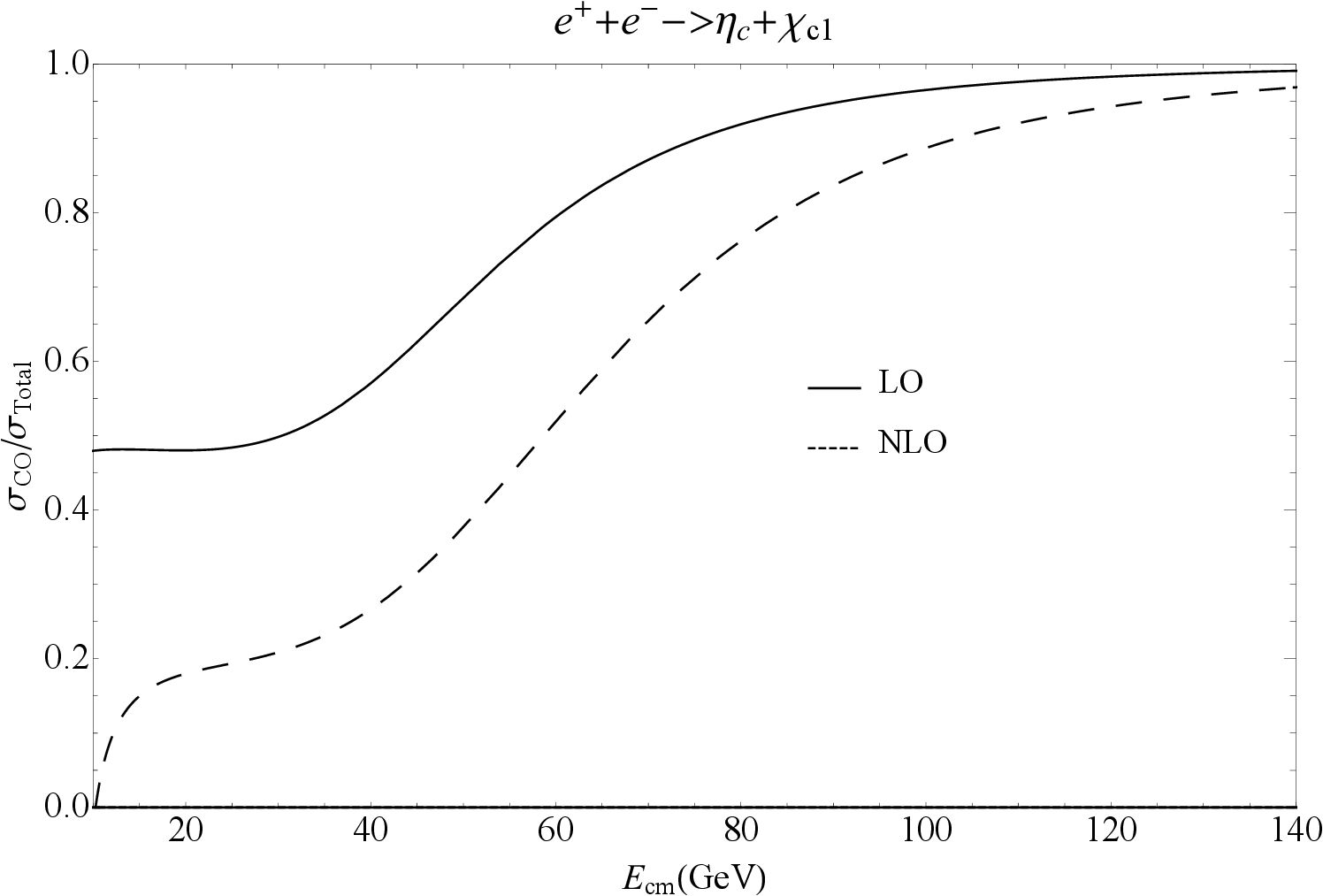}
			\includegraphics[width=0.333\textwidth]{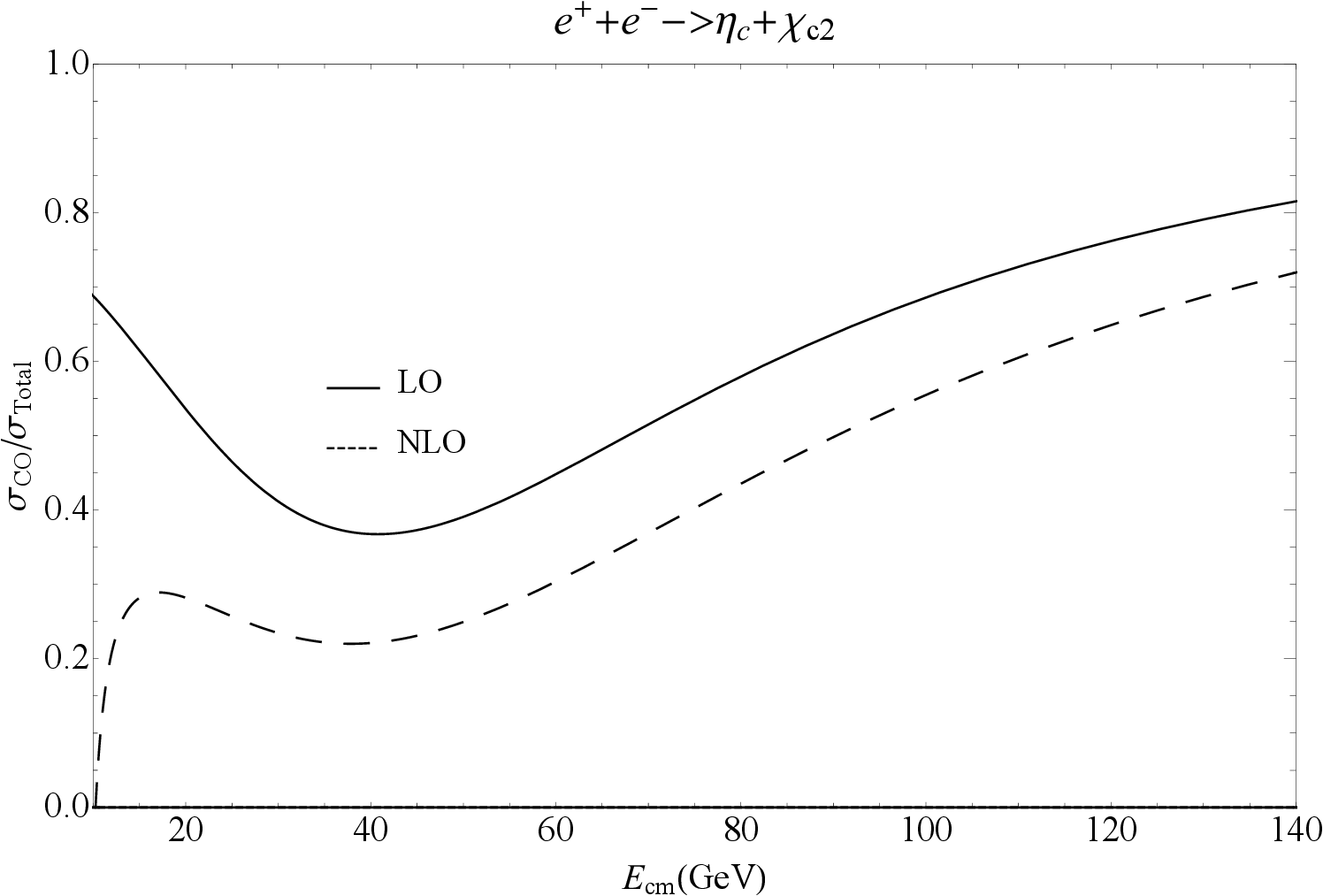}
			\includegraphics[width=0.333\textwidth]{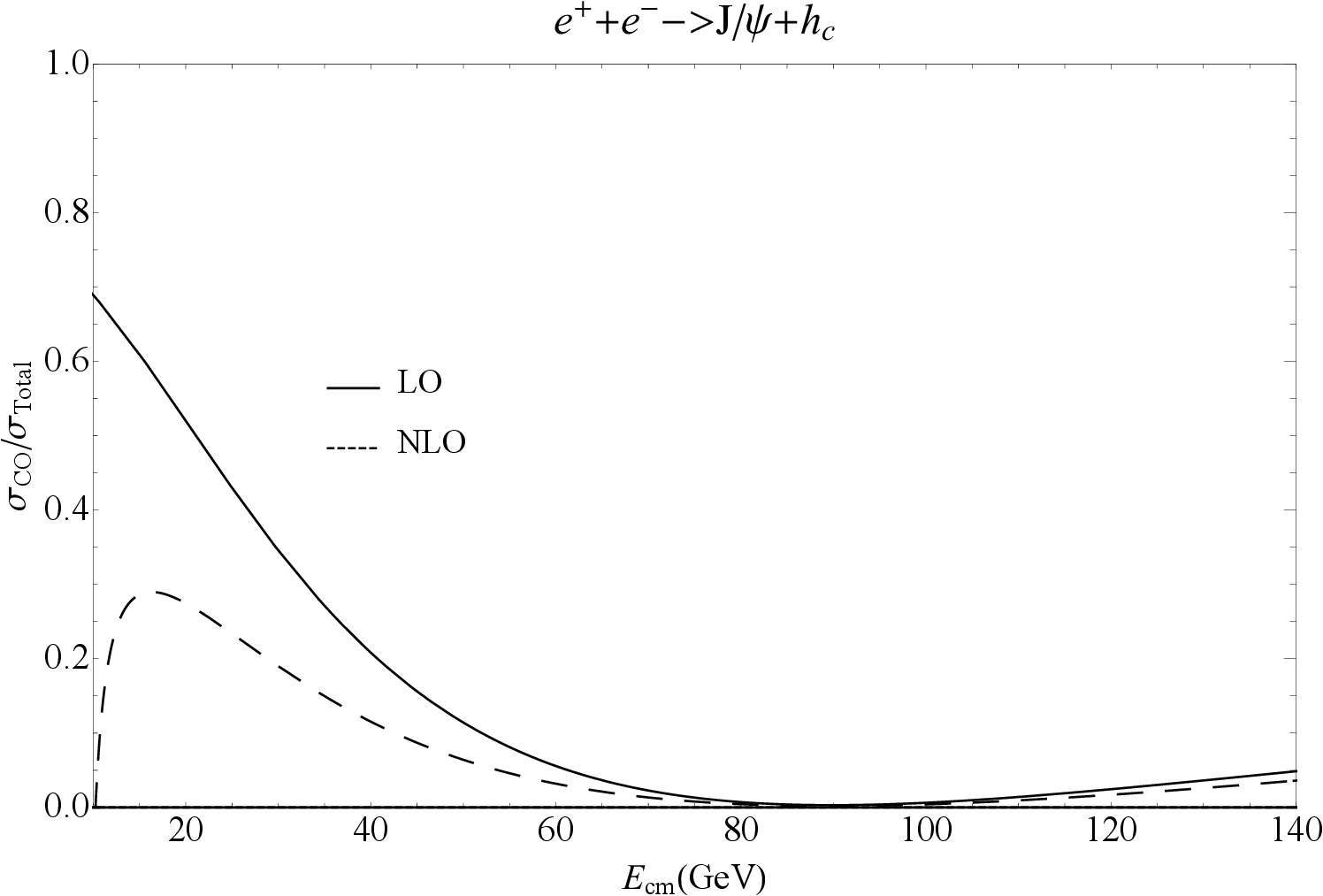}
		\end{tabular}
		\begin{tabular}{c c c }
			
			\includegraphics[width=0.333\textwidth]{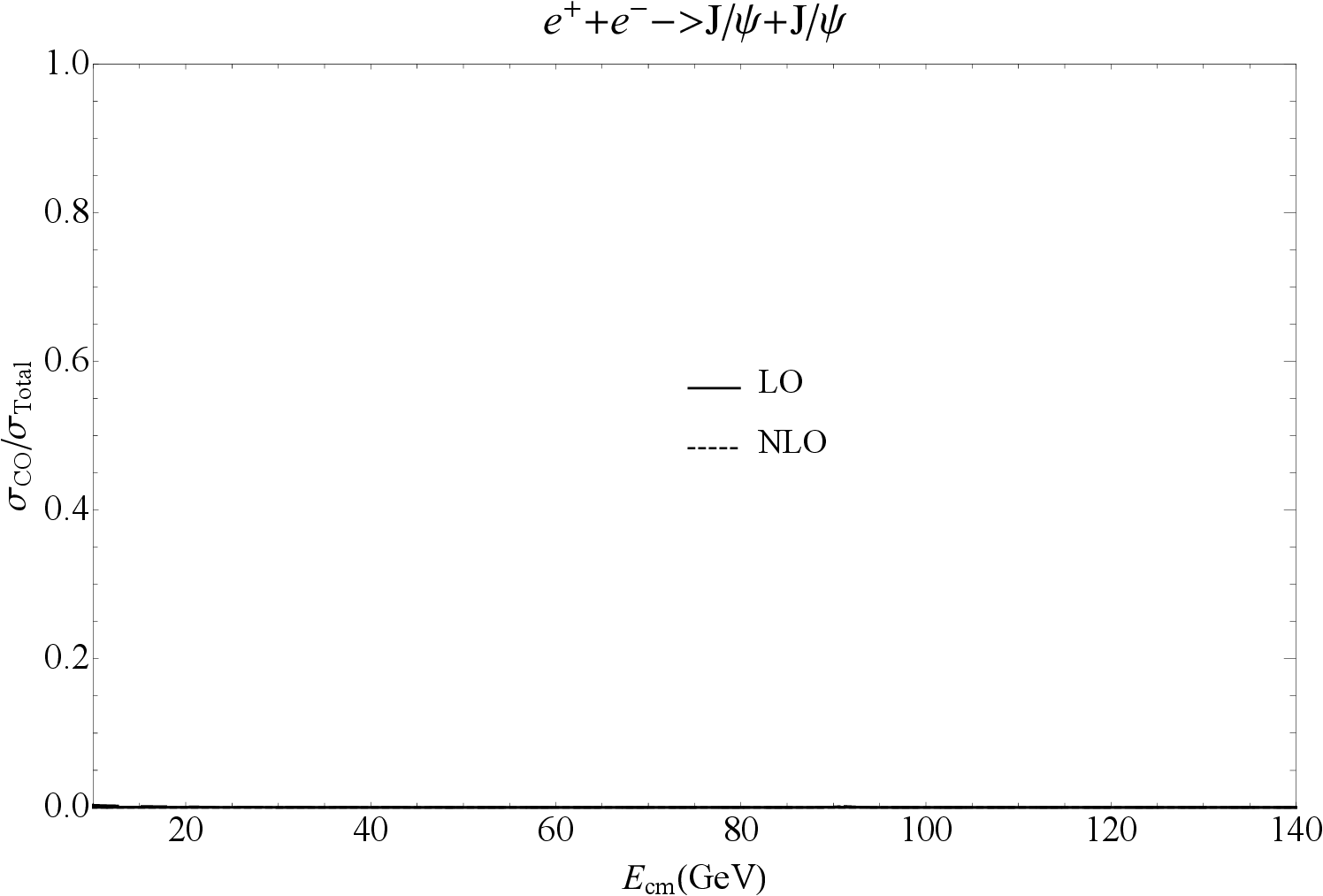}
				\includegraphics[width=0.333\textwidth]{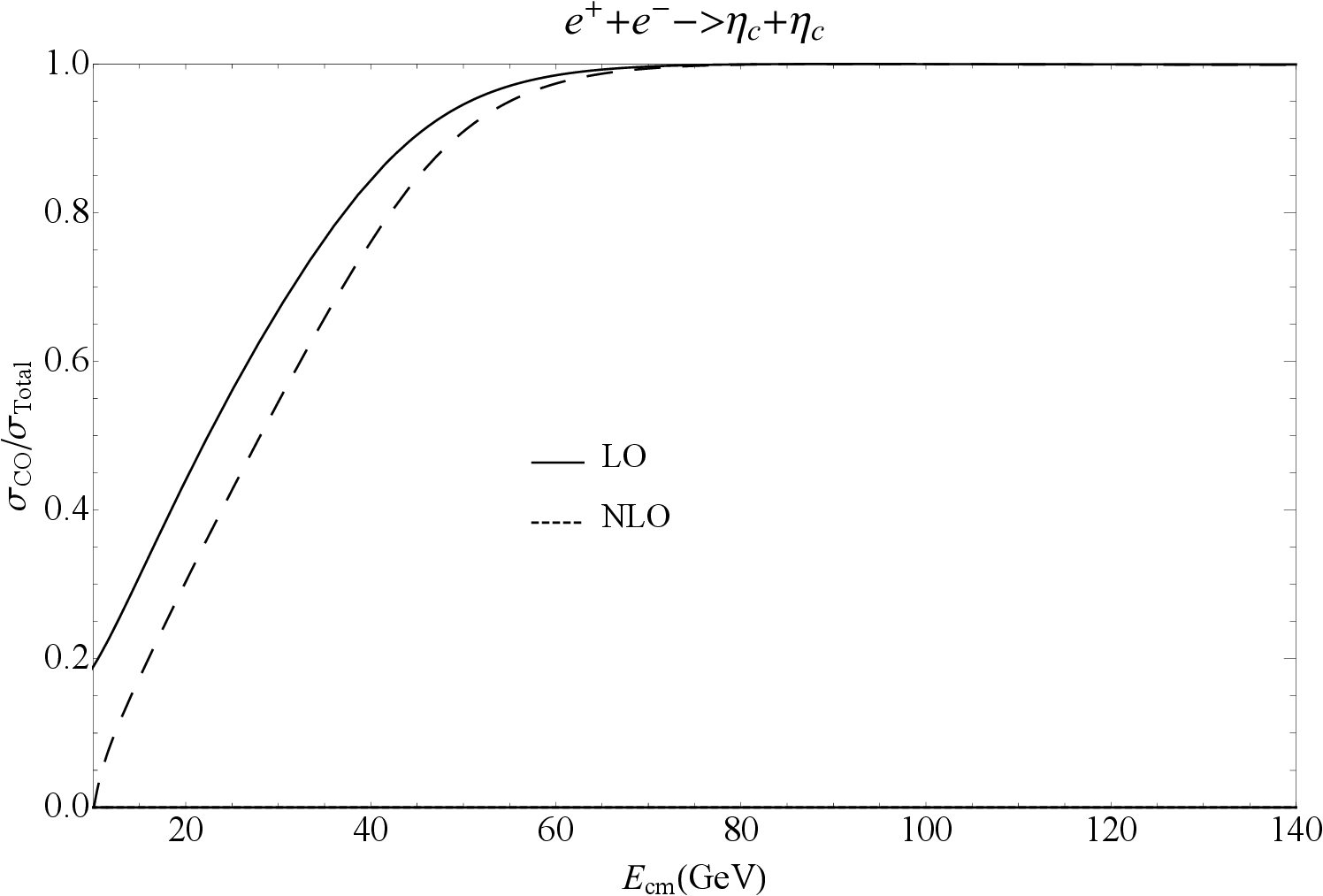}
		
		\end{tabular}
		\caption{ (Color online) The ratios ($\sigma_{CO}/\sigma_{Total}$) versus c.m. energy $(E_{cm}=\sqrt{s})$ ~for double  charmonium production. The solid line represents leading order (LO)  and the dashed line represents next-to-leading order in $v^2$ (NLO) results. }
		\label{z0ccco}
	\end{figure*}
	\FloatBarrier
\end{widetext}

	\begin{widetext}
	\begin{figure*}[htbp]
		\begin{tabular}{c c c}
			\includegraphics[width=0.333\textwidth]{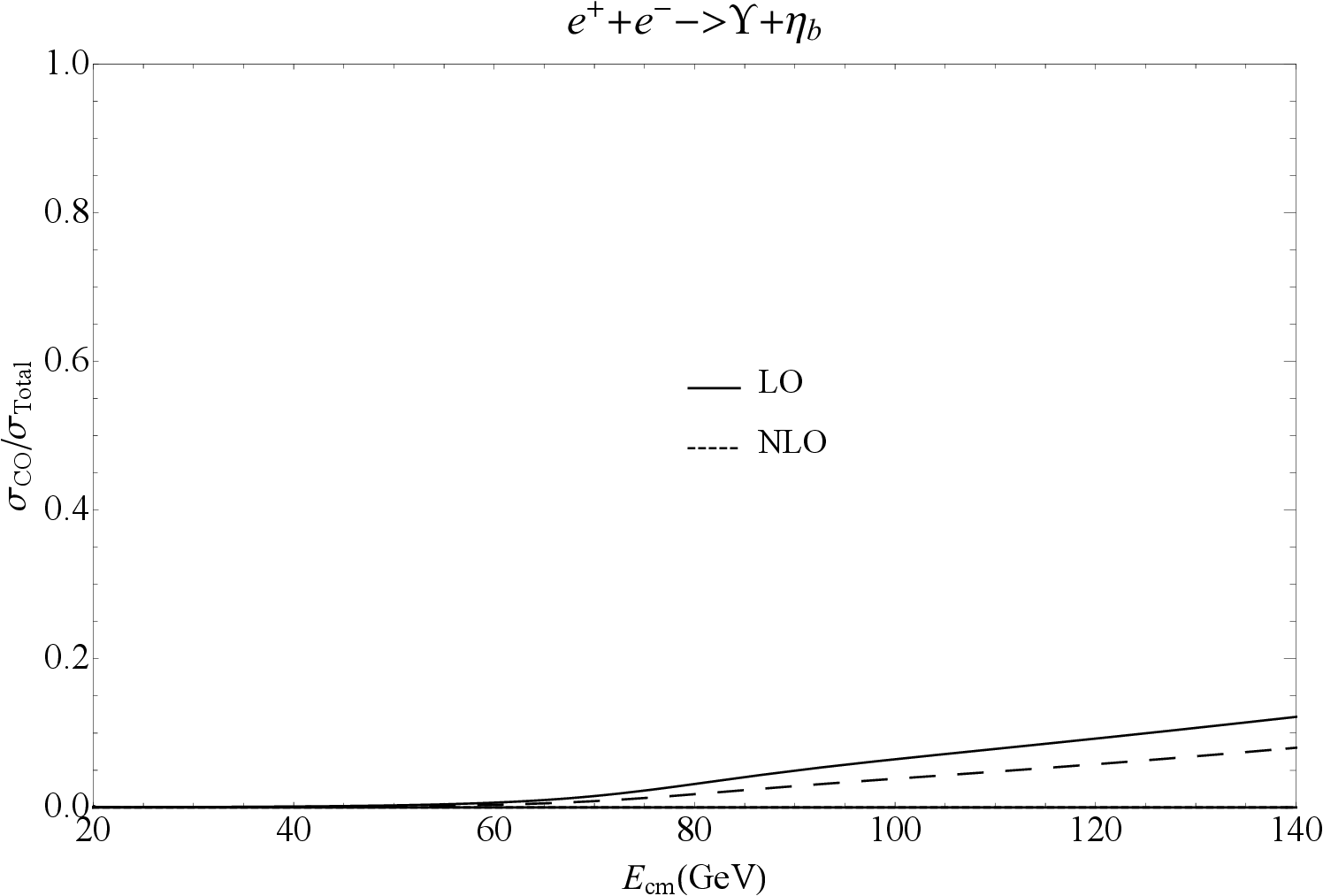}
			\includegraphics[width=0.333\textwidth]{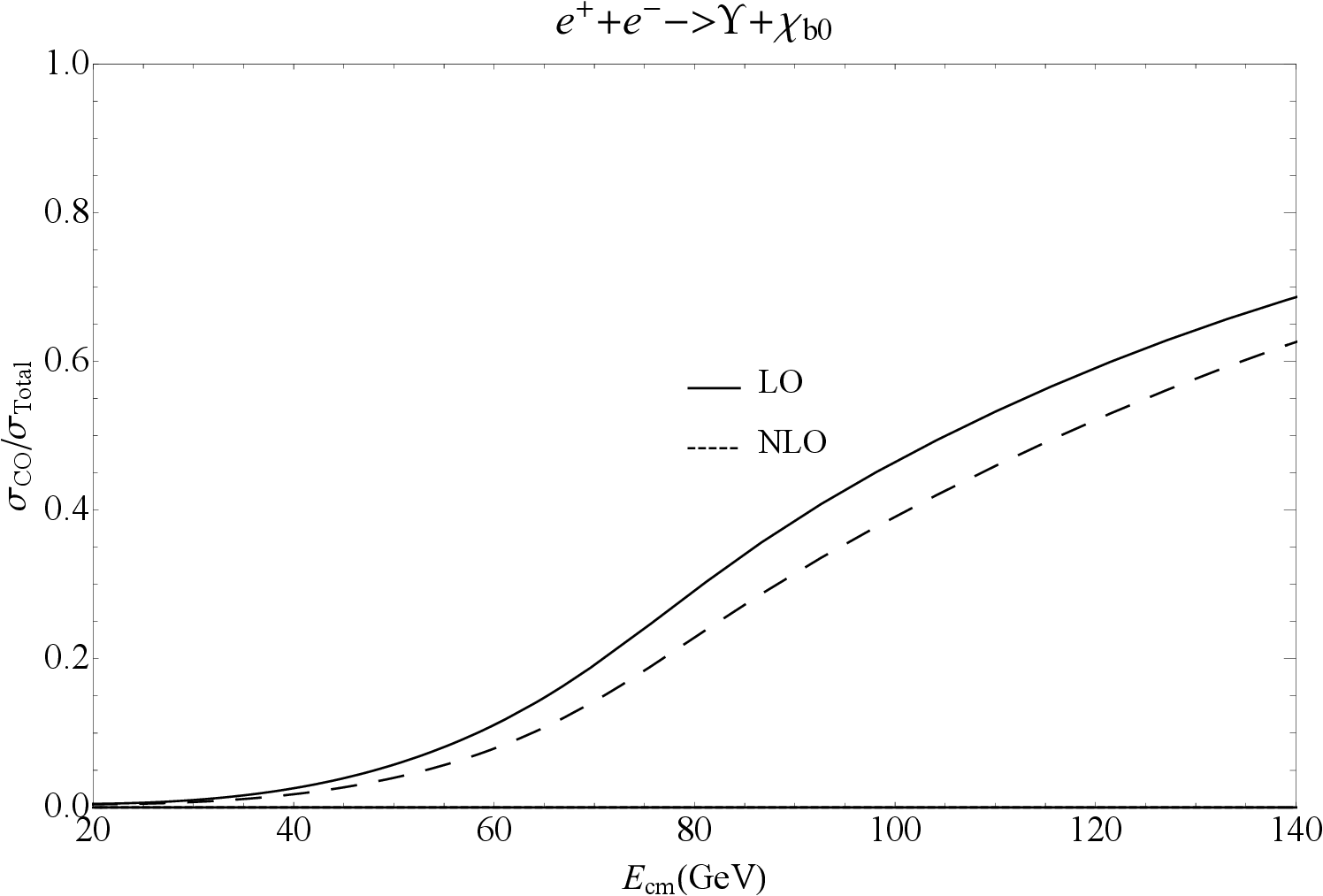}
			\includegraphics[width=0.333\textwidth]{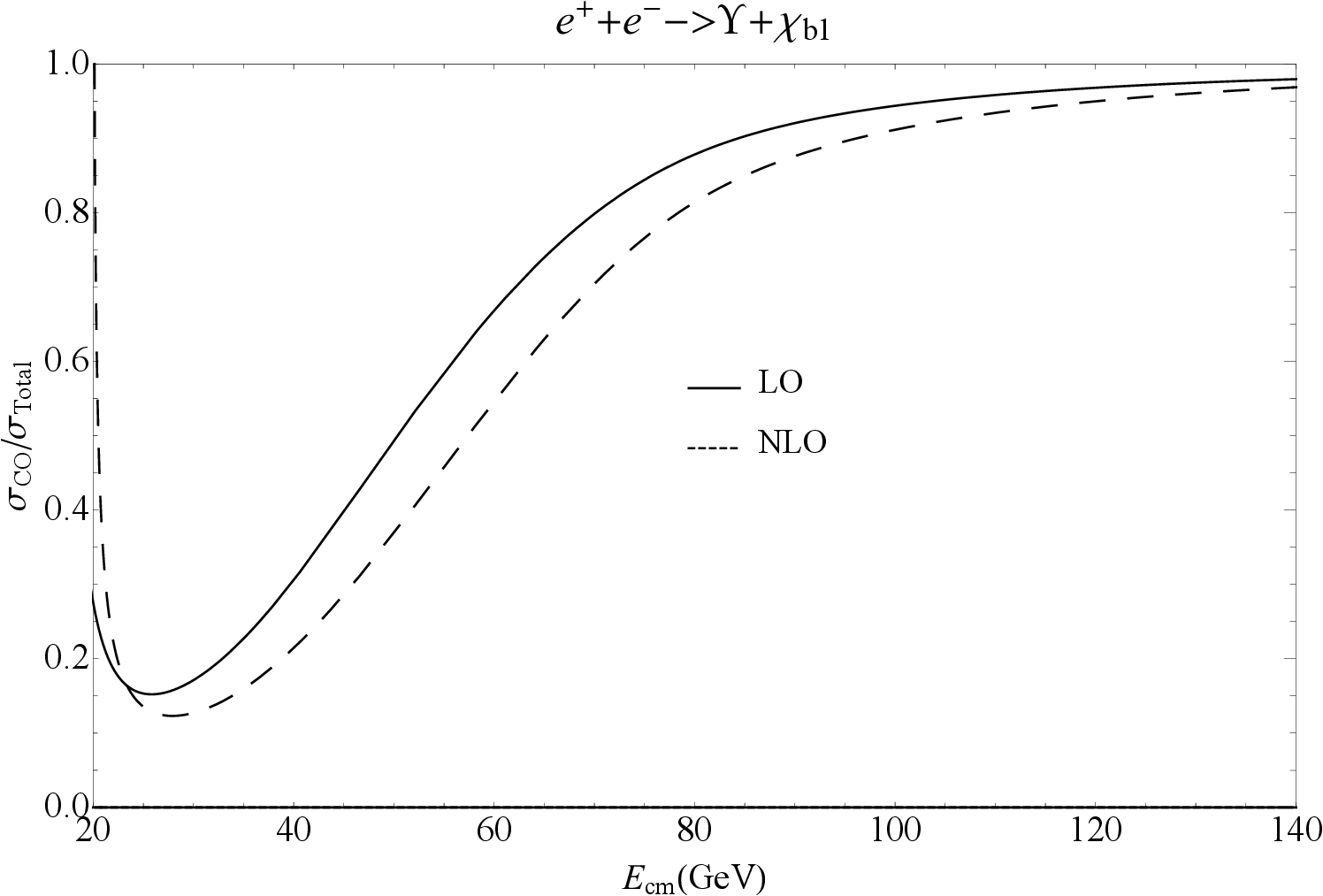}
		\end{tabular}
		\begin{tabular}{c c c}
			\includegraphics[width=0.333\textwidth]{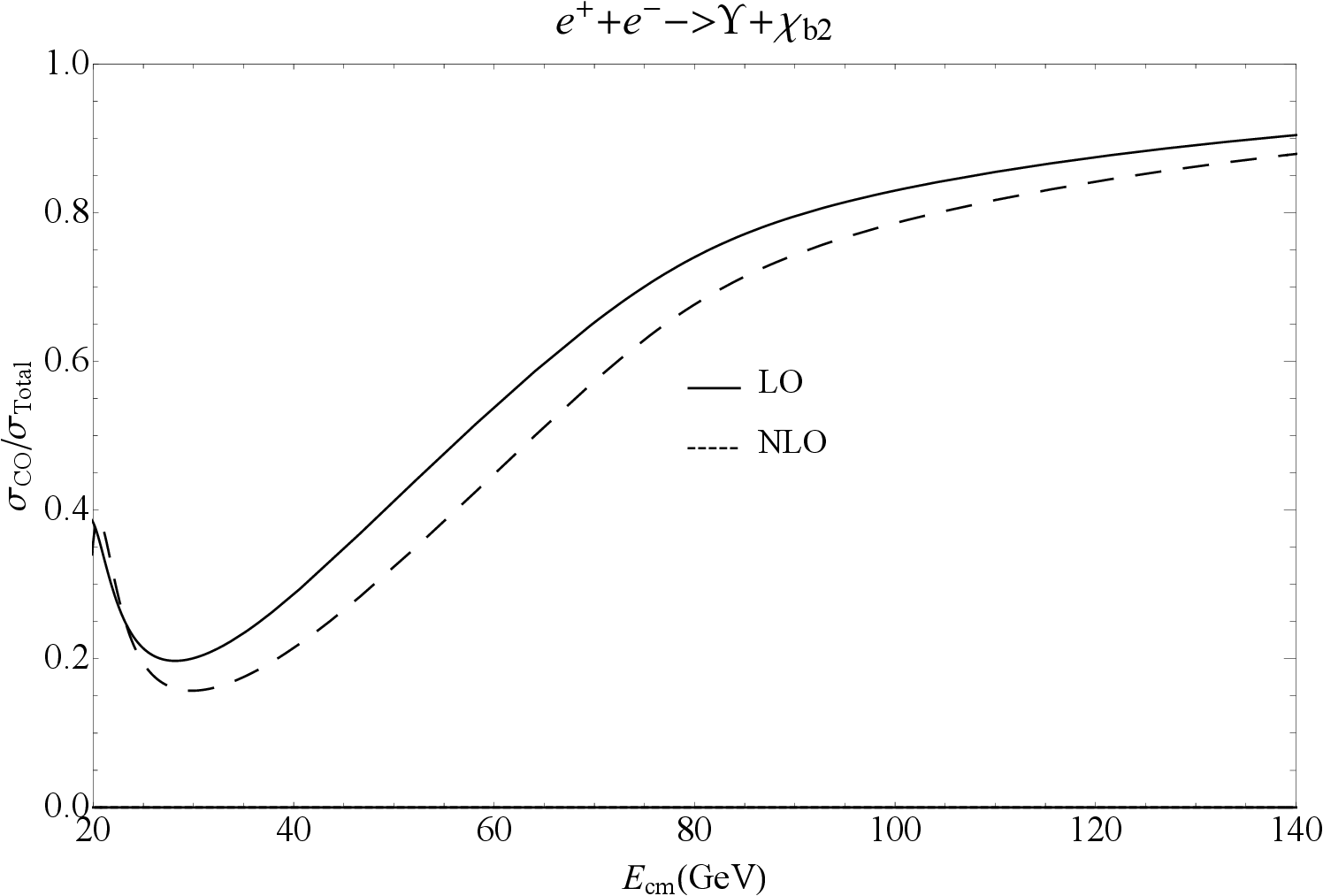}
			\includegraphics[width=0.333\textwidth]{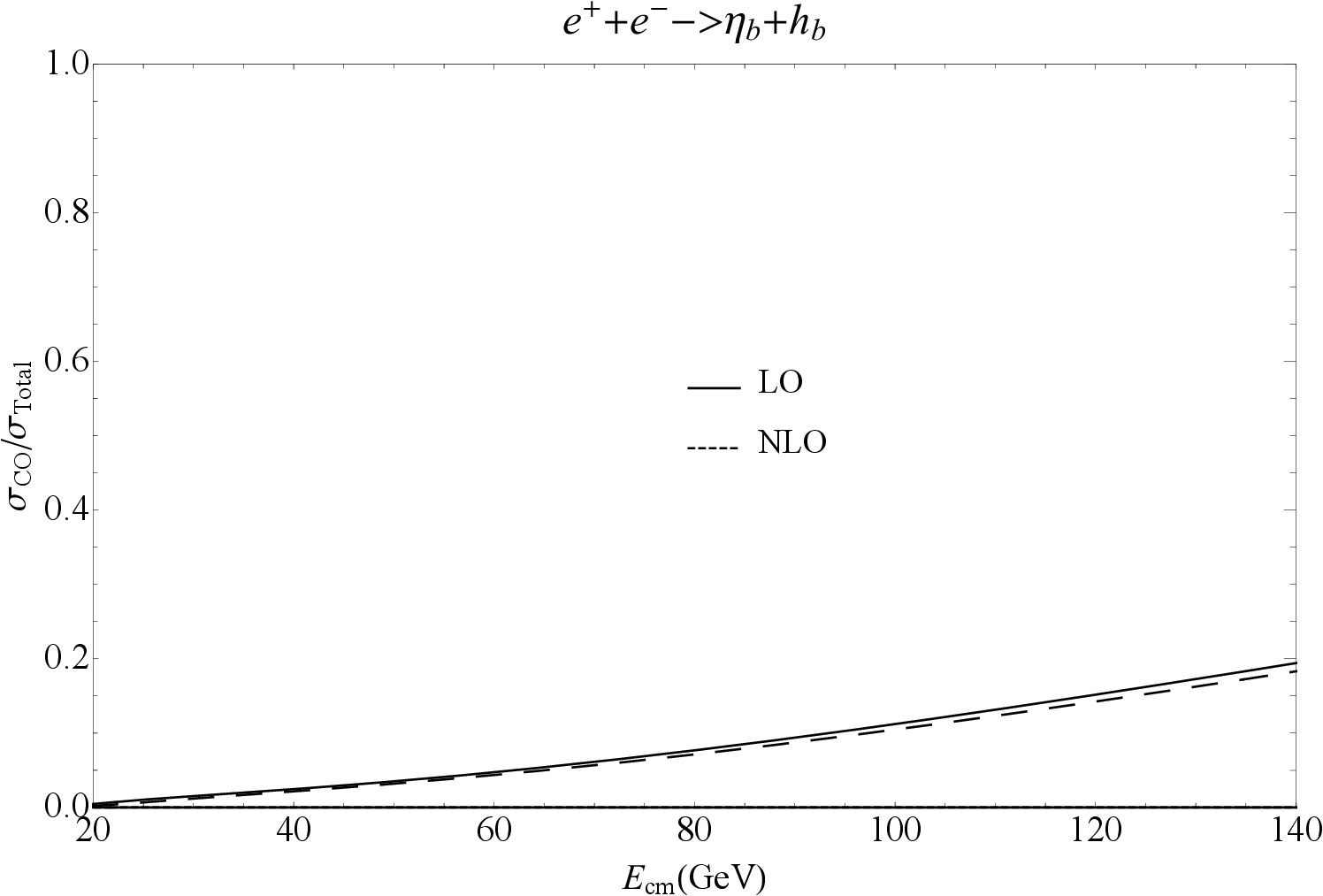}
			\includegraphics[width=0.333\textwidth]{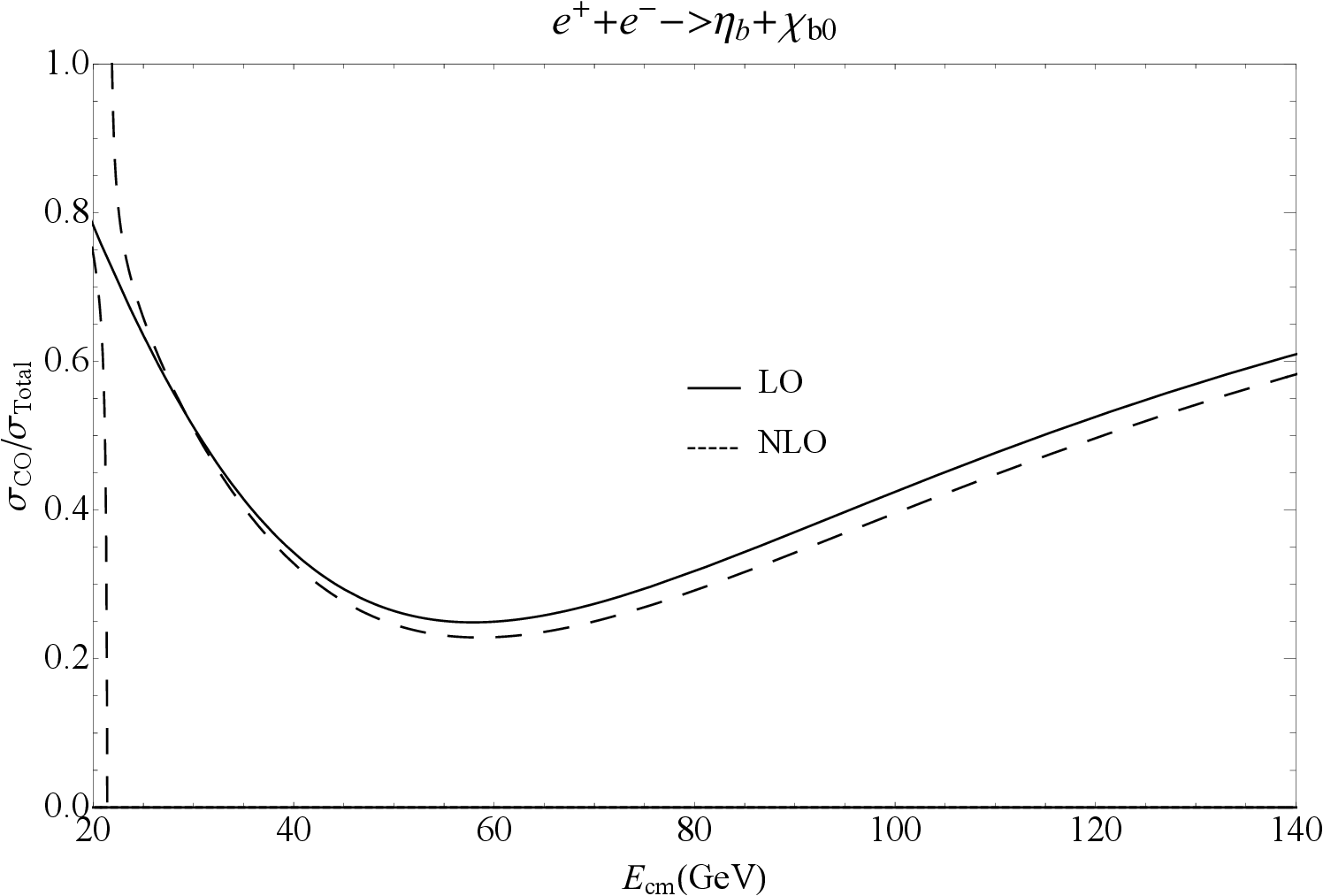}
		\end{tabular}
		\begin{tabular}{c c c}
			\includegraphics[width=0.333\textwidth]{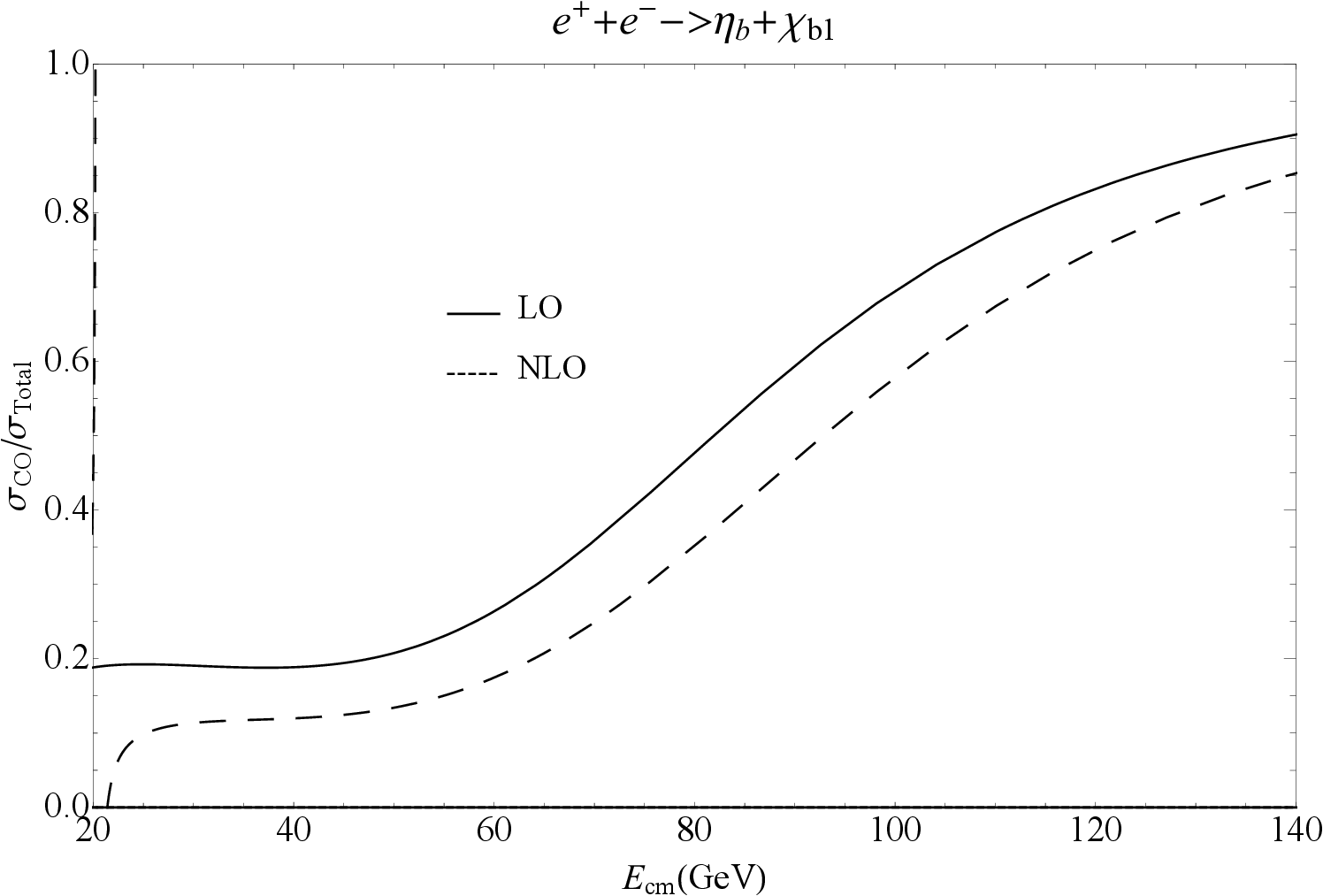}
			\includegraphics[width=0.333\textwidth]{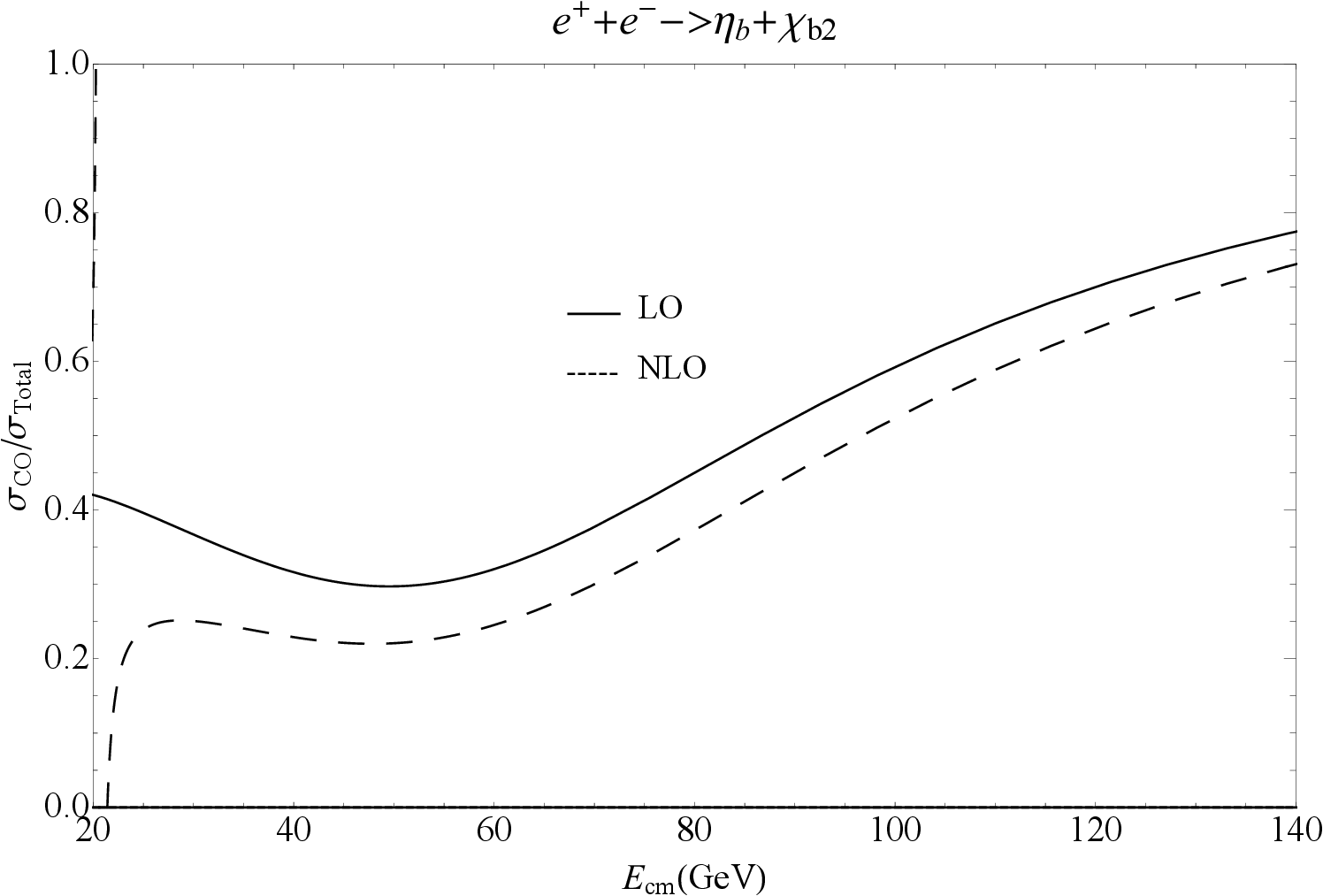}
			\includegraphics[width=0.333\textwidth]{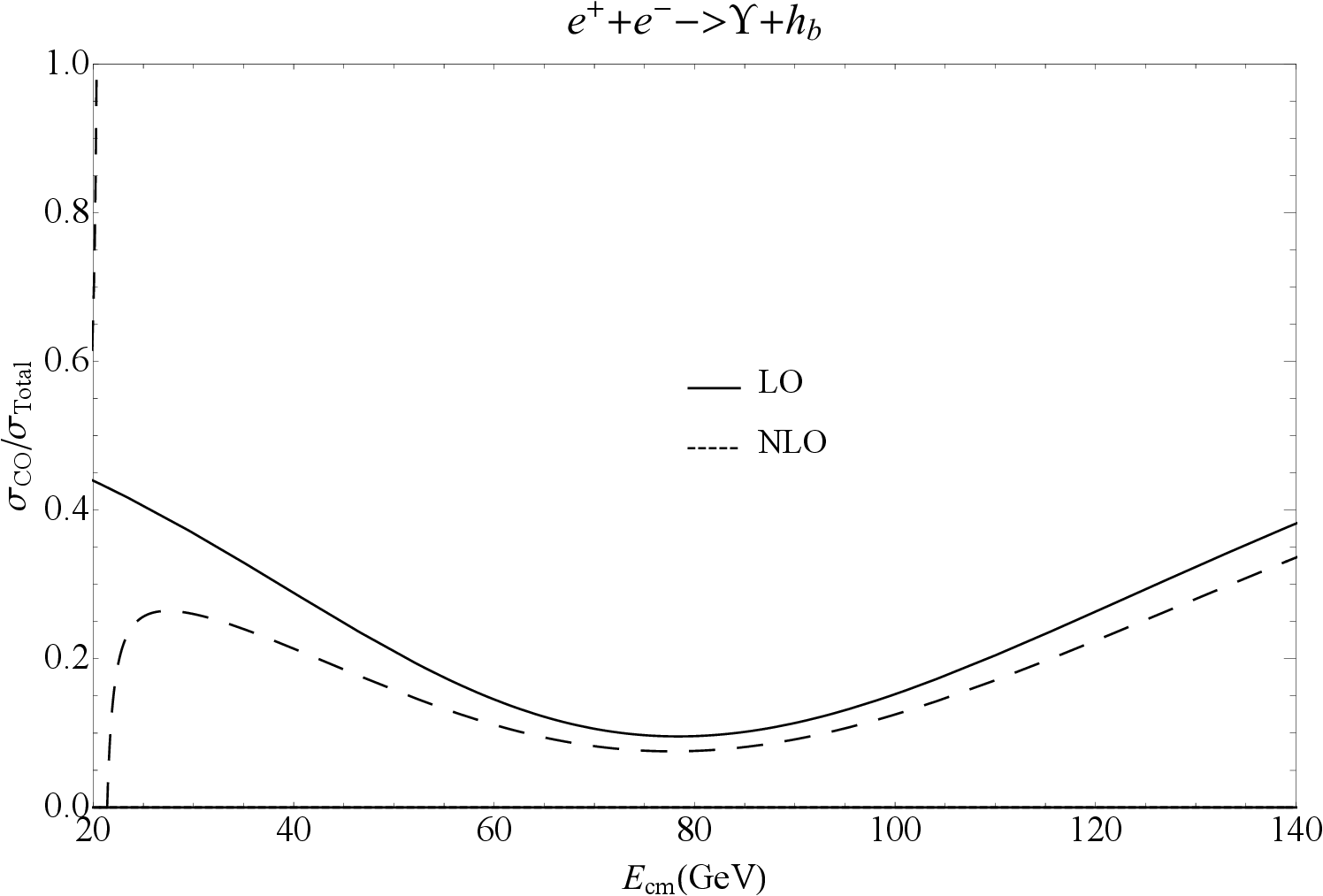}
		\end{tabular}
		\begin{tabular}{c c c }
			
			\includegraphics[width=0.333\textwidth]{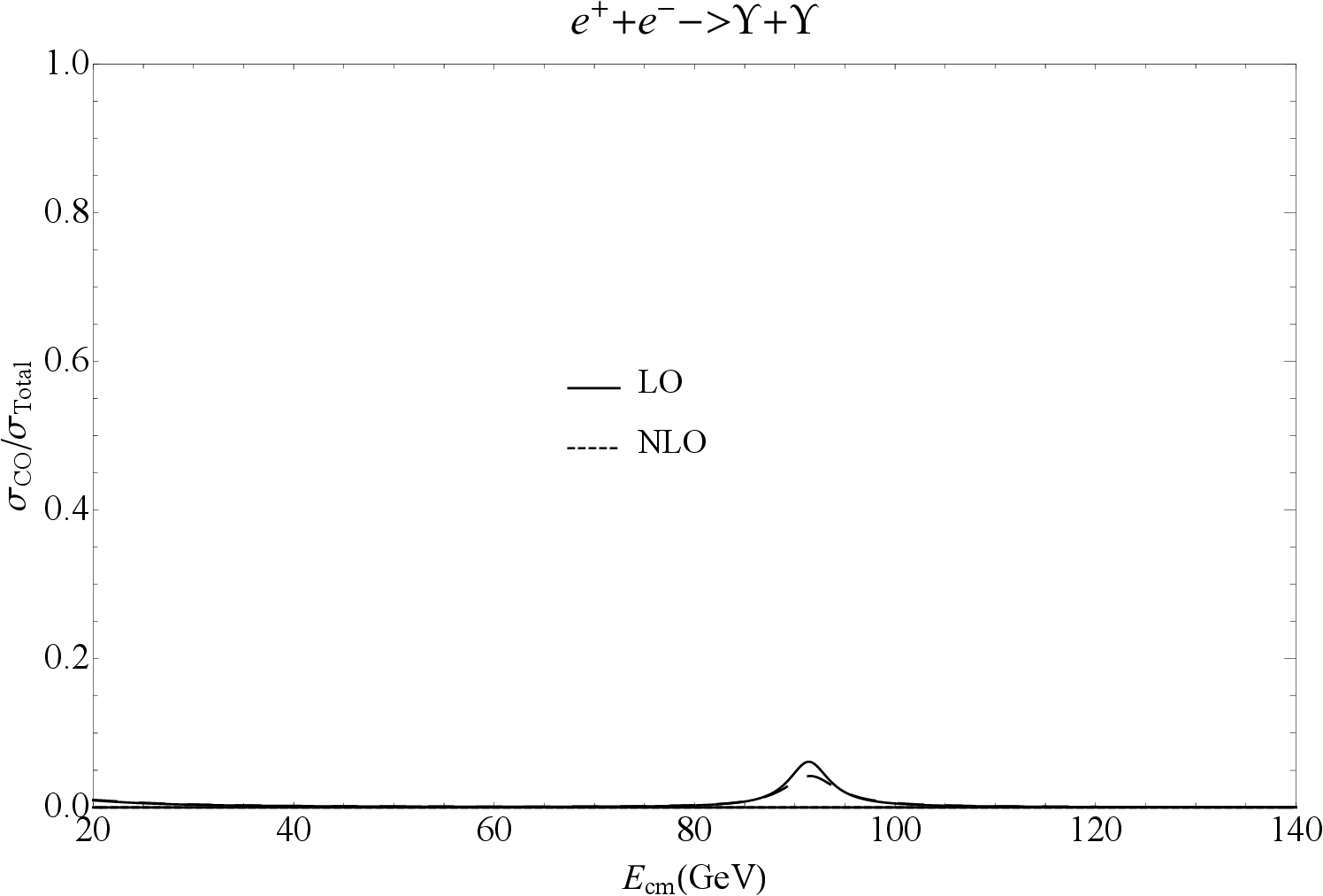}
				\includegraphics[width=0.333\textwidth]{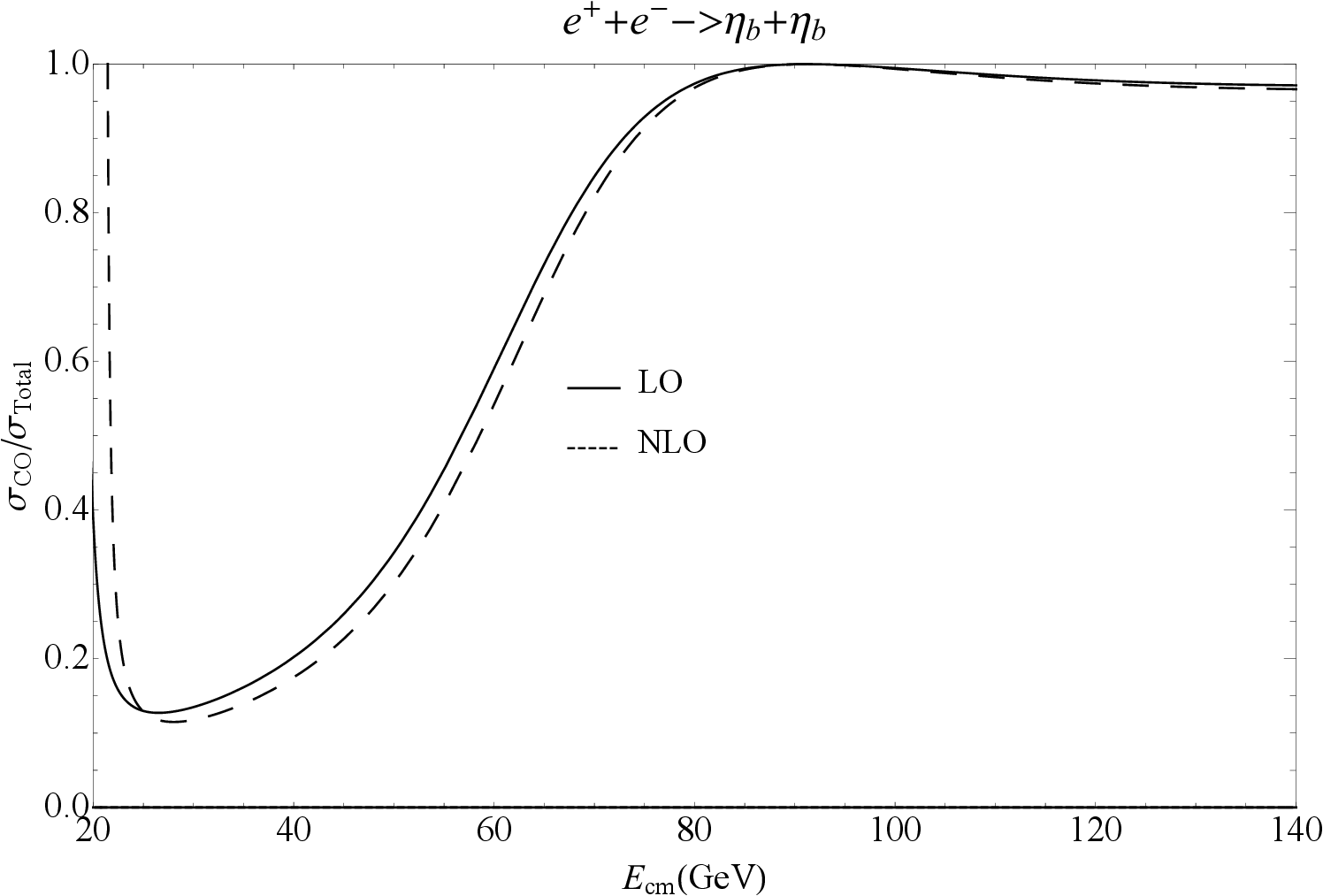}
			
		\end{tabular}
		\caption{ (Color online) The ratios ($\sigma_{CO}/\sigma_{Total}$) versus c.m. energy $(E_{cm}=\sqrt{s})$ ~for double  bottomonium production. TThe solid line represents leading order (LO)  and the dashed line represents next-to-leading order in $v^2$ (NLO) results. }
		\label{z0bbco}
	\end{figure*}
	\FloatBarrier
\end{widetext}

	\subsection{Events}

    CEPC has  two interaction points. And for its Z factory mode, the designed integrated luminosity within two years is $16 ab^{-1}$\cite{CEPC}.
     The FCC-ee, with a designed luminosity of $150 ab^{-1}$\cite{FCC}, would operate as a super Z factory in its first four years. The designed luminosity of FCC-ee is nine times that of CEPC. The estimated events for CEPC and FCC-ee are shown in Table \ref{events}. 

We use the $K$ factors of the NLO cross sections for QCD and EW channels in $\alpha_s$ from Ref. \cite{Belov} ($K_{QCD}^{(\alpha_s)}\simeq3.75, 3.9, 2.55, 2.5$, $ K_{EW}^{(\alpha_s)}\simeq1.08, 1.01, 0.775, 0.908$ for $J/\psi+\eta_c$, $J/\psi+J/\psi$, $\Upsilon+\eta_b$, $\Upsilon+\Upsilon$). Respectively, the NLO($\alpha_s$) cross sections would be obtained as $\sigma^{(\alpha_s)}_{CS(QCD)}= (0.7; 1.3; 58.9; 28.9)\times10^{-4} fb$, $\sigma^{(\alpha_s)}_{CS(EW)}=(8.7; 54.8; 1.0; 2.4)\times10^{-4} fb$\footnote{The cross sections obtained by summing the QCD and EW contributions for $\Upsilon+\eta_b$, $\Upsilon+\Upsilon$ processes are different from those in Ref.\cite{Belov}.  This is due to the adoption of different parameters, especially $\alpha_s$ and LDMEs.}, and with the interference contributions between QCD and EW channels as
	$(2.5; 8.4; 11.2; 11.1)\times10^{-4} fb$. The CO cross sections are
	$(20.9; 1.4; 1.9; 2.2)\times10^{-4} fb$. The t-channel double-photon-fragmentation QED contributions are $(0; 2673.0; 0; 9.1)\times10^{-4} fb$. Therefore the total cross section at $\mathcal{O}(v^0)$ would be $(32.7; 2738.9; 73.1; 53.8)\times10^{-4} fb$, and the final events would be   $(52, 4382, 117, 86)$ and $(491, 41083, 1096, 806)$ for the two colliders.

	\begin{table}
	\caption{ The events of double heavy quarkonia production at $\sqrt{s}=$91.1876 GeV for the CEPC and FCC-ee. In each cell, the values outside/inside the brackets are for leading order and next-to-leading order results in the $v^2$ expansions, respectively. }
	\begin{tabular}{|c|c|c| |c|c|c|}
		\hline
		\multicolumn{6}{|c|}{ CEPC($16ab^{-1}$)}\\
		\hline
		\hline
		~& CS & CS+CO & 	& CS & CS+CO  \\
		\hline
		$J/\psi+J/\psi$&4378(569)&4380(570)& $\Upsilon+\Upsilon$&55(59)&59(61) \\
		\hline
		$J/\psi+\eta_c$& 17(10)&51(22)& 	$\Upsilon+\eta_b$& 57(69)&60(71) \\
		\hline
		$J/\psi+h_{c}$& 53(32)&53(32)& 	$\Upsilon+h_{b}$& 36(33)&40(36) \\
		\hline
		$J/\psi+\chi_{c0}$& 3(2)&7(3)& 	$\Upsilon+\chi_{b0}$& 13(13)&21(19) \\
		\hline
		$J/\psi+\chi_{c1}$& 4(1)&13(5)& $\Upsilon+\chi_{b1}$& 2(2)&27(21)~ \\
		\hline
		$J/\psi+\chi_{c2}$& 5(4)&21(9) &$\Upsilon+\chi_{b2}$& 11(10)&53(41) \\
		\hline
		$\eta_c+\eta_c$& 0(0)&63(21)& 	$\eta_b+\eta_b$&0(0) &1(0) \\
		\hline
		$\eta_c+h_c$& 5(2)&7(3)	& $\eta_b+h_b$& 11(9)&13(10)\\
		\hline
		$\eta_c+\chi_{c0}$& 11(6)&19(9)& $\eta_b+\chi_{b0}$& 8(6)&13(9) \\
		\hline
		$\eta_c+\chi_{c1}$&1(2)&26(11)& $\eta_b+\chi_{b1}$&9(11)&24(21) \\
		\hline
		$\eta_c+\chi_{c2}$& 23(15)&63(30)& 	$\eta_b+\chi_{b2}$& 21(19)&46(36) \\
		\hline
		\hline
		\multicolumn{6}{|c|}{ FCC-ee($150ab^{-1}$)}\\
		\hline
		\hline
		$J/\psi+J/\psi$&41041(5335)&41061(5343)& $\Upsilon+\Upsilon$&516(552)&550(576) \\
		\hline
		$J/\psi+\eta_c$& 163(93)&481(206)& 	$\Upsilon+\eta_b$& 535(645)&564(665) \\
			\hline
		$J/\psi+h_{c}$& 498(304)&499(304)& $\Upsilon+h_{b}$& 333(305)&378(337) \\
		\hline
		$J/\psi+\chi_{c0}$& 32(21)&61(32)& 	$\Upsilon+\chi_{b0}$& 120(119)&199(176) \\
		\hline
		$J/\psi+\chi_{c1}$& 34(13)&123(45)& 	$\Upsilon+\chi_{b1}$& 20(23)&257(194) \\
		\hline
		$J/\psi+\chi_{c2}$& 48(35)&196(89)& $\Upsilon+\chi_{b2}$& 99(95)&494(380) \\
		\hline
		$\eta_c+\eta_c$& 0(0)&587(199)& $\eta_b+\eta_b$&0(0) &5(3) \\
		\hline
		$\eta_c+h_c$& 46(23)&67(31)	& $\eta_b+h_b$& 106(82)&118(91) \\
		\hline
		$\eta_c+\chi_{c0}$& 102(52)&179(80)& $\eta_b+\chi_{b0}$& 75(58)&121(89) \\
		\hline
		$\eta_c+\chi_{c1}$&12(16)&241(103)& 	$\eta_b+\chi_{b1}$&88(100)&225(193) \\
		\hline
		$\eta_c+\chi_{c2}$& 211(141)&593(285)& 	$\eta_b+\chi_{b2}$& 200(182)&428(336) \\
		\hline
	\end{tabular}
	\label{events}
\end{table}
\FloatBarrier

	\subsection{Uncertainties}
	The  sources of uncertainty include the heavy quark mass, LDMEs, renormalization scale, and the deviation of collision energy from $m_Z$. As in Ref. \cite{sunzhan} and for simplicity, we  won't discuss the LDMEs uncertainty. However, the  CO LDMEs  used in the present paper are actually moderate compared to those in some other Refs. \cite{jpsiCOLDMEs1,jpsiCOLDMEs2,universalityt2,etacCOLDMEs2,hcCOLDMEs1,hcCOLDMEs2,psianomalyt,hcCOLDMEs4,hcCOLDMEs5,hcCOLDMEs6,chicjCOLDMEs1,universalityt3,chicjCOLDMEs3}. Ref. \cite{Liao_2023} used five CS LDMEs to see the uncertainties caused by them. When we used the largest set (B.T potential model)\cite{LDLiao,LDBT}, we found that the COM  is still significant   around the $Z^0$ pole for many processes.
	
	The uncertainties caused by the renormalization scale ($\mu$) are shown in Figs. \ref{z0ccmu} and \ref{z0bbmu}. 
	
	To show the sensitivity of the total cross sections to the collision energy around the $Z^0$ peak, we calculate the total cross sections using $E_{cm}=(1\pm3\%)m_Z$. The results are given in Tables \ref{ccmz} and \ref{bbmz}, where we use the same ratios as in Ref. \cite{sunzhan}.
	\bea
	R_-=\frac{\sigma(E_{cm}=97\%m_Z)}{\sigma(E_{cm}=m_Z)}\cr
	R_+=\frac{\sigma(E_{cm}=103\%m_Z)}{\sigma(E_{cm}=m_Z)}
	\eea
The total cross section decreases to $15\%\sim20\%$ of its peak values, consistent with the results in Tables VIII and IX of Ref. \cite{sunzhan}, which studied the semi-exclusive processes.
		
	Next, to see the uncertainty caused by the heavy quark mass, we  take $m_c=(1.5\pm0.15) ~GeV, m_b=(4.7\pm0.15)~ GeV$. The results are shown in Tables \ref{ccmc} and \ref{bbmb}. For the charmonium production, the  uncertainties associated with the  $m_c=1.5\pm0.15~GeV$ variation are as follows.

(i) For the CS channel, the uncertainties are about $0.2\%$ for $\eta_c+\chi_{c1}$ production, about $16-27\%$ for $J/\psi+h_c/\chi_{c0}/\chi_{c2}, \eta_c+h_c/\chi_{c0}/\chi_{c2}/\eta_c$ production,  about $24-37\%$ for $J/\psi+\eta_c$ production,  about $28-45\%$ for $J/\psi+\chi_{c1}$ production, about $44-91\%$ for double $J/\psi$ production; 

	(ii) For total CO channels,  about $30-50\%$ for  $J/\psi+h_c/J/\psi$ production, about $32-53\%$ for all other production processes;

	 (iii) For total cross sections, it's about $18-26\%$ for $J/\psi+h_{c}$ production, about $21-32\%$ for $J/\psi+\chi_{c0}, \eta_c+h_{c}$ production,  about $23-38\%$ for $\eta_c+\chi_{c0}/\chi_{c2}$ production, about $27-46\%$ for $J/\psi+\eta_c/\chi_{c2}, \eta_c+\chi_{c1}$ production, about $31-53\%$ for $J/\psi+\chi_{c1}, \eta_c+\eta_{c}$ production, about $44-91\%$ for double $J/\psi$ production.

	 As for the bottomonium production, the  uncertainties associated with the variation $m_b=4.7\pm0.15~GeV$ are as follows.

	  (i) For the CS channel, the uncertainties are $1\%$ for $\Upsilon/\eta_b+\chi_{b1}$ production, about $2-3\%$ for $\Upsilon+\eta_b/\chi_{b0}$ production, about  $5-6\%$ for $\Upsilon+\chi_{b2}/h_b, \eta_b+\chi_{b2}$ production, about $6-8\%$ for  $\eta_b+h_{b}/\chi_{b0}/\eta_b, \Upsilon+\Upsilon$ production;

	  (ii) For  total CO channels, it's about $12-15\%$ for  all production processes;

	  (iii) For total cross sections, it's about $3\%$ for $\Upsilon+\eta_b$ production, about $6\%$ for $\Upsilon+h_b/\chi_{b0}$ production, about $6-8\%$ for $\eta_b+h_{b}/\chi_{b1}, \Upsilon+\Upsilon$ production, about $9-10\%$ for $\eta_b+\chi_{b0}/\chi_{b2}$ production, about $11-13\%$  for $\Upsilon+\chi_{b1}/\chi_{b2}$ production, about $13-15\%$ for double $\eta_b$ production. 

\subsection{$J/\psi+\eta_c$ production}

We present cross-sections of $J/\psi+\eta_c$ production at NLO$(\alpha_s,v^2)$ with uncertainties induced by the quark mass in the left panel of Fig. \ref{jpsietacuncerta}. From the figure, one can find the effects of CO channels are significant. CS cross sections are summed by those of QCD, EW channels. The $K$ factors of NLO$(\alpha_s)$ are estimated from the calculations of Ref.\cite{Belov}, i.e., $K_{QCD}=1.21 - 0.37\ln(4m_c^2/s)$, and $K_{QED}=0.67 - 0.06\ln(4m_c^2/s)$. The central lines correspond to $m_c = 1.5\,\text{GeV}$. The upper/lower bounds of the bands  correspond to $m_c = 1.35,1.65\,\text{GeV}$, respectively.
Furthermore, in an attempt to eliminate the uncertainties associated with CS LDMEs, we plot the ratios of the production rates of $J/\psi+\eta_c$ to those of double $J/\psi$ in the right panel.

\begin{widetext}
	\begin{figure*}[htbp]
		\begin{tabular}{c c c }
			\includegraphics[width=0.43333\textwidth]{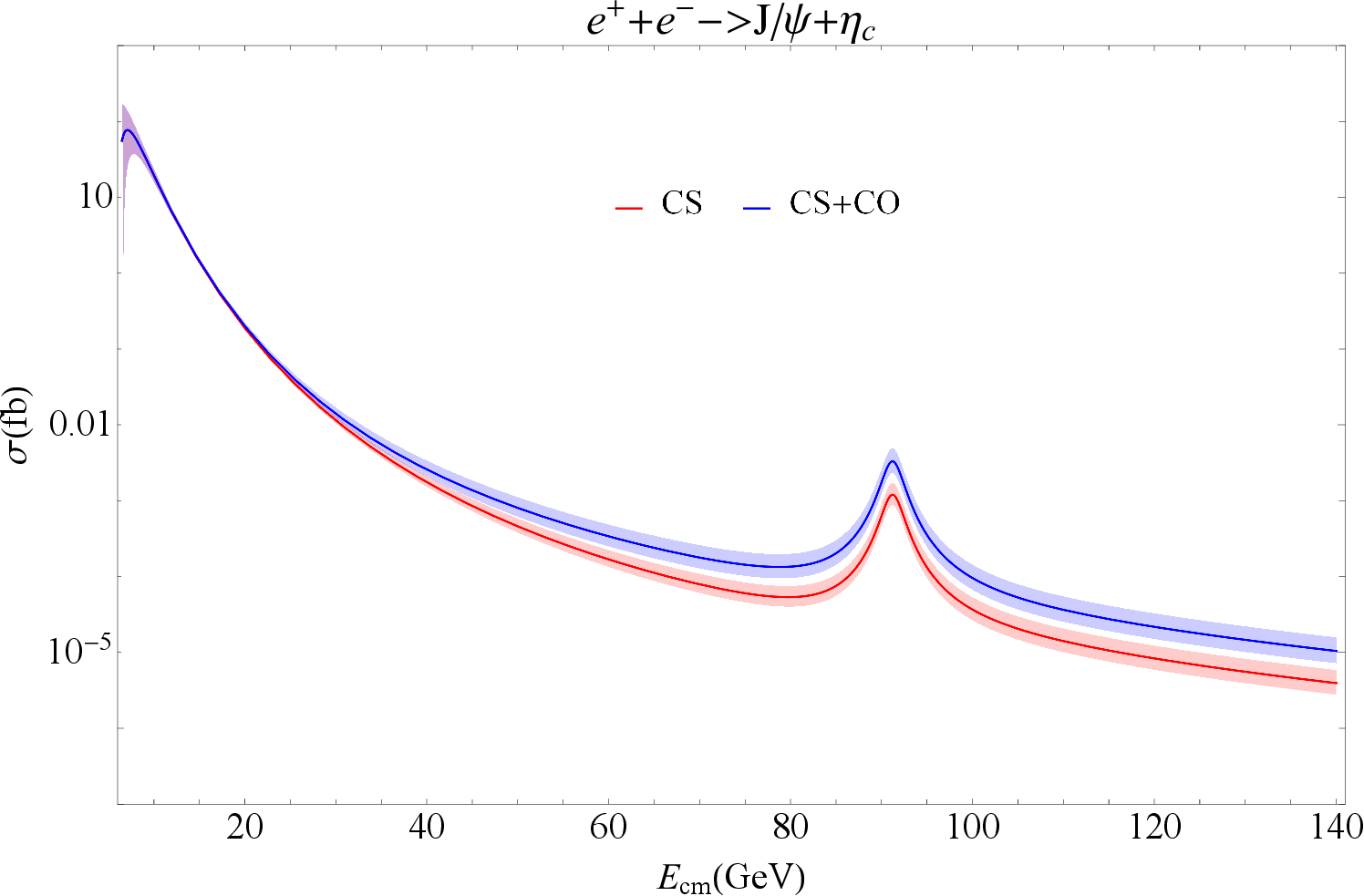}
			\includegraphics[width=0.43333\textwidth]{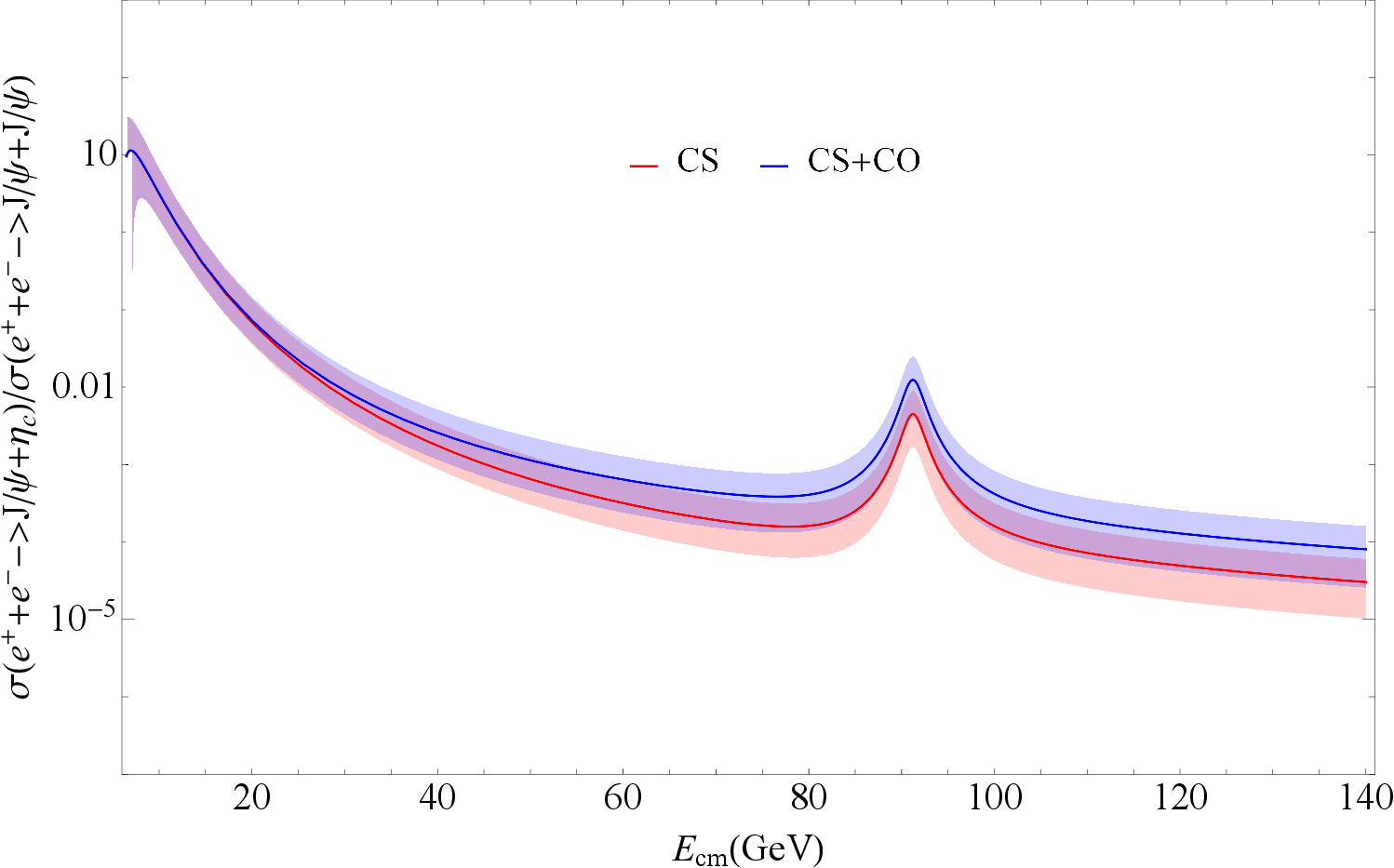}
		 	 
		\end{tabular}
		\caption{ (Color online) The cross sections with uncertainties of $J/\psi+\eta_c$ production (left panel). The ratios between the cross sections of $J/\psi+\eta_c$ production and that of double $J/\psi$ production (right panel).}
		\label{jpsietacuncerta}
	\end{figure*}
	\FloatBarrier
\end{widetext}

\section{Summary}
\label{summary}
 In this paper, we have studied the production of double heavy quarkonium  in $e^+e^-$ annihilation at the Z factories, considering the CO contributions along with CS channels up-to  $\mathcal{O}(v^2)$. 
 
 We considered CS channels including QCD and EW processes and all the CO channels at the level of tree diagrams in the calculations. The CO channels contribute substantially more than the CS channels in several processes, particularly for the combinations involving the $^3S_1^{[8]}$ state. This indicates that the processes of gluon fragmentation into the intermediate $^3S_1^{[8]}$ state might play a crucial role. It implies that studying the production of double heavy quarkonia  may also be a good way to determine the LDMEs $\langle\mathcal{O}^H(^3S_1^{[8]})\rangle$ in the future Z factory experiment with precise measurement.
 Meanwhile, the relativistic corrections are significant and the LO cross sections would be suppressed by the factors of at least $50\%$ for charmonia and $20\%$ for bottomonia, respectively. 
 With the results of cross sections, we estimate the events of double heavy quarkonium production for future experimental measurements (CEPC and FCC-ee).

 The COM is indispensable in hadron collision processes\cite{psianomalyt,universalityt2,hcCOLDMEs2,hcCOLDMEs6,universalityt3,chicjCOLDMEs3}, while in the double  charmonium production via $e^+e^-$ annihilation at B factory, the COM is negligible\cite{COM-Bfac2}. In this work, we show that the COM is significant or dominant in $e^+e^-$ annihilation at  Z factory and higher collision energy. It is expected that the COM is also significant for some other heavy quarkonia production processes (e.g. inclusive production processes, or D-wave heavy quarkonia production) at the Z factory and higher energy region. However, the COM may be negligible for $B_c$ meson production, since the gluon can't be fragmented into a $^3S_1^{[8]}$ $c\bar{b}$ quark pair. And it will  be important to continue to calculate  QCD loop corrections to these  processes to complement the study of heavy quarkonia production  at future $e^+e^-$ colliders.

 \begin{widetext}
	\begin{figure*}[htbp]
		\begin{tabular}{c c c }
			\includegraphics[width=0.333\textwidth]{ 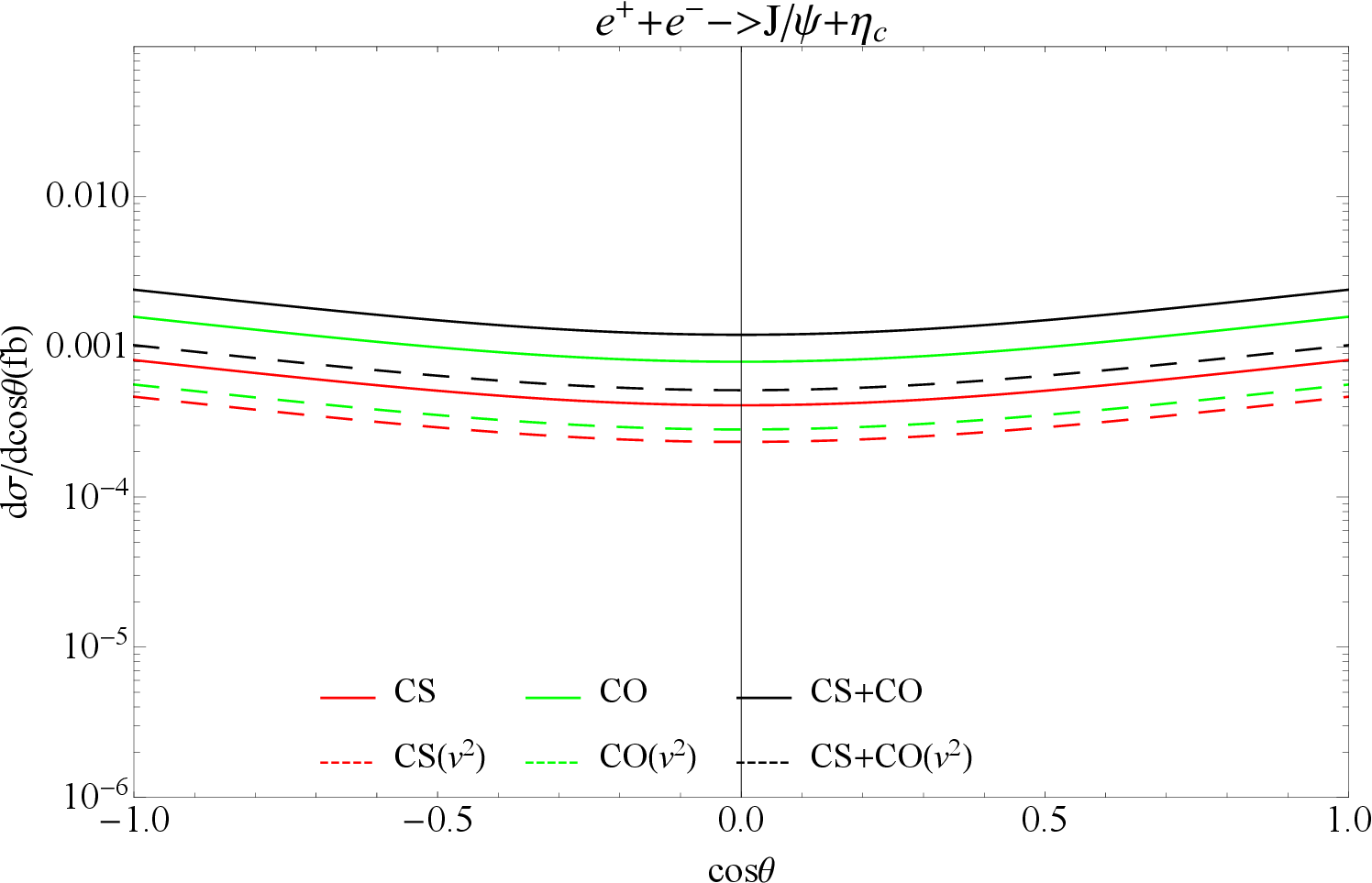}
			\includegraphics[width=0.333\textwidth]{ 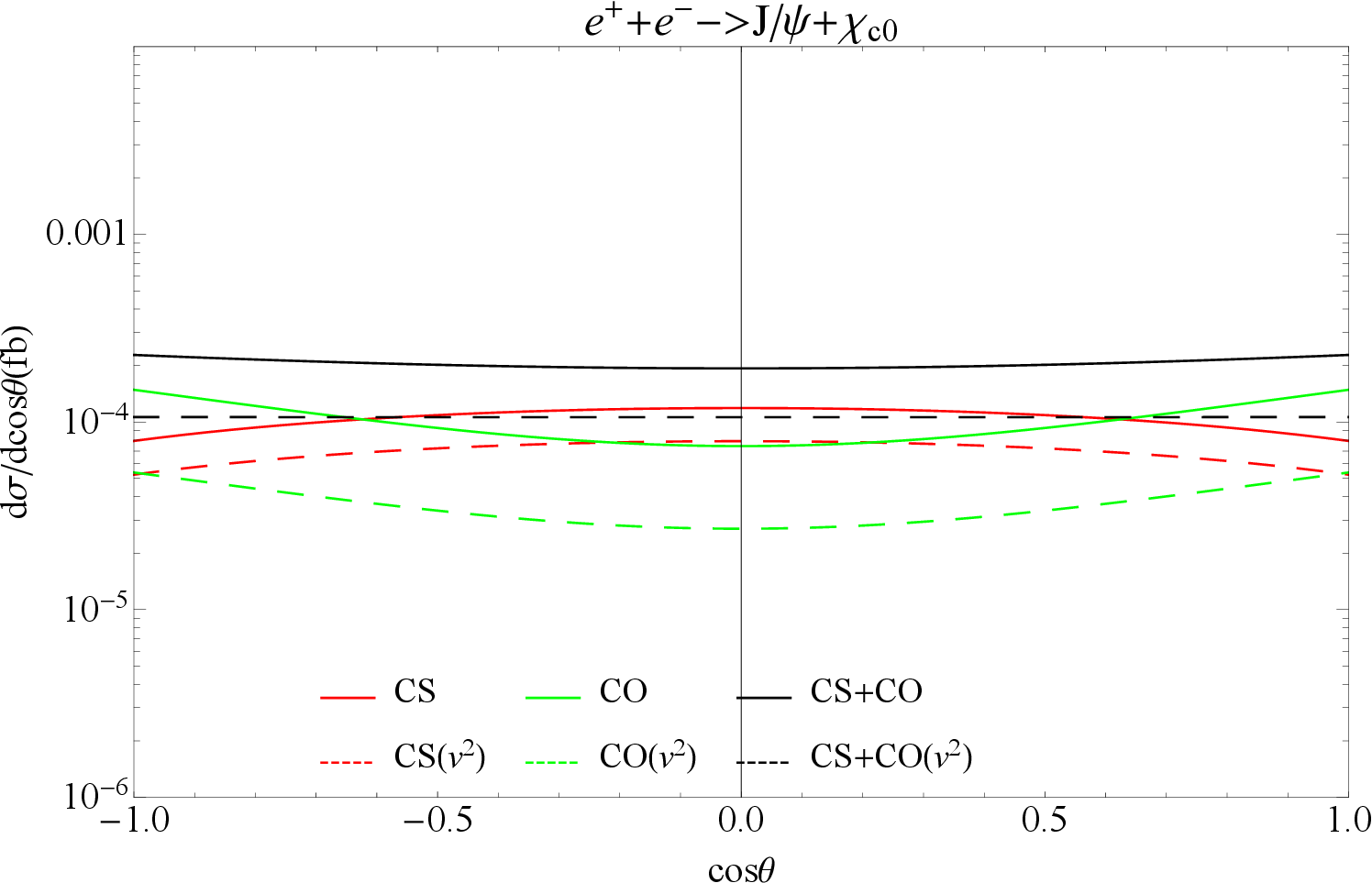}
				\includegraphics[width=0.333\textwidth]{ 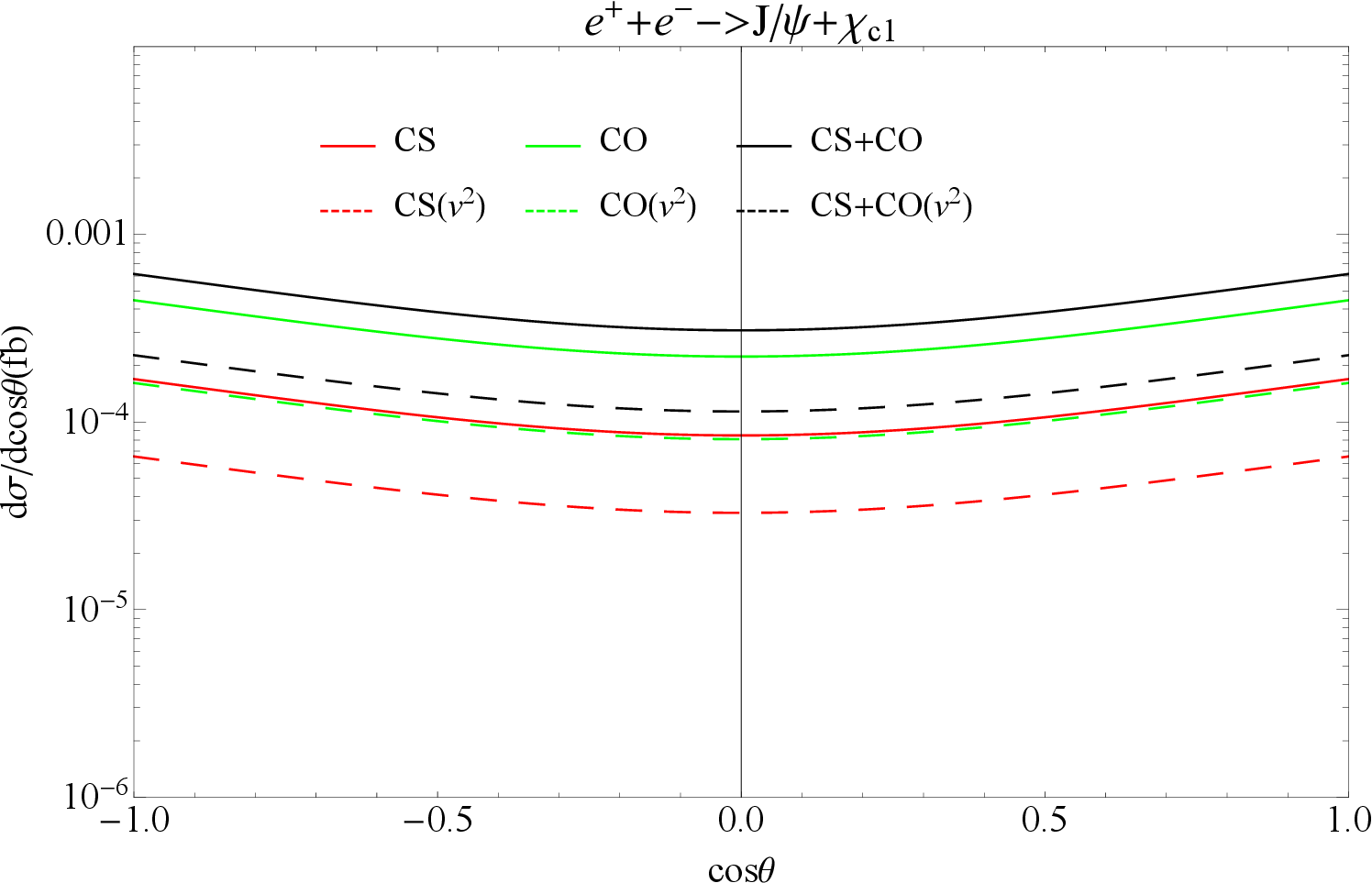}
		\end{tabular}
		\begin{tabular}{c c c }
			\includegraphics[width=0.333\textwidth]{ 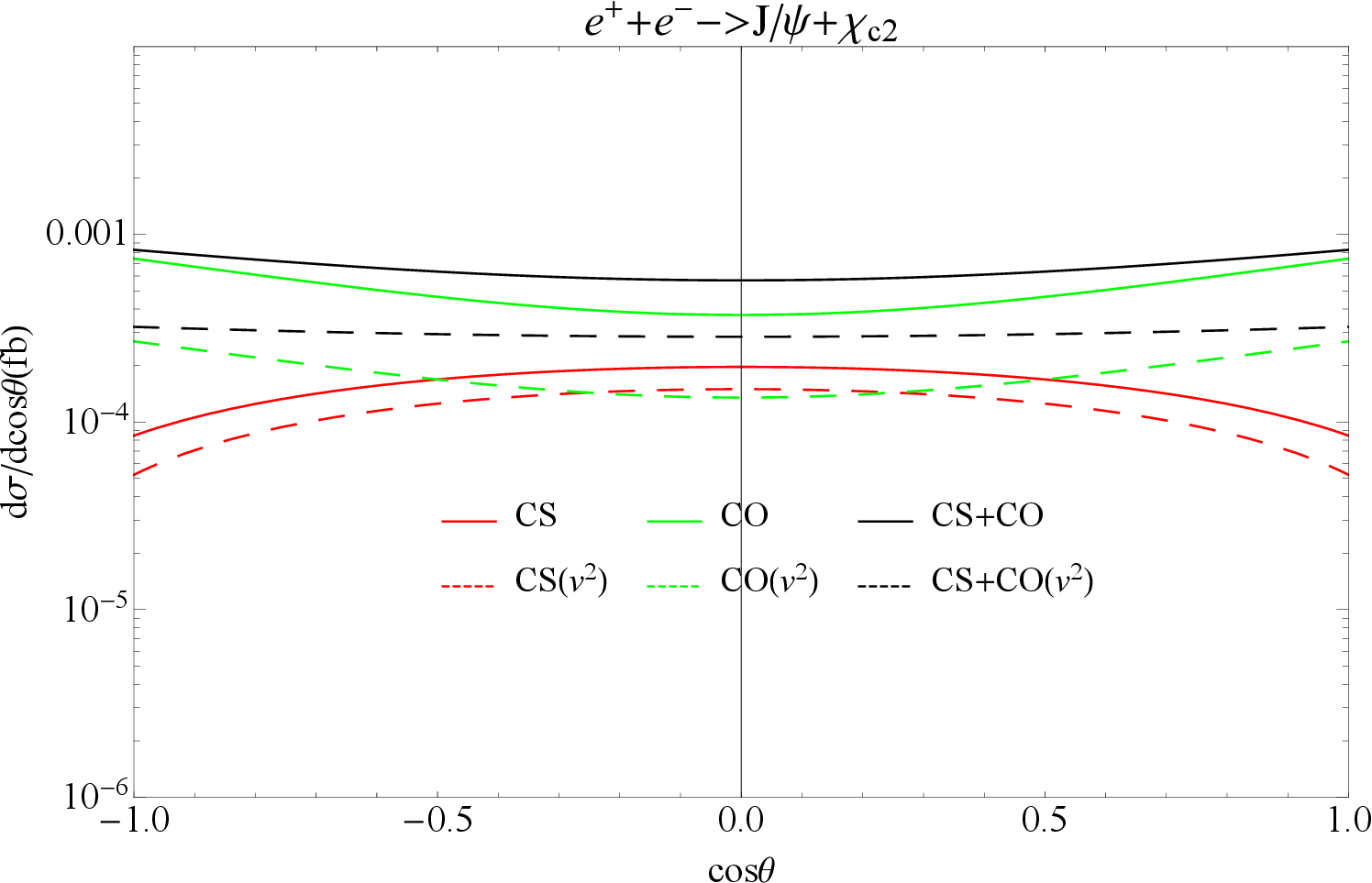}
			\includegraphics[width=0.333\textwidth]{ 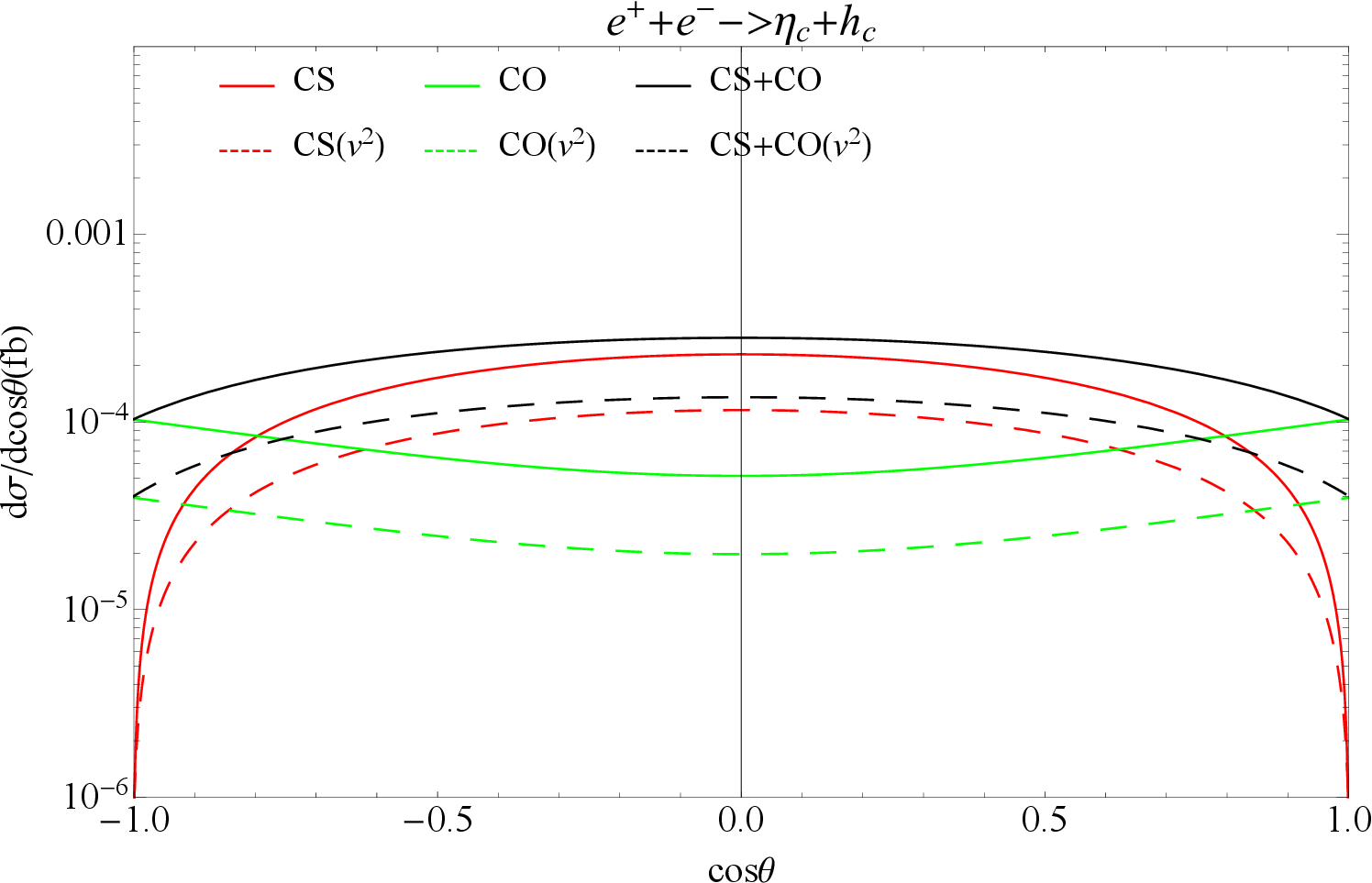}
				\includegraphics[width=0.333\textwidth]{ 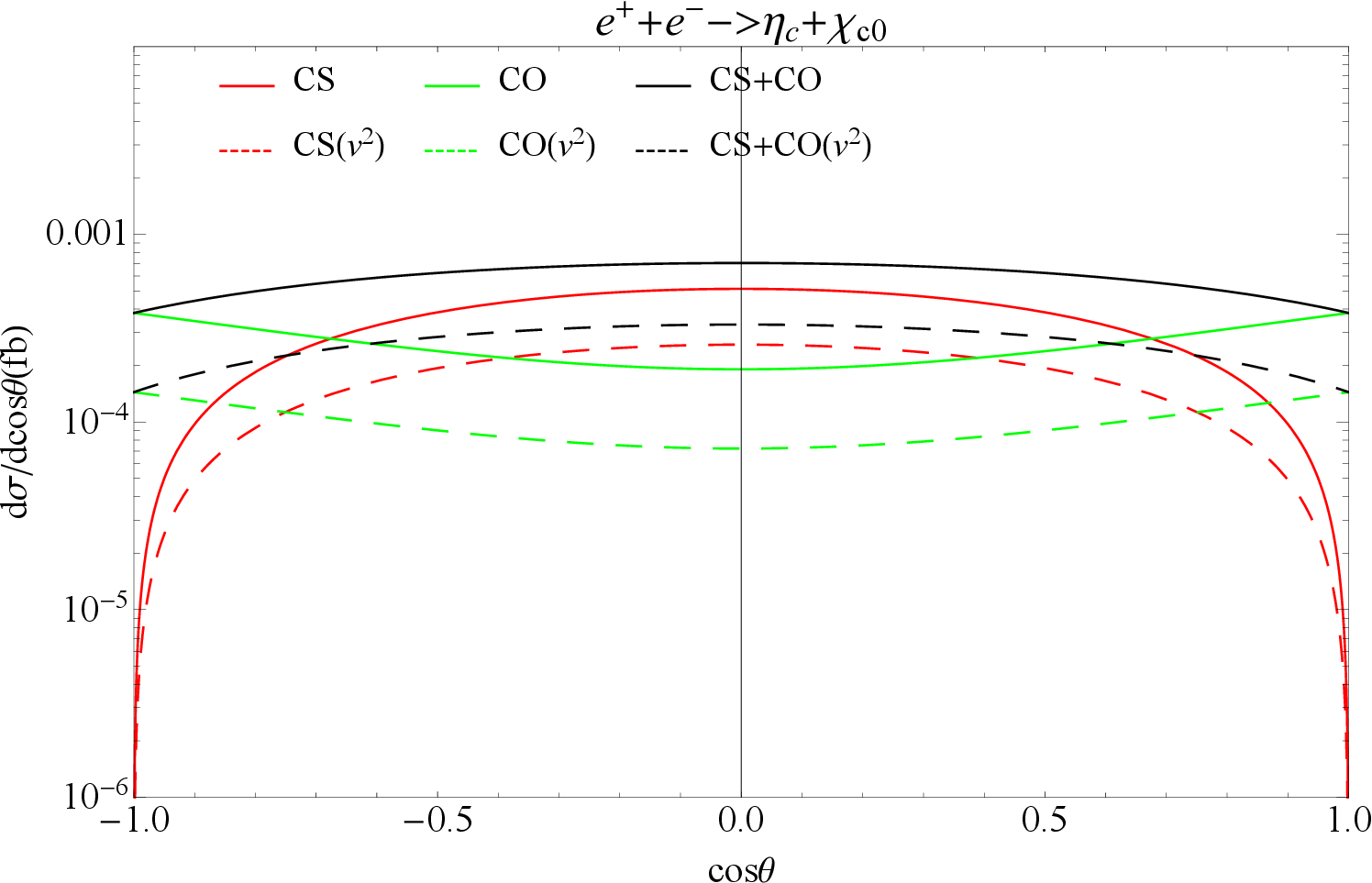}
		\end{tabular}
		\begin{tabular}{c c c }
		\includegraphics[width=0.333\textwidth]{ 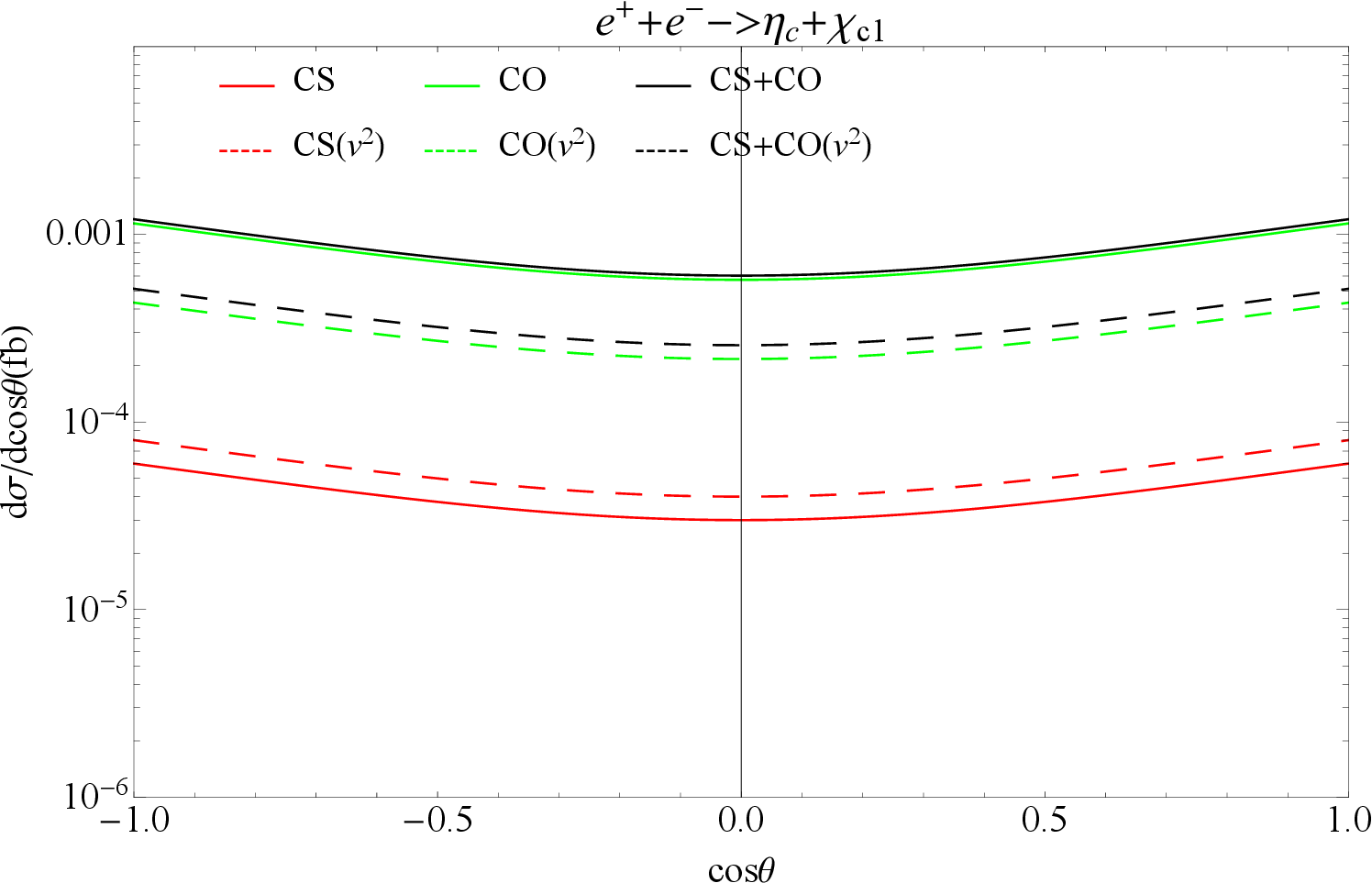}
		\includegraphics[width=0.333\textwidth]{ 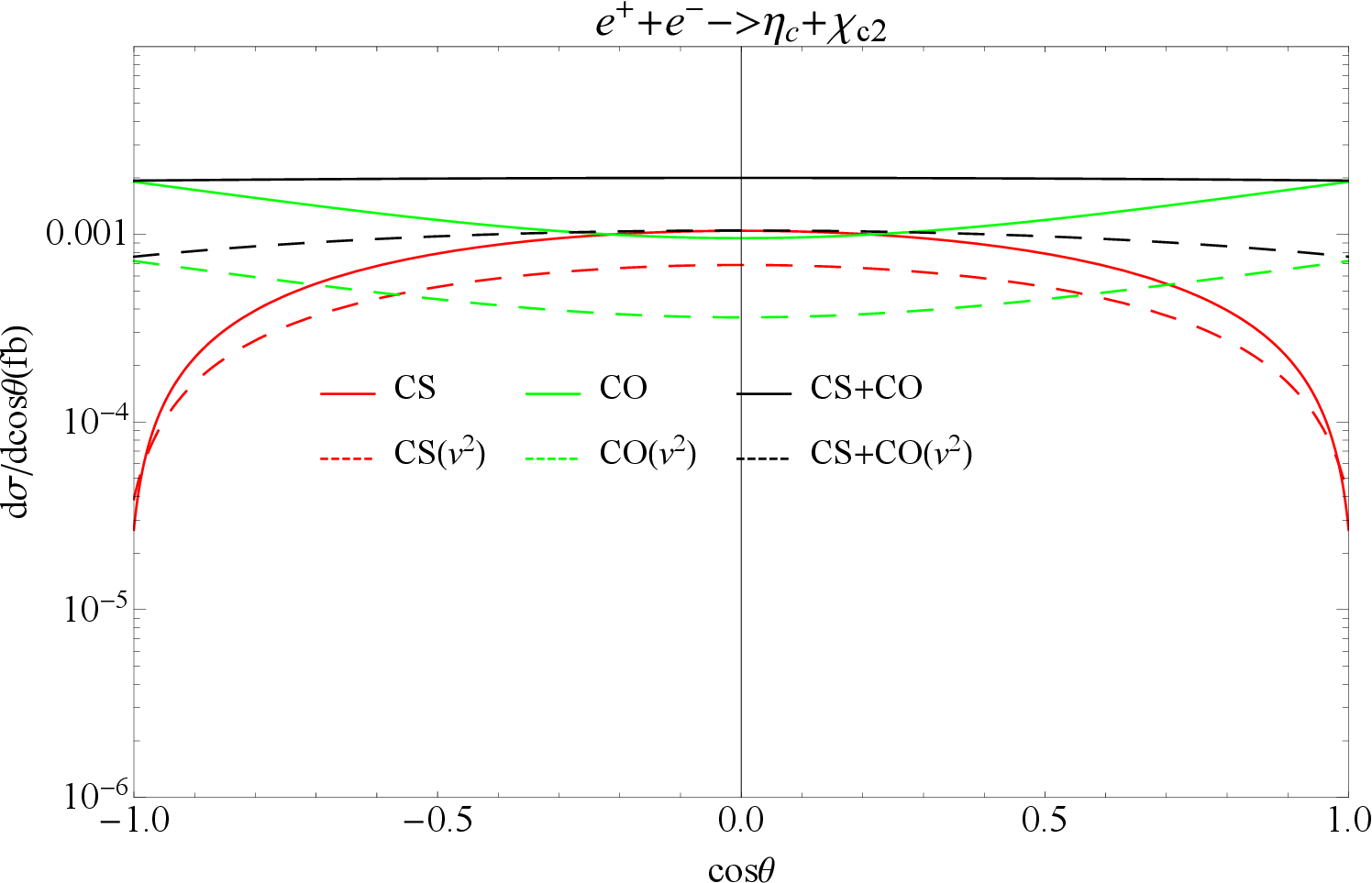}
			\includegraphics[width=0.333\textwidth]{ 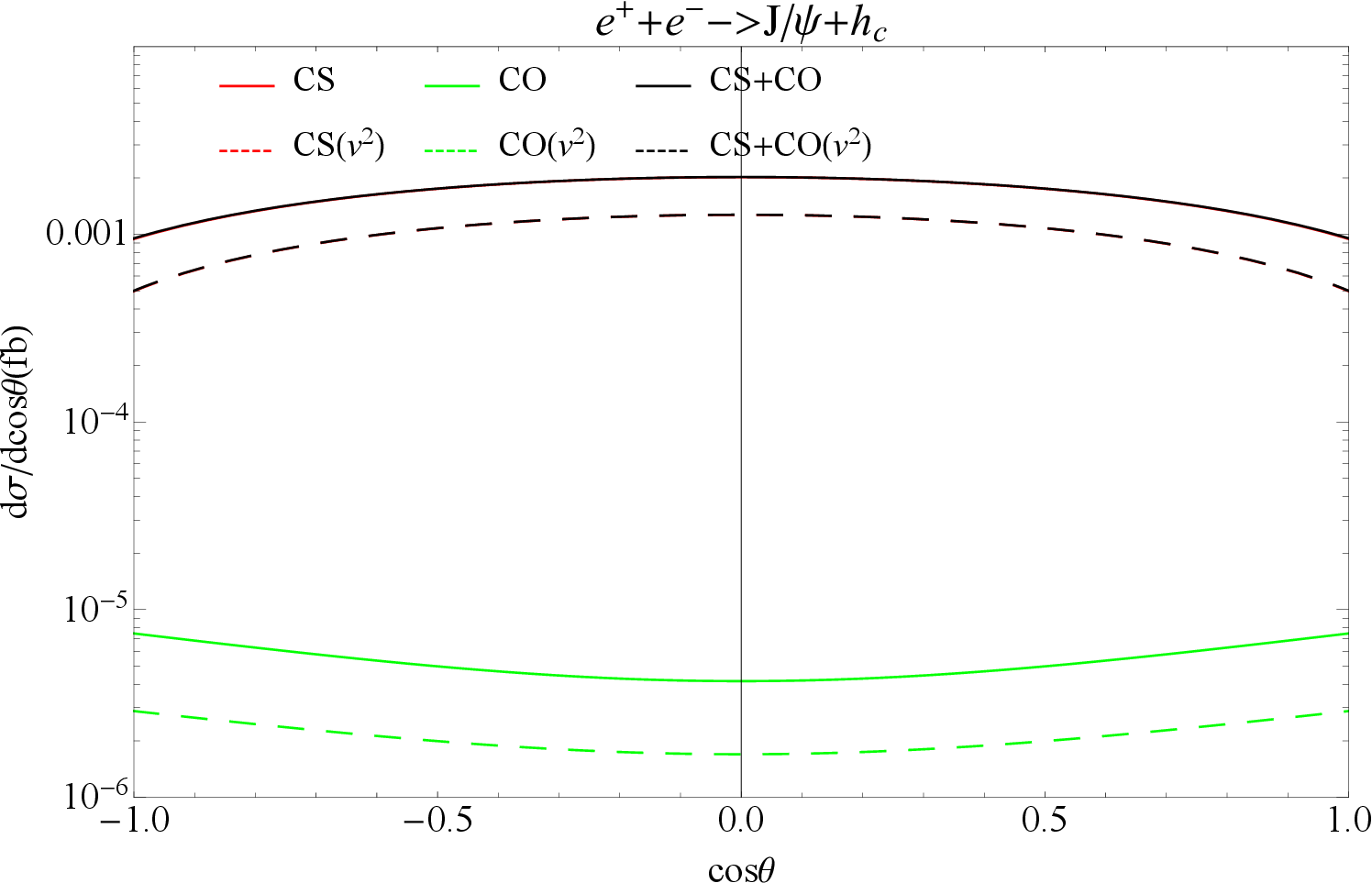}
		\end{tabular}
		\begin{tabular}{c c c}
		\includegraphics[width=0.333\textwidth]{ 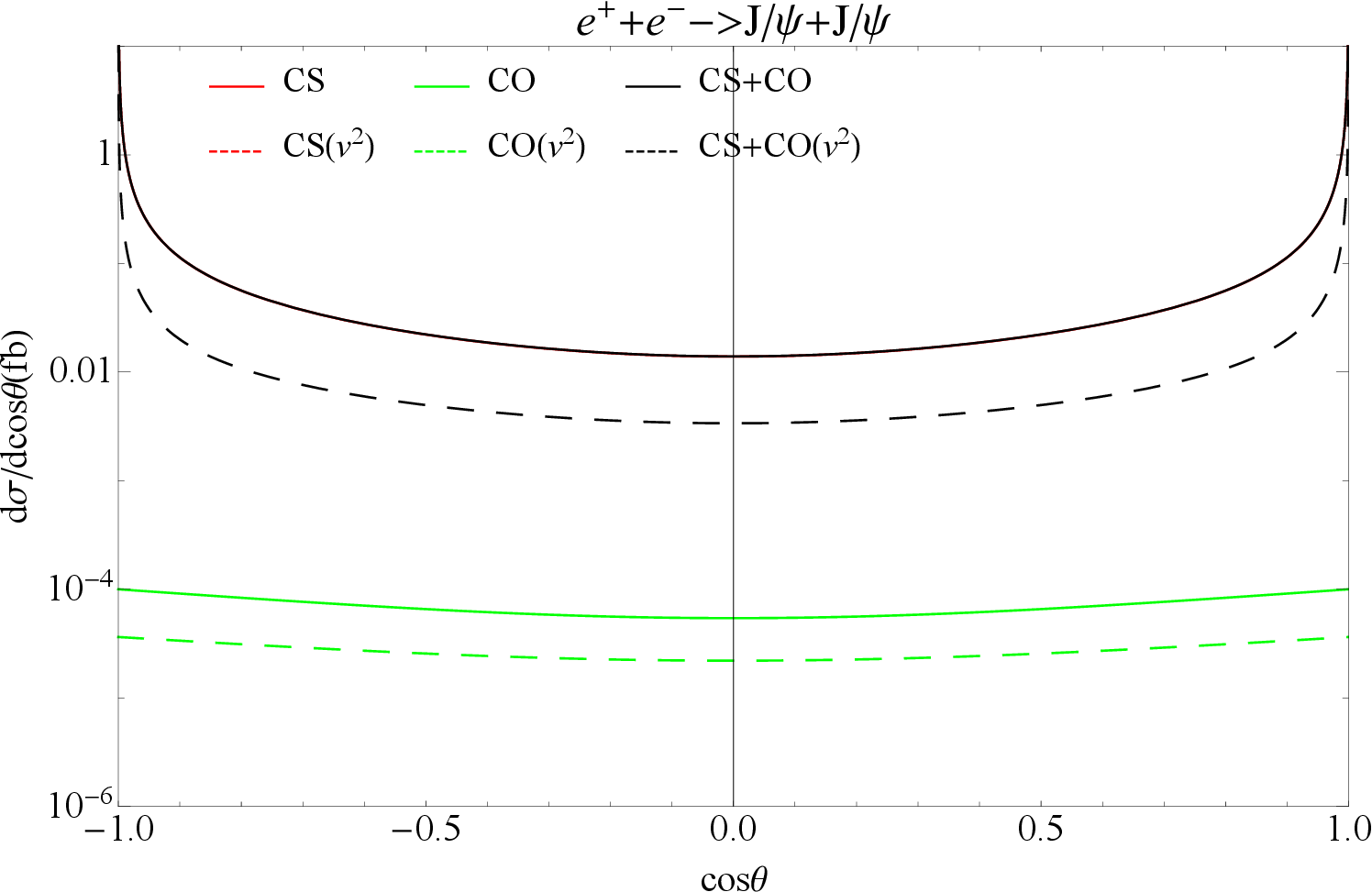}
			\includegraphics[width=0.333\textwidth]{ 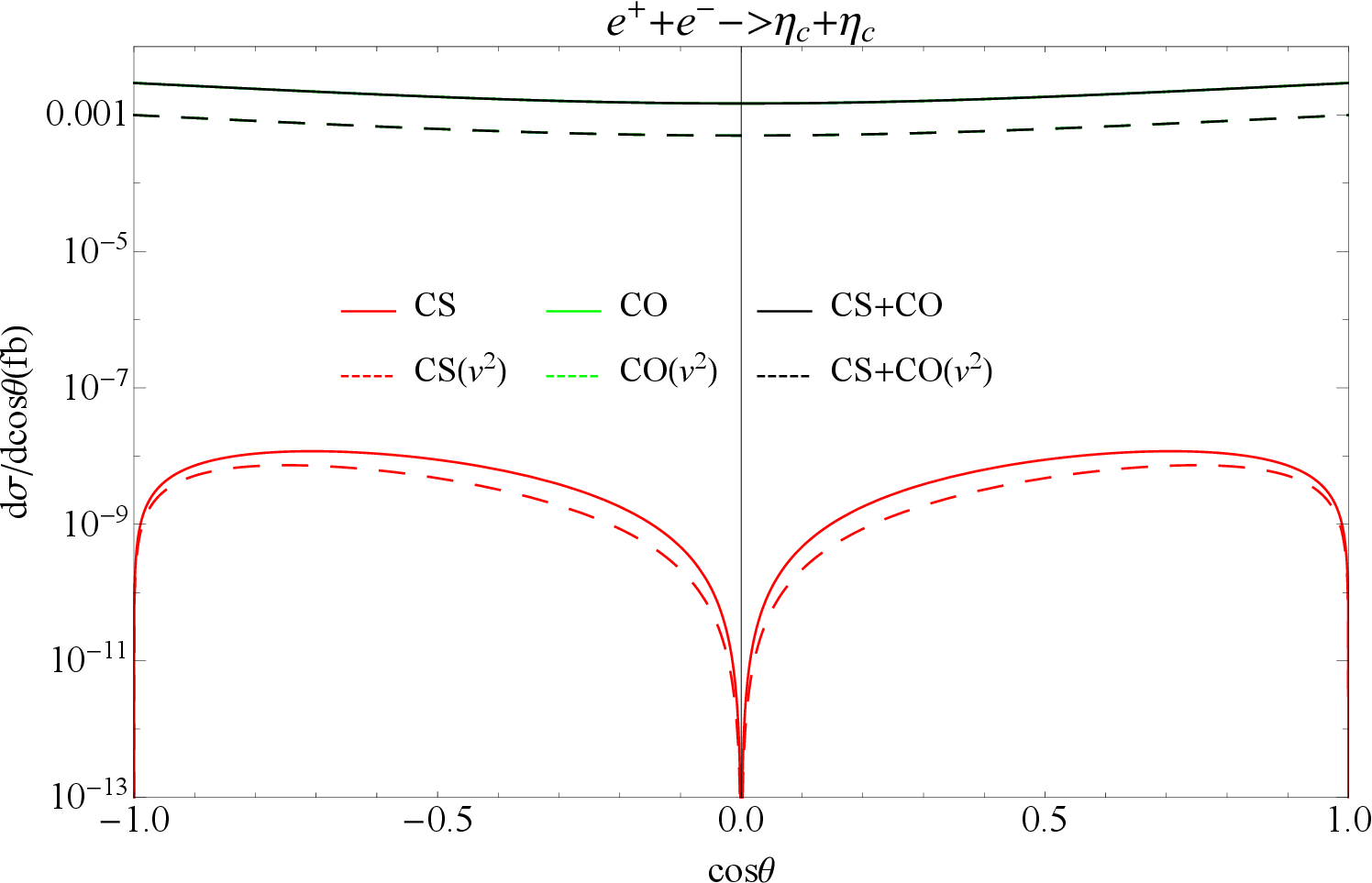}
		\end{tabular}
		\caption{ (Color online) The differential cross sections  $d\sigma/d\cos\theta$ ~for double  charmonium production at $\sqrt{s}$=$m_Z$. The solid line represents leading order (LO)  and the dashed line represents next-to-leading order in $v^2$ (NLO) results. The red line represents the CS channel, the green line represents the total CO channels and the black line represents the sum of  CS and CO. }
		\label{z0cccos}
	\end{figure*}
		\FloatBarrier
\end{widetext}

\begin{widetext}
	\begin{figure*}[htbp]
		\begin{tabular}{c c c}
			\includegraphics[width=0.333\textwidth]{ 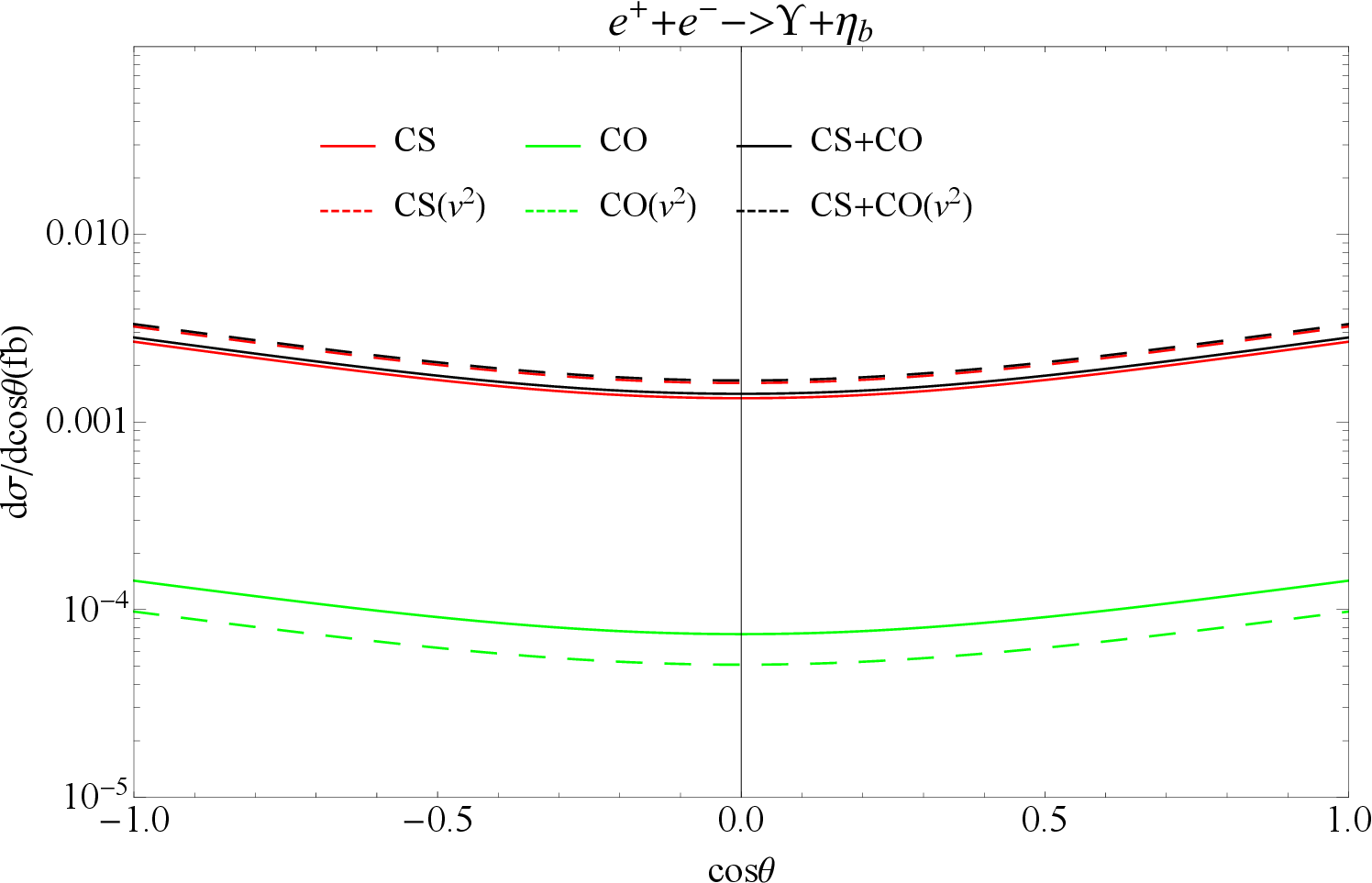}
			\includegraphics[width=0.333\textwidth]{ 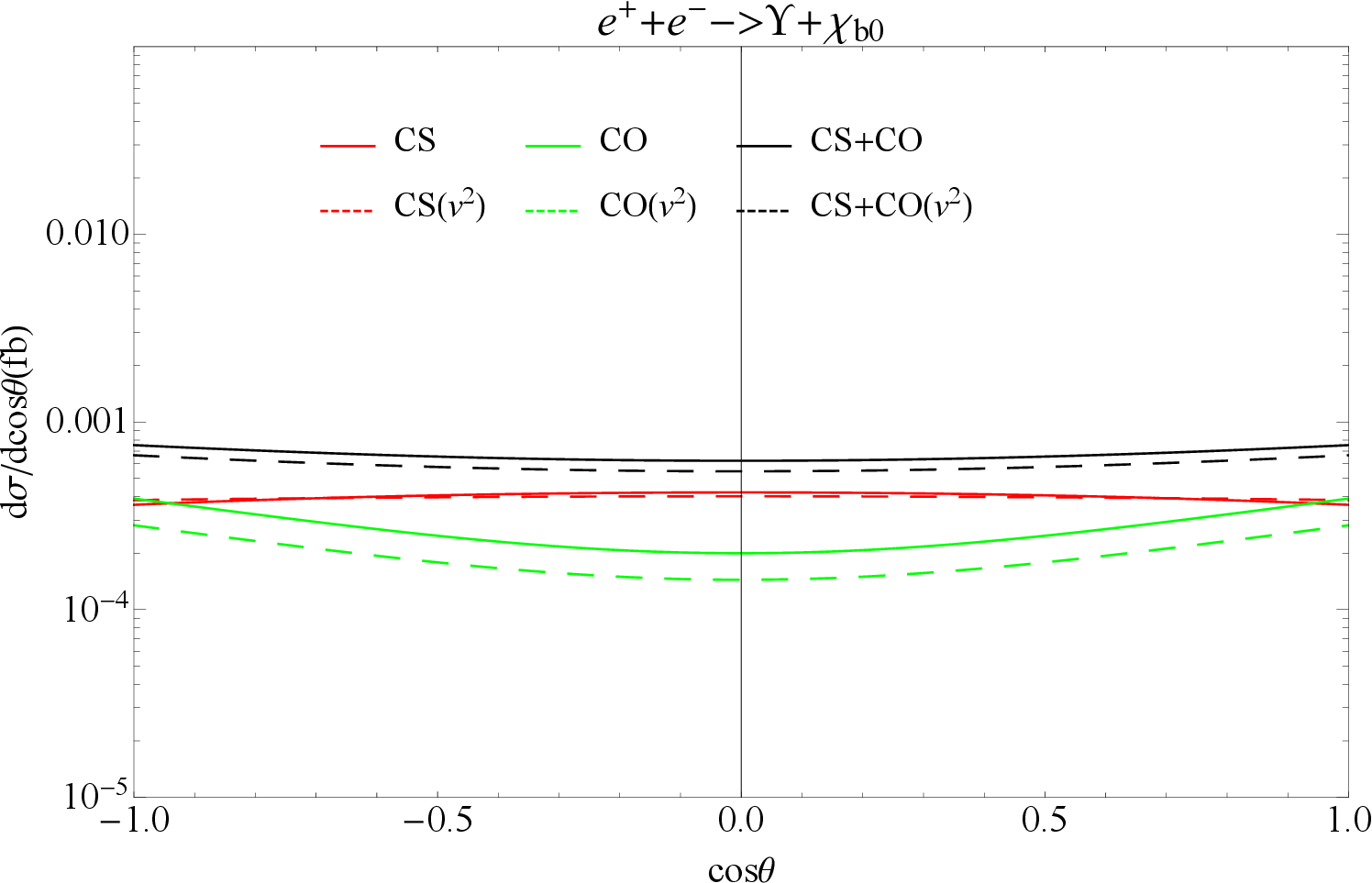}
			\includegraphics[width=0.333\textwidth]{ 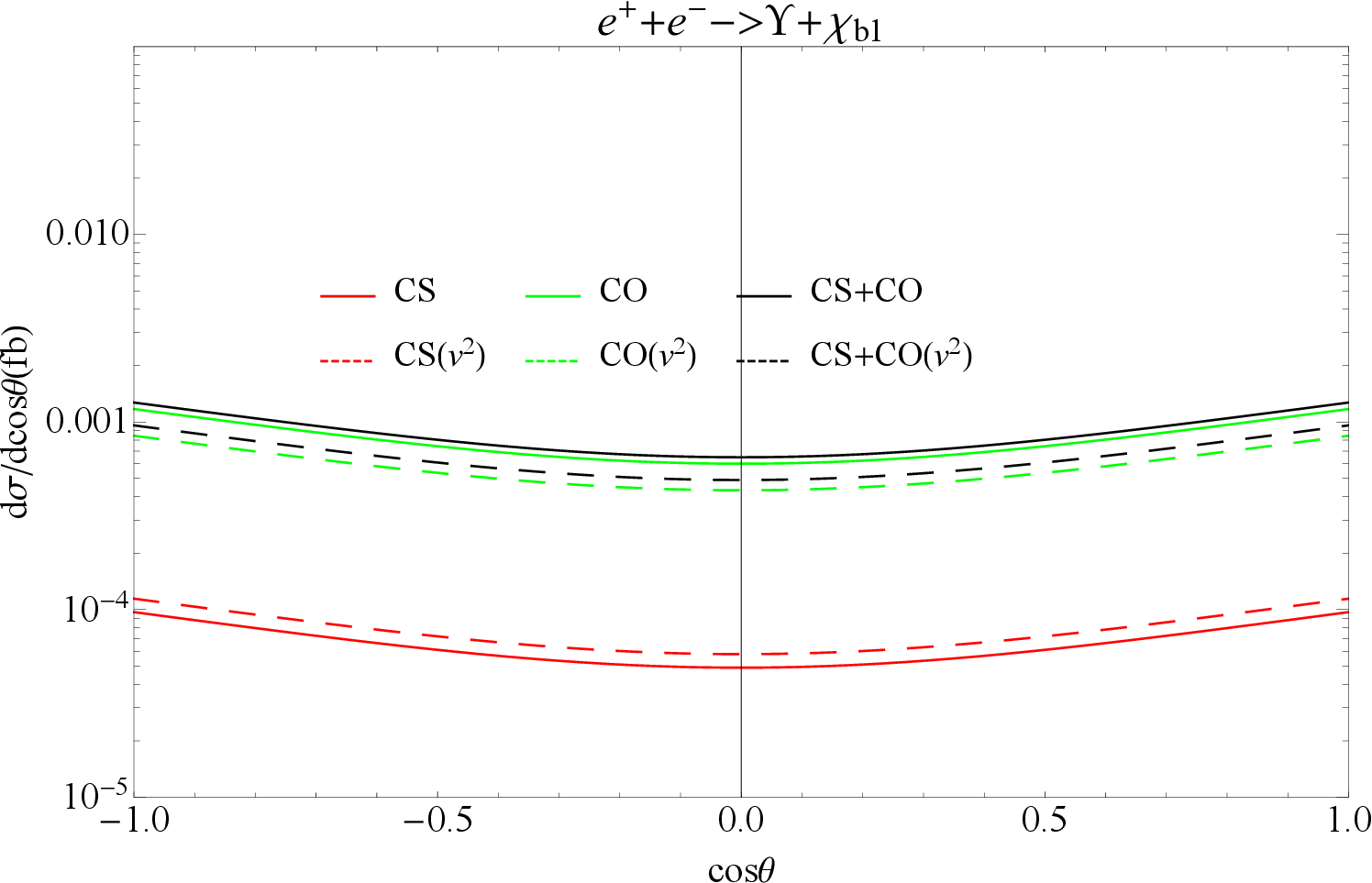}
		\end{tabular}
		\begin{tabular}{c c c}
			\includegraphics[width=0.333\textwidth]{ 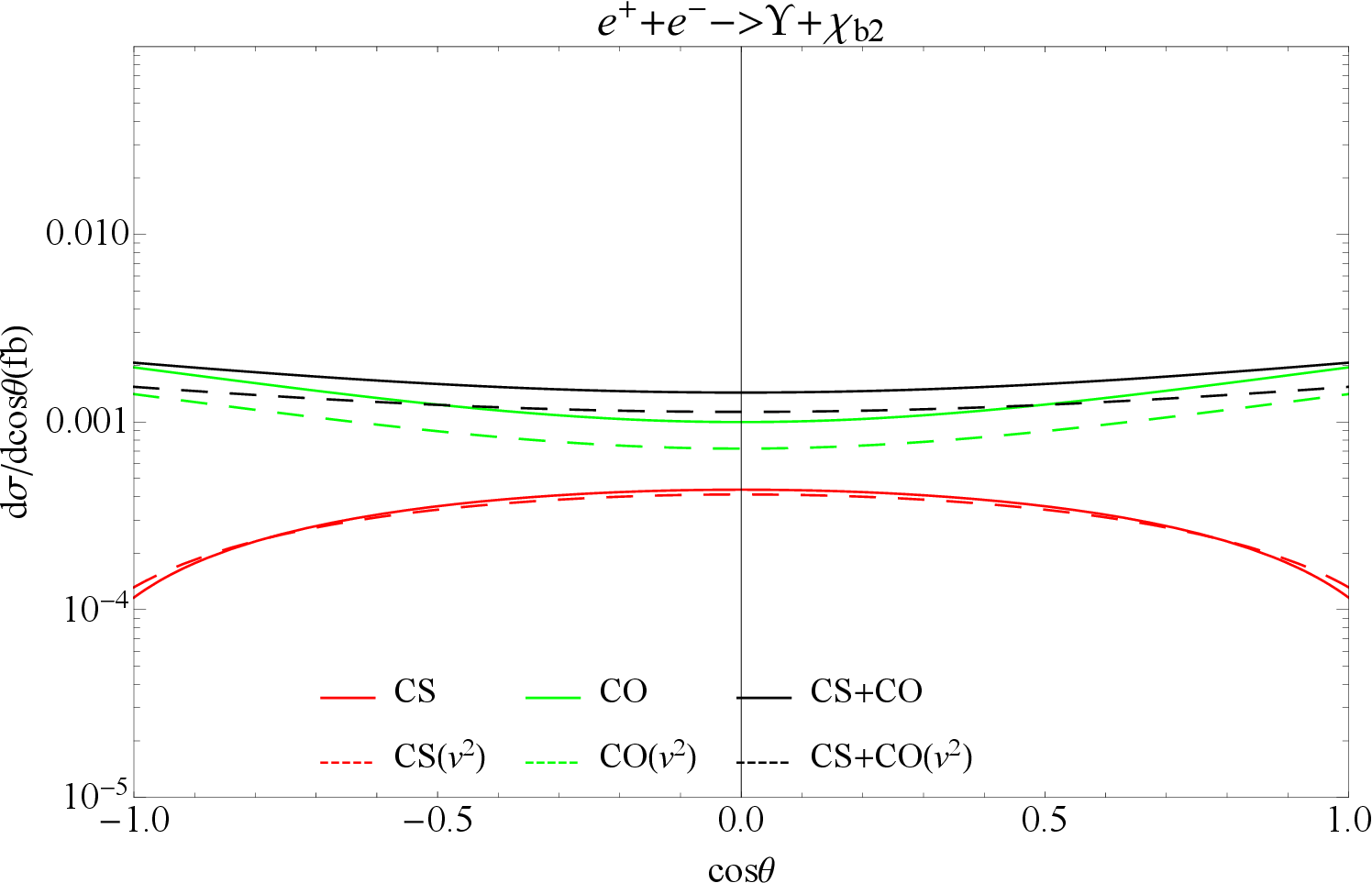}
			\includegraphics[width=0.333\textwidth]{ 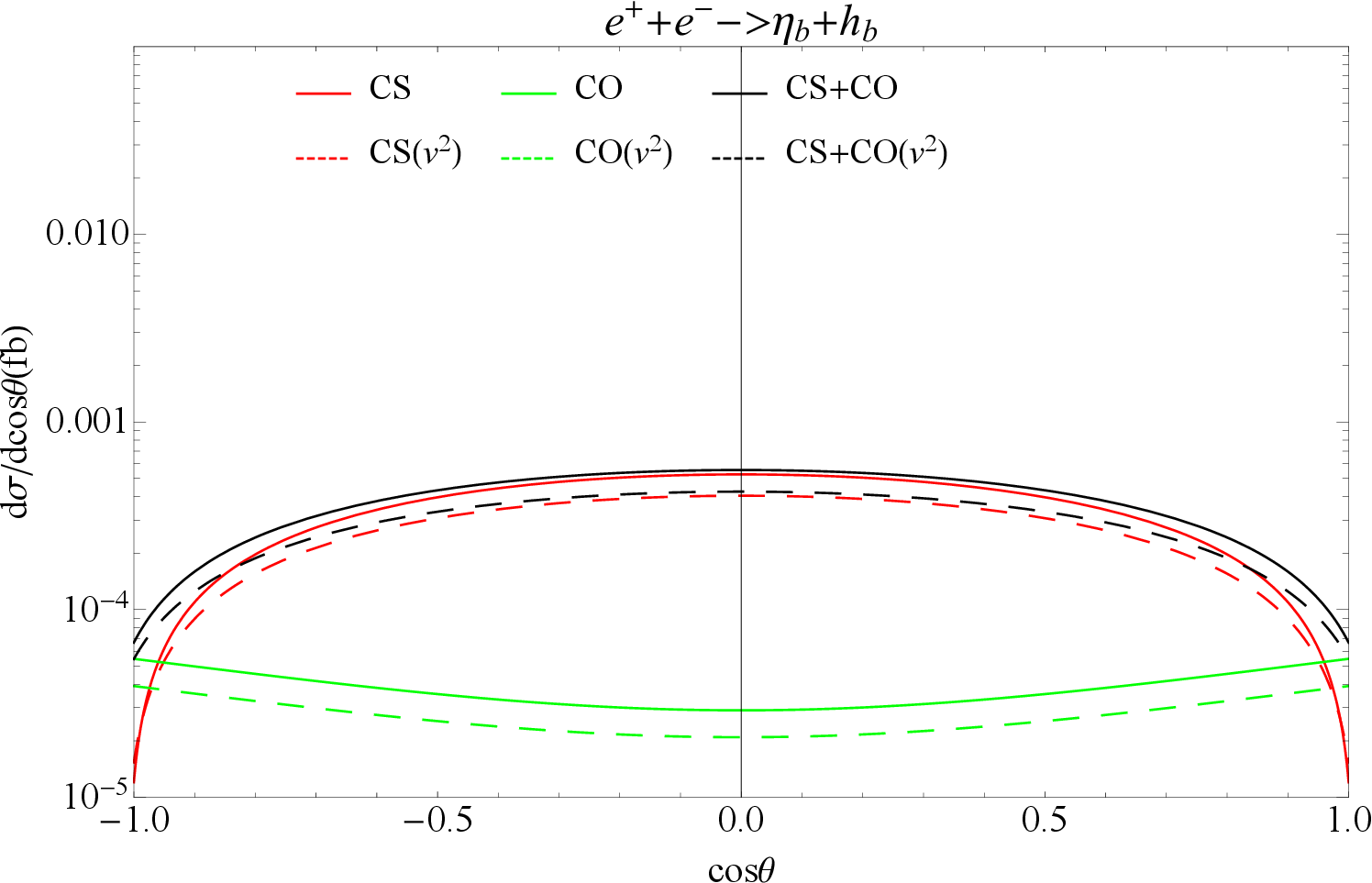}
			\includegraphics[width=0.333\textwidth]{ 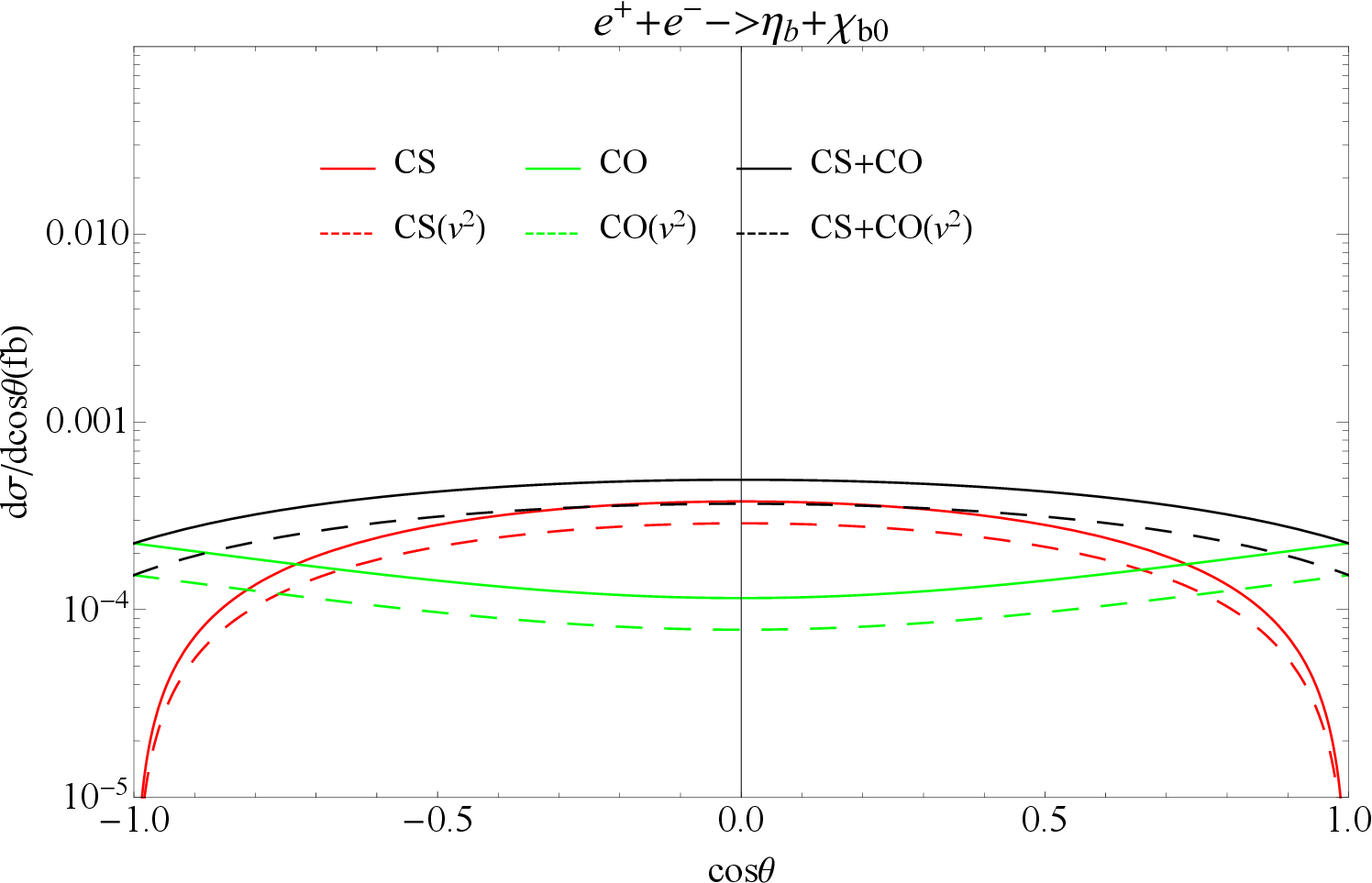}
		\end{tabular}
		\begin{tabular}{c c c}
			\includegraphics[width=0.333\textwidth]{ 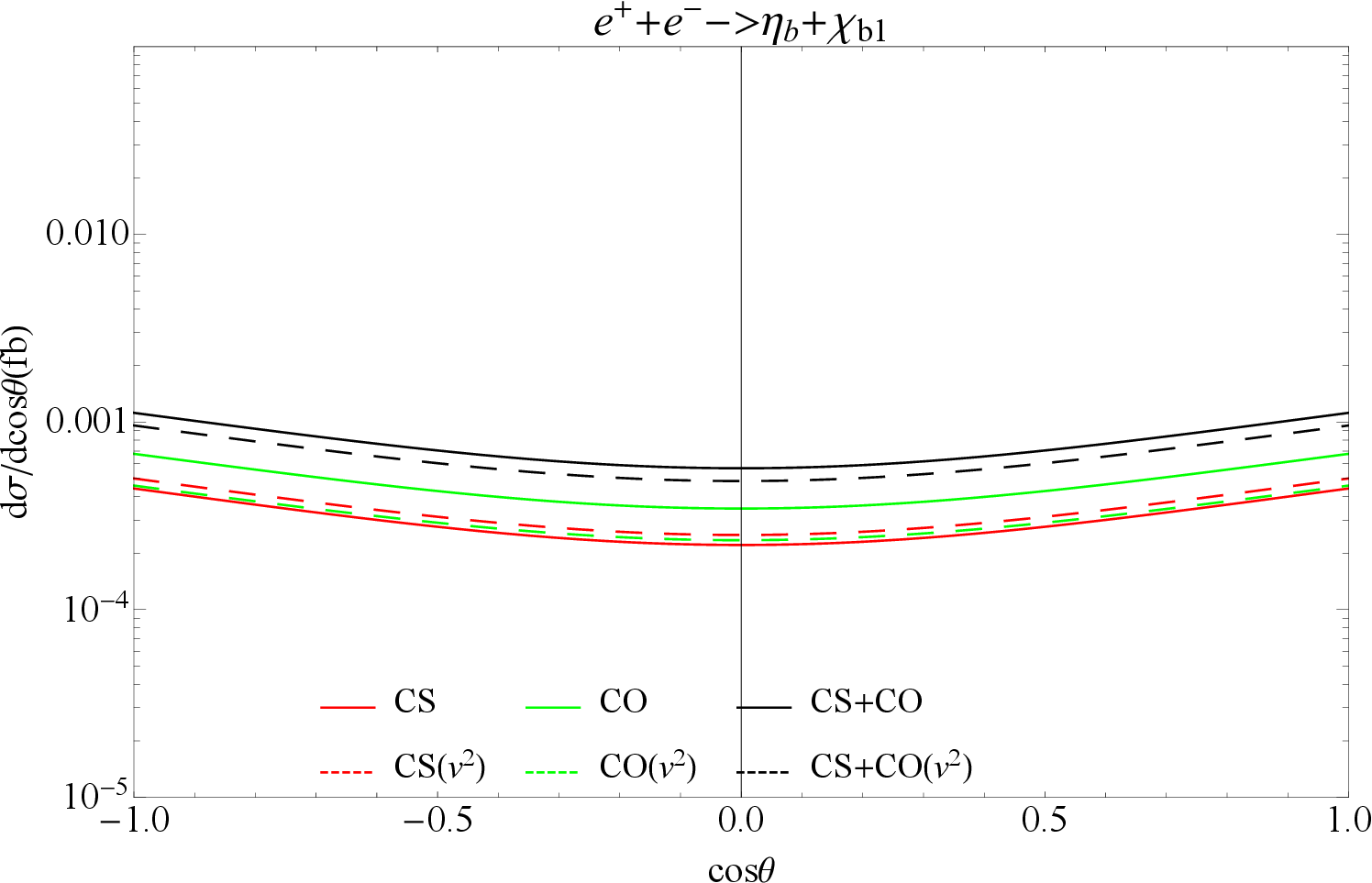}
			\includegraphics[width=0.333\textwidth]{ 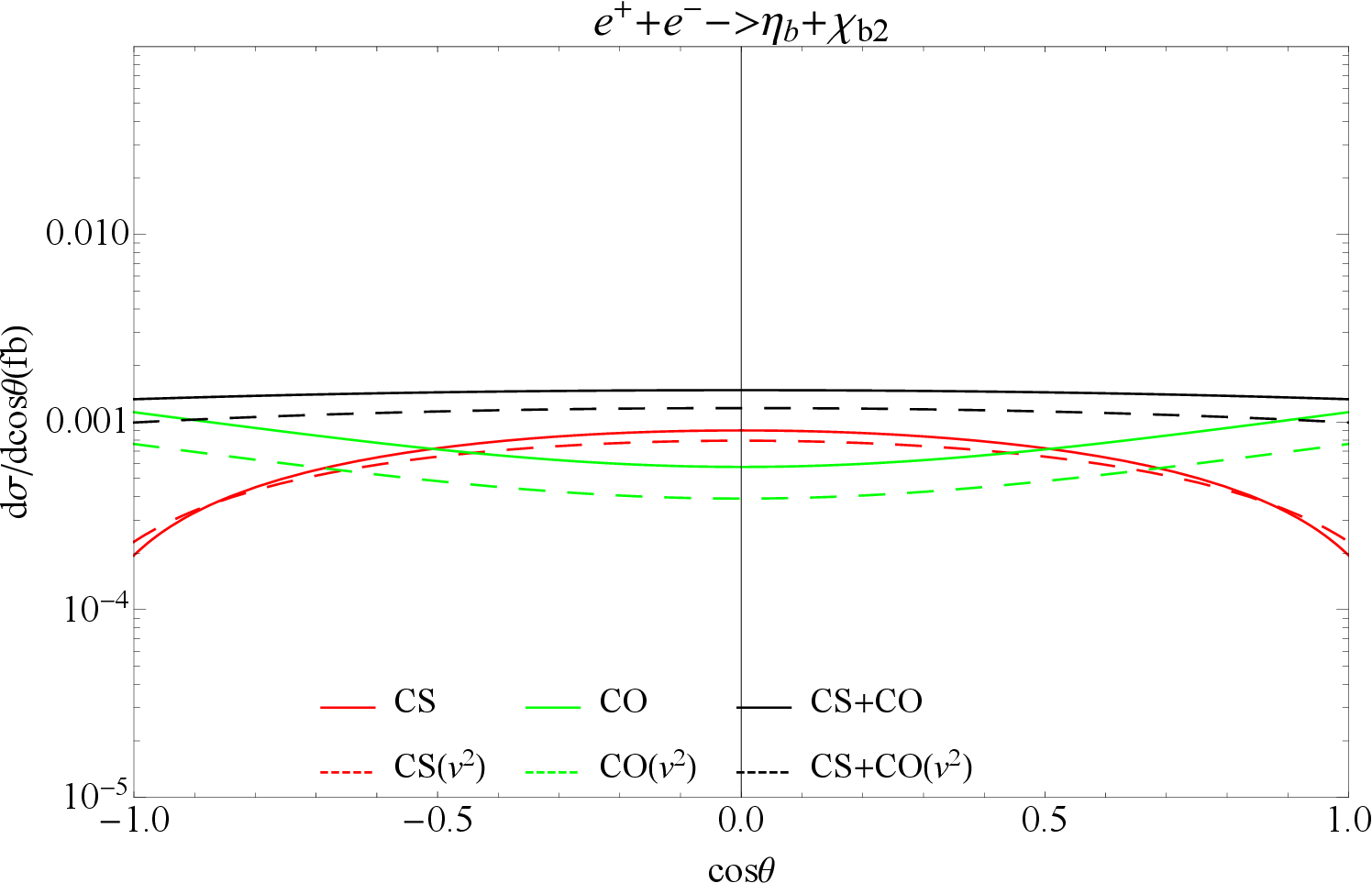}
			\includegraphics[width=0.333\textwidth]{ 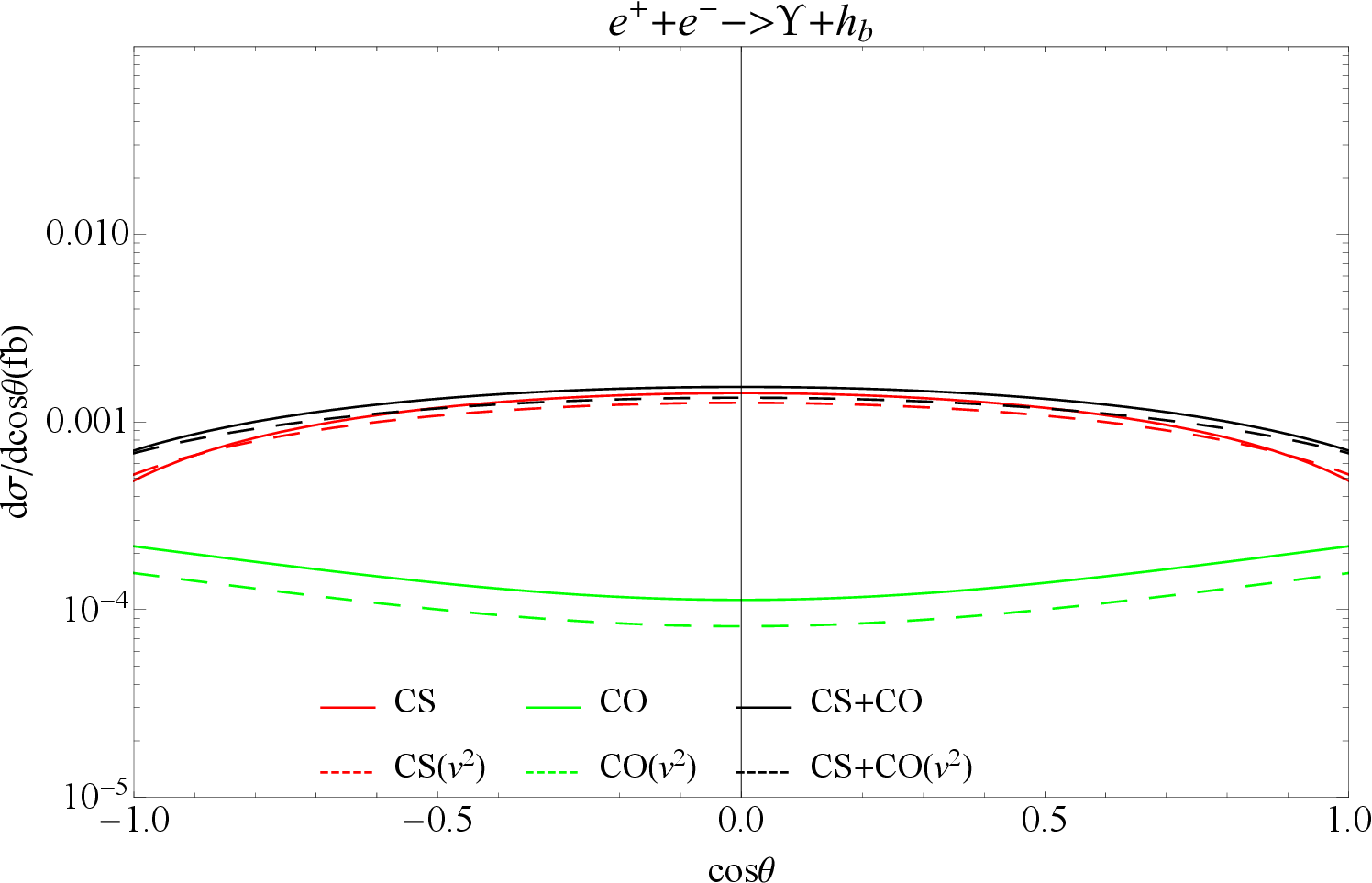}
		\end{tabular}
		\begin{tabular}{c c c }
			
			\includegraphics[width=0.333\textwidth]{ 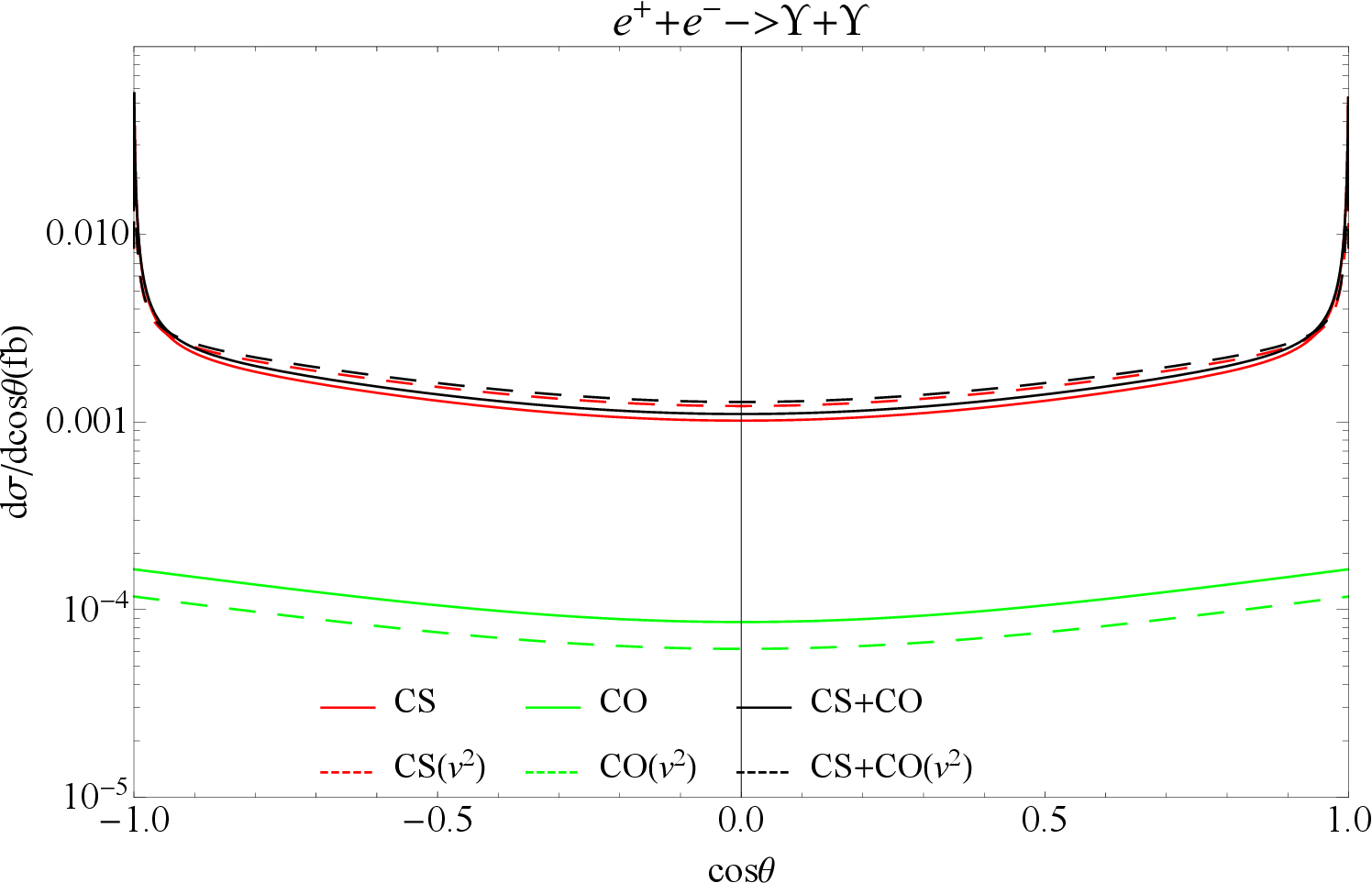}
				\includegraphics[width=0.333\textwidth]{ 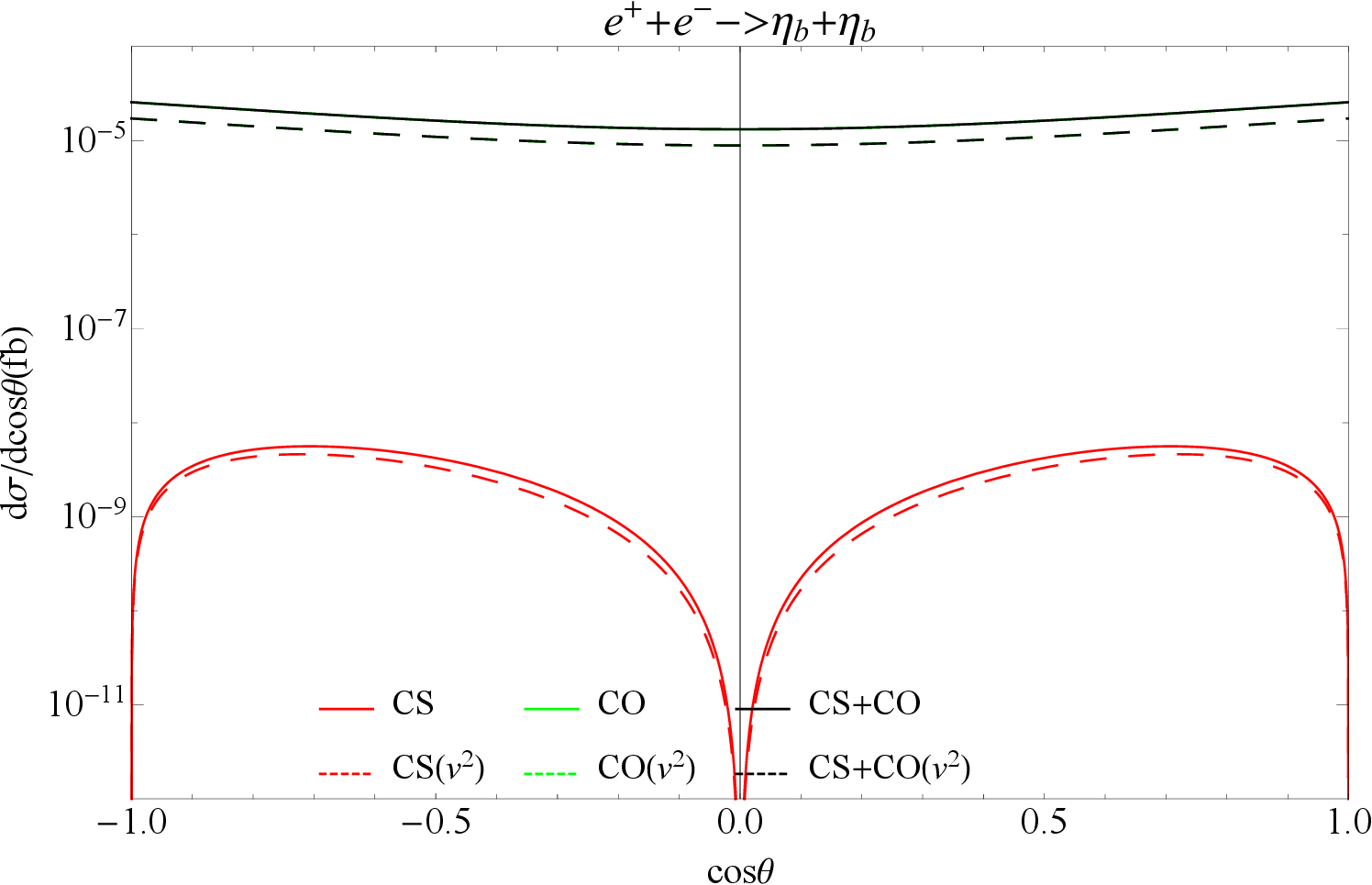}
			
		\end{tabular}
		\caption{ (Color online) The differential cross sections  $d\sigma/d\cos\theta$ ~for double bottomonium production at $\sqrt{s}$=$m_Z$. The solid line represents leading order (LO)  and the dashed line represents next-to-leading order in $v^2$ (NLO) results. The red line represents the CS channel, the green line represents the total CO channels and the black line represents the sum of  CS and CO. }
		\label{z0bbcos}
	\end{figure*}
	\FloatBarrier
\end{widetext}

\begin{widetext}
	\begin{figure*}[htbp]
		\begin{tabular}{c c c }
			\includegraphics[width=0.333\textwidth]{ 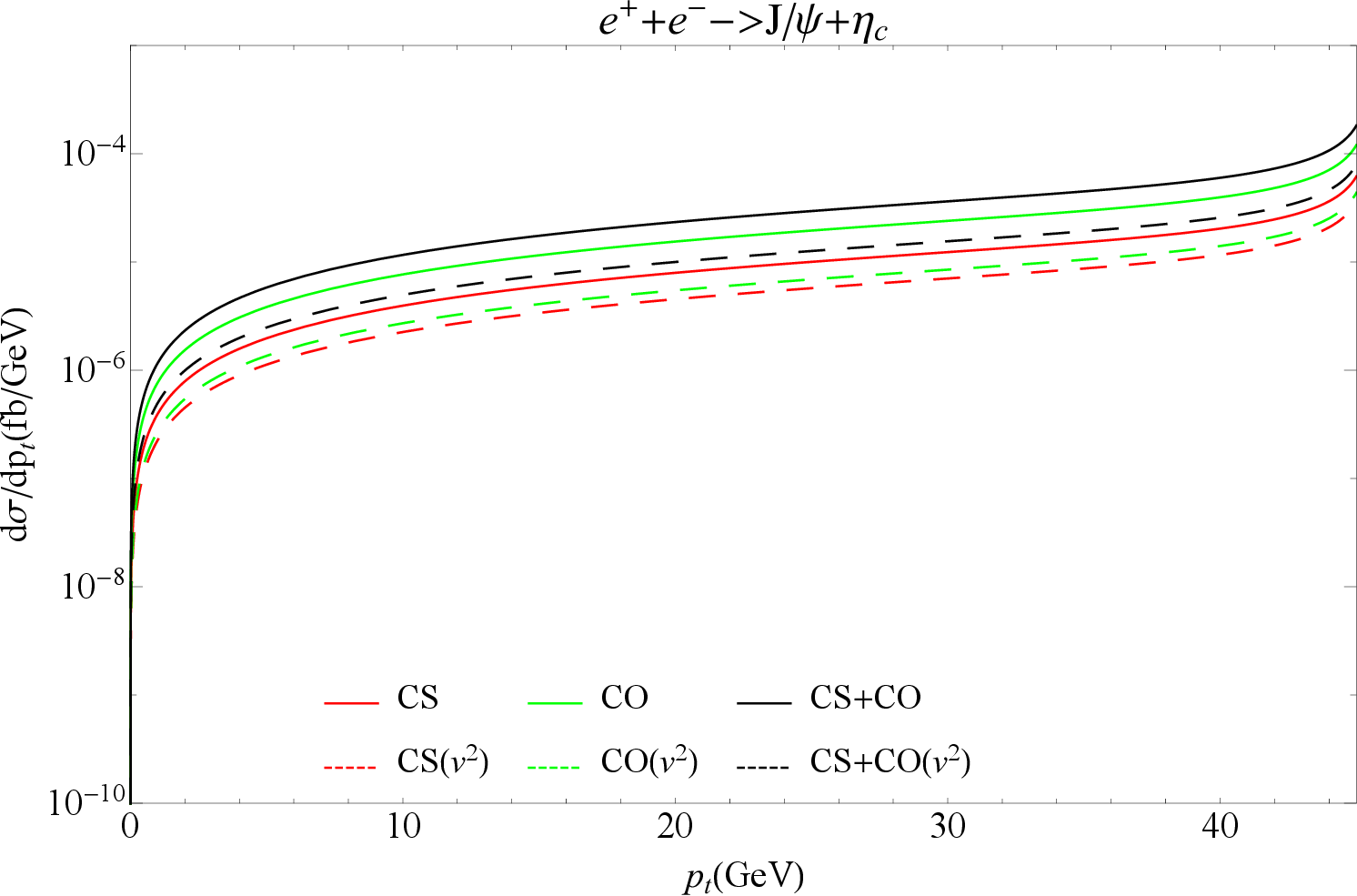}
			\includegraphics[width=0.333\textwidth]{ 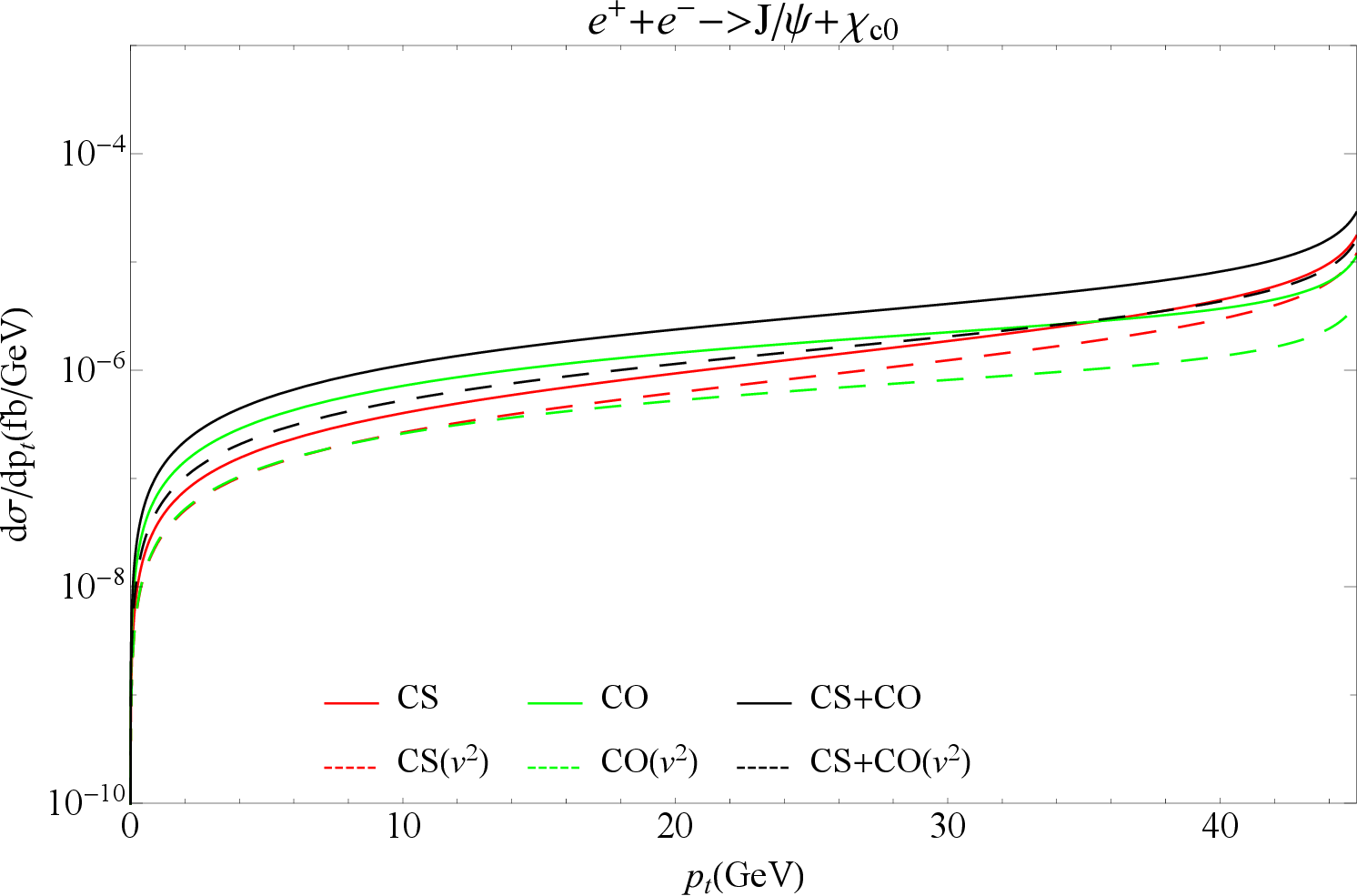}
			\includegraphics[width=0.333\textwidth]{ 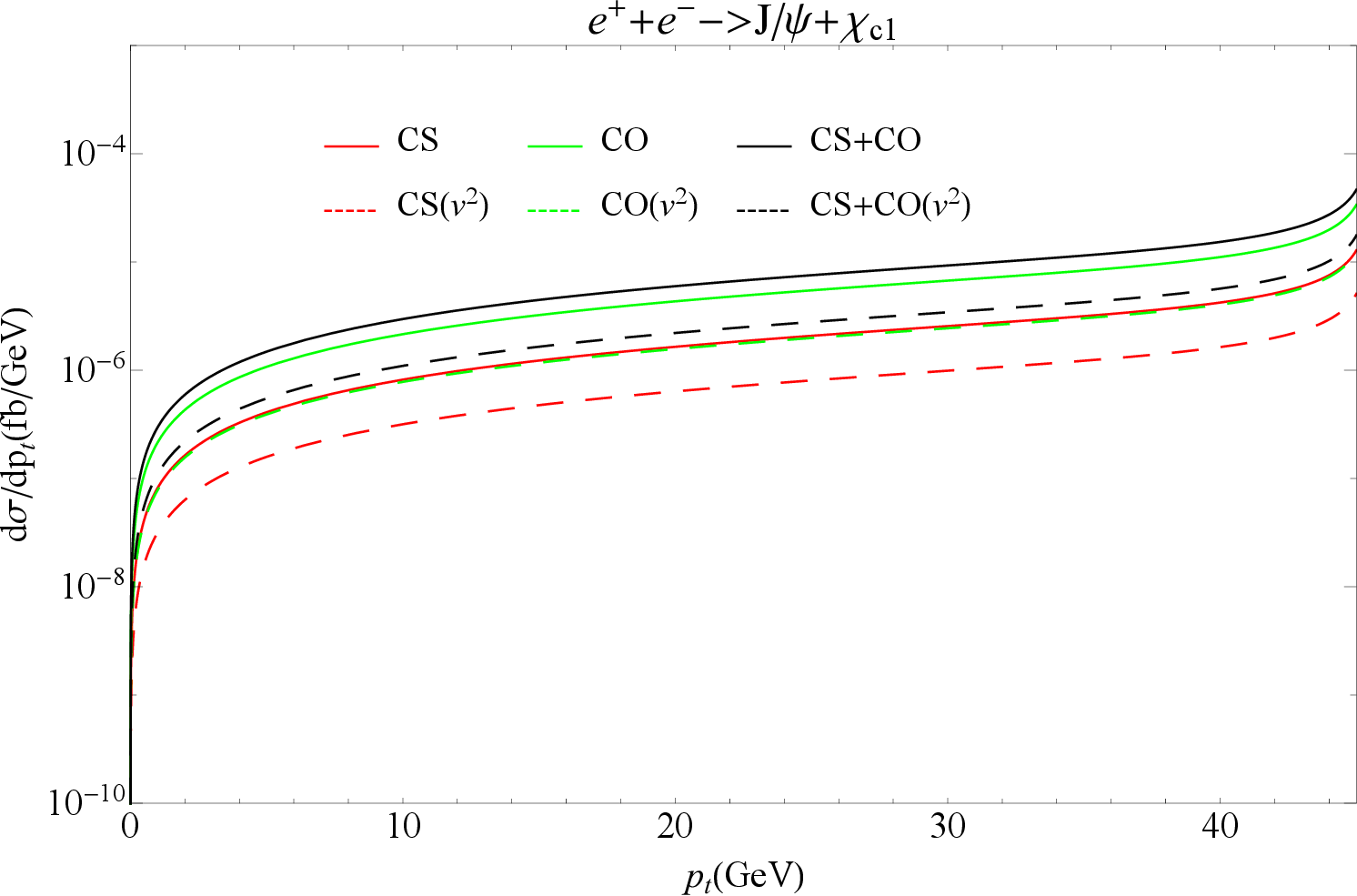}
		\end{tabular}
		\begin{tabular}{c c c }
			\includegraphics[width=0.333\textwidth]{ 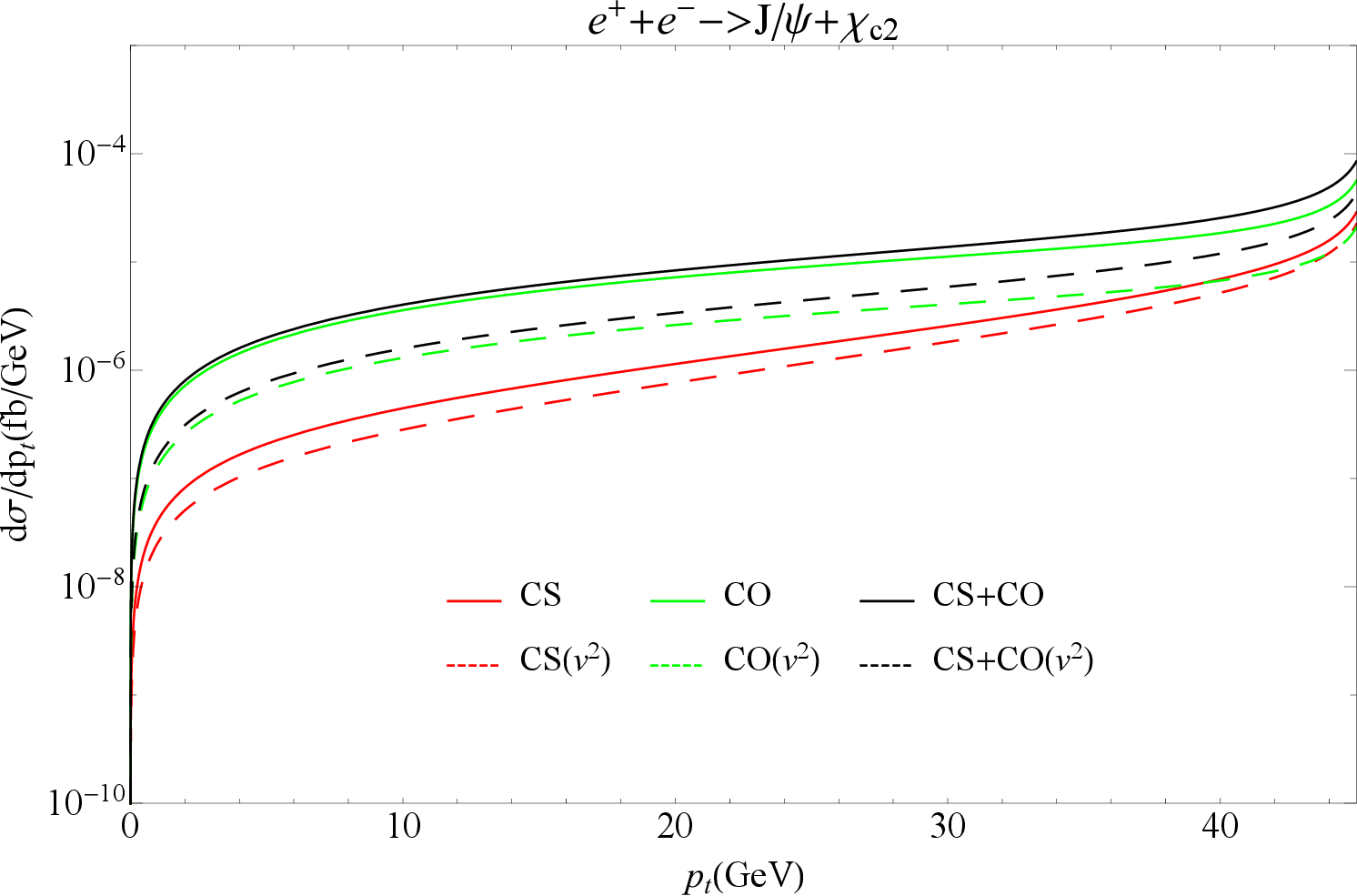}
			\includegraphics[width=0.333\textwidth]{ 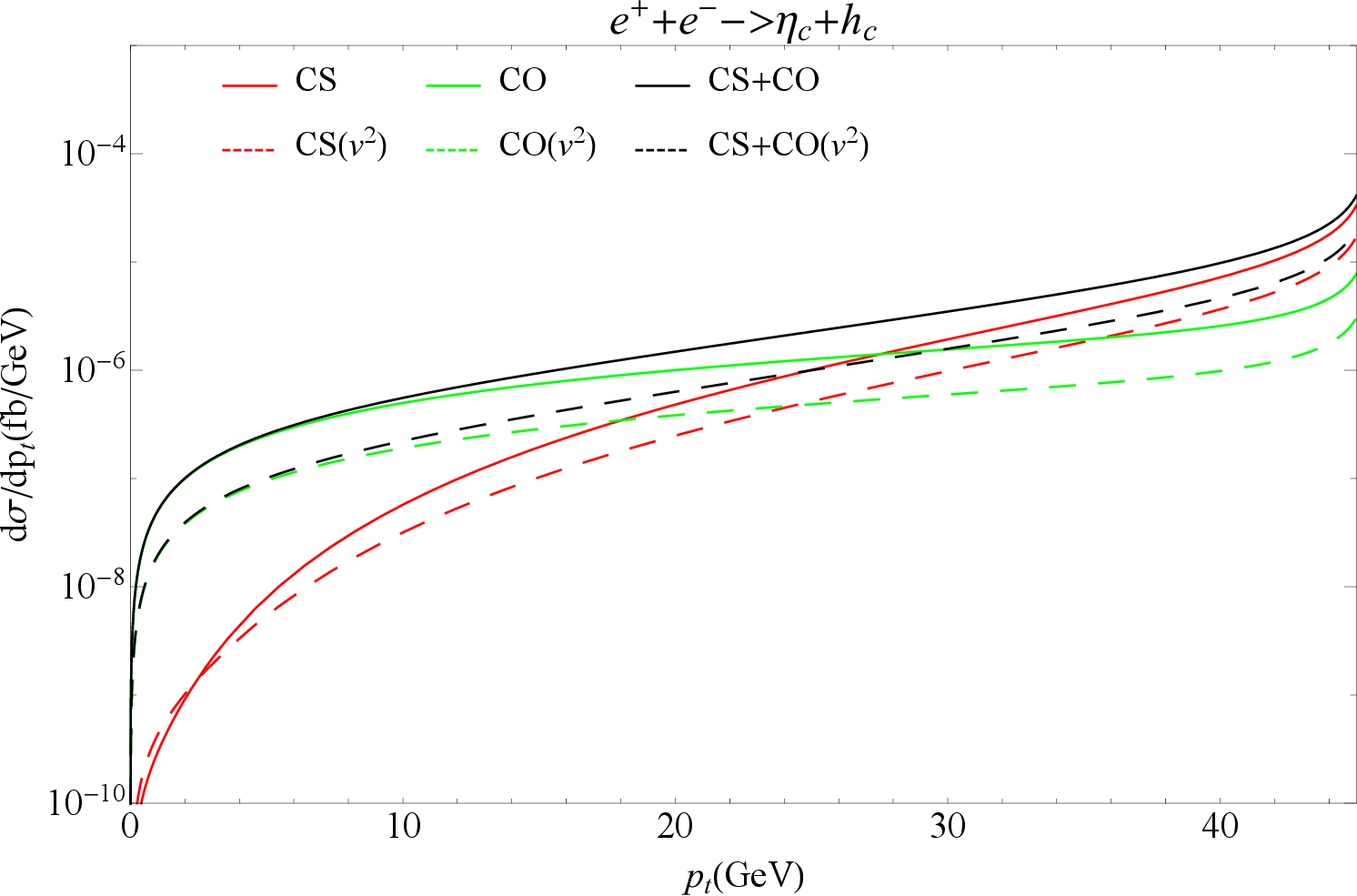}
			\includegraphics[width=0.333\textwidth]{ 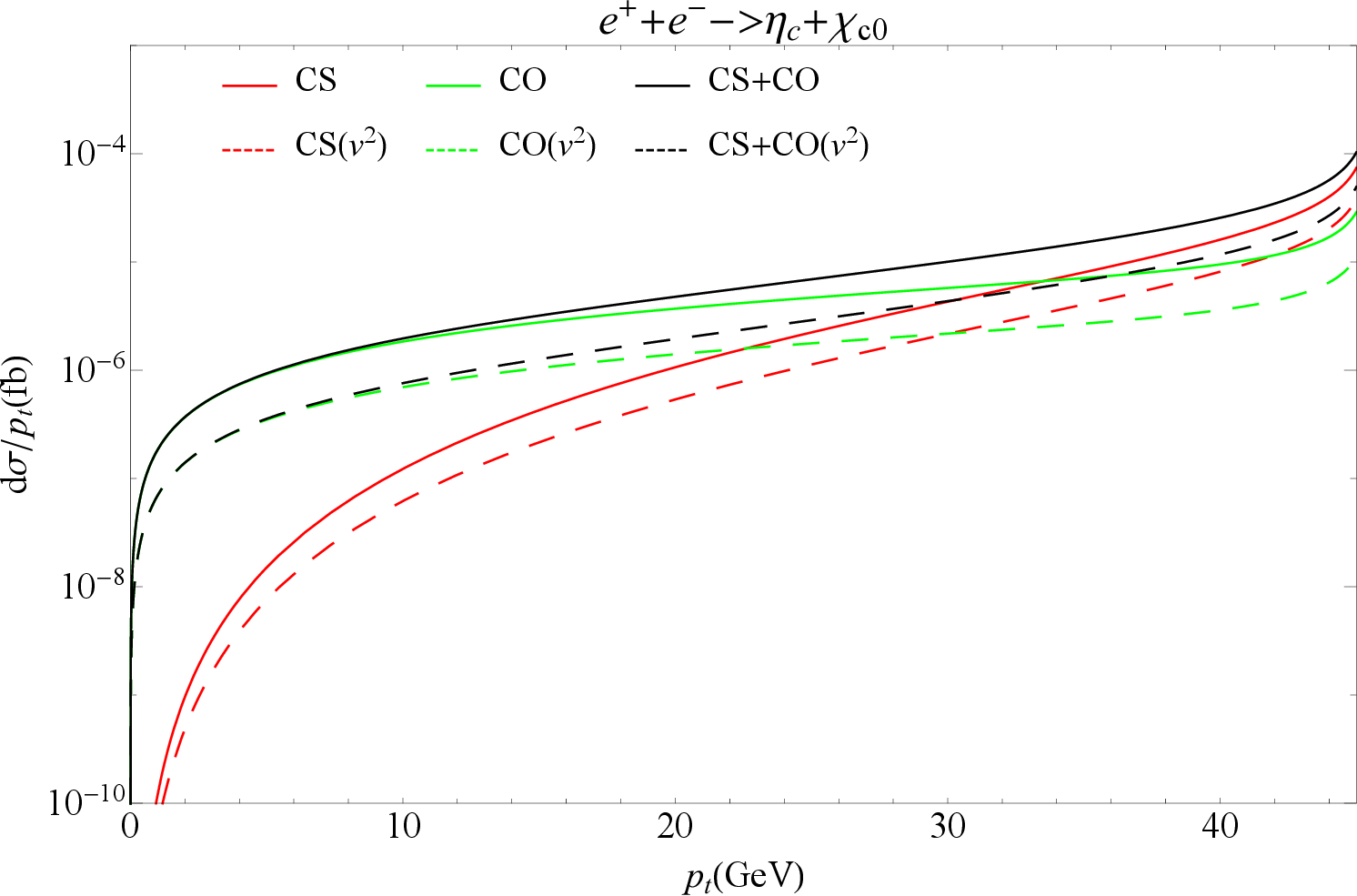}
		\end{tabular}
		\begin{tabular}{c c c }
			\includegraphics[width=0.333\textwidth]{ 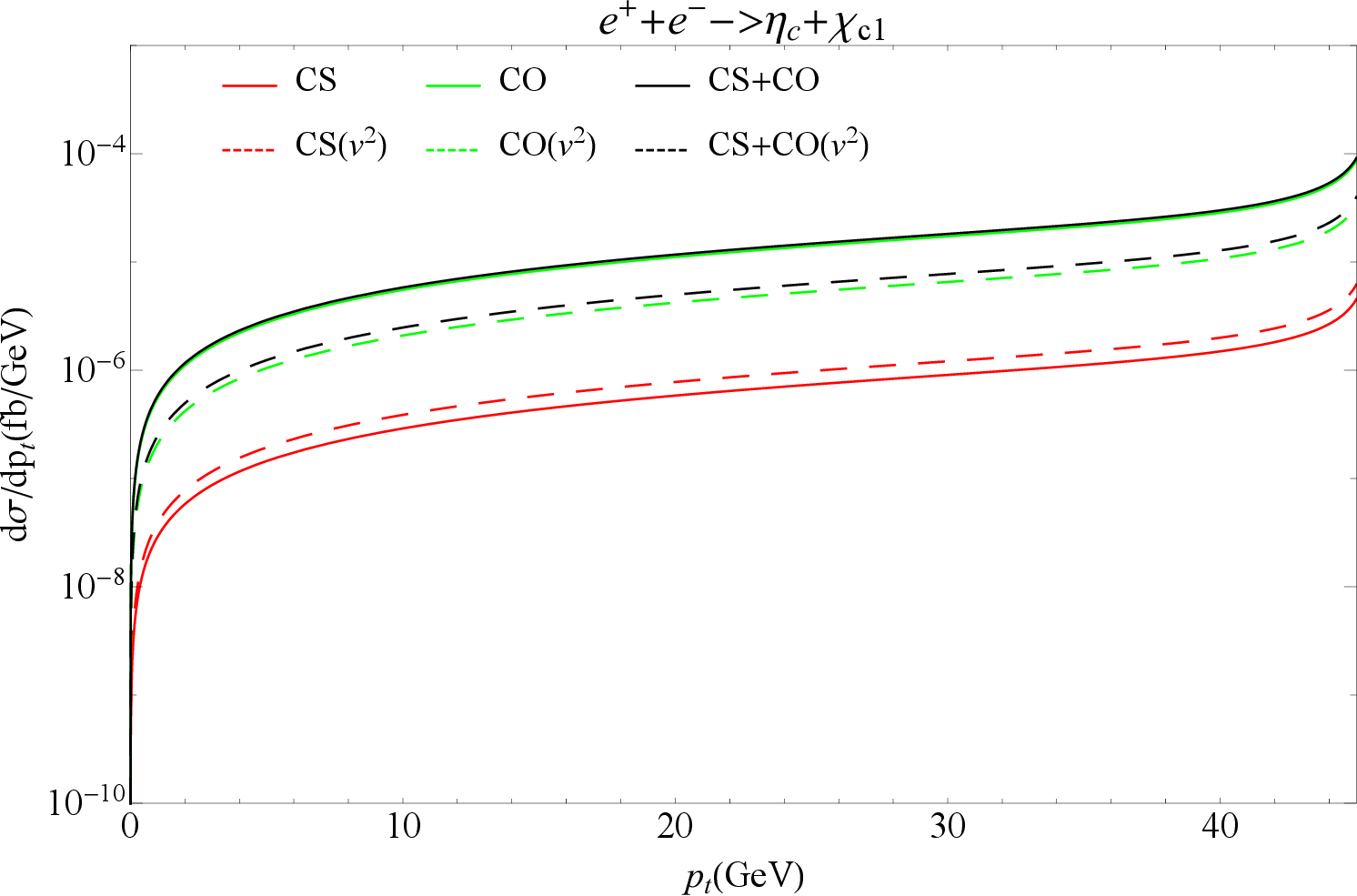}
			\includegraphics[width=0.333\textwidth]{ 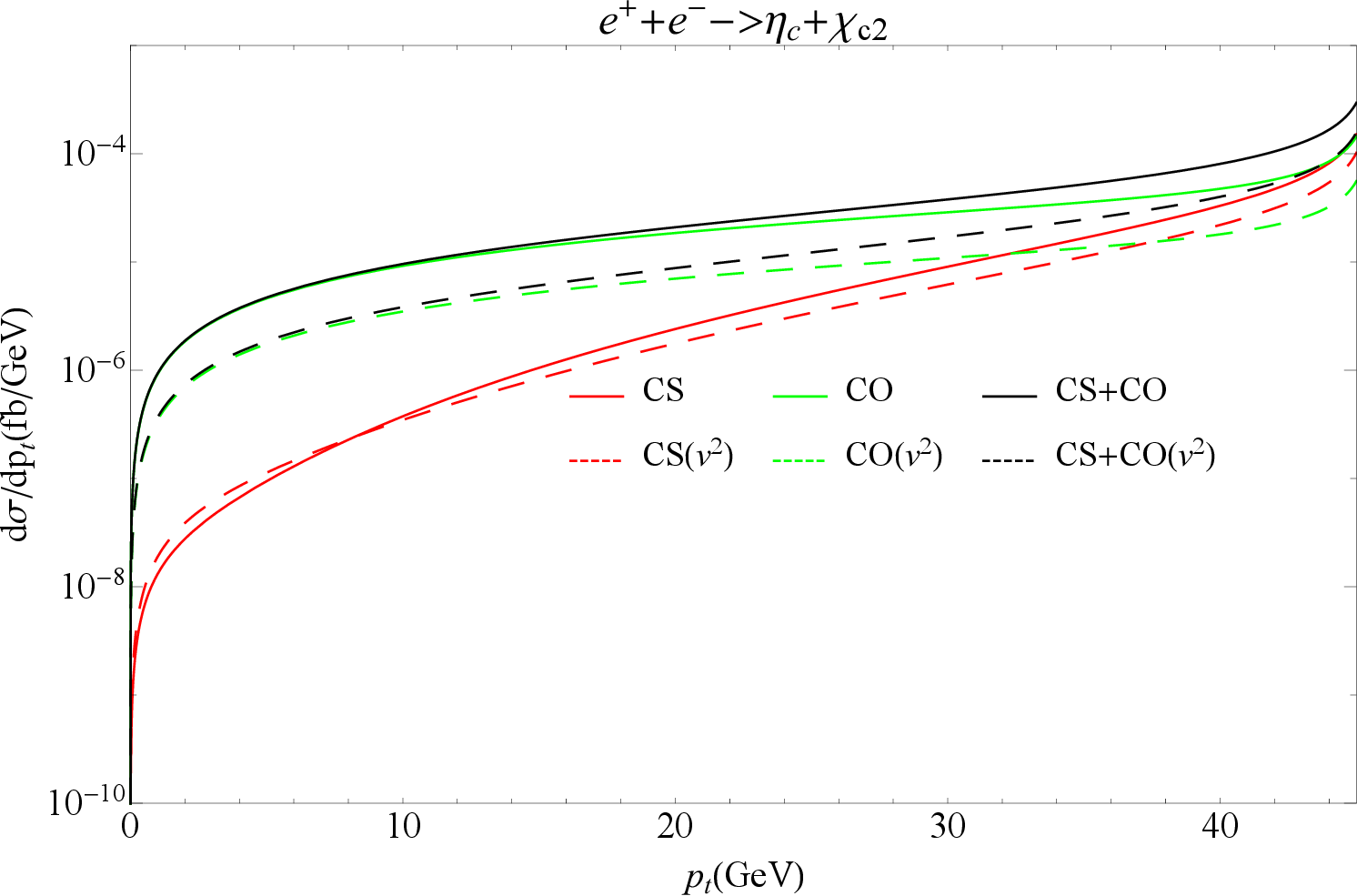}
			\includegraphics[width=0.333\textwidth]{ 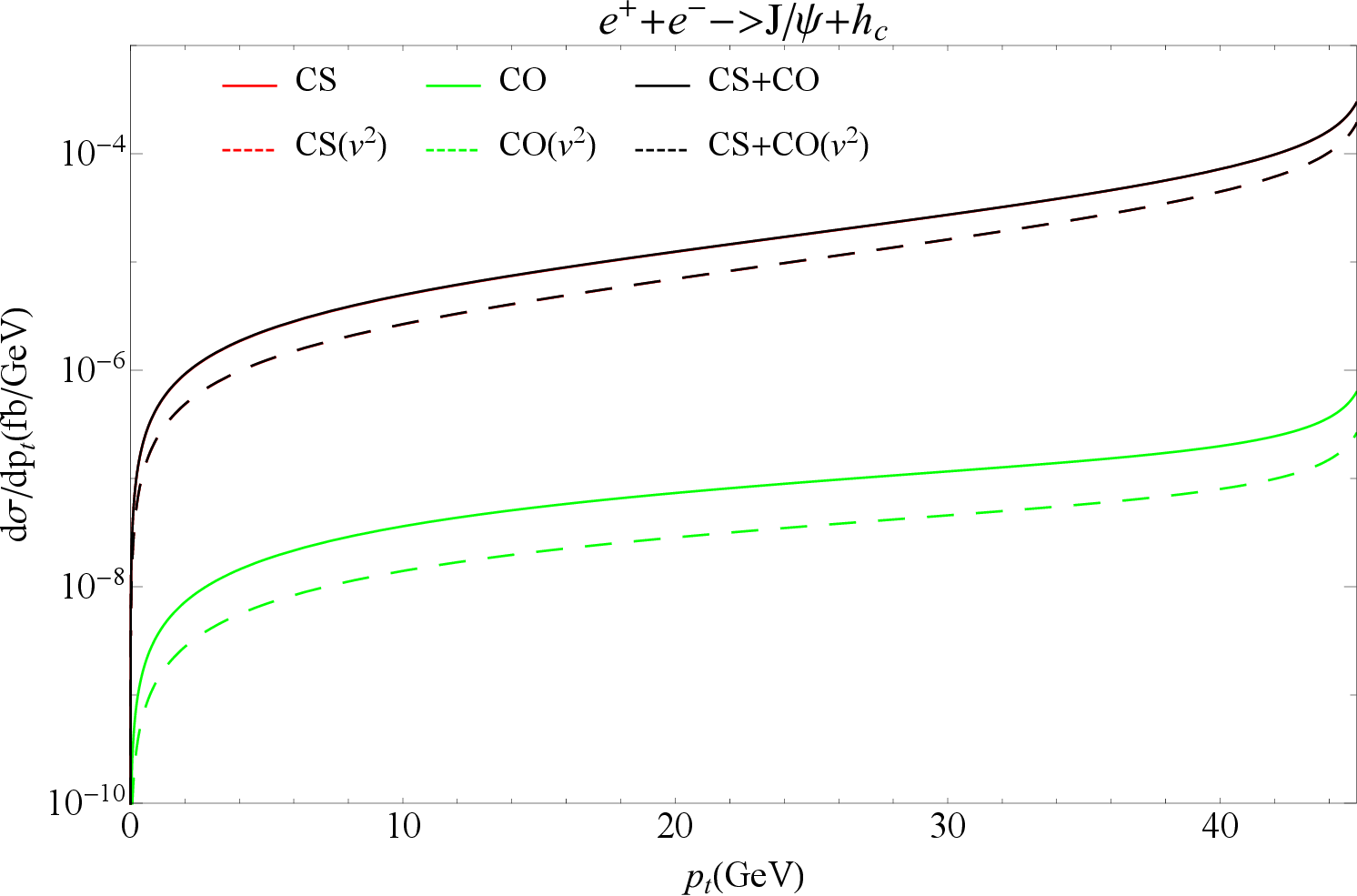}
		\end{tabular}
		\begin{tabular}{c c c}
			\includegraphics[width=0.333\textwidth]{ 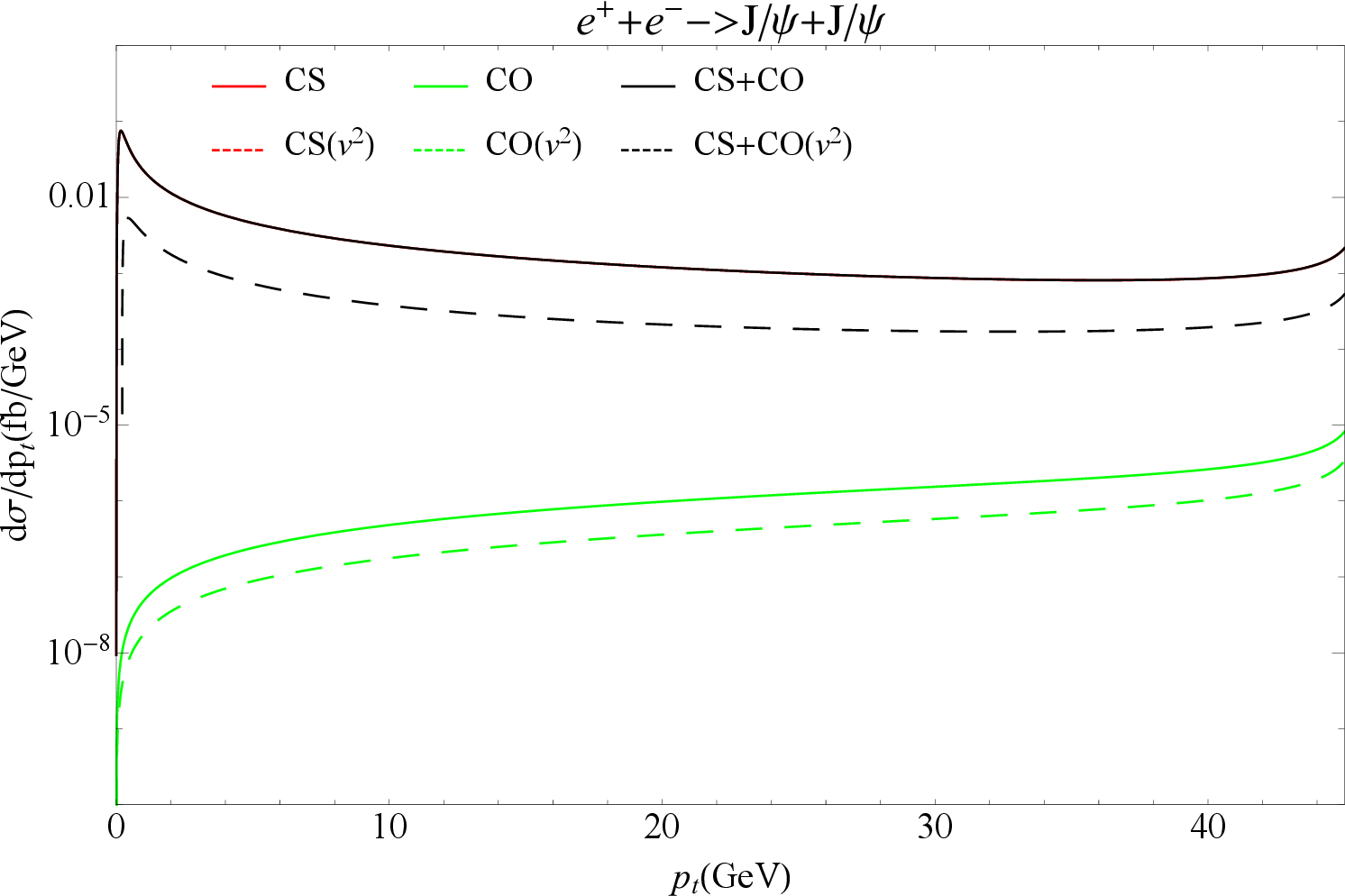}
				\includegraphics[width=0.333\textwidth]{ 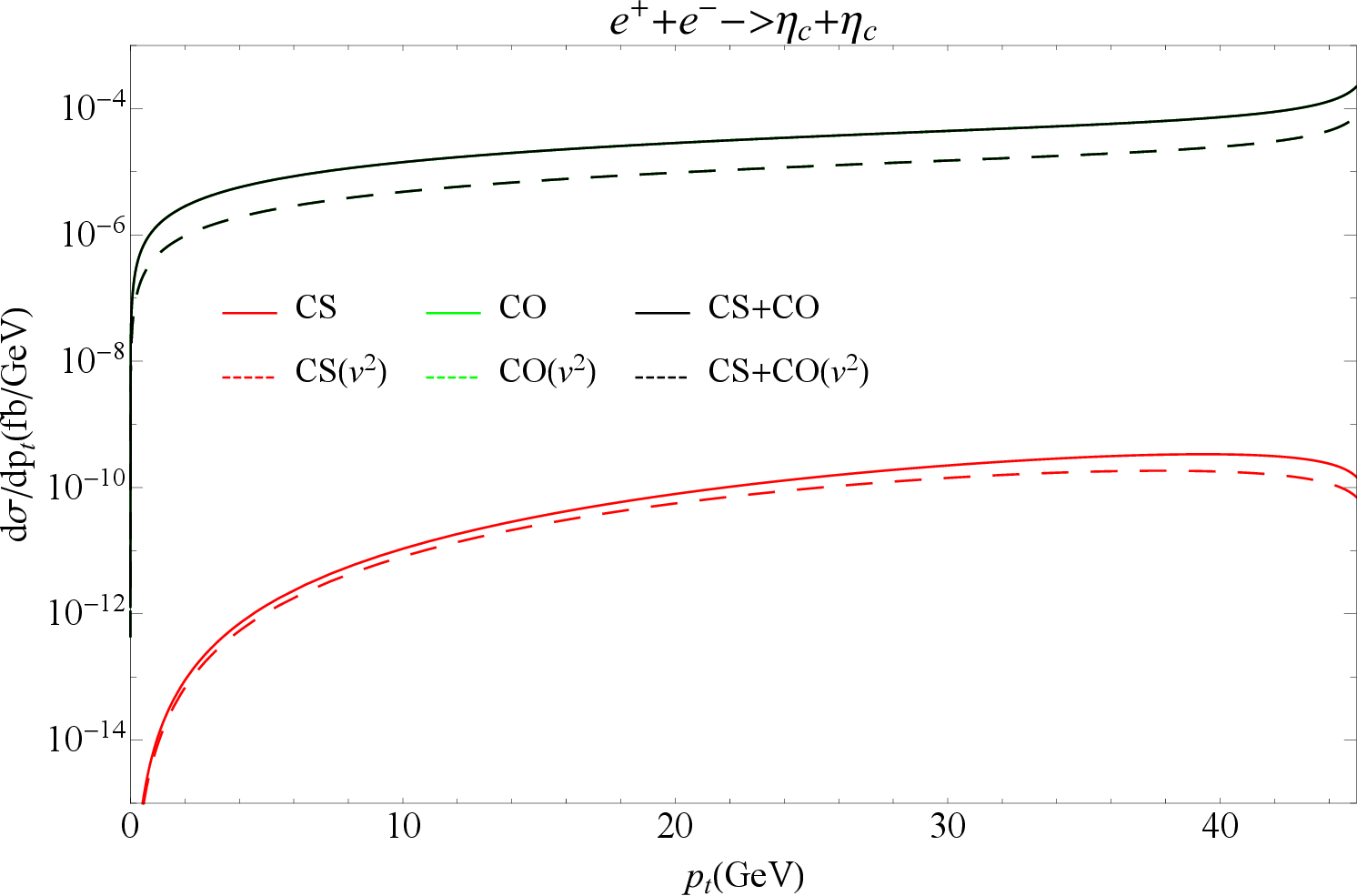}
			
		\end{tabular}
		\caption{ (Color online) The differential cross sections ($d\sigma/dp_t$) for double  charmonium production  at $\sqrt{s}$=$m_Z$. The solid line represents leading order (LO)  and the dashed line represents next-to-leading order in $v^2$ (NLO) results. The red line represents the CS channel, the green line represents the total CO channels and the black line represents the sum of  CS and CO. }
		\label{ccpt}
	\end{figure*}
		\FloatBarrier
\end{widetext}

	\begin{widetext}
	\begin{figure*}[htbp]
		\begin{tabular}{c c c }
			\includegraphics[width=0.333\textwidth]{ 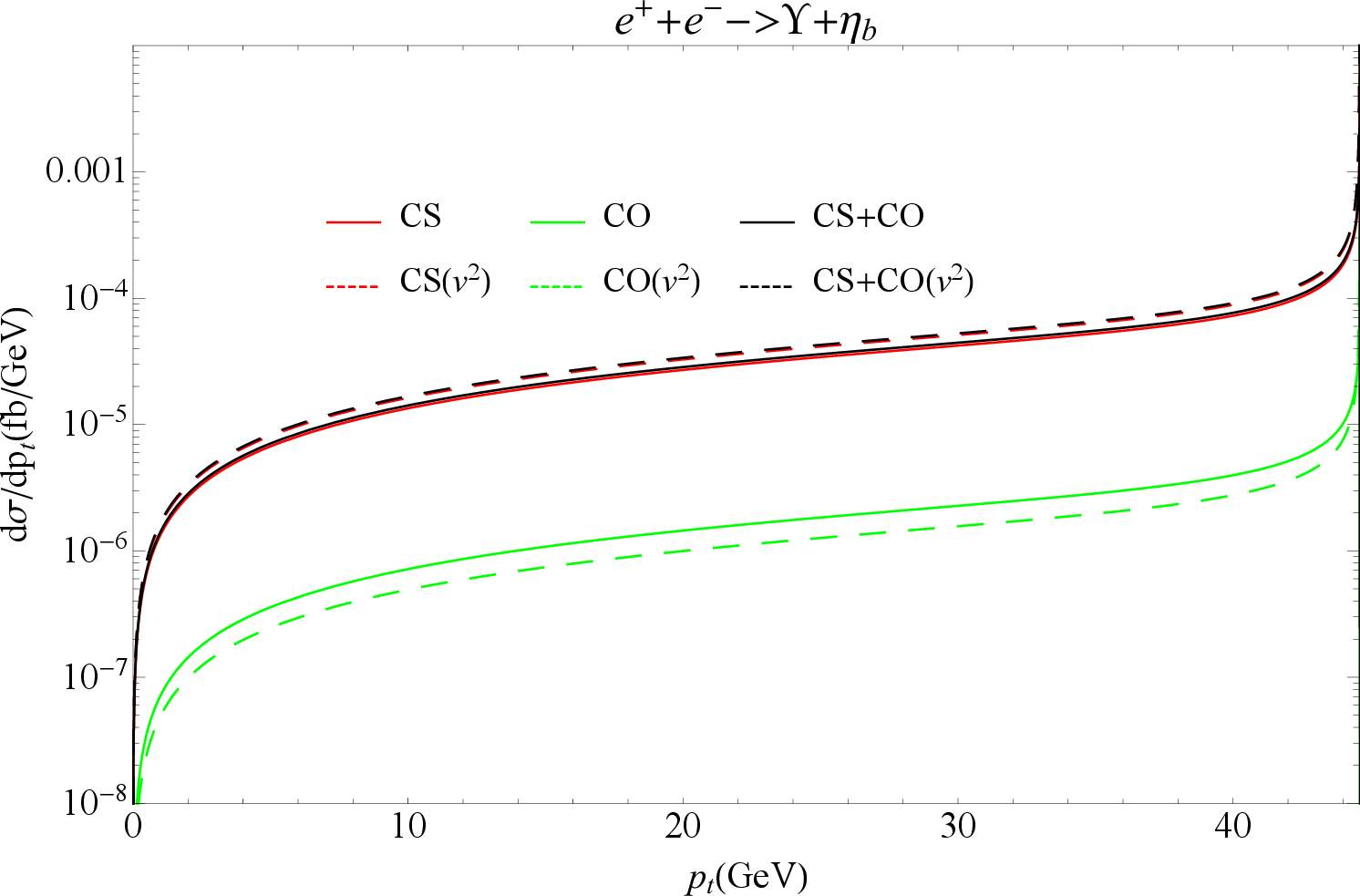}
			\includegraphics[width=0.333\textwidth]{ 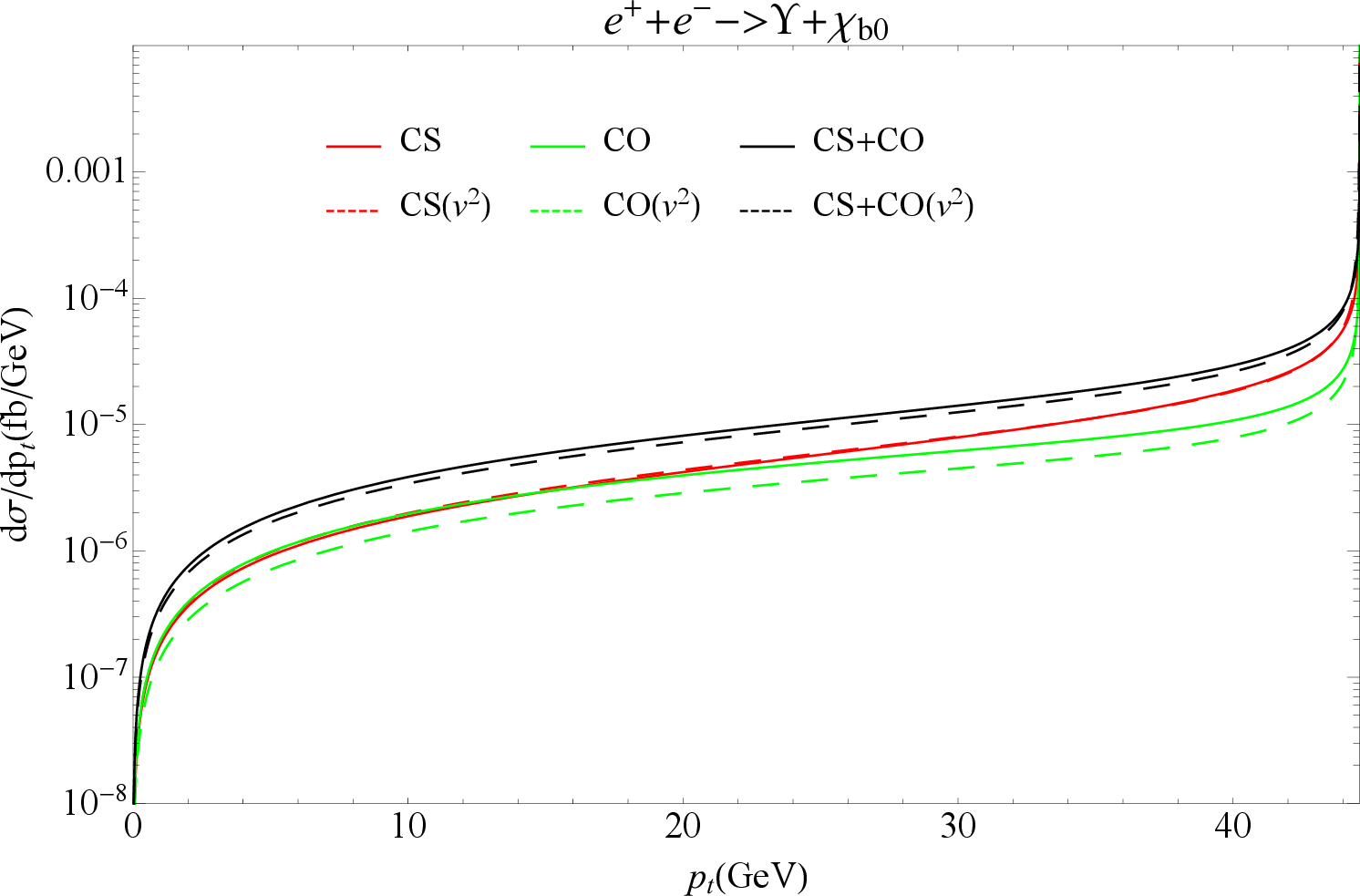}
			\includegraphics[width=0.333\textwidth]{ 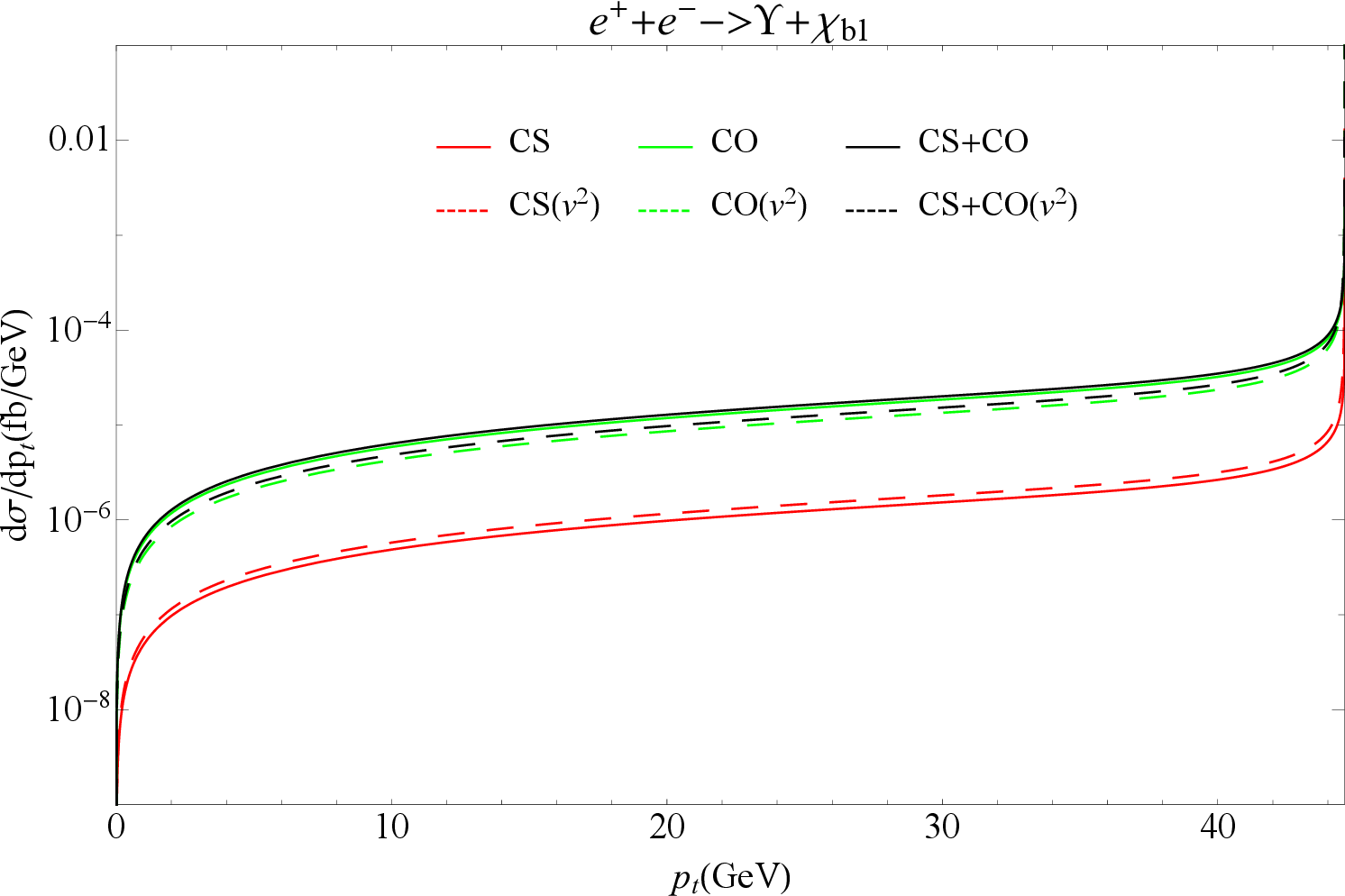}
		\end{tabular}
		\begin{tabular}{c c c }
			\includegraphics[width=0.333\textwidth]{ 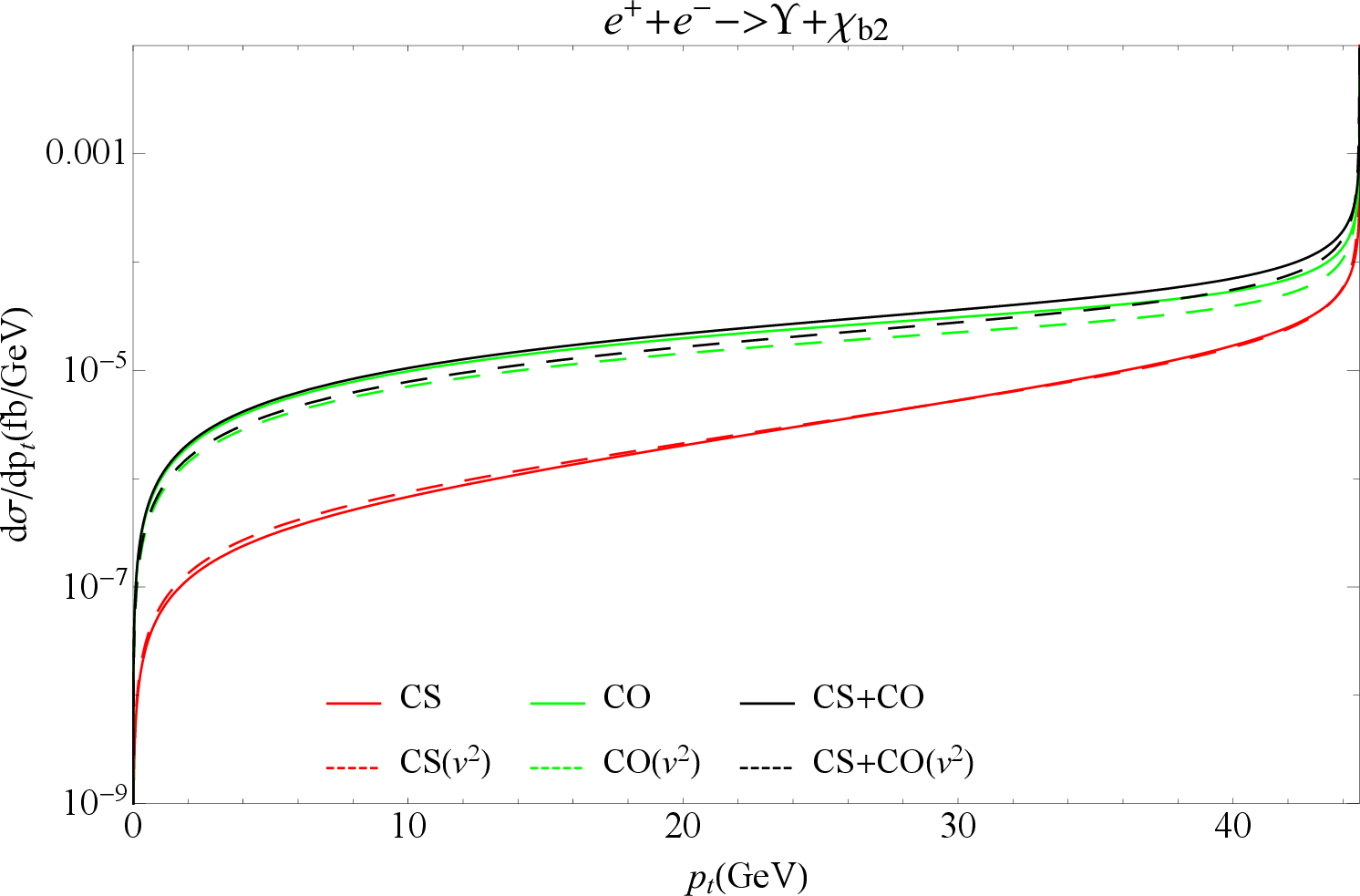}
			\includegraphics[width=0.333\textwidth]{ 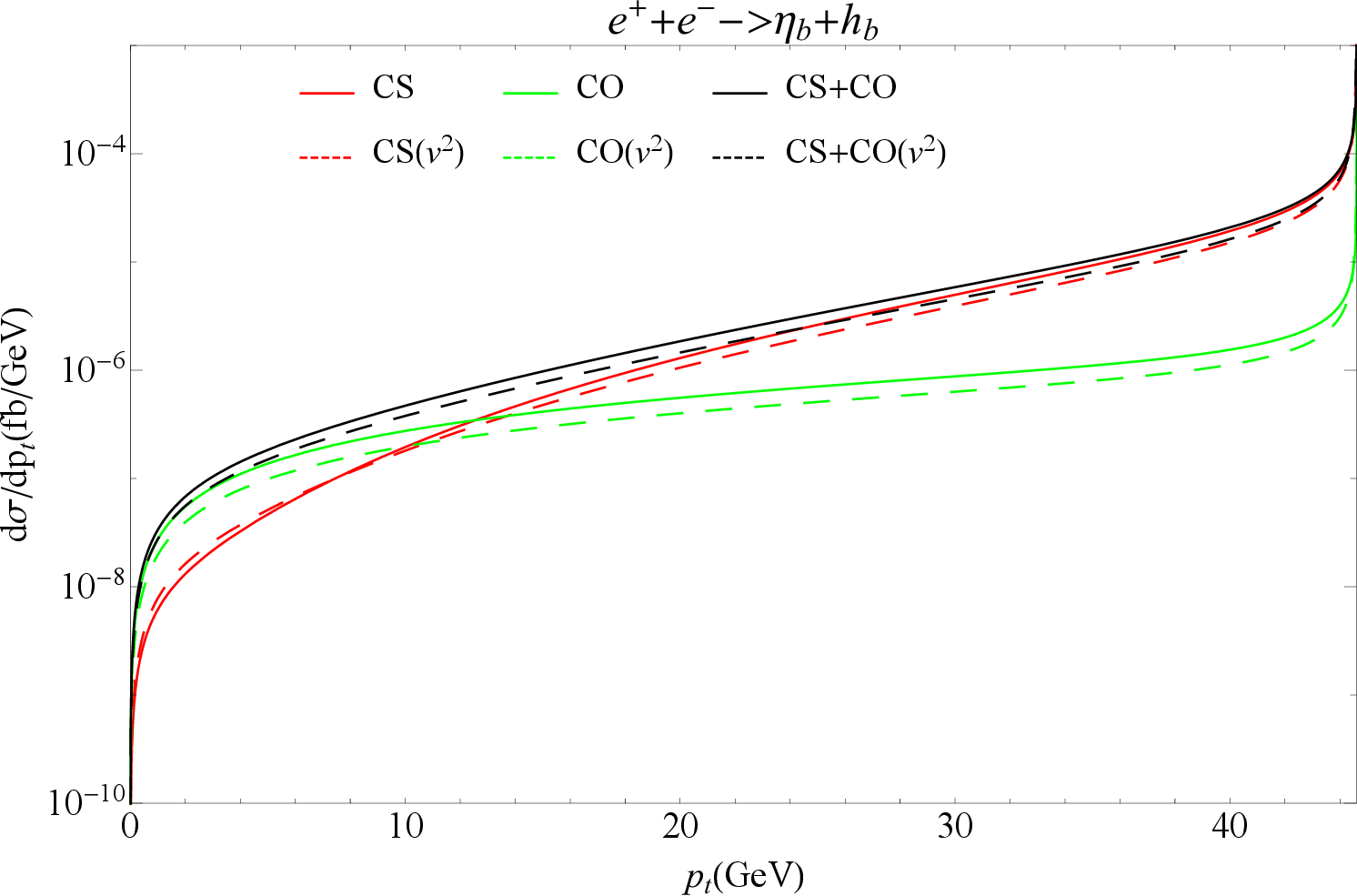}
			\includegraphics[width=0.333\textwidth]{ 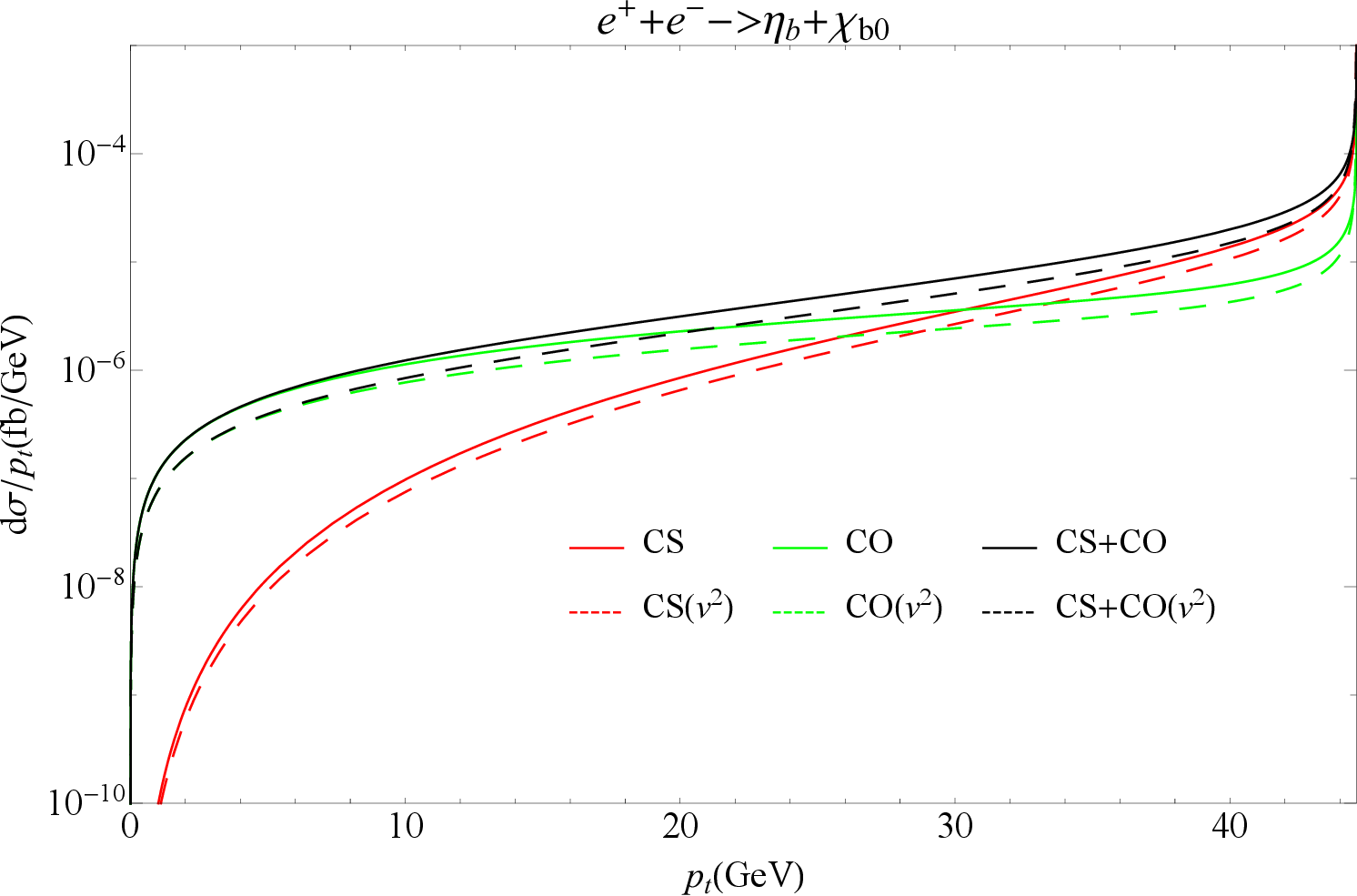}
		\end{tabular}
		\begin{tabular}{c c c }
			\includegraphics[width=0.333\textwidth]{ 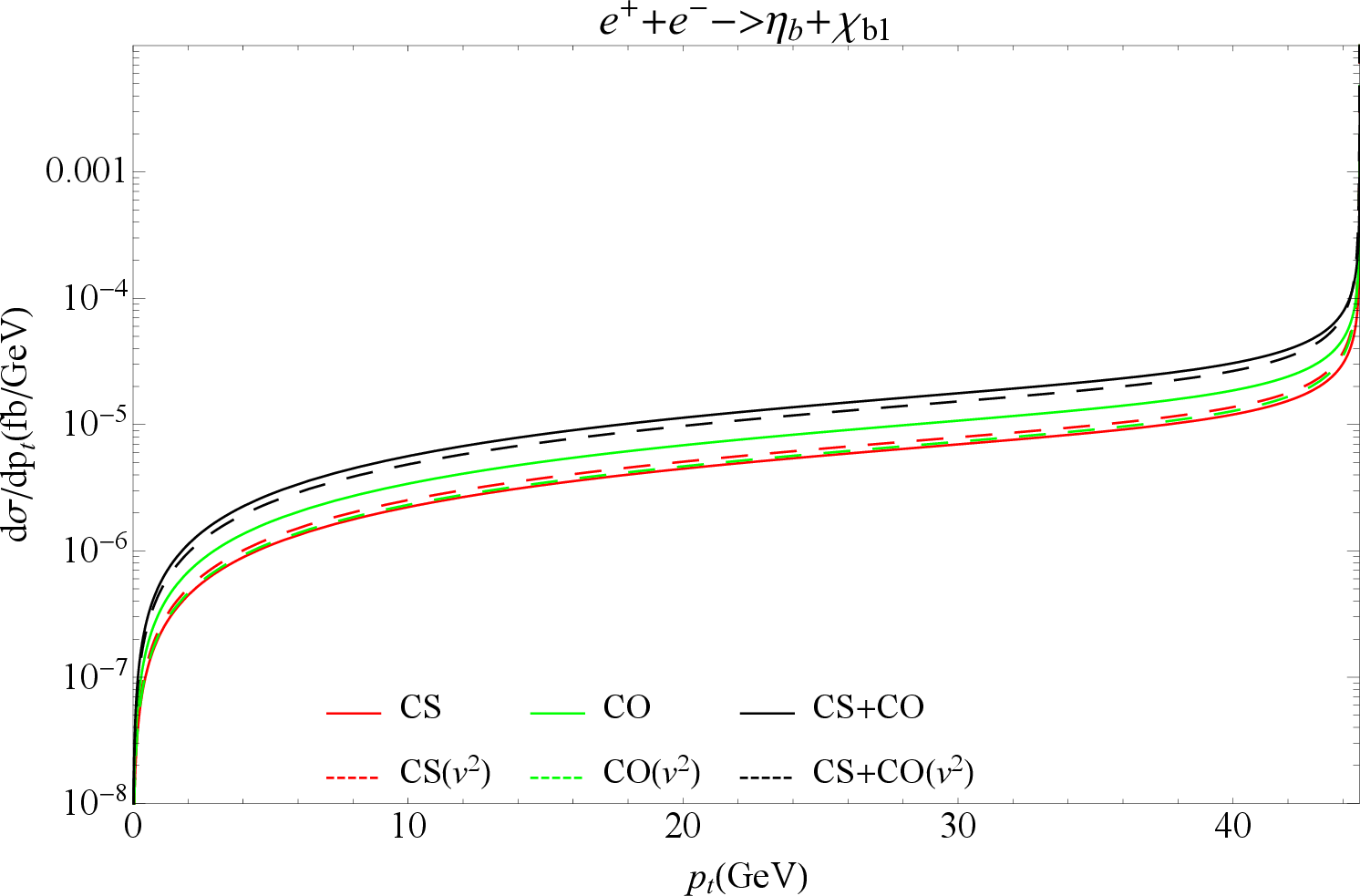}
			\includegraphics[width=0.333\textwidth]{ 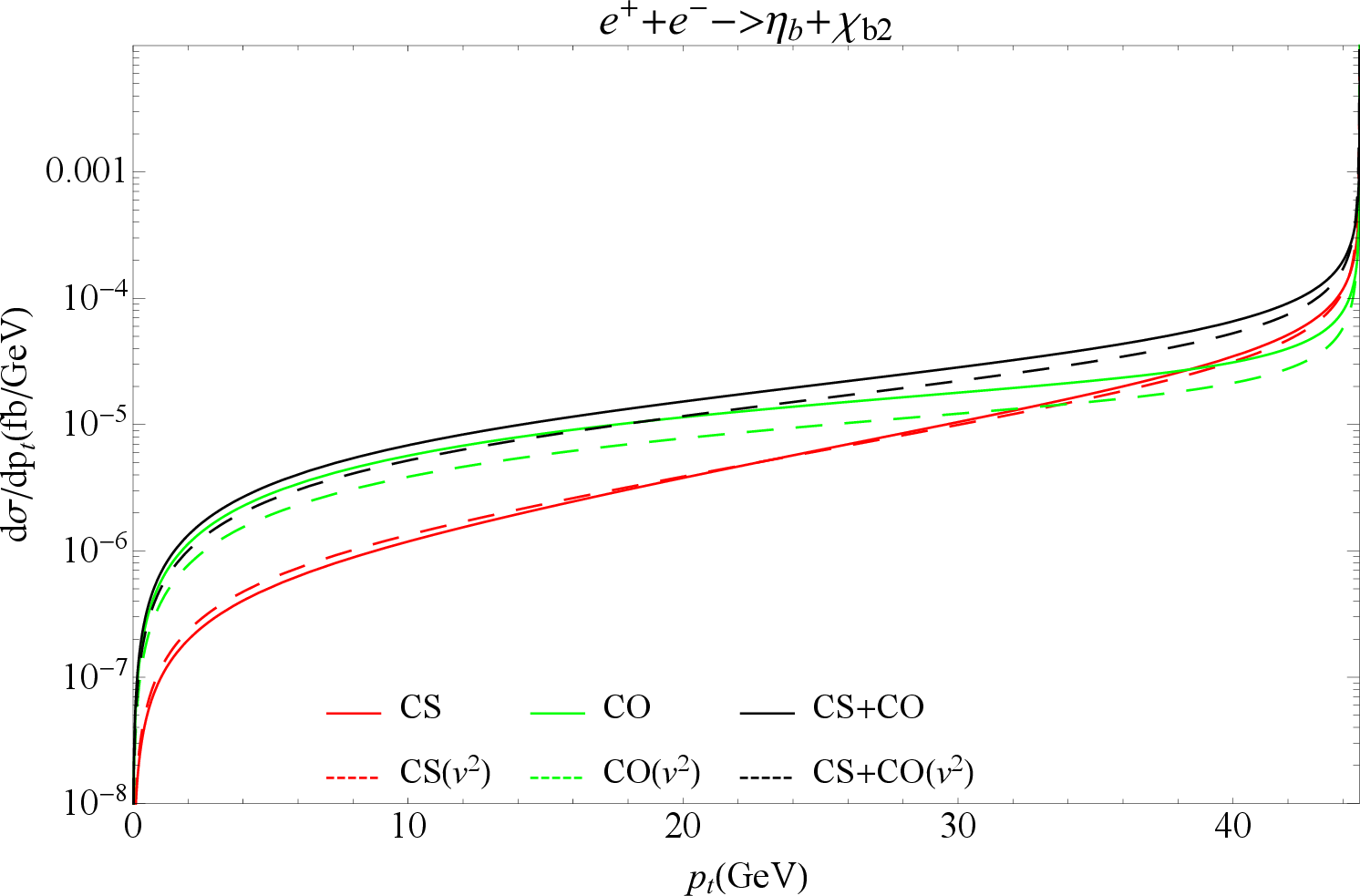}
			\includegraphics[width=0.333\textwidth]{ 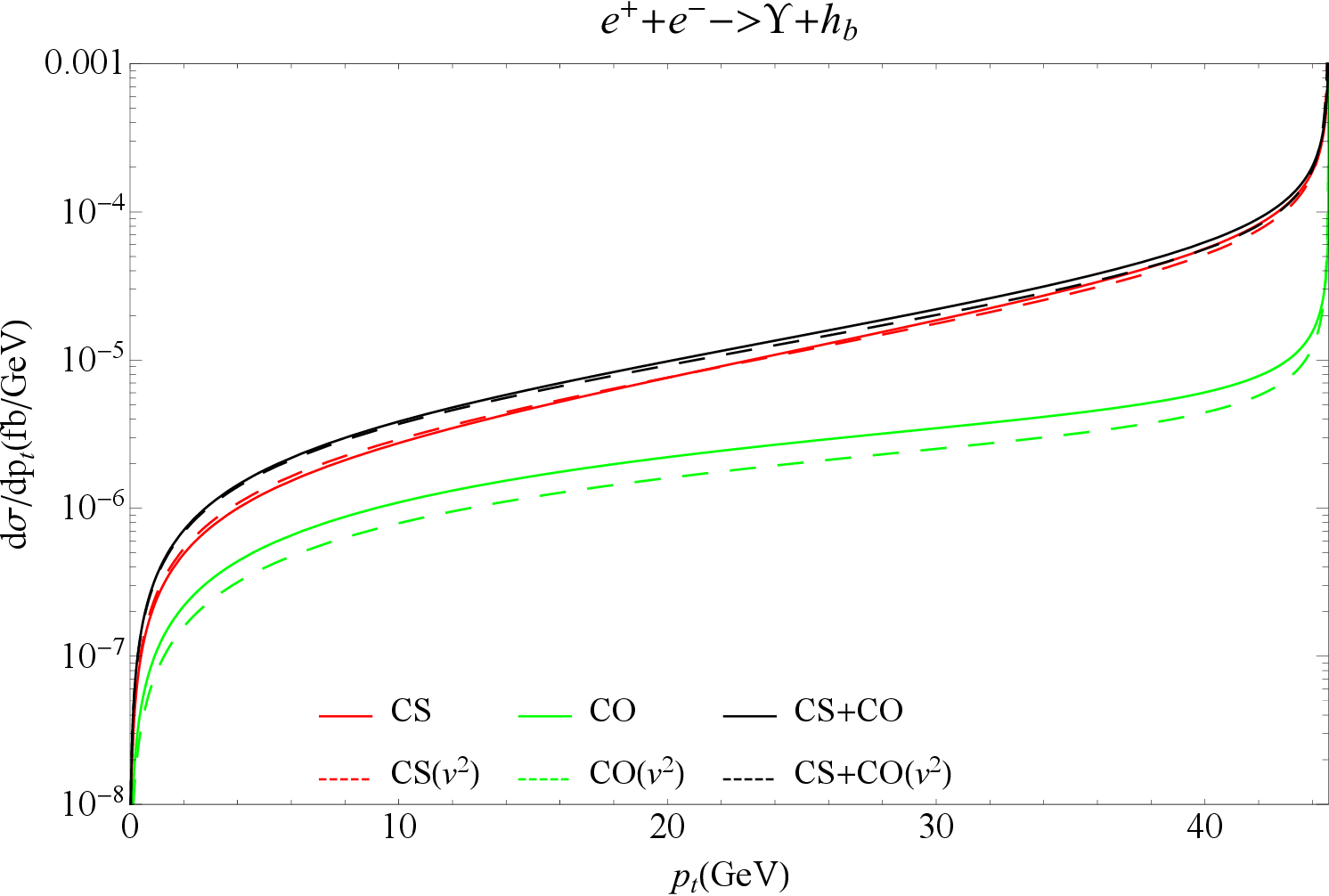}
		\end{tabular}
		\begin{tabular}{c c c}
			\includegraphics[width=0.333\textwidth]{ 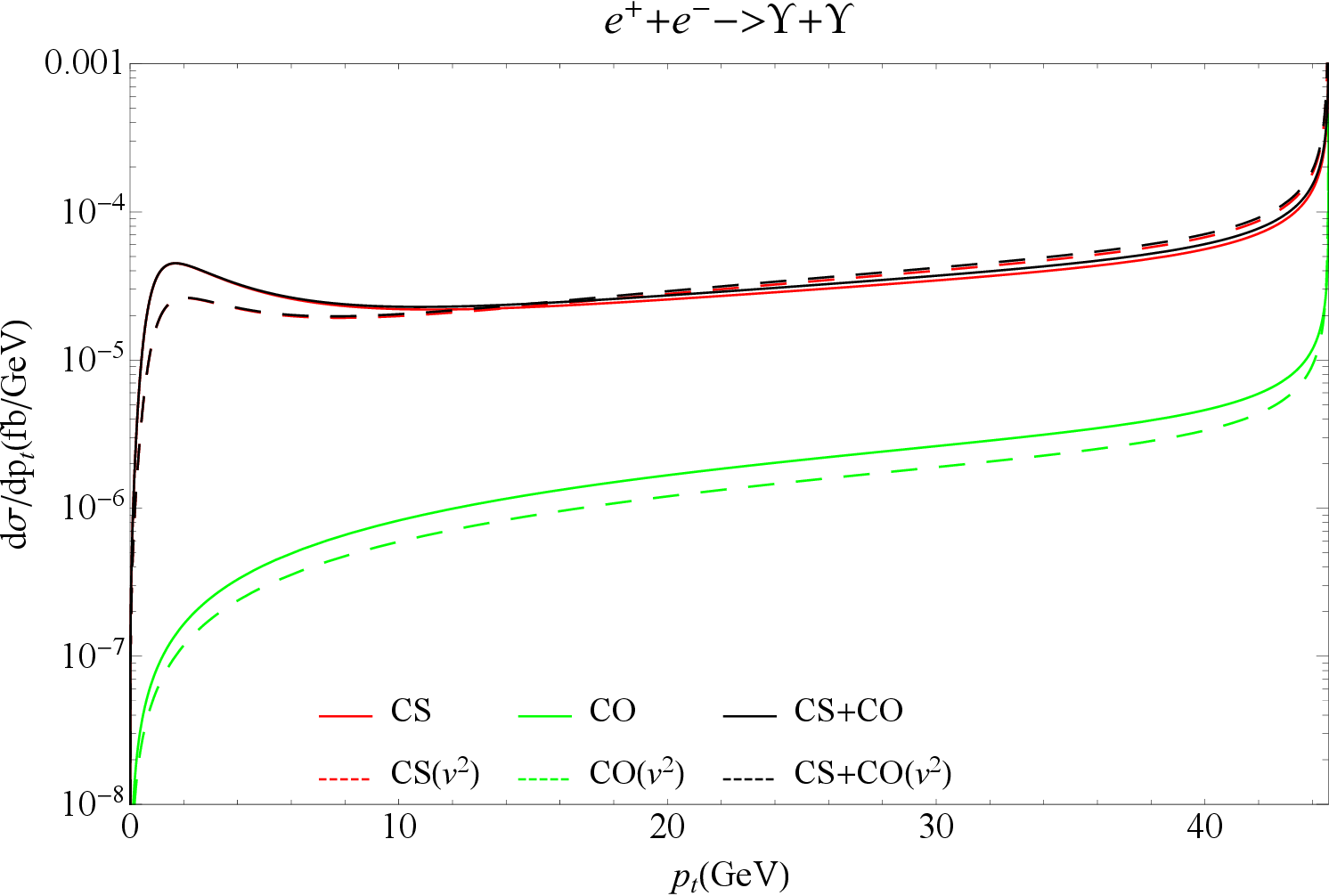}
				\includegraphics[width=0.333\textwidth]{ 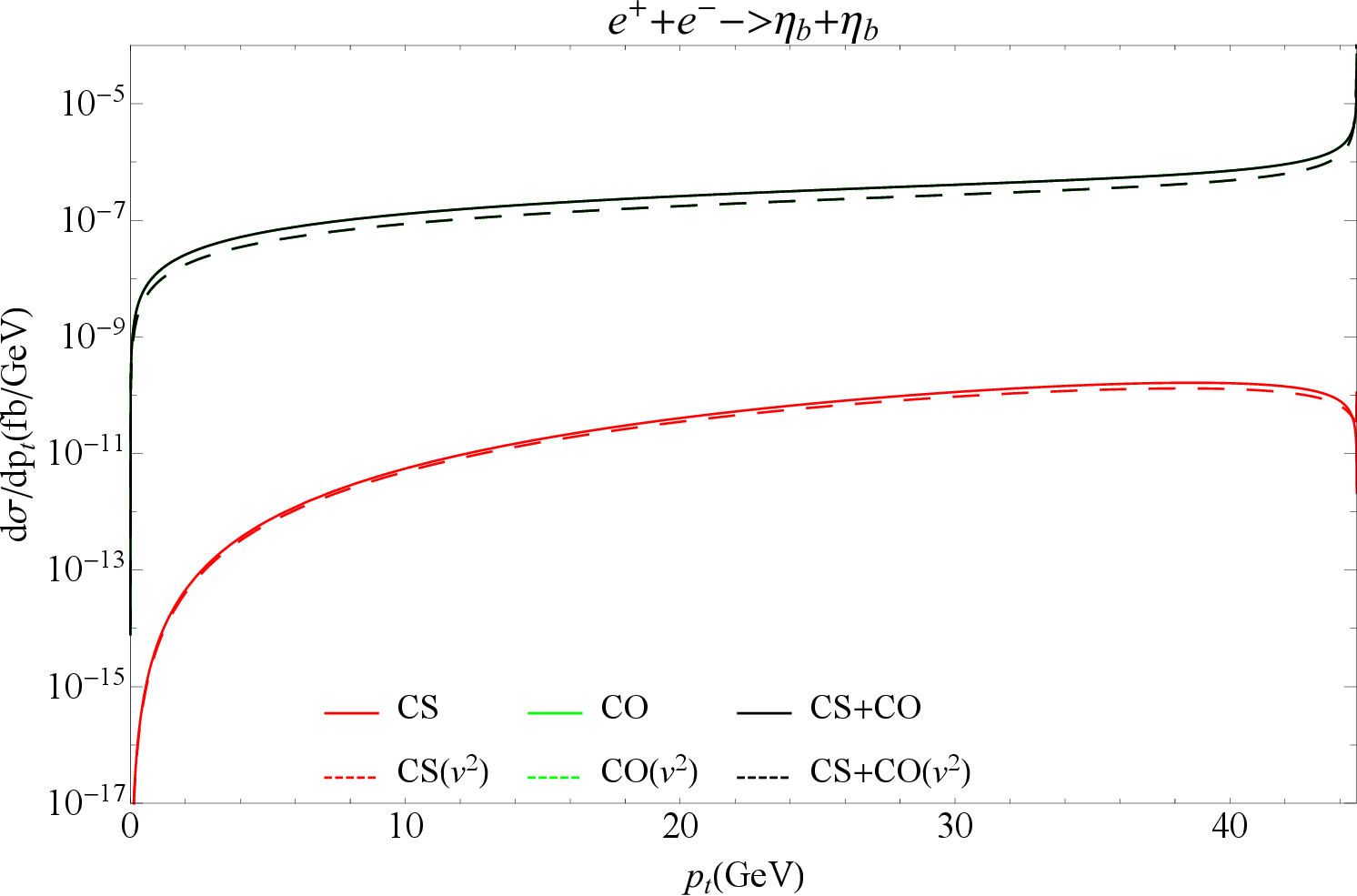}
		\end{tabular}
		\caption{ (Color online) The differential cross section ($d\sigma/dp_t$) for double  bottomonium production  at $\sqrt{s}$=$m_Z$. The solid line represents leading order (LO)  and the dashed line represents next-to-leading order in $v^2$ (NLO) results. The red line represents the CS channel, the green line represents the total CO channels and the black line represents the sum of  CS and CO. }
		\label{bbpt}
	\end{figure*}
	\FloatBarrier
\end{widetext}

\begin{widetext}
	\begin{figure*}[htbp]
		\begin{tabular}{c c c }
			\includegraphics[width=0.333\textwidth]{ 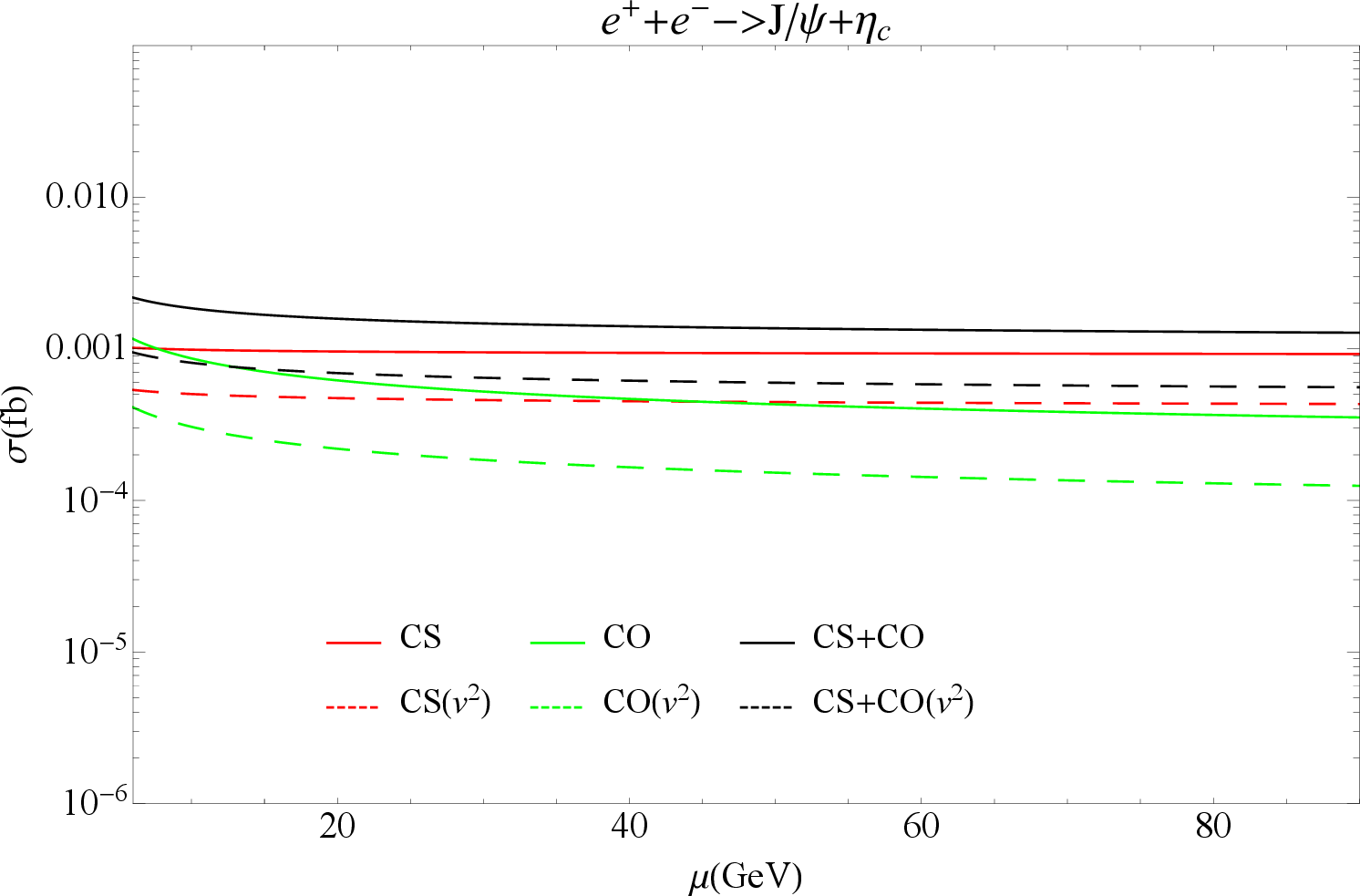}
			\includegraphics[width=0.333\textwidth]{ 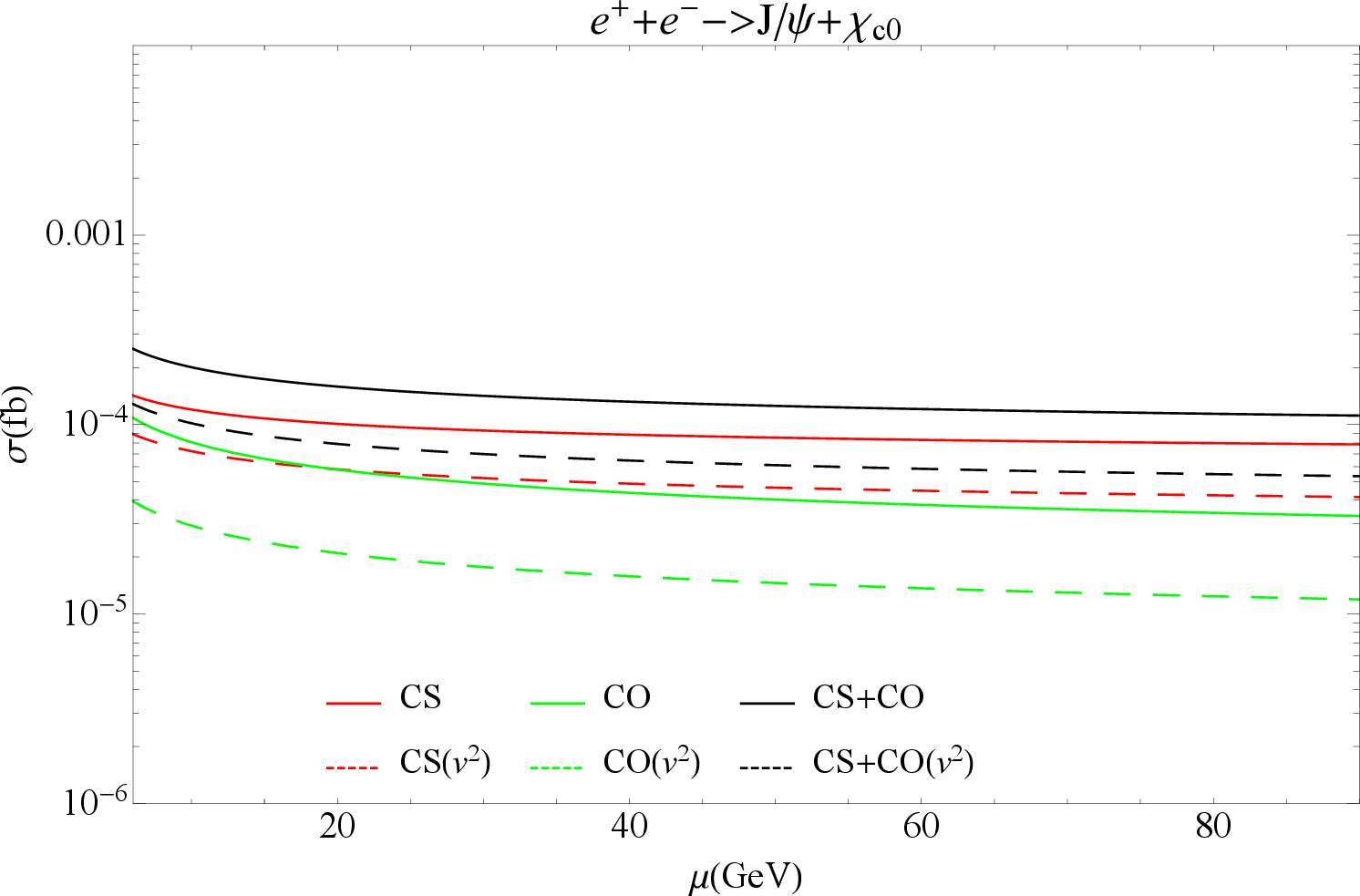}
			\includegraphics[width=0.333\textwidth]{ 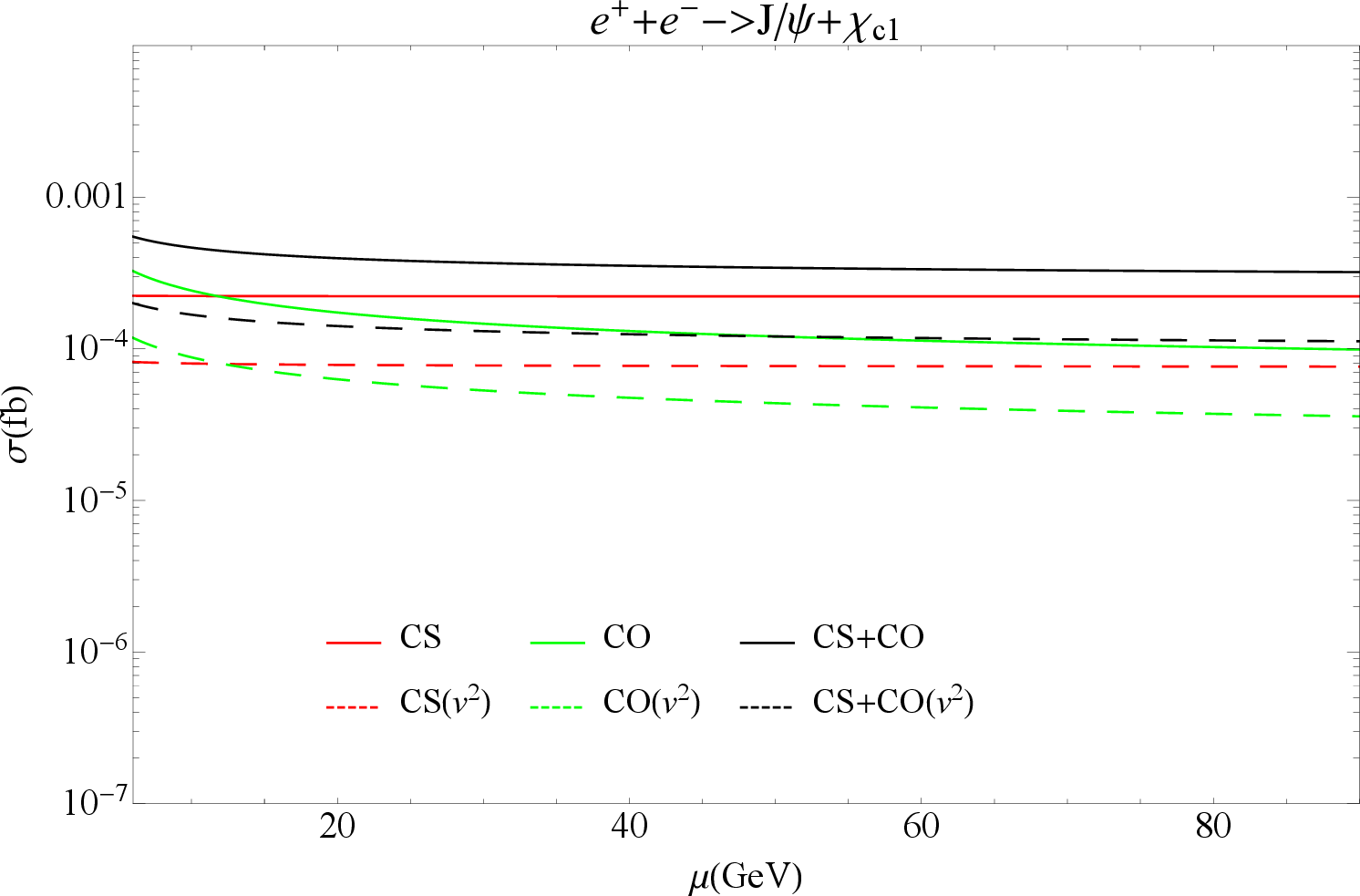}
		\end{tabular}
		\begin{tabular}{c c c }
			\includegraphics[width=0.333\textwidth]{ 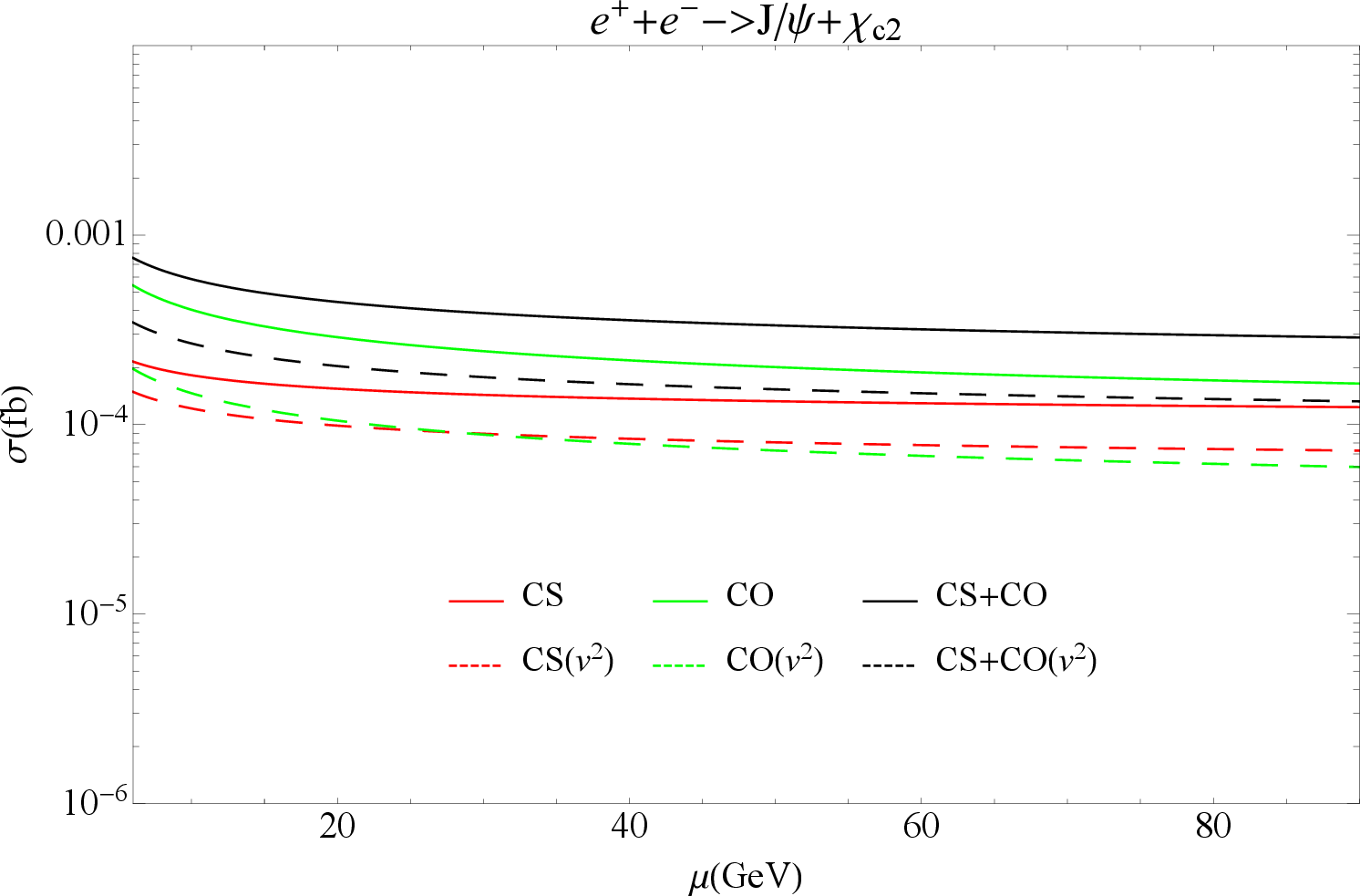}
			\includegraphics[width=0.333\textwidth]{ 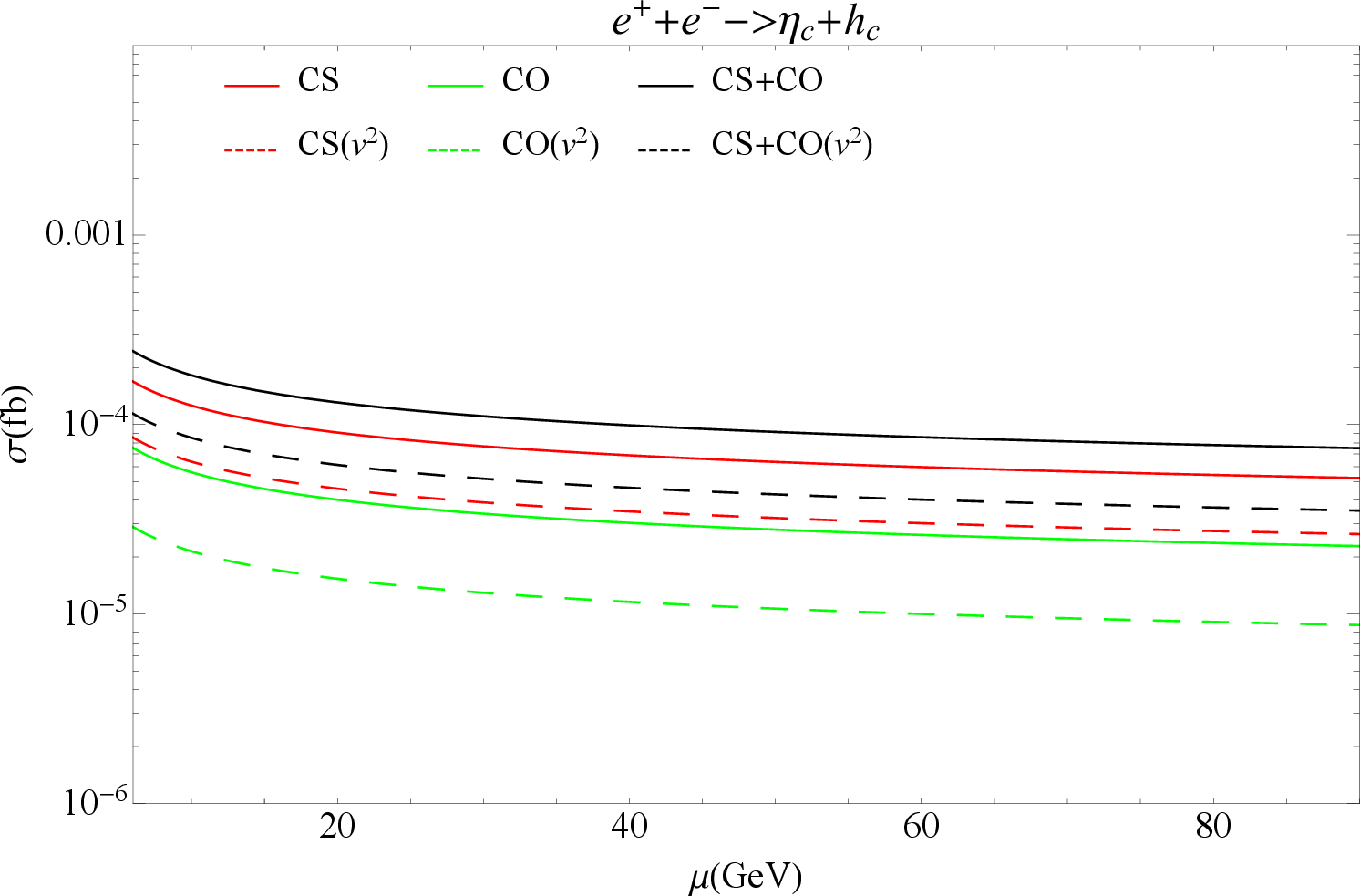}
			\includegraphics[width=0.333\textwidth]{ 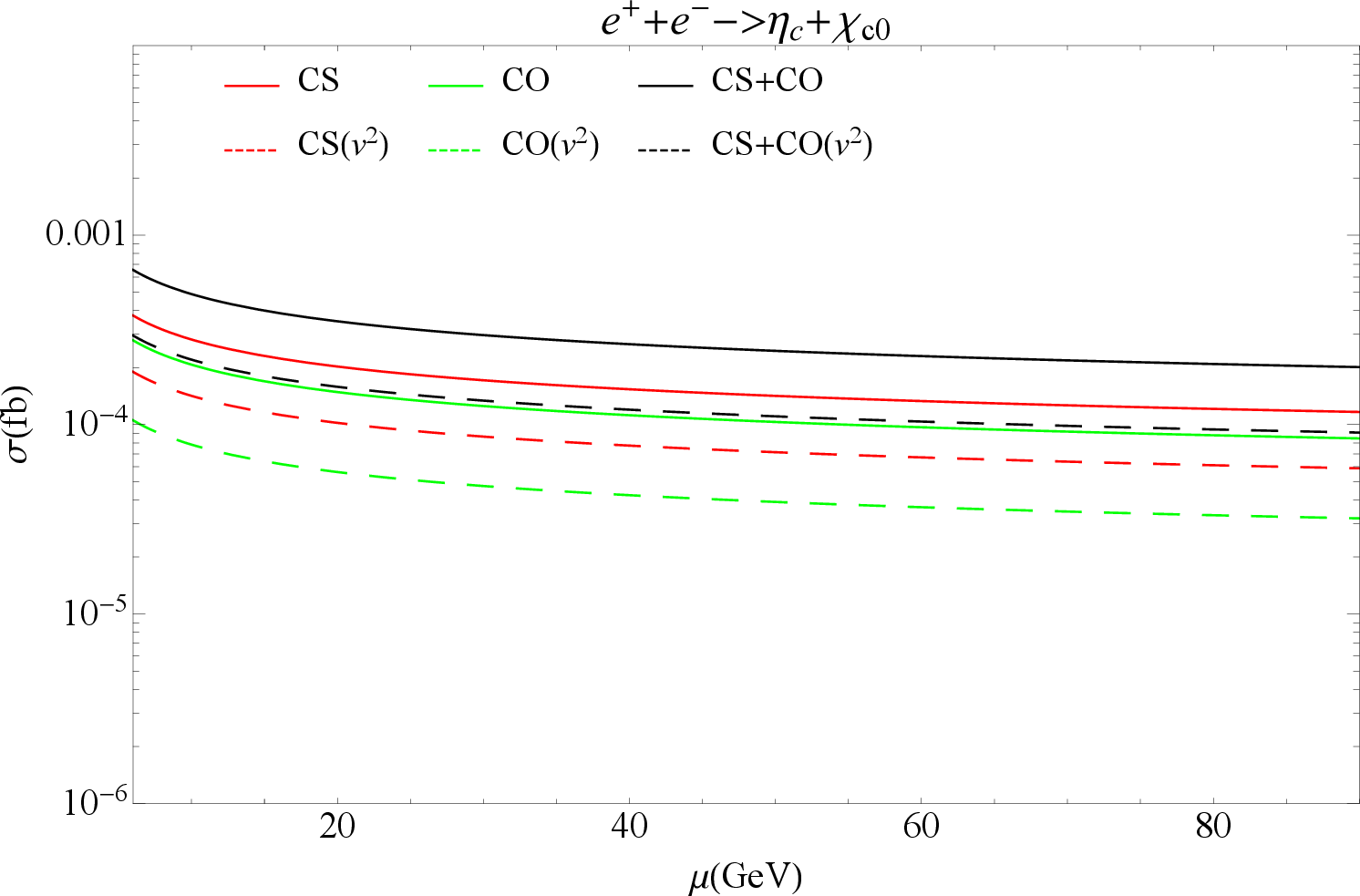}
		\end{tabular}
		\begin{tabular}{c c c }
			\includegraphics[width=0.333\textwidth]{ 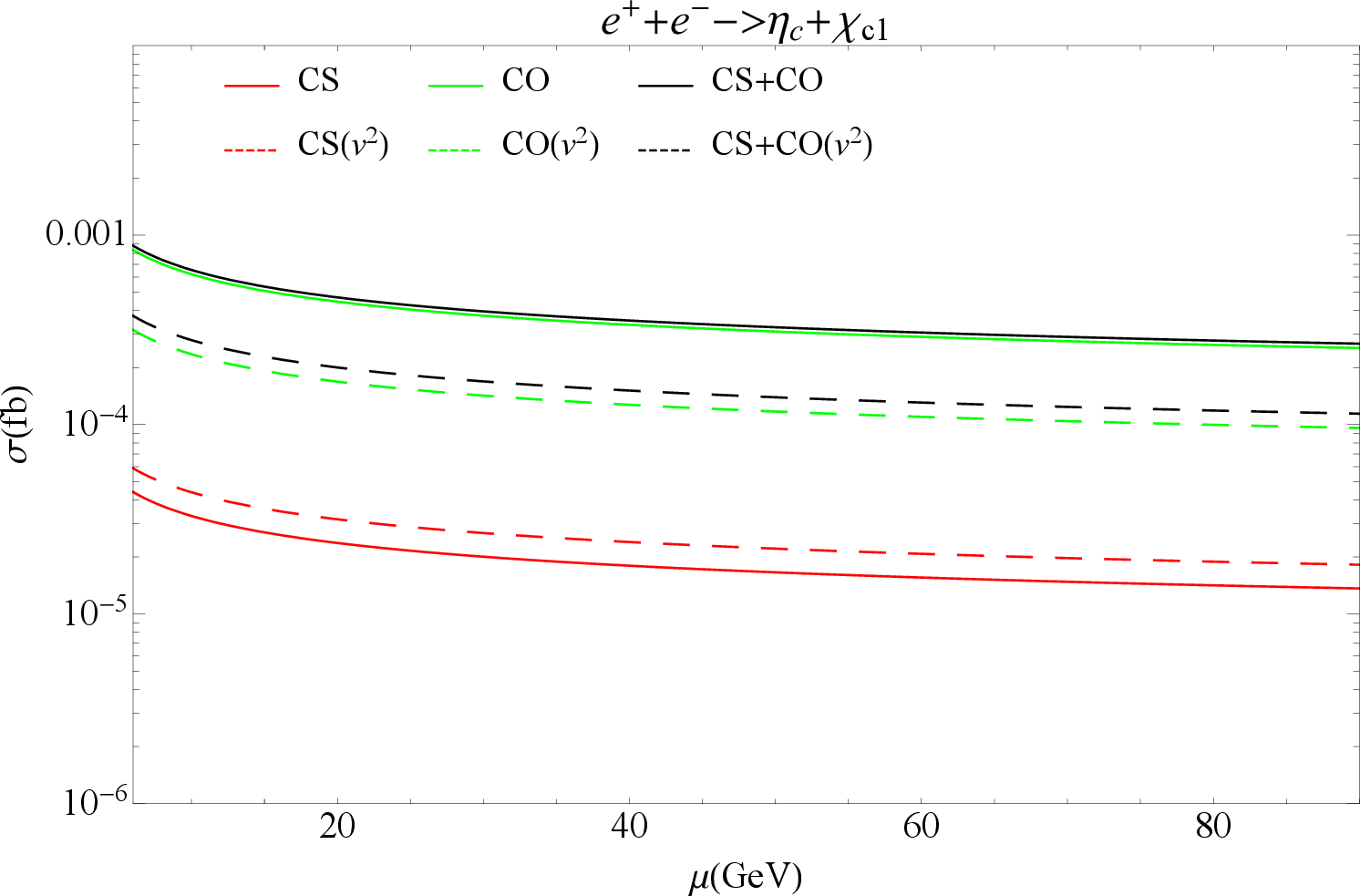}
			\includegraphics[width=0.333\textwidth]{ 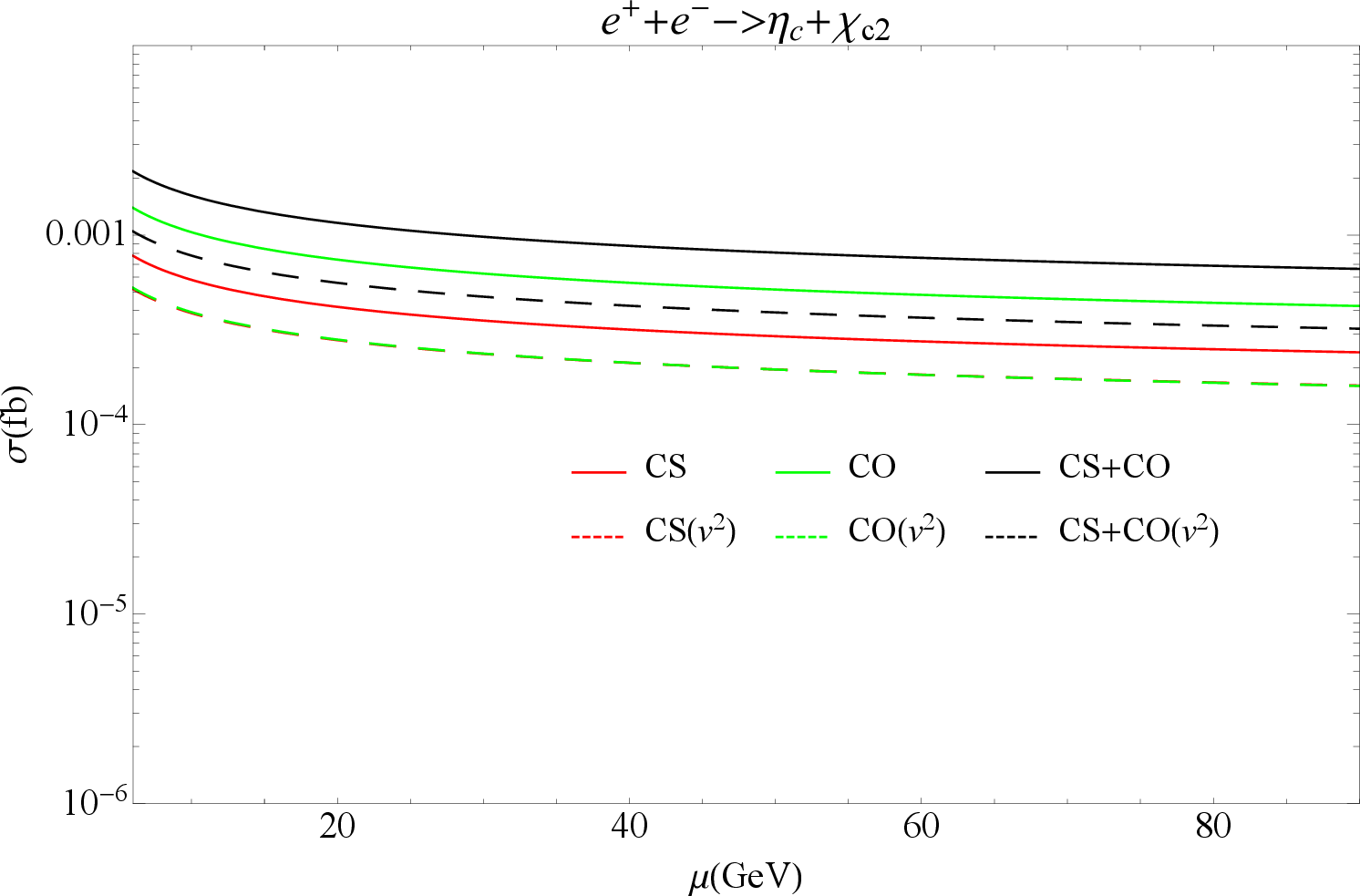}
			\includegraphics[width=0.333\textwidth]{ 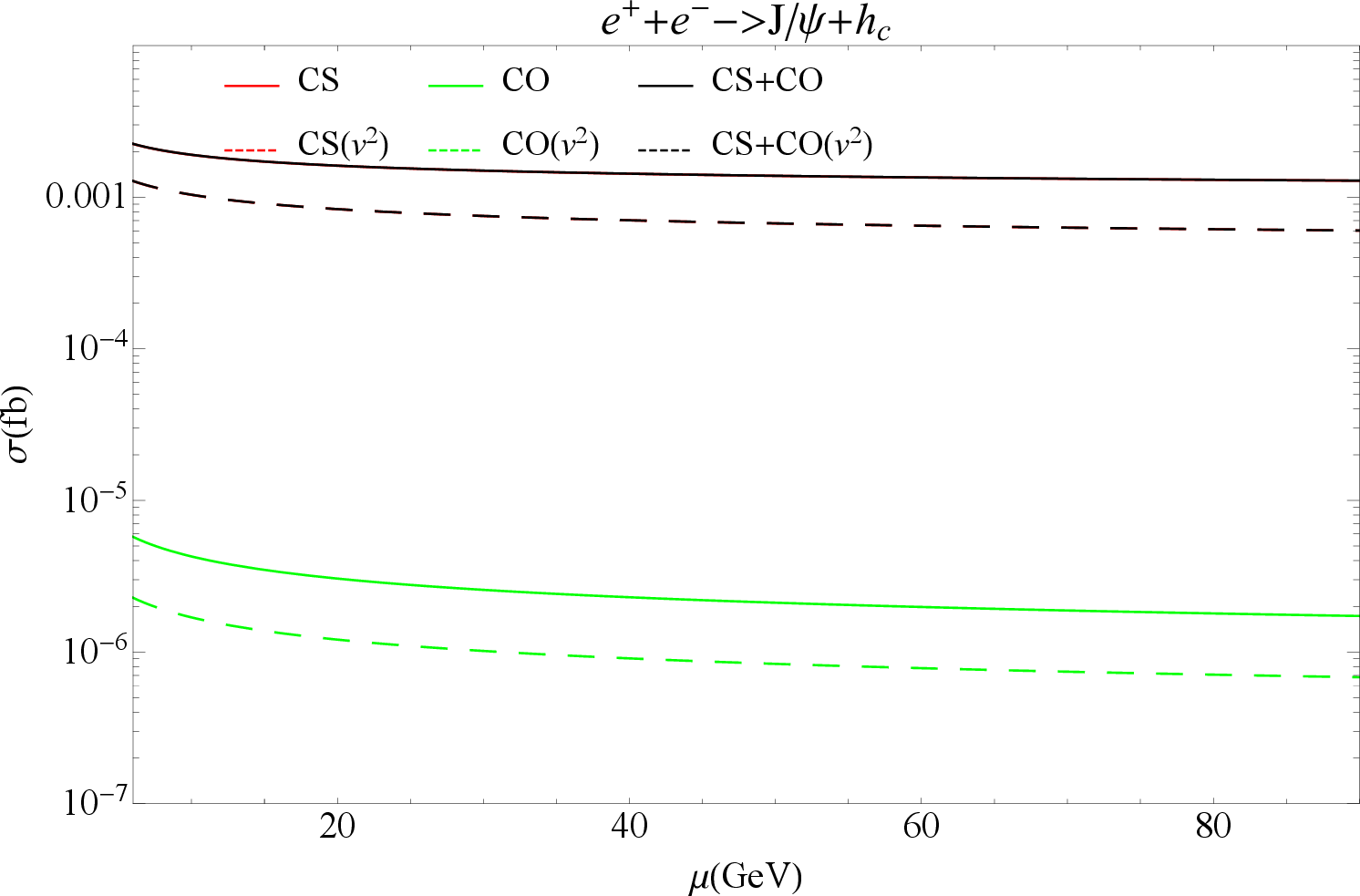}
		\end{tabular}
		\begin{tabular}{c c c}
			\includegraphics[width=0.333\textwidth]{ 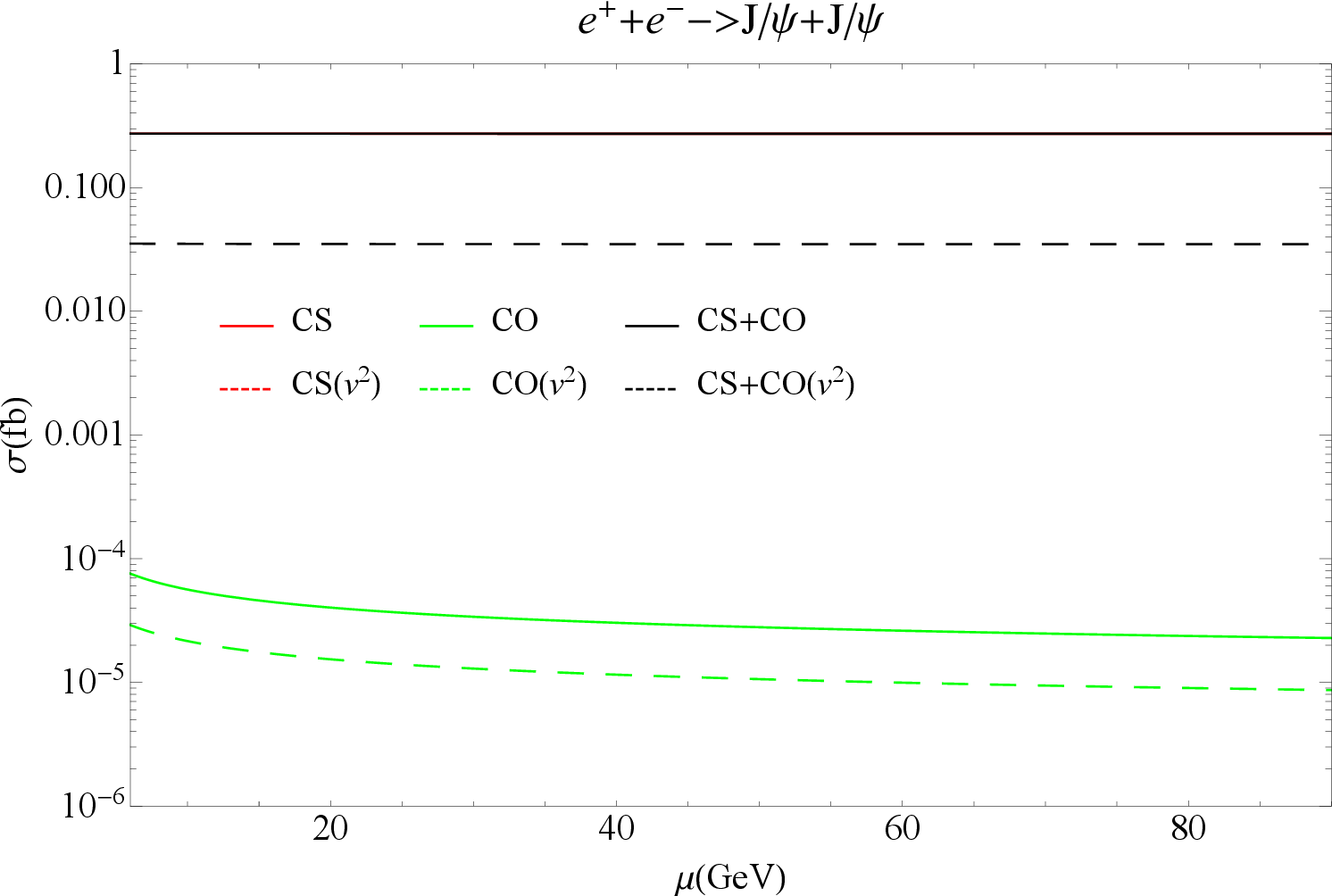}	
				\includegraphics[width=0.333\textwidth]{ 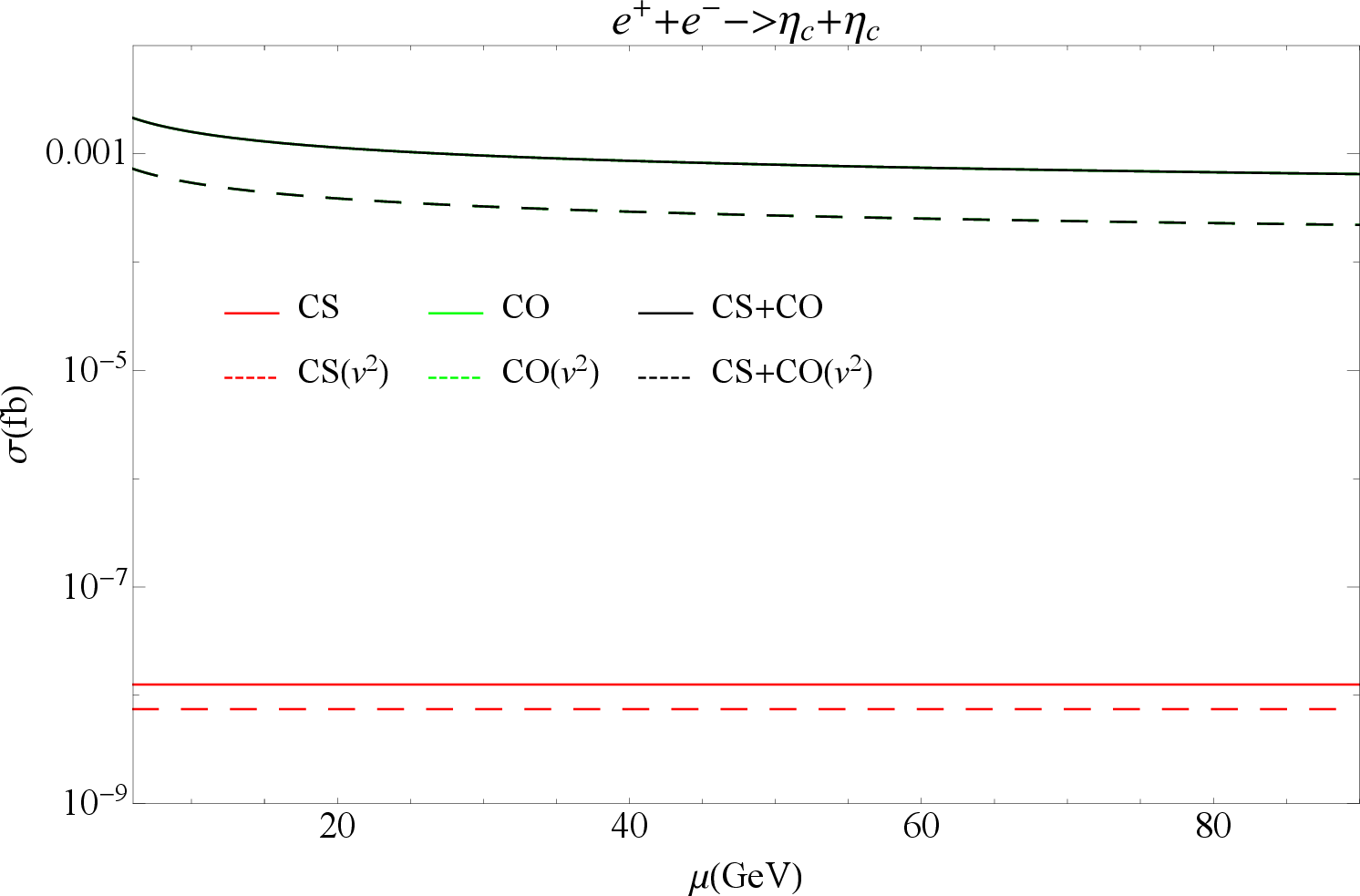}
		\end{tabular}
		\caption{ (Color online) Cross sections ($\sigma$) versus renormalization scale ($\mu$)~for double  charmonium production  at $\sqrt{s}$=$m_Z$. The solid line represents LO  and dashed line represents NLO($v^2$) result. The solid line represents leading order (LO)  and the dashed line represents next-to-leading order in $v^2$ (NLO) results. The red line represents the CS channel, the green line represents the total CO channels and the black line represents the sum of  CS and CO. }
		\label{z0ccmu}
	\end{figure*}
		\FloatBarrier
\end{widetext}

\begin{widetext}
	\begin{figure*}[htbp]
		\begin{tabular}{c c c }
			\includegraphics[width=0.333\textwidth]{ 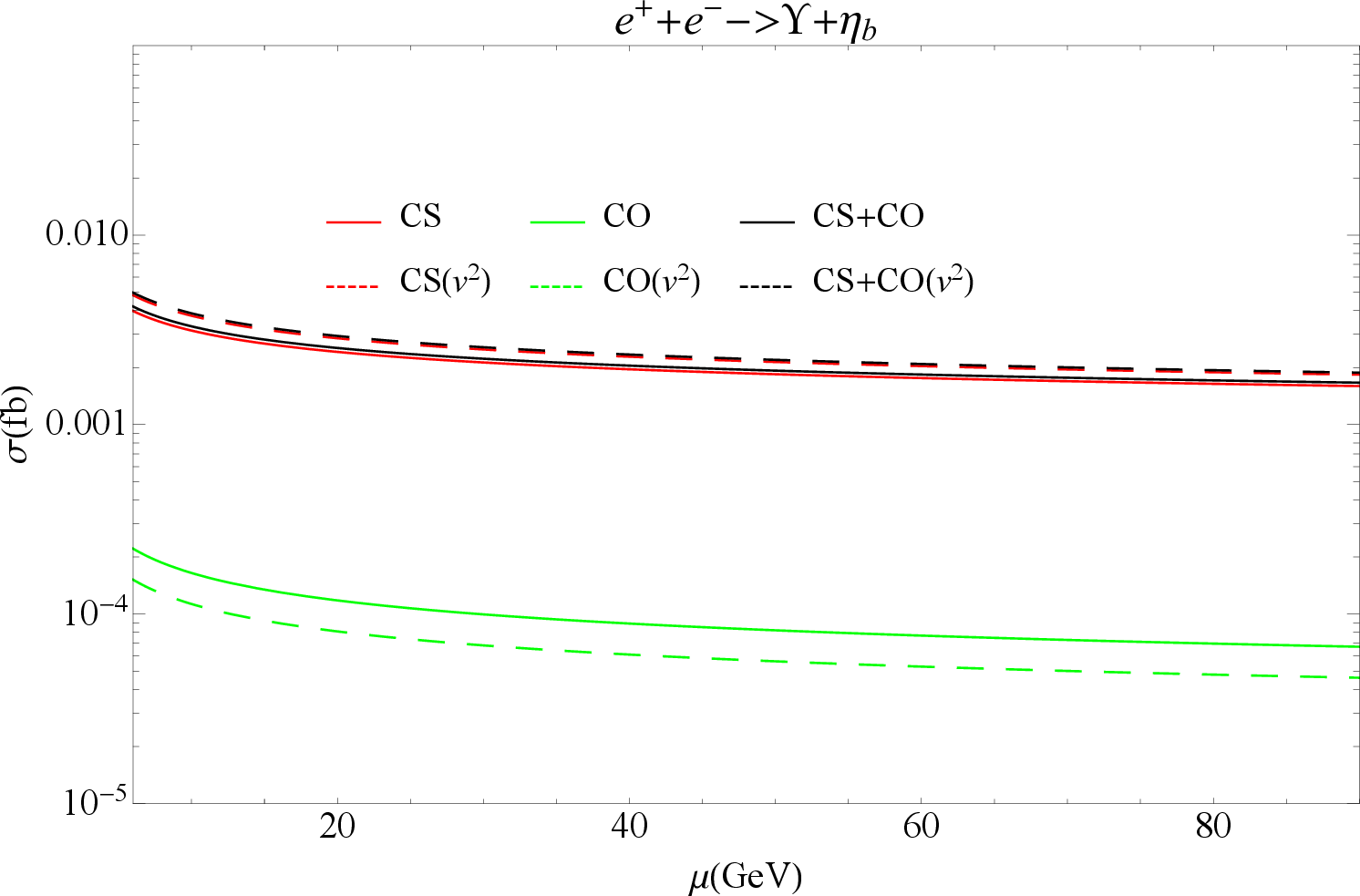}
			\includegraphics[width=0.333\textwidth]{ 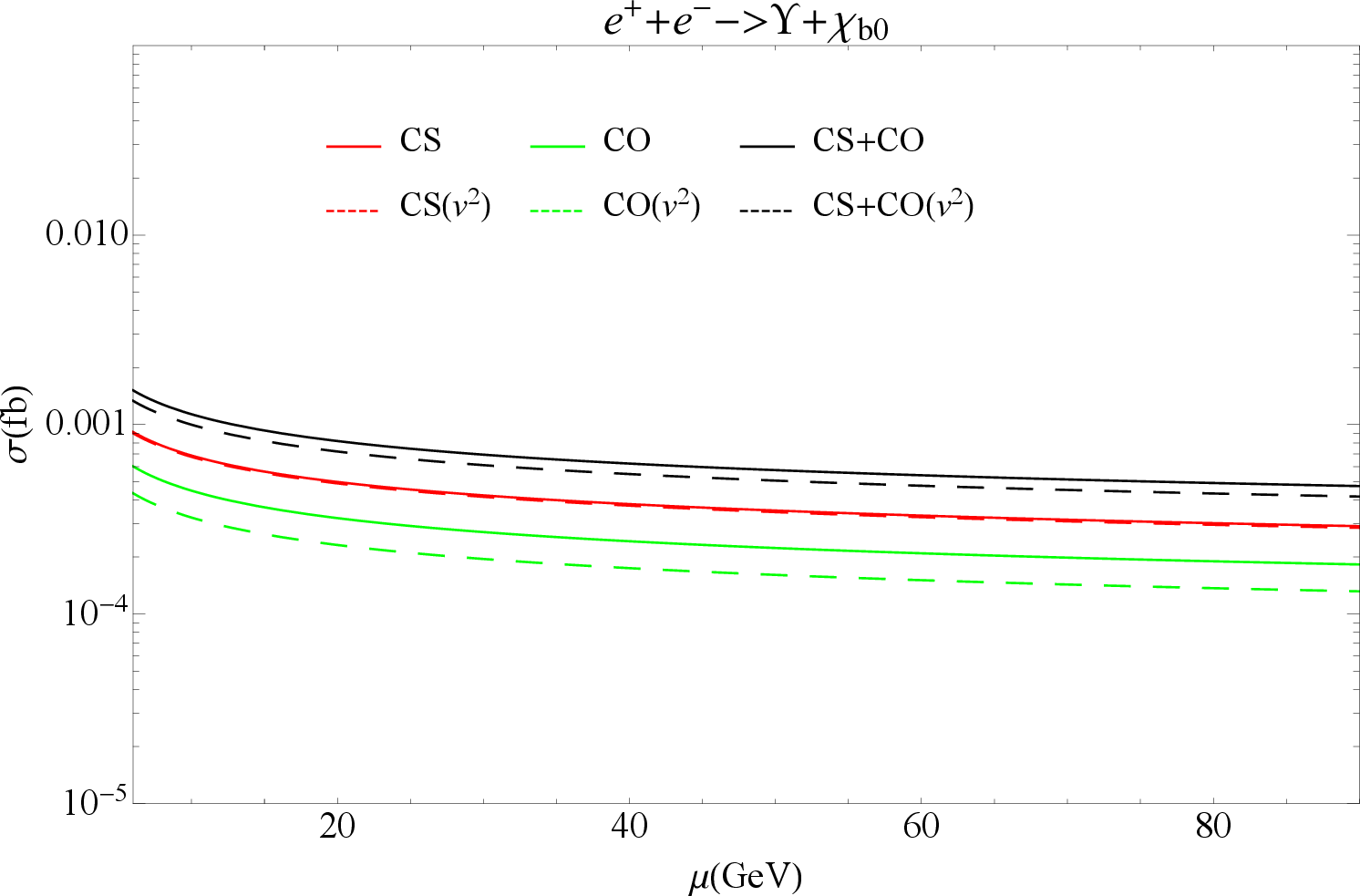}
			\includegraphics[width=0.333\textwidth]{ 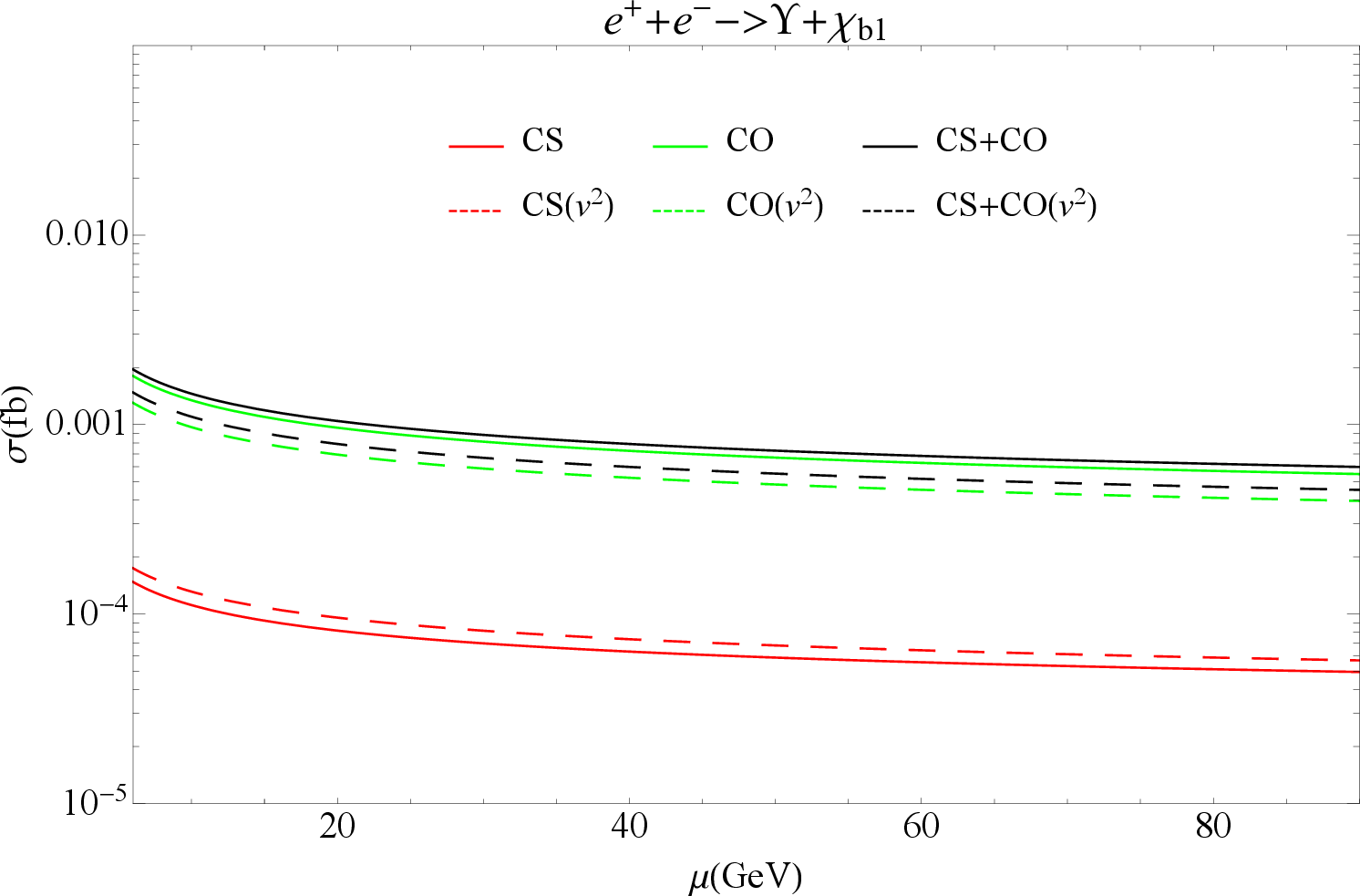}
		\end{tabular}
		\begin{tabular}{c c c }
			\includegraphics[width=0.333\textwidth]{ 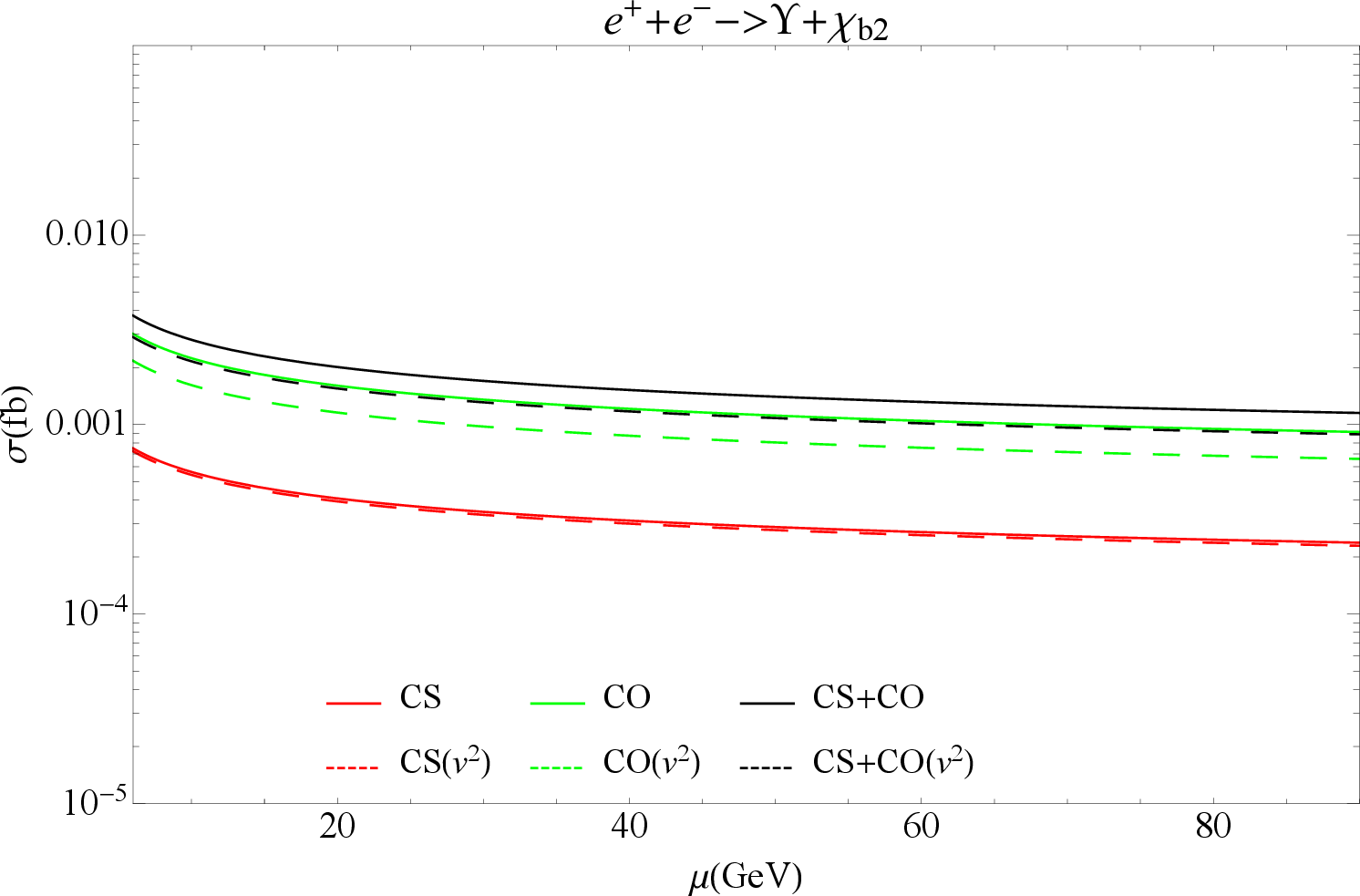}
			\includegraphics[width=0.333\textwidth]{ 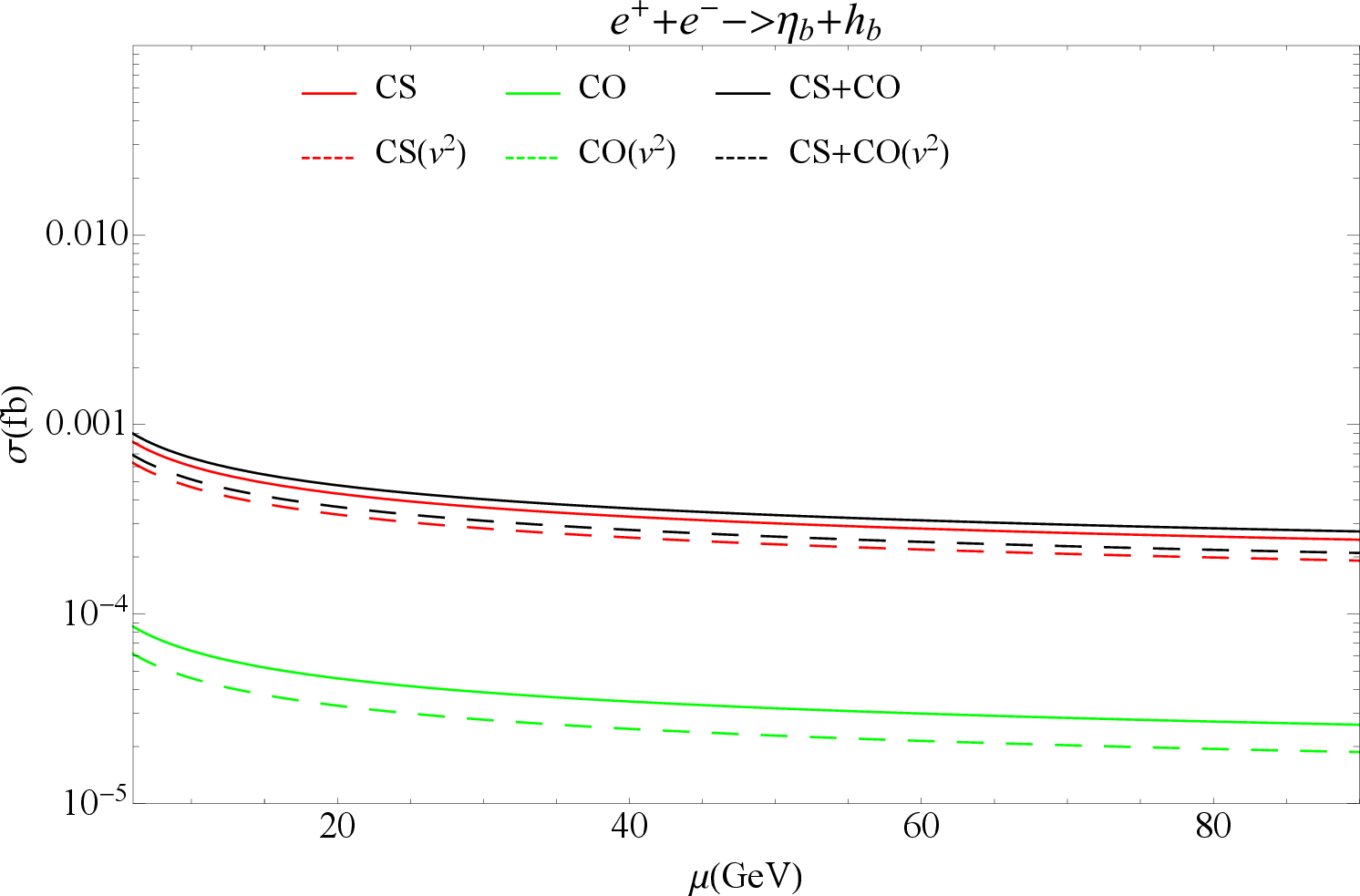}
			\includegraphics[width=0.333\textwidth]{ 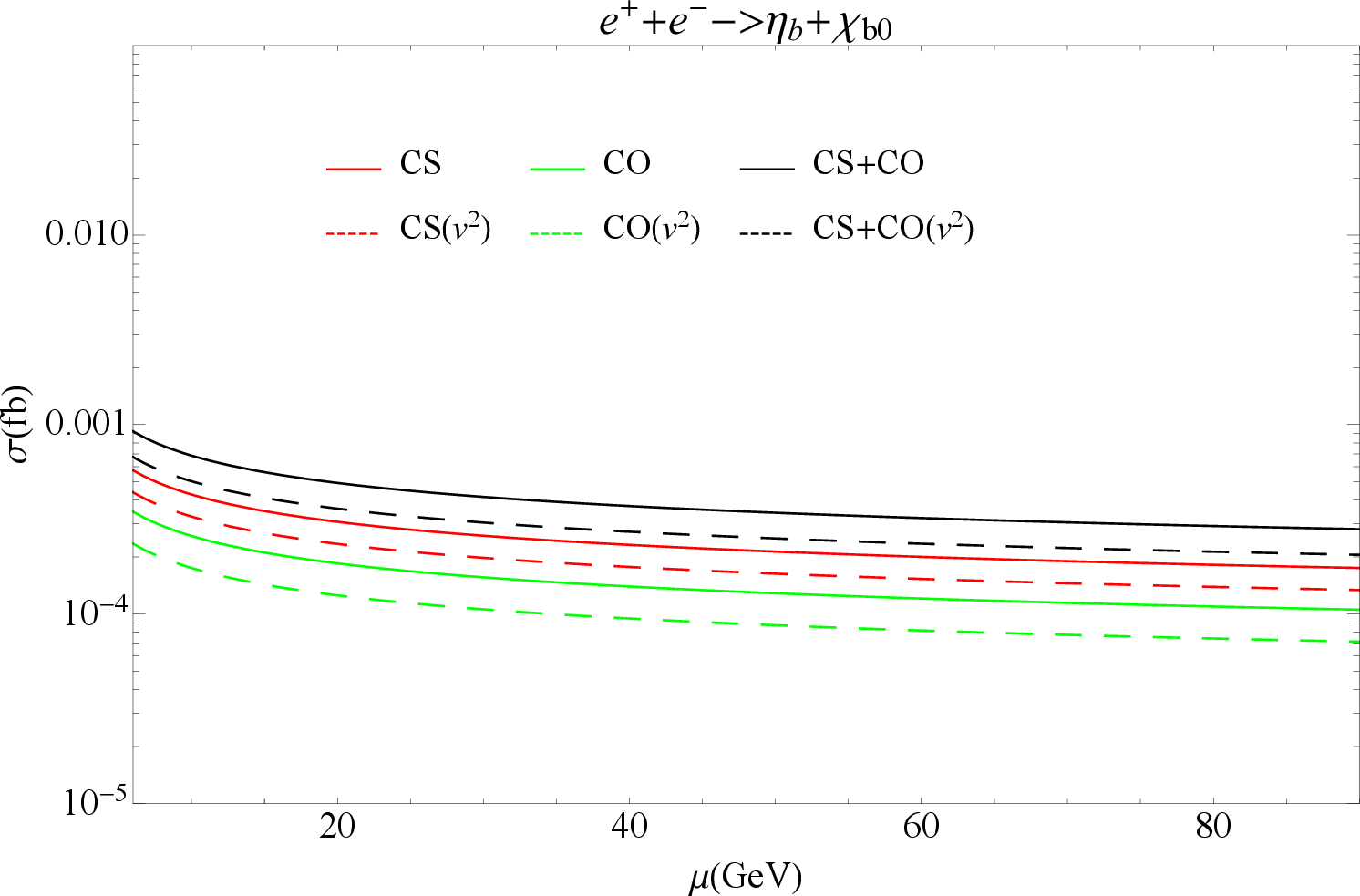}
		\end{tabular}
		\begin{tabular}{c c c }
			\includegraphics[width=0.333\textwidth]{ 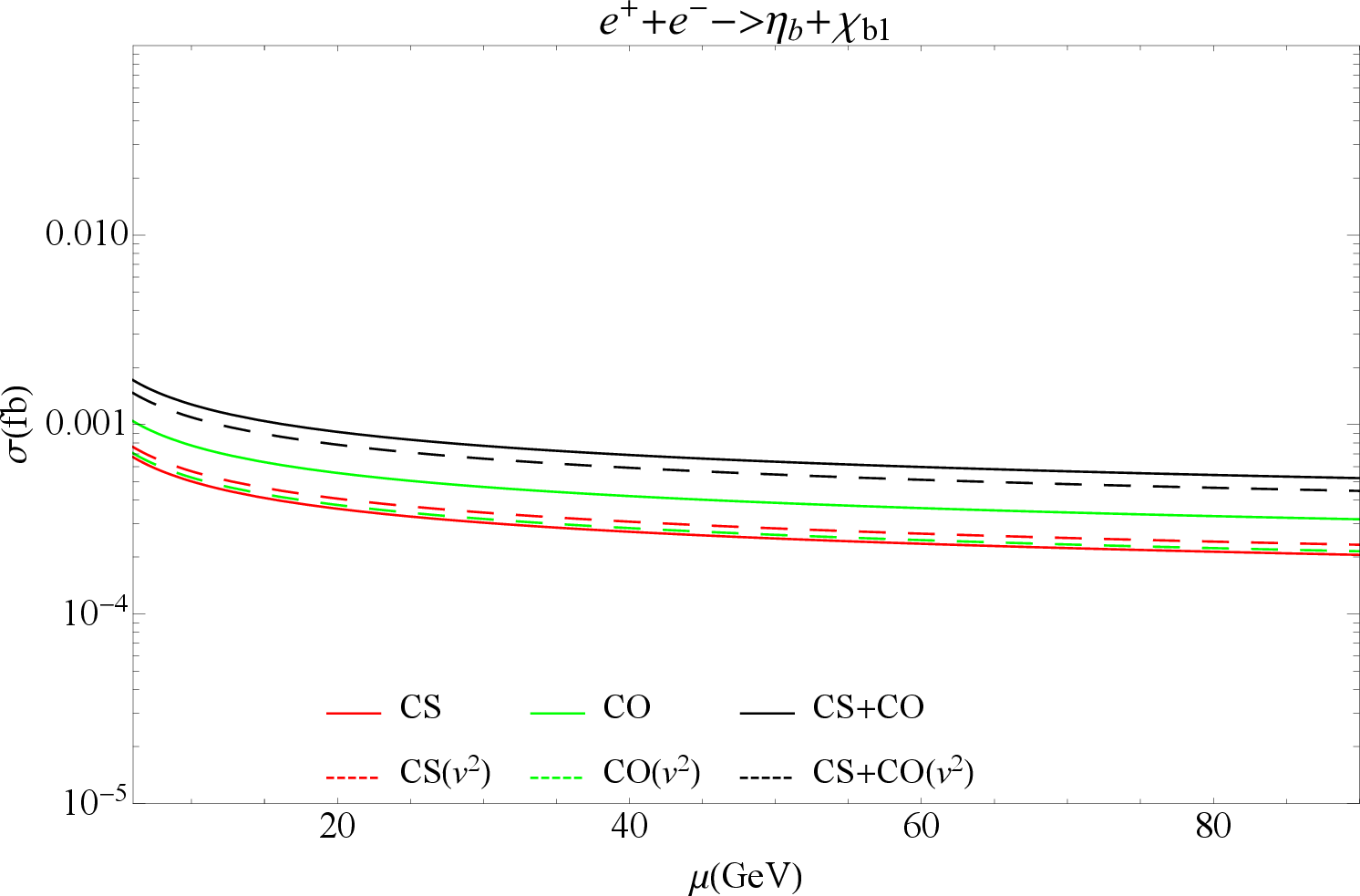}
			\includegraphics[width=0.333\textwidth]{ 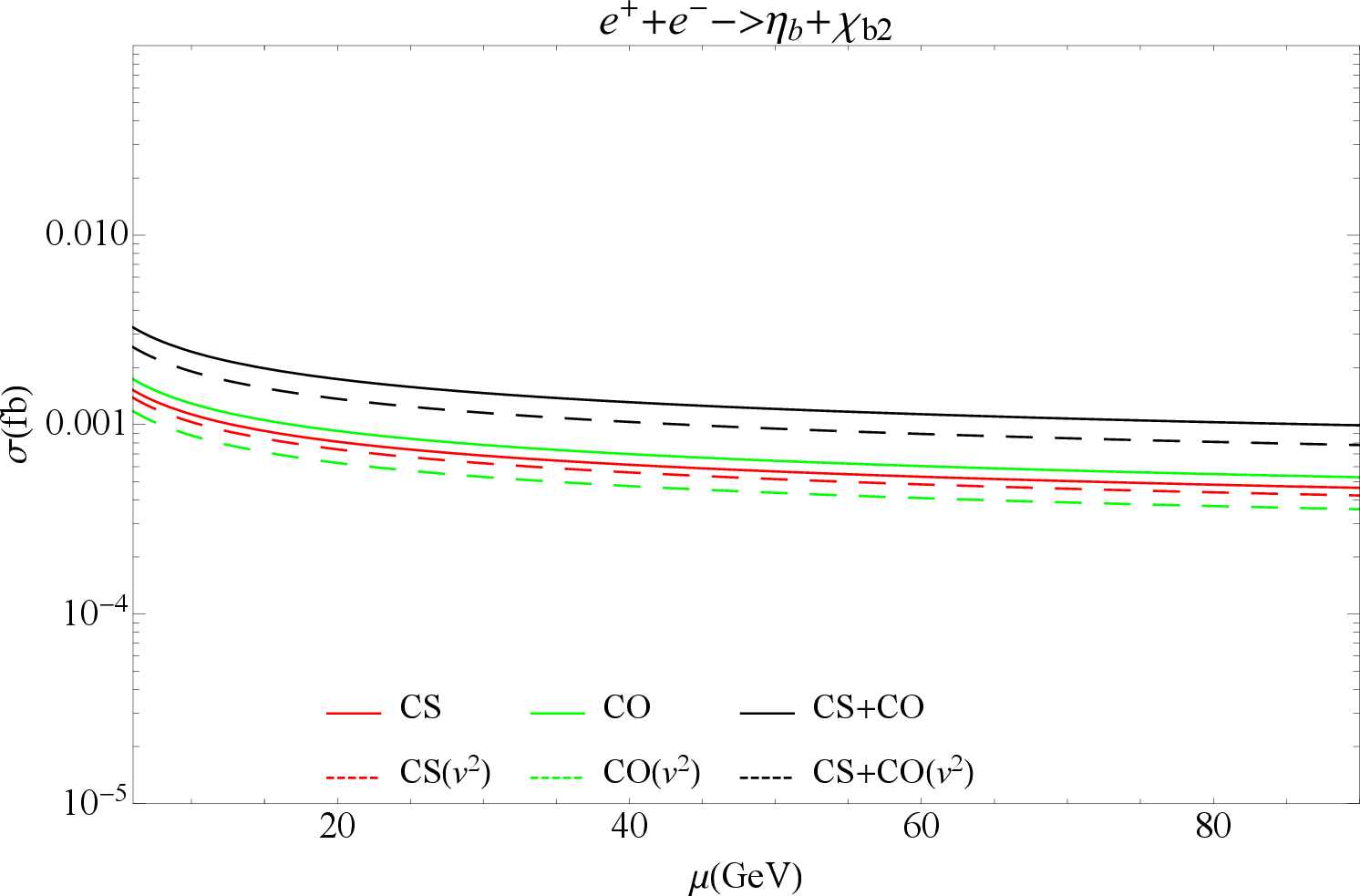}
			\includegraphics[width=0.333\textwidth]{ 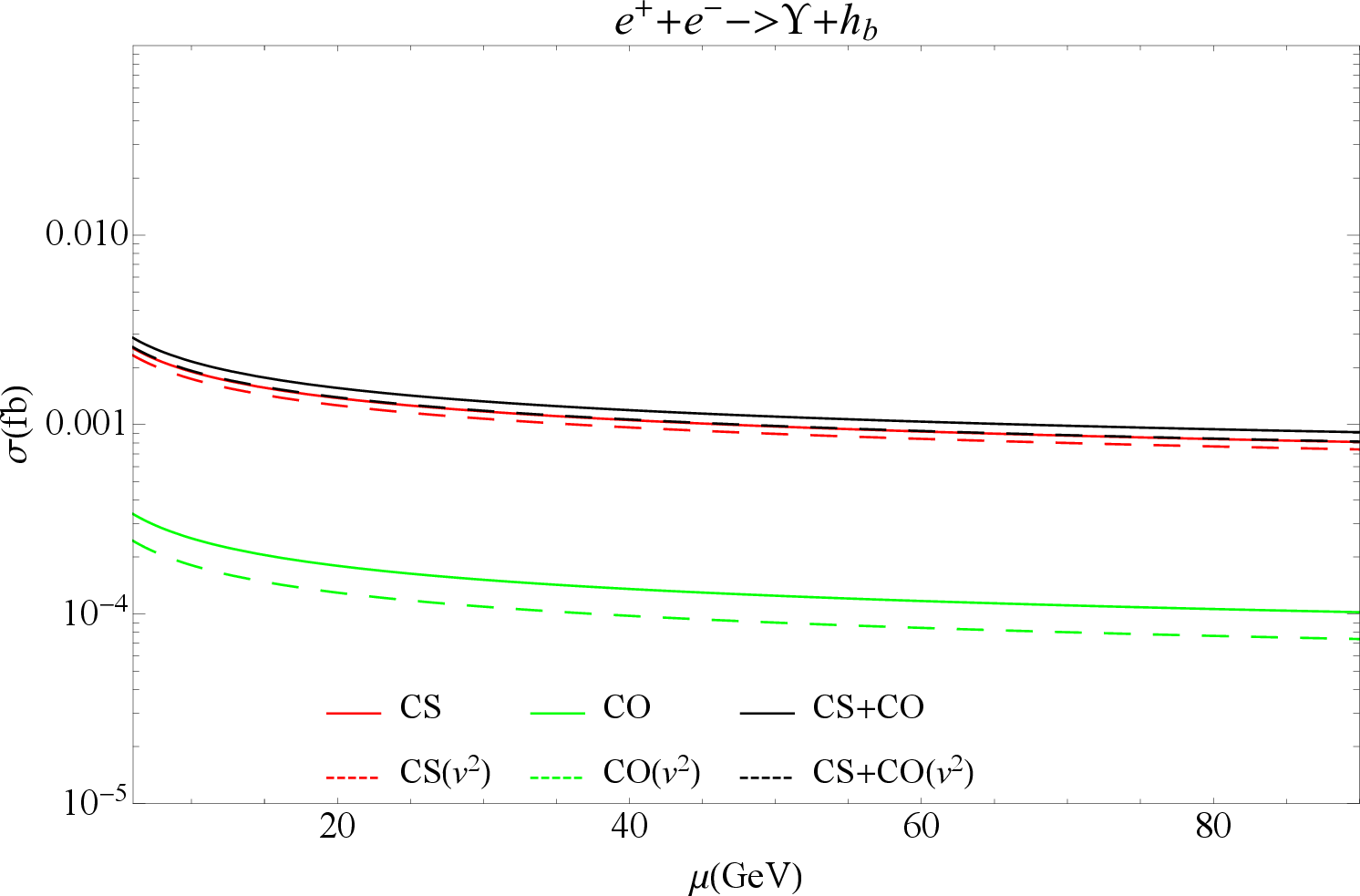}
		\end{tabular}
		\begin{tabular}{c c c}
			\includegraphics[width=0.333\textwidth]{ 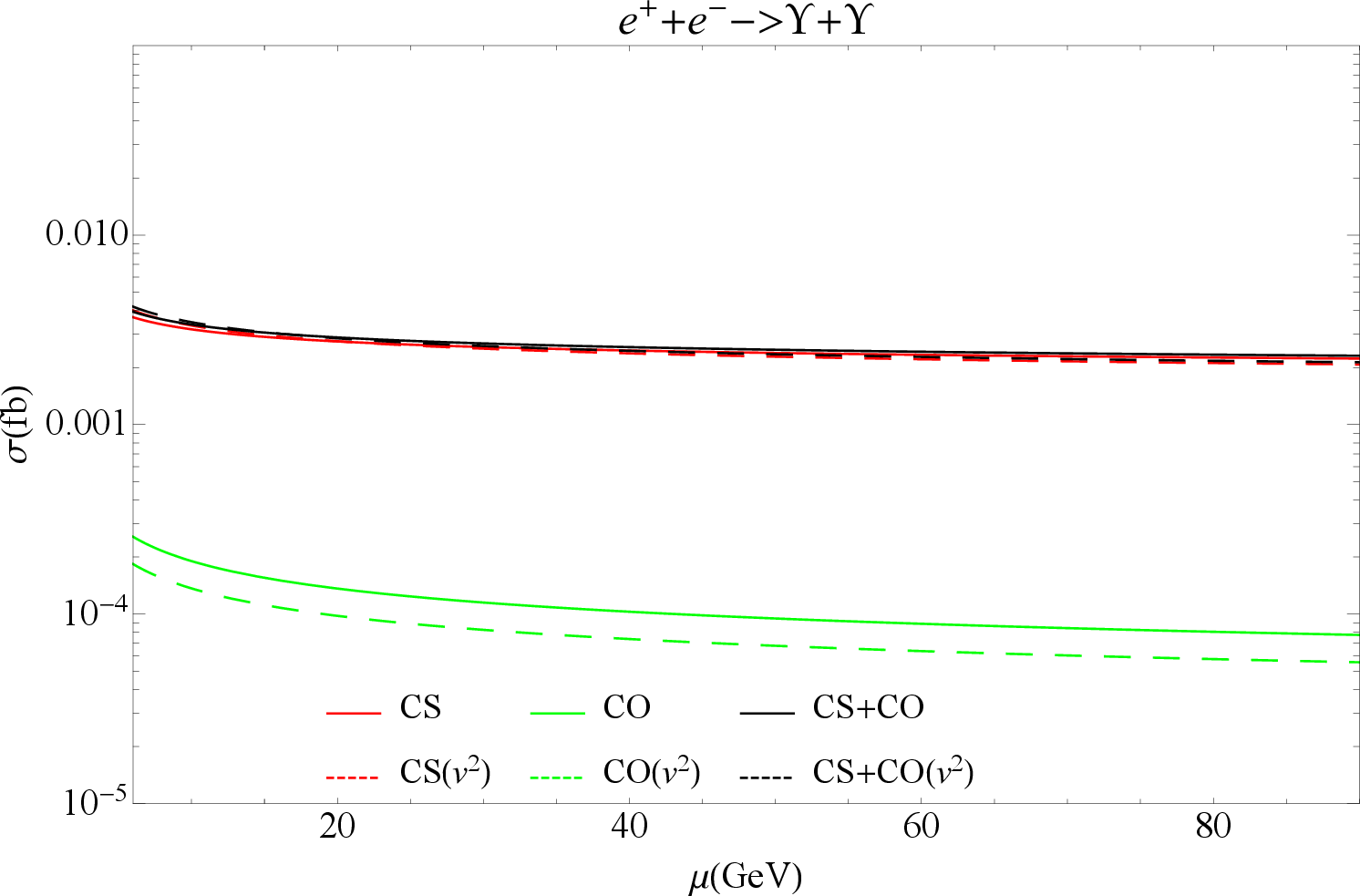}		
				\includegraphics[width=0.333\textwidth]{ 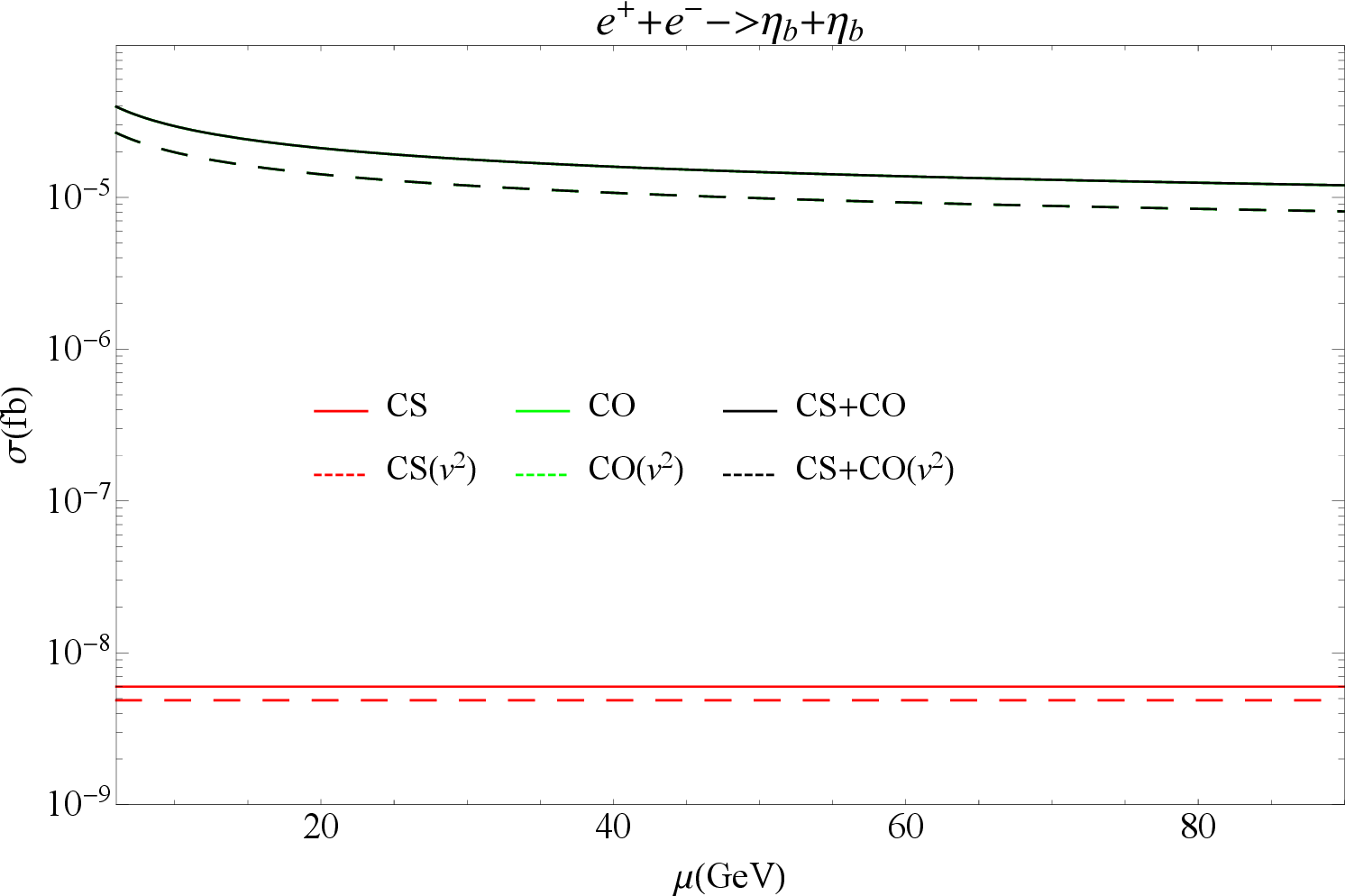}		
		\end{tabular}
		\caption{ (Color online) Cross section ($\sigma$) versus renormalization scale ($\mu$)~for double  bottomonium production  at $\sqrt{s}$=$m_Z$.  The solid line represents leading order (LO)  and the dashed line represents next-to-leading order in $v^2$ (NLO) results. The red line represents the CS channel, the green line represents the total CO channels and the black line represents the sum of  CS and CO. }
		\label{z0bbmu}
	\end{figure*}
	\FloatBarrier
\end{widetext}

\begin{table}
	\caption{Total cross sections $(units: \times 10^{-4}~fb)$ up to $\mathcal{O}(v^2)$  for the double charmonia production, with varying values of $E_{cm}$. The ratio $R_{\pm}$ show how the cross sections are changed with varying $E_{cm}$.
		}
	\begin{tabular}{|c|c|c|c| |c|c|c| |c|c|c|}
		\hline
		~	&\multicolumn{3}{c| |}{CS } & \multicolumn{3}{c| |}{CO } & \multicolumn{3}{c|}{Total }\\
		\hline
		\hline
		~&$97\%m_Z$&$103\%m_Z$&\Bigg(\makecell{$R_-$\\$R_+$}\Bigg)&$97\%m_Z$&$103\%m_Z$&\Bigg(\makecell{$R_-$\\$R_+$}\Bigg)&$97\%m_Z$&$103\%m_Z$&\Bigg(\makecell{$R_-$\\$R_+$}\Bigg)\\
		\hline
		\hline
	$J/\psi+\eta_c$& 1.2&1.1&\Bigg(\makecell{$19\%$\\$18\%$}\Bigg) &1.4&1.4 &\Bigg(\makecell{$18\%$\\$18\%$}\Bigg)&2.6 &2.5&\Bigg(\makecell{$19\%$\\$18\%$}\Bigg) \\
		\hline
	$J/\psi+h_c$& 3.8&3.2&\Bigg(\makecell{$19\%$\\$16\%$}\Bigg) &0.0&0.0 &\Bigg(\makecell{$19\%$\\$19\%$}\Bigg)&3.8 &~3.2~&\Bigg(\makecell{$19\%$\\$16\%$}\Bigg) \\
	\hline
		$J/\psi+\chi_{c0}$& 0.3&0.2&\Bigg(\makecell{$20\%$\\$18\%$}\Bigg) &0.1&0.1 &\Bigg(\makecell{$18\%$\\$19\%$}\Bigg)&0.4 &0.4&\Bigg(\makecell{$19\%$\\$18\%$}\Bigg) \\
	\hline
		$J/\psi+\chi_{c1}$& 0.2&0.2&\Bigg(\makecell{$19\%$\\$19\%$}\Bigg) &0.4&0.4 &\Bigg(\makecell{$18\%$\\$19\%$}\Bigg)&0.6 &0.6&\Bigg(\makecell{$19\%$\\$19\%$}\Bigg) \\
	\hline
		$J/\psi+\chi_{c2}$& 0.5&0.4&\Bigg(\makecell{$20\%$\\$18\%$}\Bigg) &~0.7~&0.7 &\Bigg(\makecell{$18\%$\\$19\%$}\Bigg)&1.1 &~1.1~&\Bigg(\makecell{$19\%$\\$18\%$}\Bigg) \\
	\hline
		$\eta_c+h_c$& 0.3&0.3&\Bigg(\makecell{$20\%$\\$18\%$}\Bigg) &~0.1~&0.1 &\Bigg(\makecell{$19\%$\\$19\%$}\Bigg)&0.4 &~0.4~&\Bigg(\makecell{$19\%$\\$18\%$}\Bigg) \\
	\hline
		$\eta_c+\chi_{c0}$& 0.6&0.5&\Bigg(\makecell{$19\%$\\$16\%$}\Bigg) &~0.3~&0.3 &\Bigg(\makecell{$18\%$\\$17\%$}\Bigg)&1.0 &0.9&\Bigg(\makecell{$18\%$\\$16\%$}\Bigg) \\
	\hline
		$\eta_c+\chi_{c1}$& 0.2&~0.2~&\Bigg(\makecell{$20\%$\\$15\%$}\Bigg) &~1.0~&1.0 &\Bigg(\makecell{$18\%$\\$17\%$}\Bigg)&1.2 &~1.1~&\Bigg(\makecell{$18\%$\\$17\%$}\Bigg) \\
	\hline
		$\eta_c+\chi_{c2}$& 1.8&1.5&\Bigg(\makecell{$19\%$\\$16\%$}\Bigg) &~1.7~&1.6 &\Bigg(\makecell{$18\%$\\$17\%$}\Bigg)&3.5 &~3.1~&\Bigg(\makecell{$18\%$\\$16\%$}\Bigg) \\
	\hline
		$J/\psi+J/\psi$& 669.4&620.0&\Bigg(\makecell{$94\%$\\$87\%$}\Bigg) &~0.2~&0.2 &\Bigg(\makecell{$19\%$\\$18\%$}\Bigg)&670.0 &620.0&\Bigg(\makecell{$94\%$\\$87\%$}\Bigg) \\
		\hline
	$\eta_c+\eta_c$& 0.0&0.0&\Bigg(\makecell{$120\%$\\$84\%$}\Bigg)&~4.7~&4.5 &\Bigg(\makecell{$18\%$\\$17\%$}\Bigg)&4.7 & 4.5&\Bigg(\makecell{$18\%$\\$17\%$}\Bigg) \\
		\hline
	\end{tabular}
	\label{ccmz}
\end{table}
\FloatBarrier

\begin{table}
	\caption{Total cross sections $(units: \times 10^{-4}~fb)$ up to $\mathcal{O}(v^2)$  for the double bottomonia production, with varying values of $E_{cm}$. The ratio $R_{\pm}$ show how the cross sections are changed with varying $E_{cm}$.
		}
	\begin{tabular}{|c|c|c|c| |c|c|c| |c|c|c|}
		\hline
		~	&\multicolumn{3}{c| |}{CS } & \multicolumn{3}{c| |}{CO } & \multicolumn{3}{c|}{Total }\\
		\hline
		\hline
		~&$97\%m_Z$&$103\%m_Z$&\Bigg(\makecell{$R_-$\\$R_+$}\Bigg)&$97\%m_Z$&$103\%m_Z$&\Bigg(\makecell{$R_-$\\$R_+$}\Bigg)&$97\%m_Z$&$103\%m_Z$&\Bigg(\makecell{$R_-$\\$R_+$}\Bigg)\\
		\hline
		\hline
		$\Upsilon+\eta_b$& 8.3&6.7&\Bigg(\makecell{$19\%$\\$16\%$}\Bigg) &~0.2~&0.2 &\Bigg(\makecell{$17\%$\\$17\%$}\Bigg)&8.6 &~6.9~&\Bigg(\makecell{$19\%$\\$16\%$}\Bigg) \\
			\hline
		$\Upsilon+h_b$& 3.9&3.2&\Bigg(\makecell{$19\%$\\$16\%$}\Bigg) &~0.4~&0.4 &\Bigg(\makecell{$17\%$\\$17\%$}\Bigg)&4.2 &~3.6~&\Bigg(\makecell{$19\%$\\$16\%$}\Bigg) \\
		\hline
		$\Upsilon+\chi_{b0}$& 1.5&1.2&\Bigg(\makecell{$19\%$\\$16\%$}\Bigg) &~0.7~&0.7 &\Bigg(\makecell{$17\%$\\$17\%$}\Bigg)&2.2 &~1.9~&\Bigg(\makecell{$19\%$\\$16\%$}\Bigg) \\
		\hline
		$\Upsilon+\chi_{b1}$& 0.3&~0.2~&\Bigg(\makecell{$20\%$\\$15\%$}\Bigg) &~2.0~&2.0 &\Bigg(\makecell{$17\%$\\$17\%$}\Bigg)&2.3 &~2.2~&\Bigg(\makecell{$18\%$\\$17\%$}\Bigg) \\
		\hline
		$\Upsilon+\chi_{b2}$& 1.2&1.0&\Bigg(\makecell{$19\%$\\$16\%$}\Bigg) &~3.3~&3.3 &\Bigg(\makecell{$17\%$\\$17\%$}\Bigg)&4.5 &~4.3~&\Bigg(\makecell{$18\%$\\$17\%$}\Bigg) \\
		\hline
		$\eta_b+h_b$& 1.0&0.9&\Bigg(\makecell{$18\%$\\$16\%$}\Bigg) &~0.1~&0.1 &\Bigg(\makecell{$17\%$\\$17\%$}\Bigg)&1.1 &~1.0~&\Bigg(\makecell{$18\%$\\$16\%$}\Bigg) \\
		\hline
		$\eta_b+\chi_{b0}$& 0.7&~0.6~&\Bigg(\makecell{$19\%$\\$16\%$}\Bigg) &~0.4~&0.4 &\Bigg(\makecell{$17\%$\\$17\%$}\Bigg)&1.1 &~1.0~&\Bigg(\makecell{$18\%$\\$16\%$}\Bigg) \\
		\hline
		$\eta_b+\chi_{b1}$& 1.3&1.0&\Bigg(\makecell{$20\%$\\$15\%$}\Bigg) &~1.1~&1.1 &\Bigg(\makecell{$17\%$\\$17\%$}\Bigg)&2.4 &~2.1~&\Bigg(\makecell{$19\%$\\$16\%$}\Bigg) \\
		\hline
		$\eta_b+\chi_{b2}$& 2.3&1.9&\Bigg(\makecell{$19\%$\\$16\%$}\Bigg) &~1.8~&1.8 &\Bigg(\makecell{$17\%$\\$17\%$}\Bigg)&4.1 &~3.7~&\Bigg(\makecell{$18\%$\\$16\%$}\Bigg) \\
		\hline
		$\Upsilon+\Upsilon$& 12.5&9.6&\Bigg(\makecell{$34\%$\\$26\%$}\Bigg) &~0.3~&0.3 &\Bigg(\makecell{$18\%$\\$17\%$}\Bigg)&12.8 &~9.9~&\Bigg(\makecell{$18\%$\\$17\%$}\Bigg) \\
		\hline
			$\eta_b+\eta_b$&  0.0&0.0&\Bigg(\makecell{$119\%$\\$84\%$}\Bigg)&0.0&0.0 &\Bigg(\makecell{$17\%$\\$17\%$}\Bigg)&0.0 &0.0&\Bigg(\makecell{$17\%$\\$17\%$}\Bigg) \\
		\hline
	\end{tabular}
	\label{bbmz}
\end{table}
\FloatBarrier

\begin{table}
	\caption{Total cross sections ($units: \times10^{-4}~fb$) up to $\mathcal{O}(v^2)$  for the double charmonia production at $\sqrt{s}=m_Z$, with varying values of $m_c$. The uncertainties in  each the third column are the deviations from the central values corresponding to $m_c=1.5 GeV$.
		}
	\begin{tabular}{|c|c|c|c| |c|c|c| |c|c|c|}
		\hline
		~	&\multicolumn{3}{c| |}{CS} & \multicolumn{3}{c| |}{CO} & \multicolumn{3}{c|}{Total}\\
		\hline
		\hline
		$m_c$&$1.35 GeV$&$1.65 GeV$&uncertainty&$1.35GeV$&$1.65GeV$&uncertainty&$1.35GeV$&$1.65GeV$&uncertainty\\
		\hline
		\hline
		$J/\psi+\eta_c$& 8.510&4.732&\makecell{$+2.300$\\$-1.477$}&11.47&5.102 &\makecell{$+3.971$\\$-2.397$}&19.98 &9.835&\makecell{$+6.271$\\$-3.874$} \\
		\hline
		$J/\psi+h_c$& 25.41&~16.57~&\makecell{$+5.167$\\$-3.677$} &0.063&0.029 &\makecell{$+0.021$\\$-0.013$}&25.48 &16.60&\makecell{$+5.188$\\$-3.690$} \\
		\hline
		$J/\psi+\chi_{c0}$& 1.707&1.184&\makecell{$+0.305$\\$-0.218$}&1.099&0.489 &\makecell{$+0.380$\\$-0.230$}&2.806 &1.673&\makecell{$+0.685$\\$-0.448$} \\
		\hline
		$J/\psi+\chi_{c1}$& 1.271&0.631&\makecell{$+0.397$\\$-0.243$} &3.296&1.467 &\makecell{$+1.141$\\$-0.689$}&4.566 &2.098&\makecell{$+1.537$\\$-0.931$} \\
		\hline
		$J/\psi+\chi_{c2}$& 2.967&1.902&\makecell{$+0.624$\\$-0.440$} &5.493&2.444 &\makecell{$+1.901$\\$-1.148$}&8.460 &4.347&\makecell{$+2.525$\\$-1.588$} \\
		\hline
		$\eta_c+h_c$& 1.916&1.275&\makecell{$+0.368$\\$-0.273$} &0.805&0.358 &\makecell{$+0.279$\\$-0.168$}&2.721 &1.633&\makecell{$+0.647$\\$-0.441$} \\
		\hline
		$\eta_c+\chi_{c0}$& 4.268&2.833&\makecell{$+0.825$\\$-0.610$} &2.947&1.306 &\makecell{$+1.024$\\$-0.617$}&7.215 &4.139&\makecell{$+1.848$\\$-1.227$} \\
		\hline
		$\eta_c+\chi_{c1}$& 1.069&1.064&\makecell{$+0.002$\\$-0.003$} &8.840&3.918 &\makecell{$+3.071$\\$-1.852$}&9.909 &4.982&\makecell{$+3.073$\\$-1.855$} \\
		\hline
		$\eta_c+\chi_{c2}$& 11.51&7.860&\makecell{$+2.097$\\$-1.552$} &14.73&6.529 &\makecell{$+5.118$\\$-3.087$}&26.24 &14.39&\makecell{$+7.216$\\$-4.639$}\\
		\hline
		$J/\psi+J/\psi$& 679.0&199.8&\makecell{$+323.3$\\$-155.9$} &0.796&0.375 &\makecell{$+0.262$\\$-0.160$}&679.8 &200.2&\makecell{$+323.6$\\$-156.0$} \\
		\hline
			$\eta_c+\eta_c$&  0.000& 0.000& \makecell{$+0.000$\\$-0.000$} &20.35&9.005 &\makecell{$+7.076$\\$-4.266$}&  20.35& 9.006&  \makecell{$+7.076$\\$-4.266$}\\
		\hline
	\end{tabular}
	\label{ccmc}
\end{table}
\FloatBarrier

\begin{table}
	\caption{Total cross sections ($units: \times10^{-4}~fb$) up to $\mathcal{O}(v^2)$   for the double bottomonia production at $\sqrt{s}=m_Z$, with varying values of $m_b$. The uncertainties in  each the third column are the deviations from the central values corresponding to $m_b=4.7 GeV$.
	}
	\begin{tabular}{|c|c|c|c| |c|c|c| |c|c|c|}
		\hline
		~	&\multicolumn{3}{c| |}{CS} & \multicolumn{3}{c| |}{CO} & \multicolumn{3}{c|}{Total}\\
		\hline
		\hline
		$m_b$&$4.55 GeV$&$4.85 GeV$&uncertainty&$4.55GeV$&$4.85GeV$&uncertainty&$4.55GeV$&$4.85GeV$&uncertainty\\
		\hline
		\hline
		$\Upsilon+\eta_b$& 44.12&42.00&\makecell{$+1.105$\\$-1.014$}&1.526&1.162 &\makecell{$+0.197$\\$-0.167$}&45.64 &43.16&\makecell{$+1.302$\\$-1.181$} \\
			\hline
		$\Upsilon+h_b$& 21.46&19.37&\makecell{$+1.093$\\$-0.998$} &2.436&1.870 &\makecell{$+0.306$\\$-0.260$}&23.89 &21.24&\makecell{$+1.399$\\$-1.257$} \\
		\hline
		$\Upsilon+\chi_{b0}$& 8.112&7.721&\makecell{$+0.204$\\$-0.187$}&4.354&3.329 &\makecell{$+0.554$\\$-0.471$}&12.47 &11.05&\makecell{$+0.758$\\$-0.658$} \\
		\hline
		$\Upsilon+\chi_{b1}$& 1.554
		&1.516&\makecell{$+0.020$\\$-0.019$} &13.06&9.987 &\makecell{$+1.662$\\$-1.412$}&14.61 &11.50&\makecell{$+1.681$\\$-1.430$} \\
		\hline
		$\Upsilon+\chi_{b2}$& 6.716&6.021&\makecell{$+0.364$\\$-0.332$} &21.77&16.65 &\makecell{$+2.770$\\$-2.353$}&28.48 &22.67&\makecell{$+3.134$\\$-2.684$} \\
		\hline
		$\eta_b+h_b$& 5.915&5.112&\makecell{$+0.421$\\$-0.382$} &0.618&0.473 &\makecell{$+0.078$\\$-0.066$}&6.532 &5.586&\makecell{$+0.499$\\$-0.448$} \\
		\hline
		$\eta_b+\chi_{b0}$& 4.138&3.577&\makecell{$+0.294$\\$-0.267$} &2.372&1.795 &\makecell{$+0.312$\\$-0.265$}&6.510 &5.373&\makecell{$+0.606$\\$-0.531$} \\
		\hline
		$\eta_b+\chi_{b1}$& 6.707&6.620&\makecell{$+0.043$\\$-0.044$} &7.115&5.385 &\makecell{$+0.937$\\$-0.794$}&13.82 &12.01&\makecell{$+0.980$\\$-0.838$} \\
		\hline
		$\eta_b+\chi_{b2}$& 12.78&11.53&\makecell{$+0.658$\\$-0.601$} &11.86&8.975 &\makecell{$+1.561$\\$-1.323$}&24.64 &20.50&\makecell{$+2.219$\\$-1.923$}\\
		\hline
		$\Upsilon+\Upsilon$& 39.49&34.55&\makecell{$+2.671$\\$-2.266$} &1.835&1.413 &\makecell{$+0.228$\\$-0.194$}&41.32 &35.96&\makecell{$+2.899$\\$-2.459$} \\
		\hline
			$\eta_b+\eta_b$&  0.000&0.000 & \makecell{$+0.000$\\$-0.000$} &0.268&0.203 &\makecell{$+0.035$\\$-0.030$}& 0.268 &0.203 &  \makecell{$+0.035$\\$-0.030$}\\
		\hline
	\end{tabular}
	\label{bbmb}
\end{table}
\FloatBarrier

\section{Appendix. A}
\label{appdA}

The ratios of $\sigma_{NLO(v^2)}/\sigma_{LO}$ (i.e., the $K$ factors) as a function of $\sqrt{s}$ are shown as follows.

\begin{widetext}
	\begin{figure*}[htbp]
		\begin{tabular}{c c c}
			\includegraphics[width=0.333\textwidth]{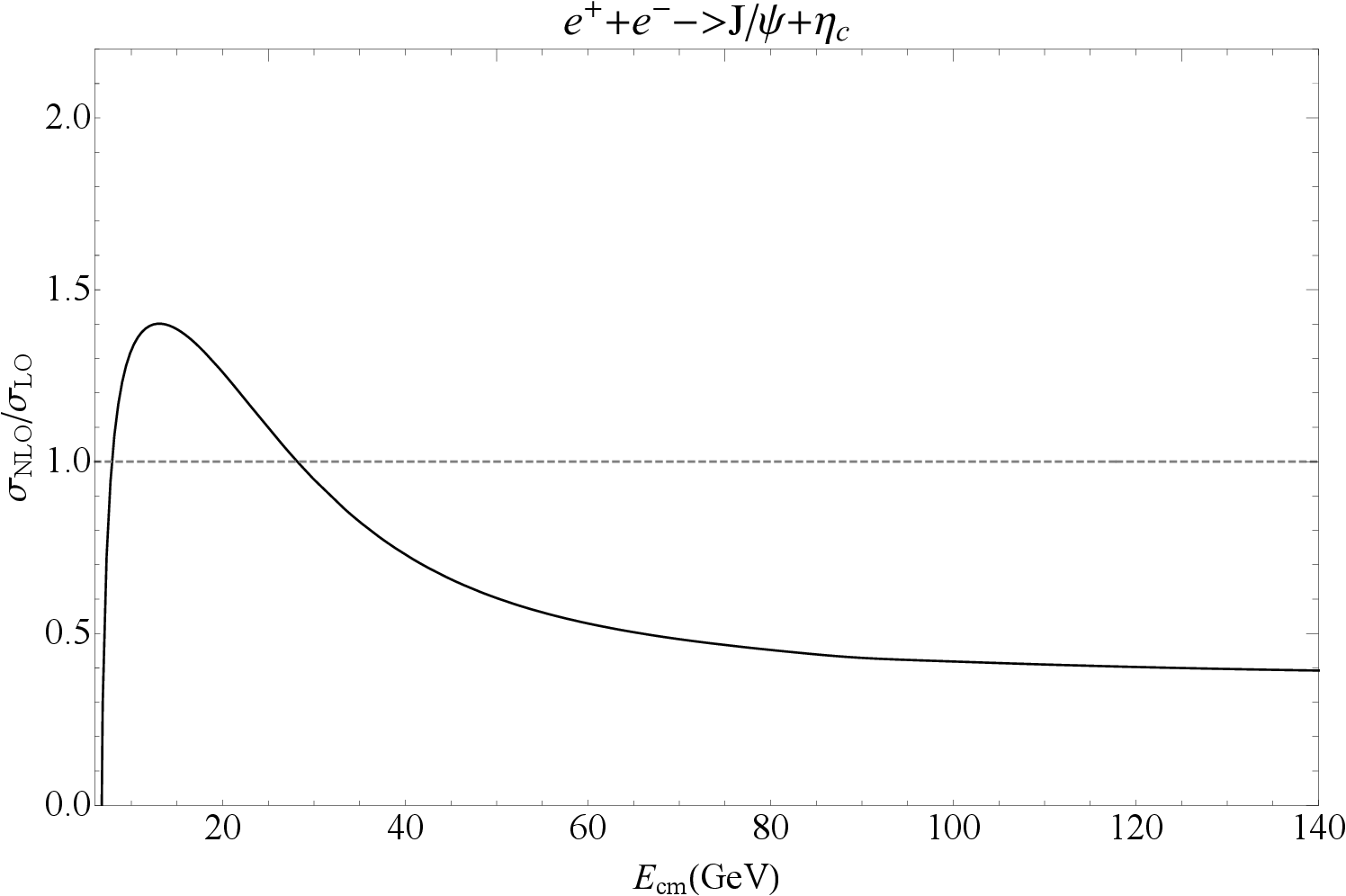}
			\includegraphics[width=0.333\textwidth]{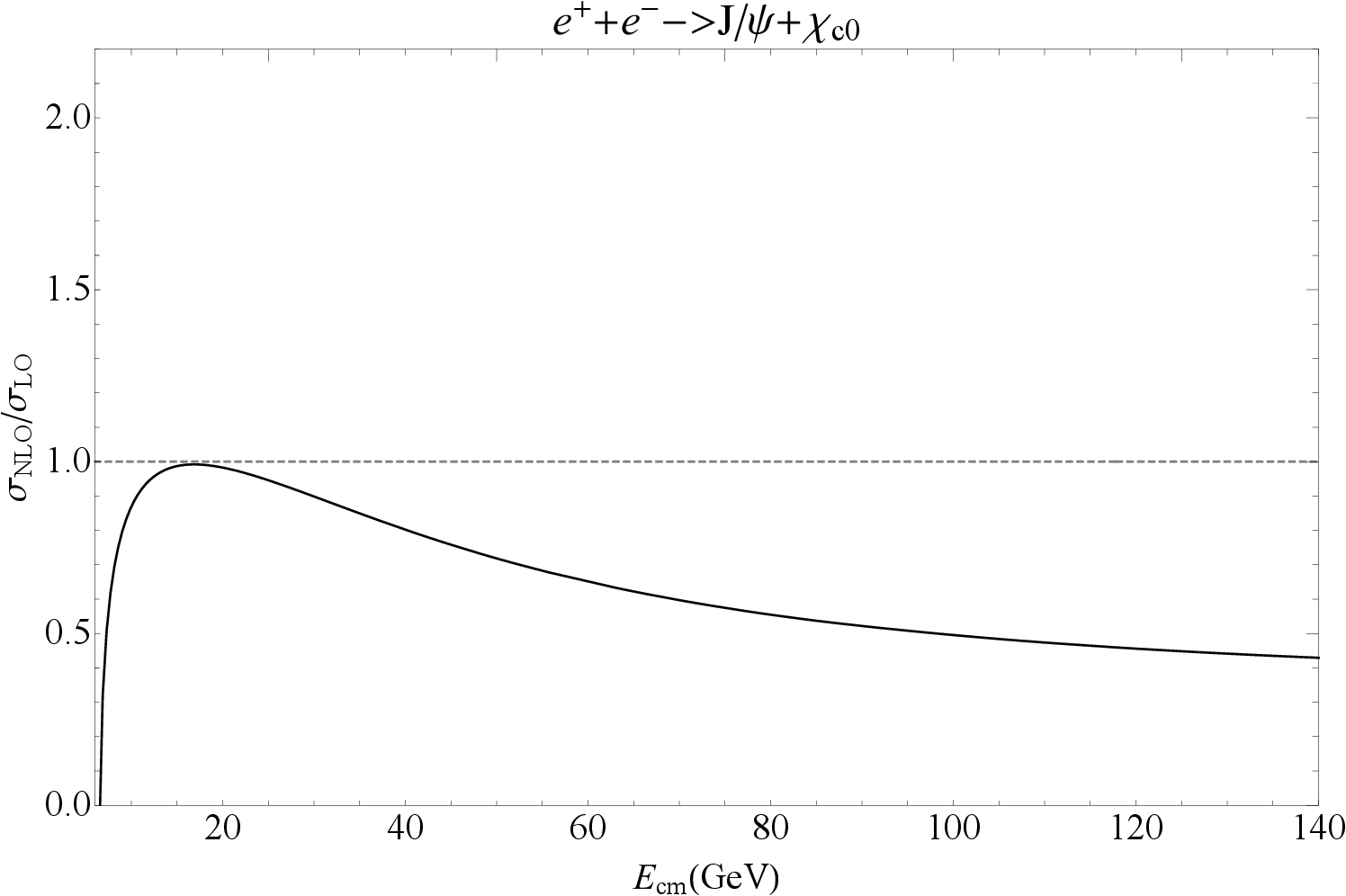}
			\includegraphics[width=0.333\textwidth]{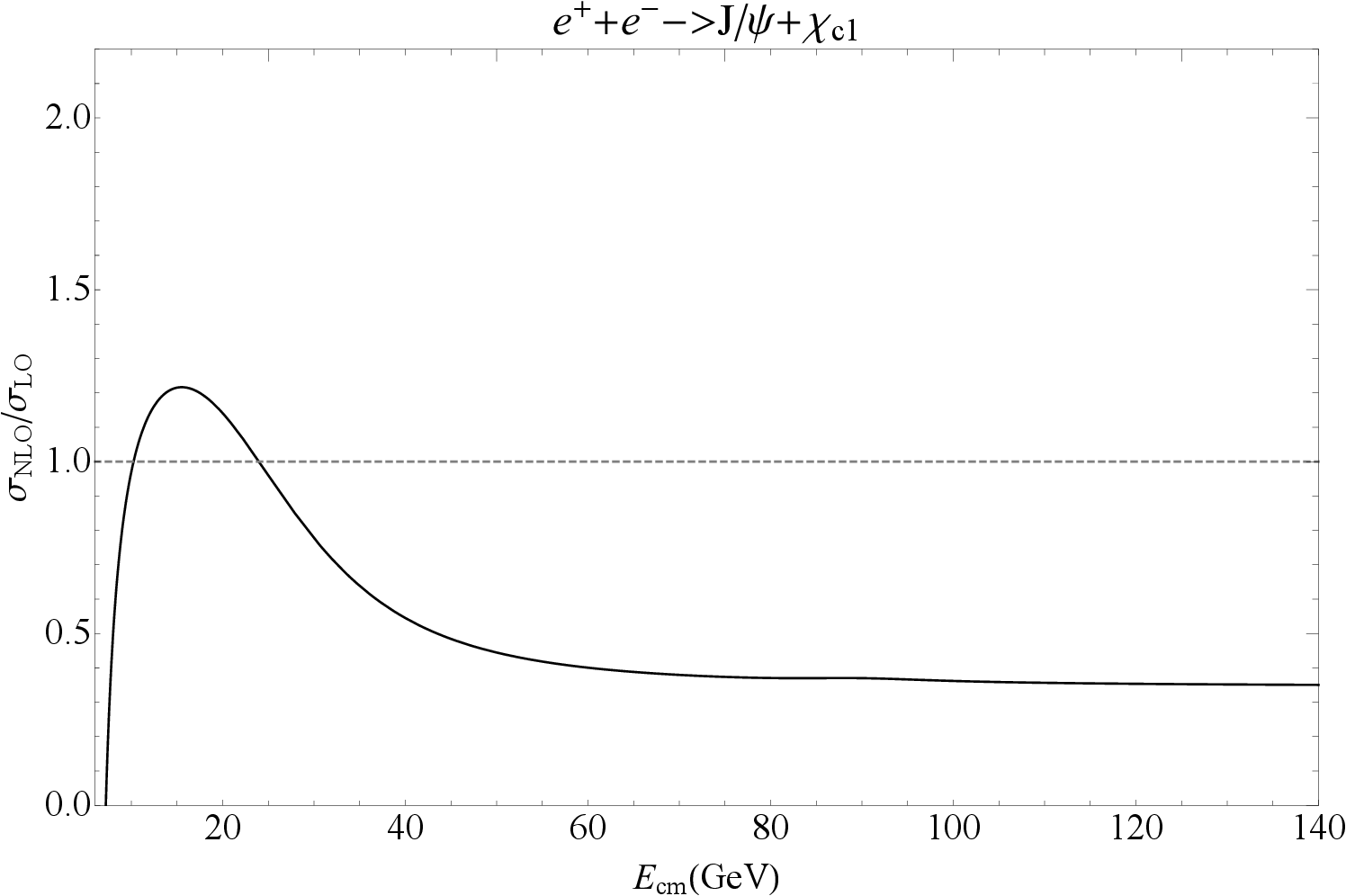}
		\end{tabular}
		\begin{tabular}{c c c}	
			\includegraphics[width=0.333\textwidth]{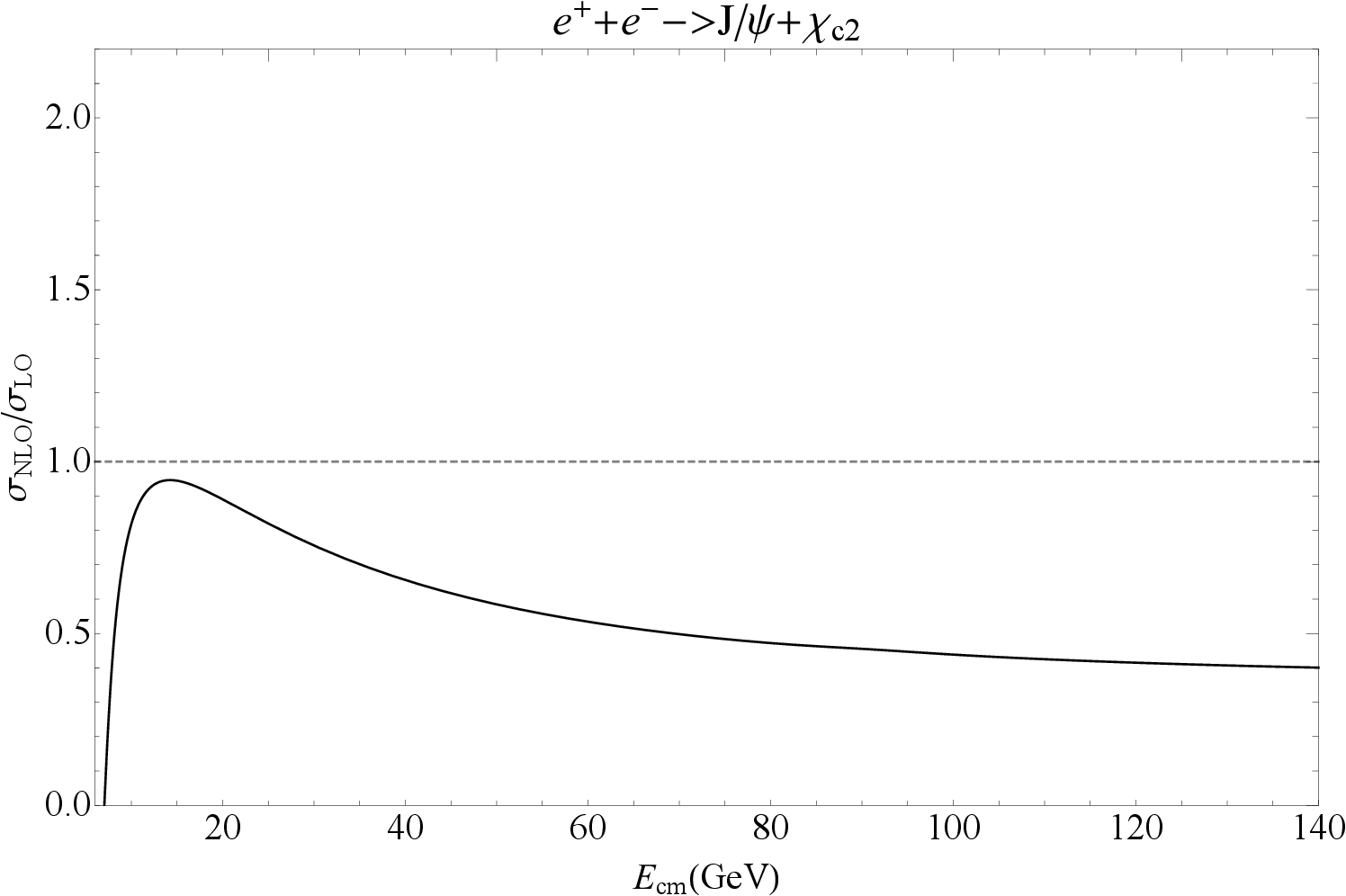}
			\includegraphics[width=0.333\textwidth]{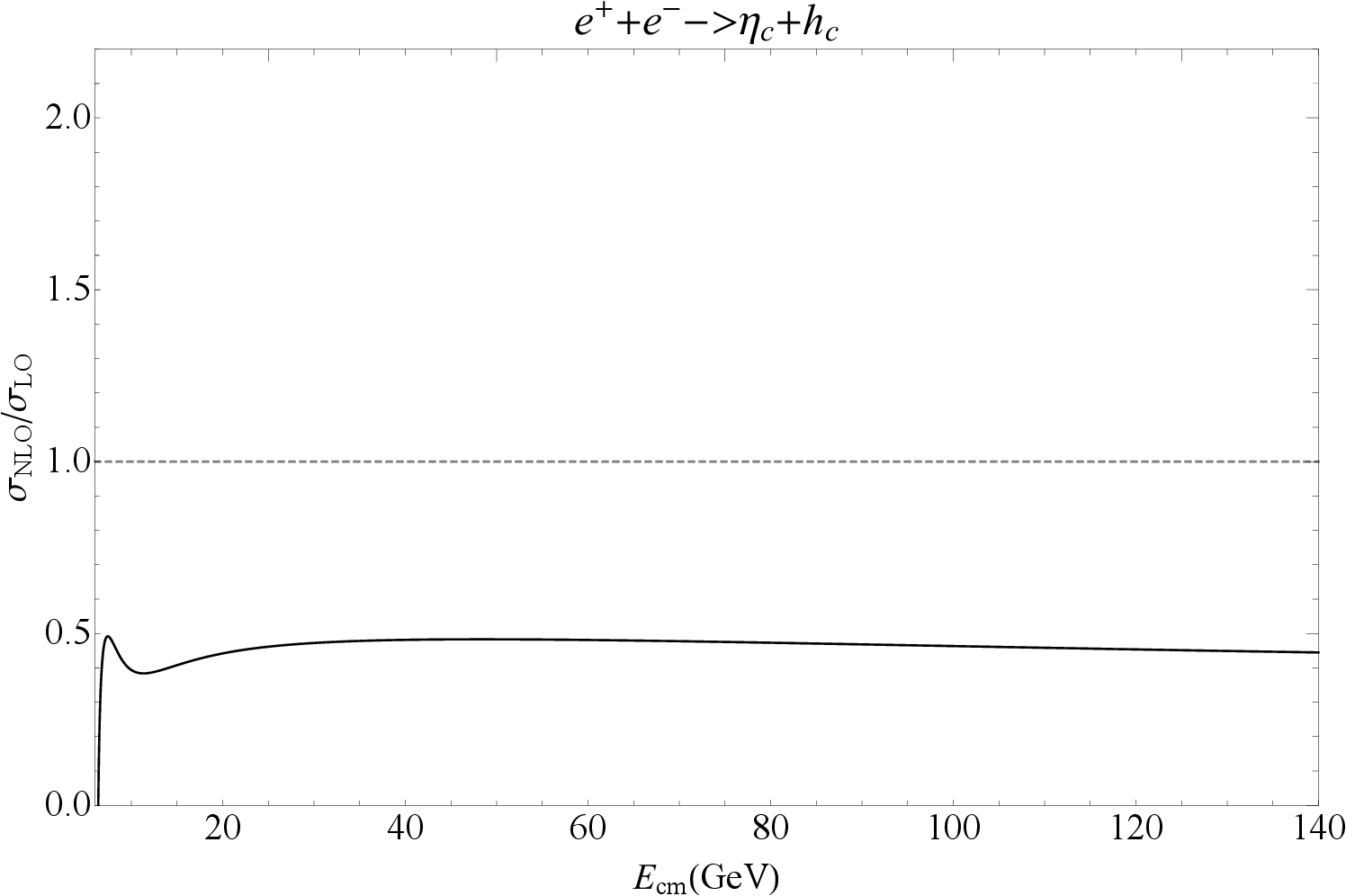}
			\includegraphics[width=0.333\textwidth]{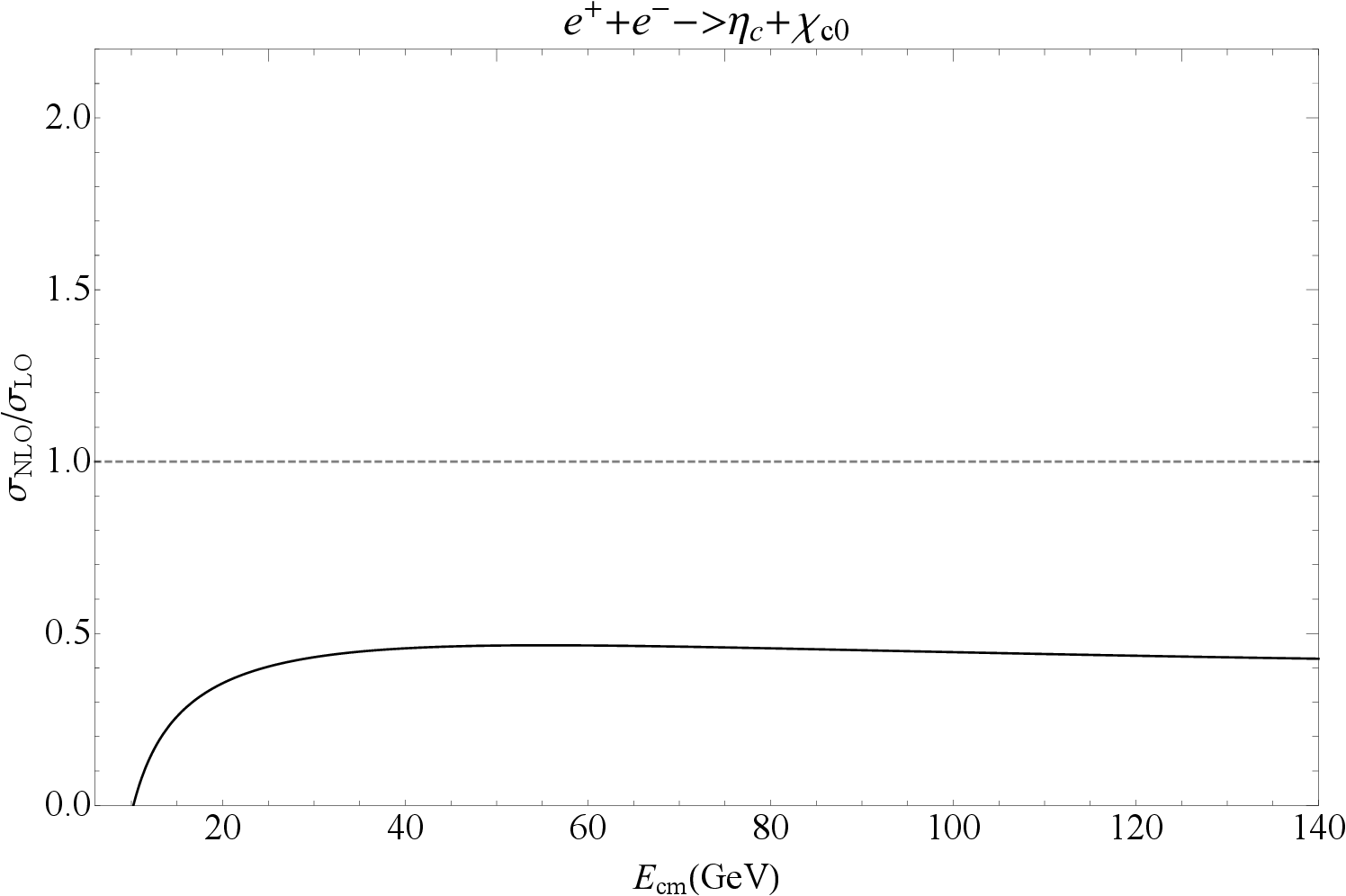}
		\end{tabular}
		\begin{tabular}{c c c}
			\includegraphics[width=0.333\textwidth]{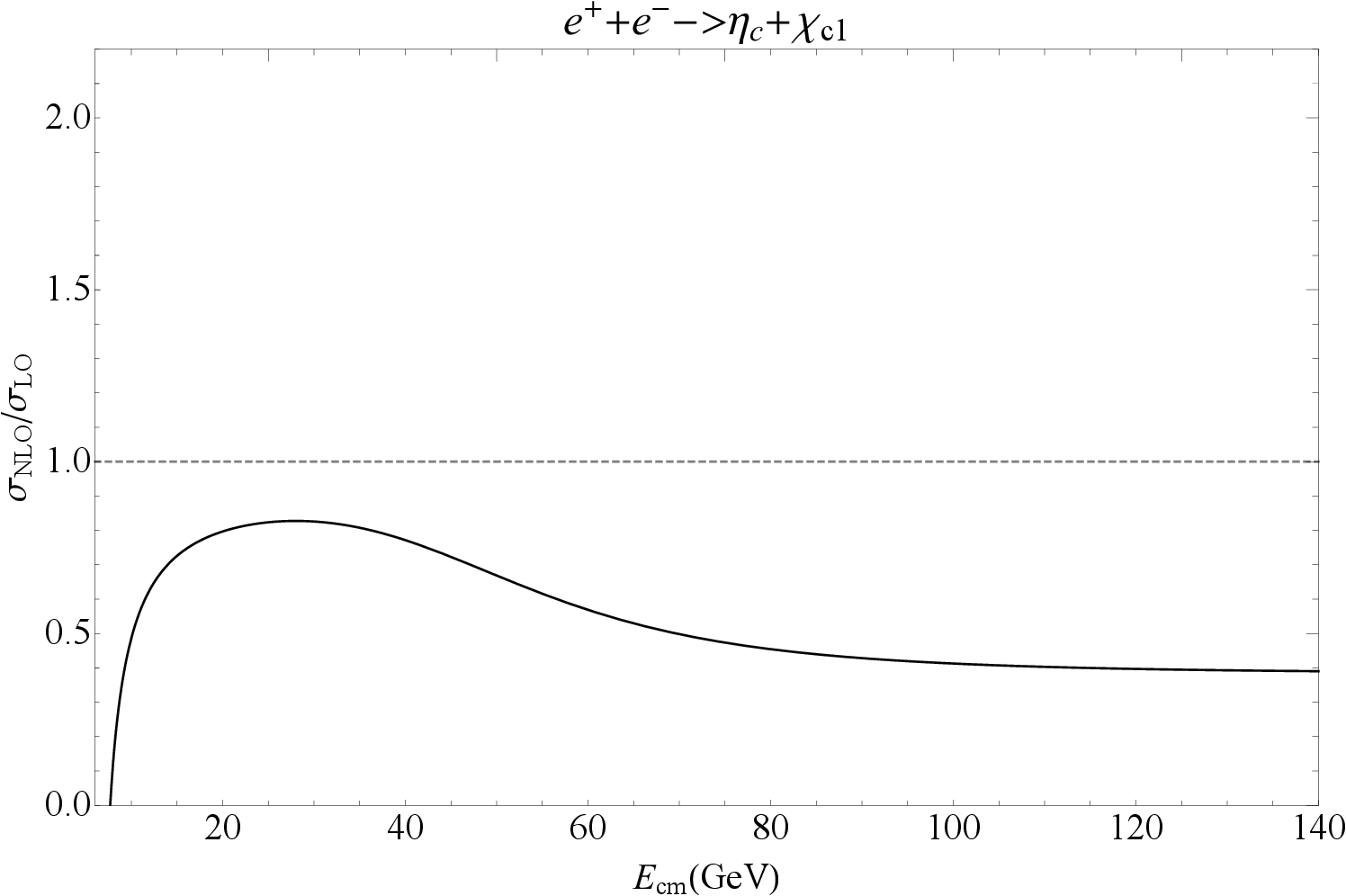}
			\includegraphics[width=0.333\textwidth]{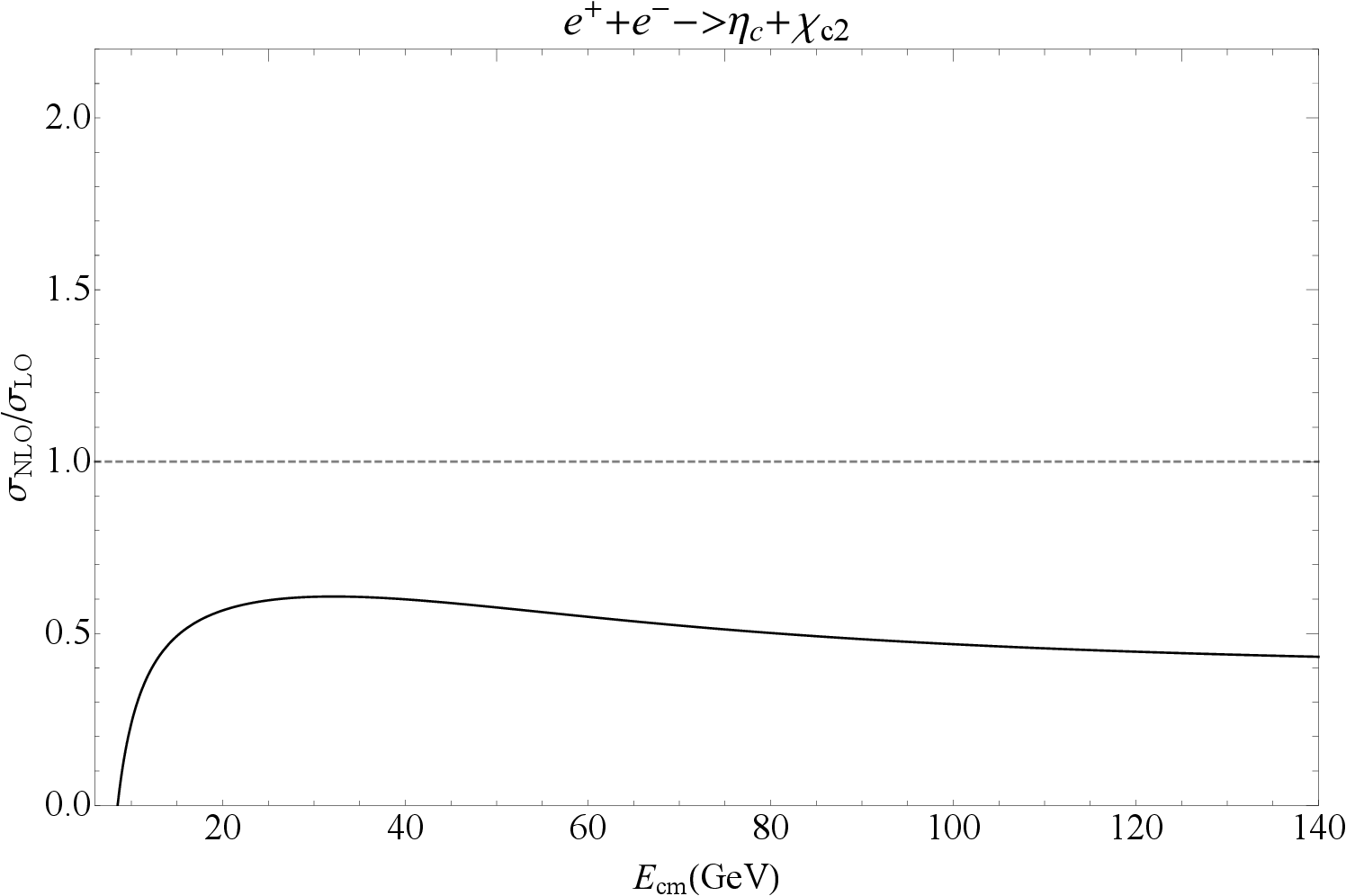}
			\includegraphics[width=0.333\textwidth]{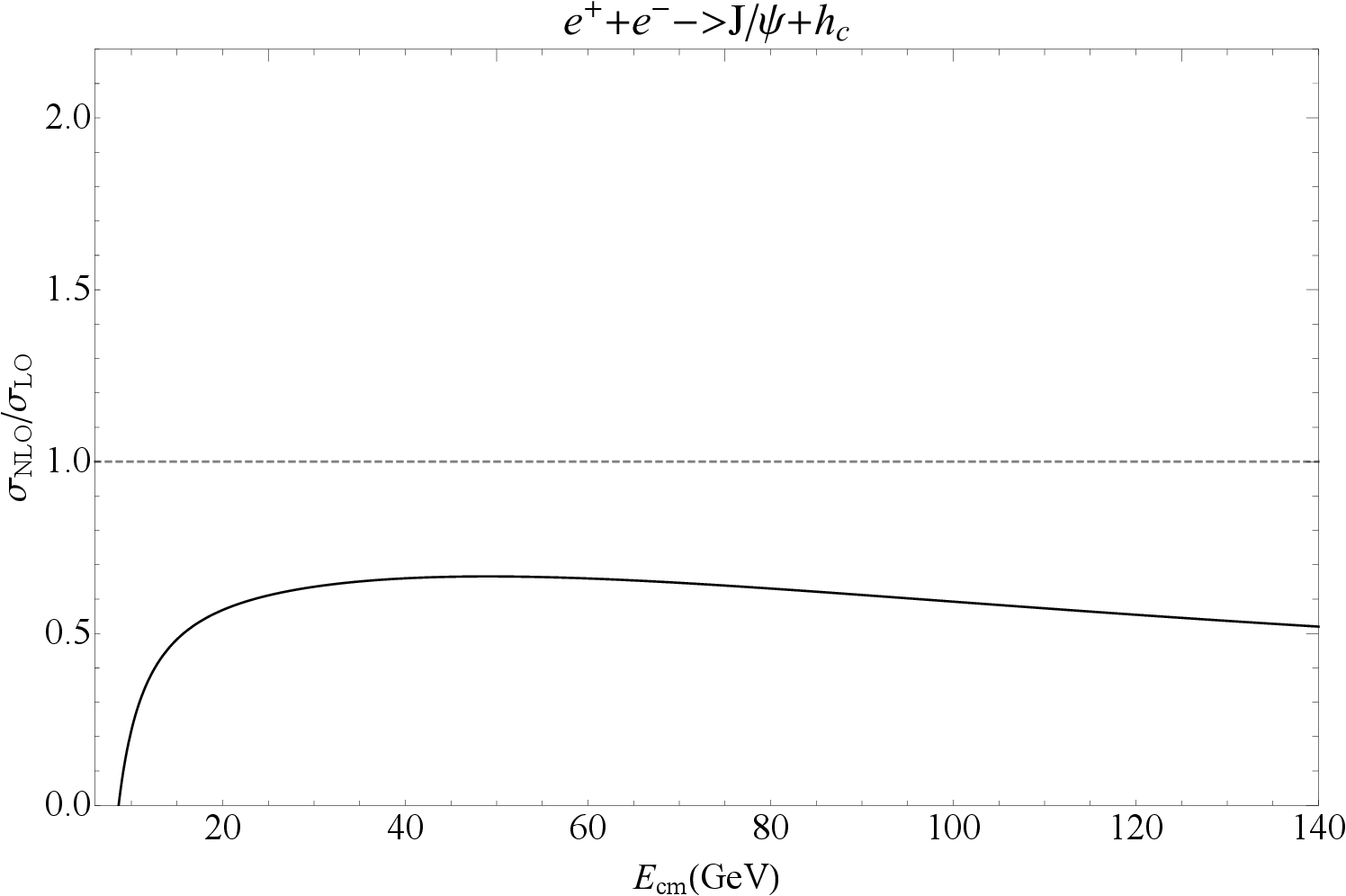}
		\end{tabular}
		\begin{tabular}{c c c }
			\includegraphics[width=0.333\textwidth]{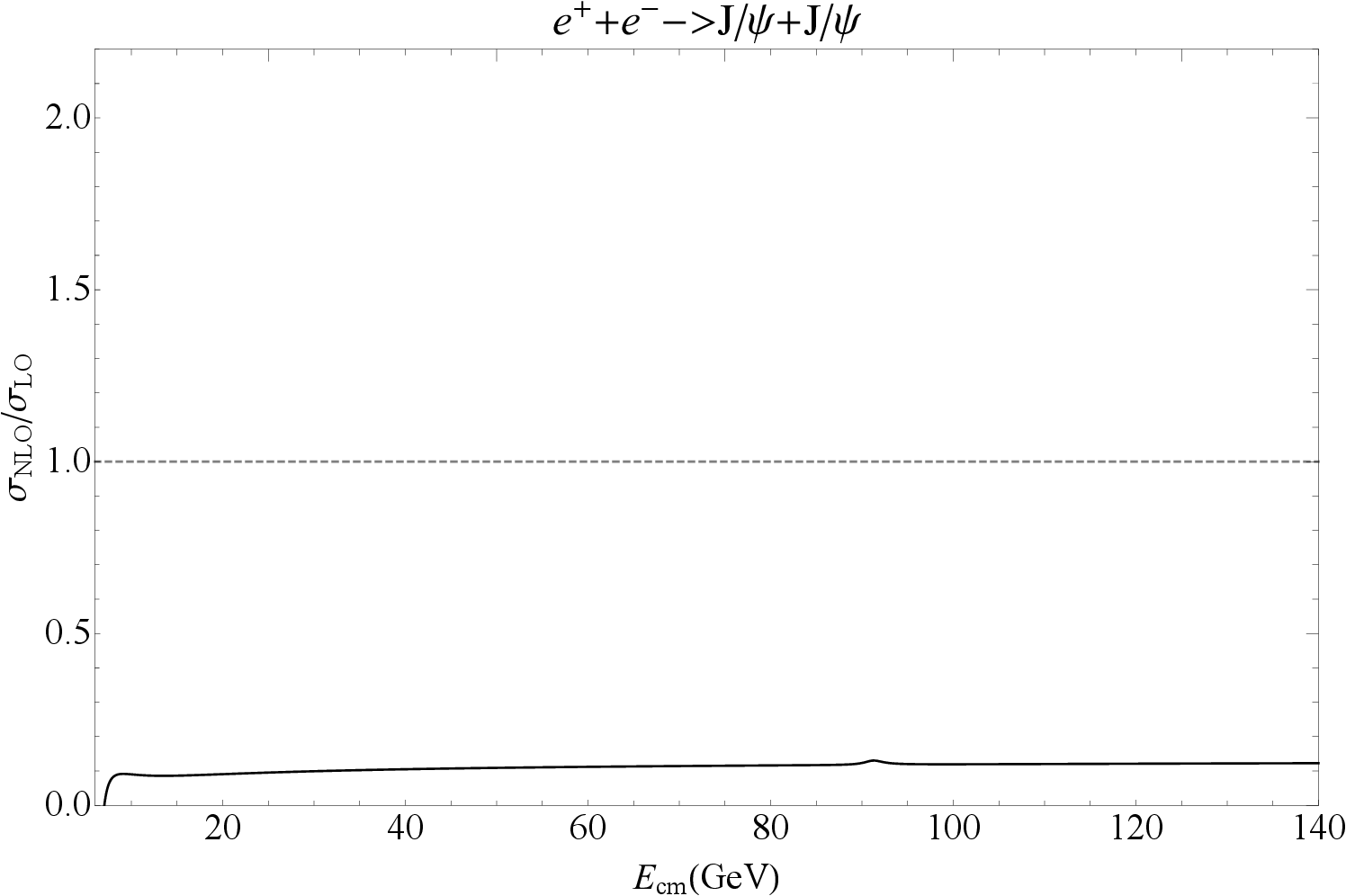}
				\includegraphics[width=0.333\textwidth]{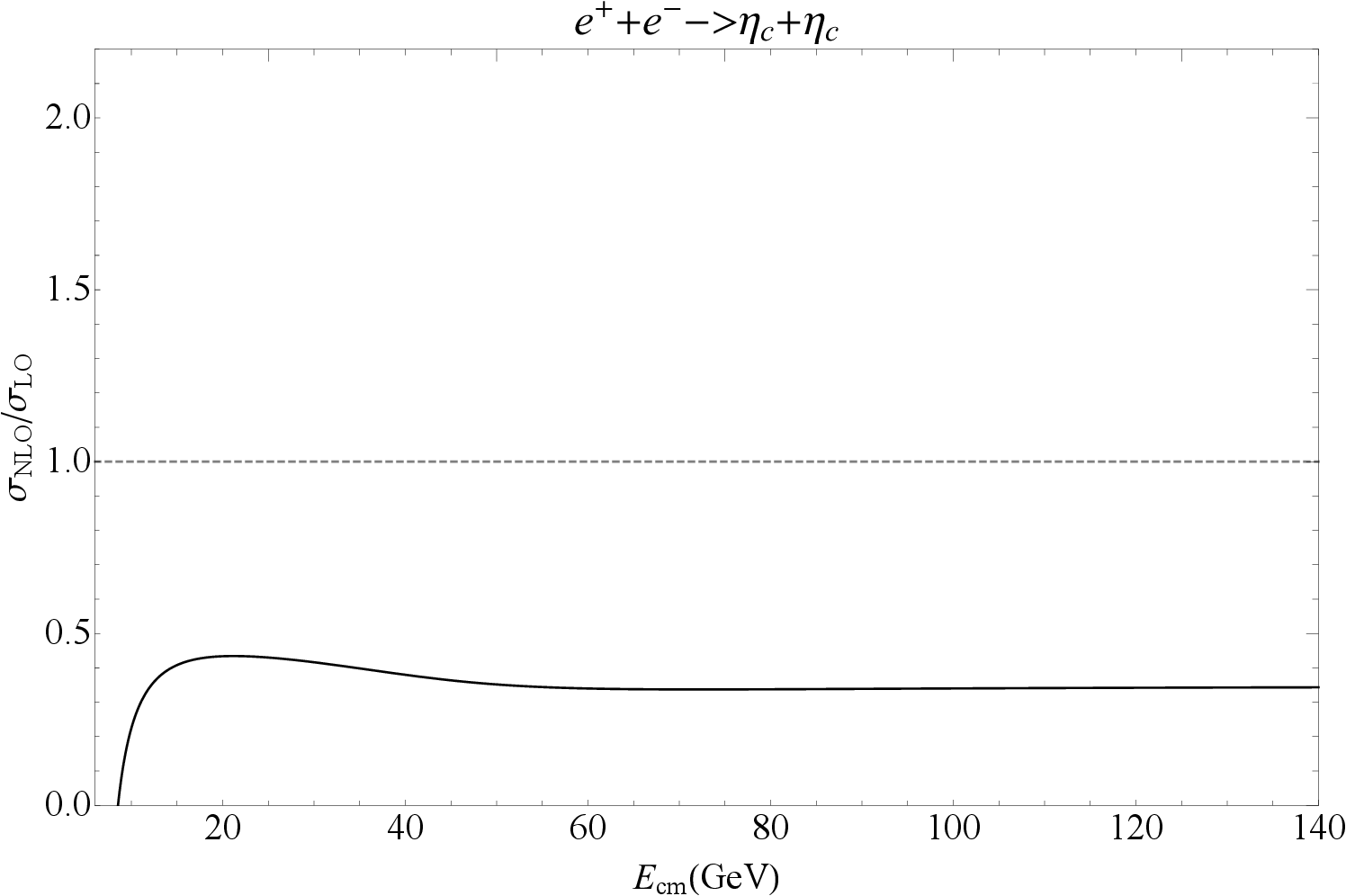}
		\end{tabular}
		\caption{ (Color online) The $K$ factors for next-to-leading order cross sections in $v^2$ as a function of c.m. energy for the production of double charmonium. }
		\label{z0cck}
	\end{figure*}
	\FloatBarrier
\end{widetext}

	\begin{widetext}
	\begin{figure*}[htbp]
		\begin{tabular}{c c c }
			\includegraphics[width=0.333\textwidth]{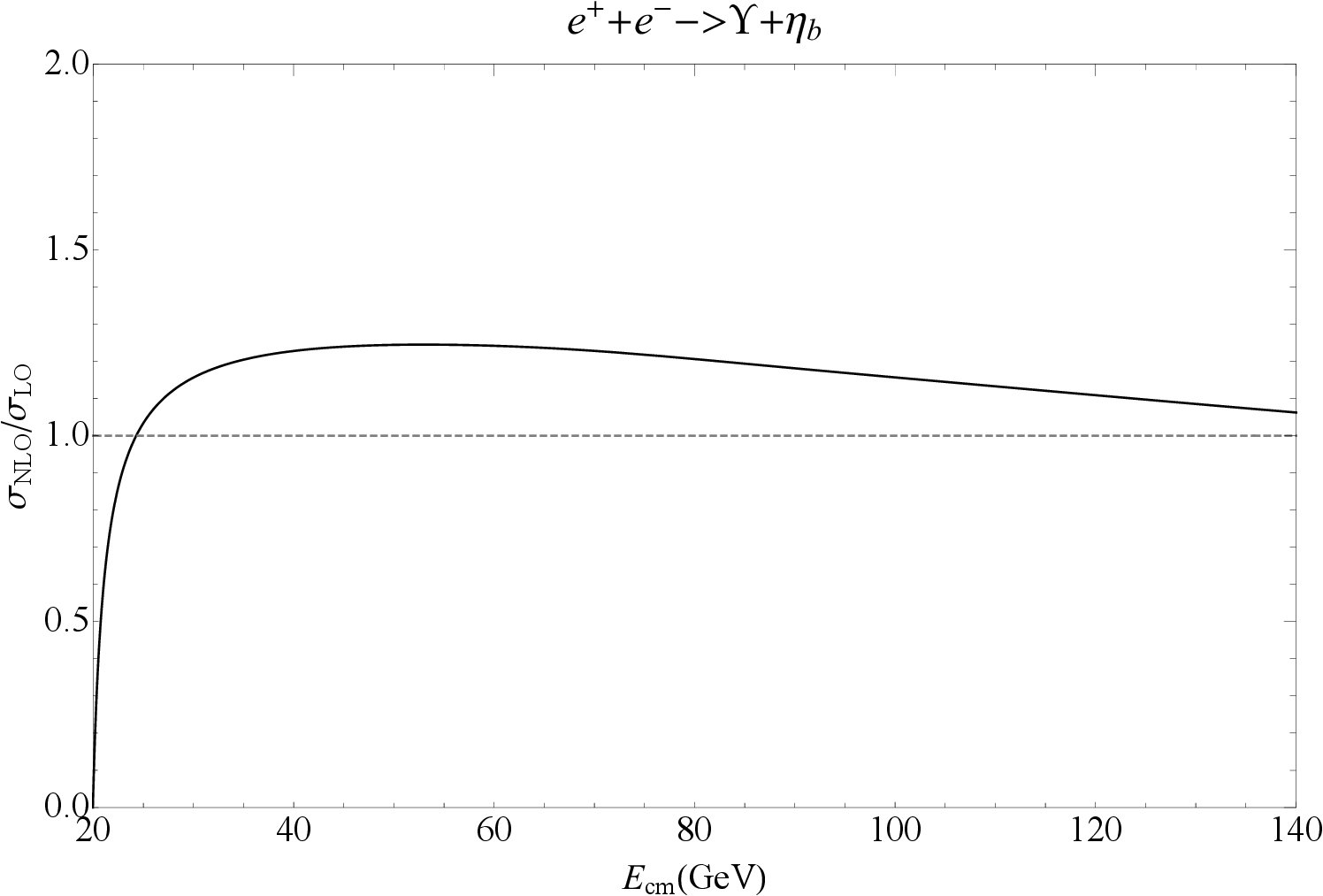}
			\includegraphics[width=0.333\textwidth]{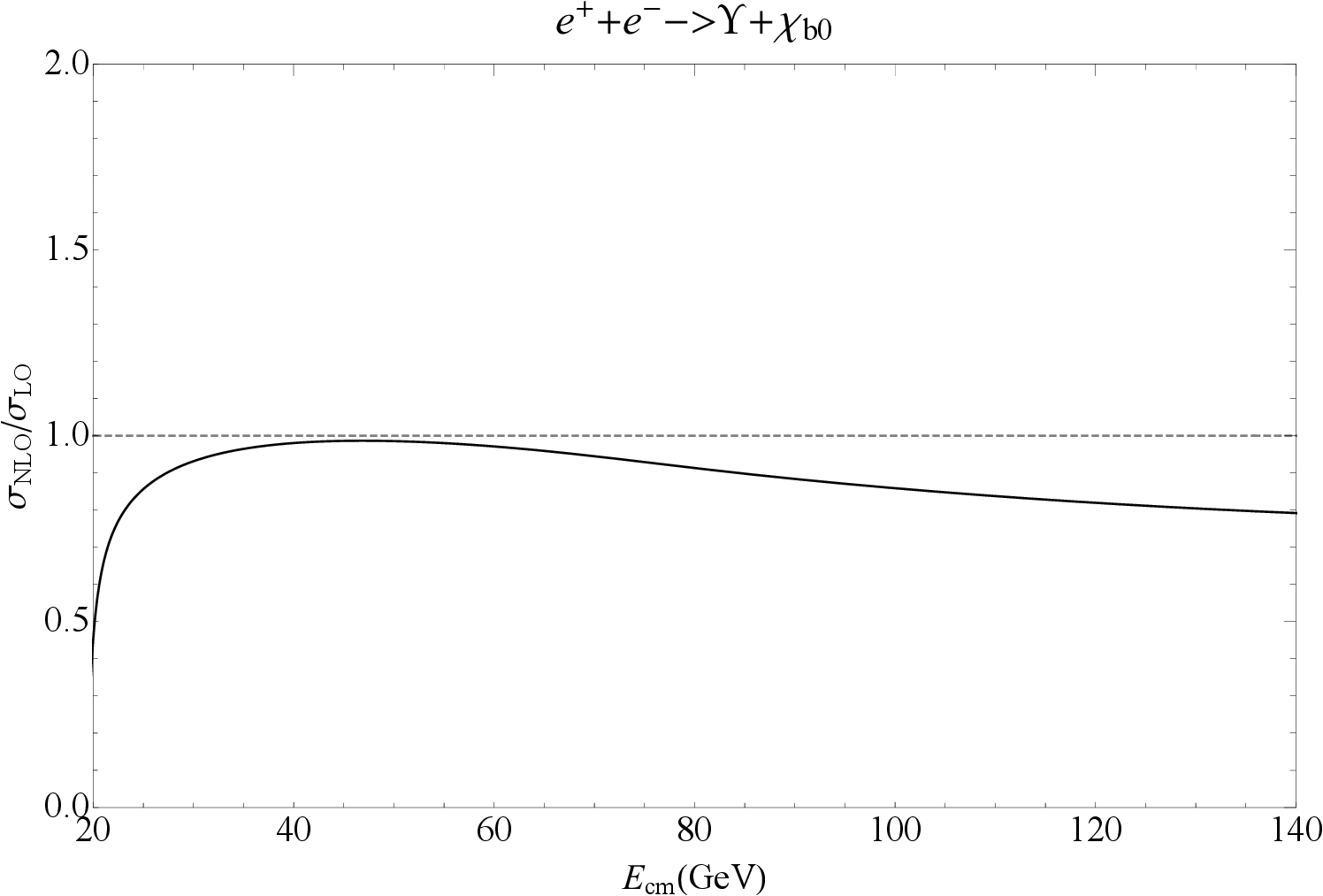}
			\includegraphics[width=0.333\textwidth]{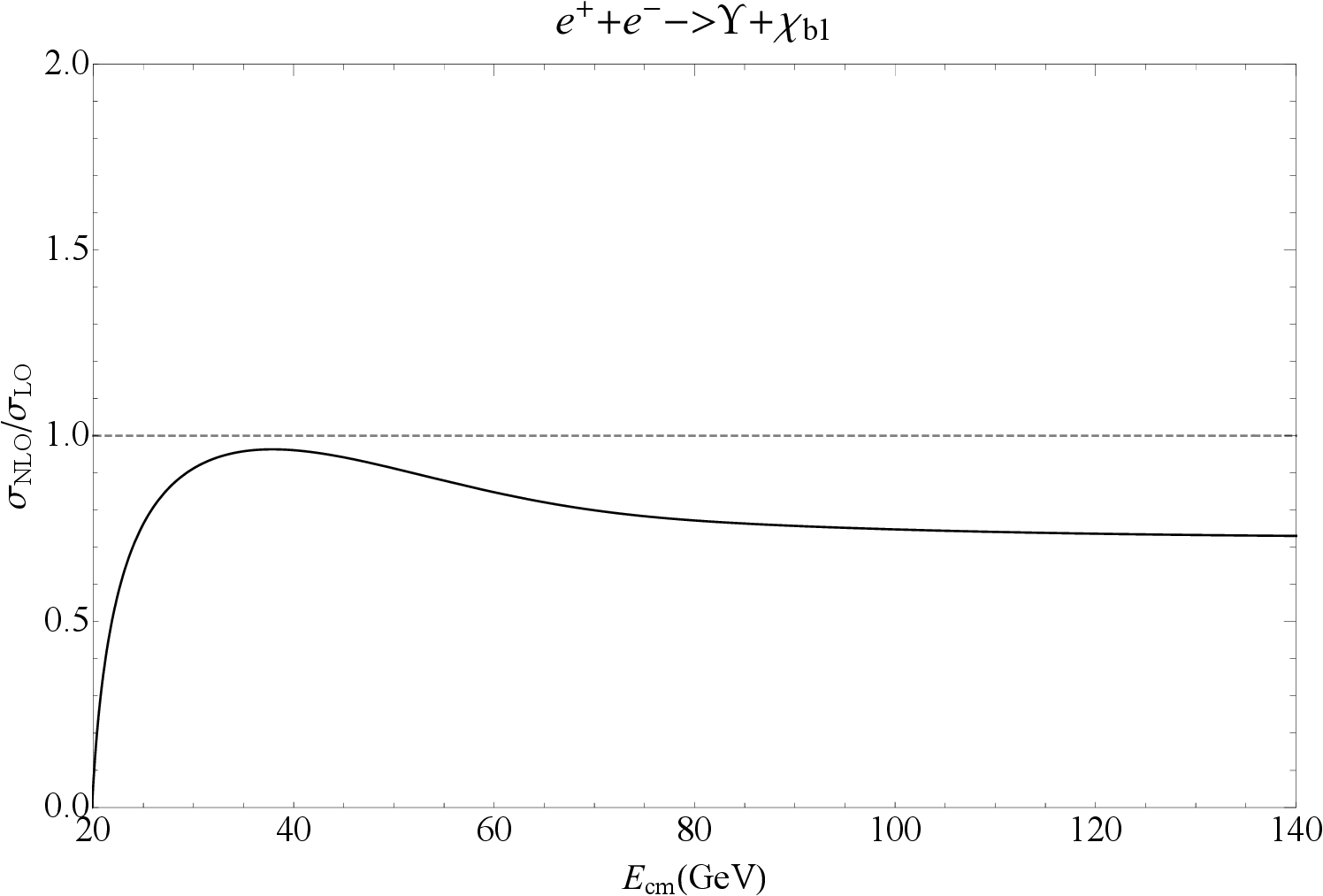}
		\end{tabular}
		\begin{tabular}{c c c }
			\includegraphics[width=0.333\textwidth]{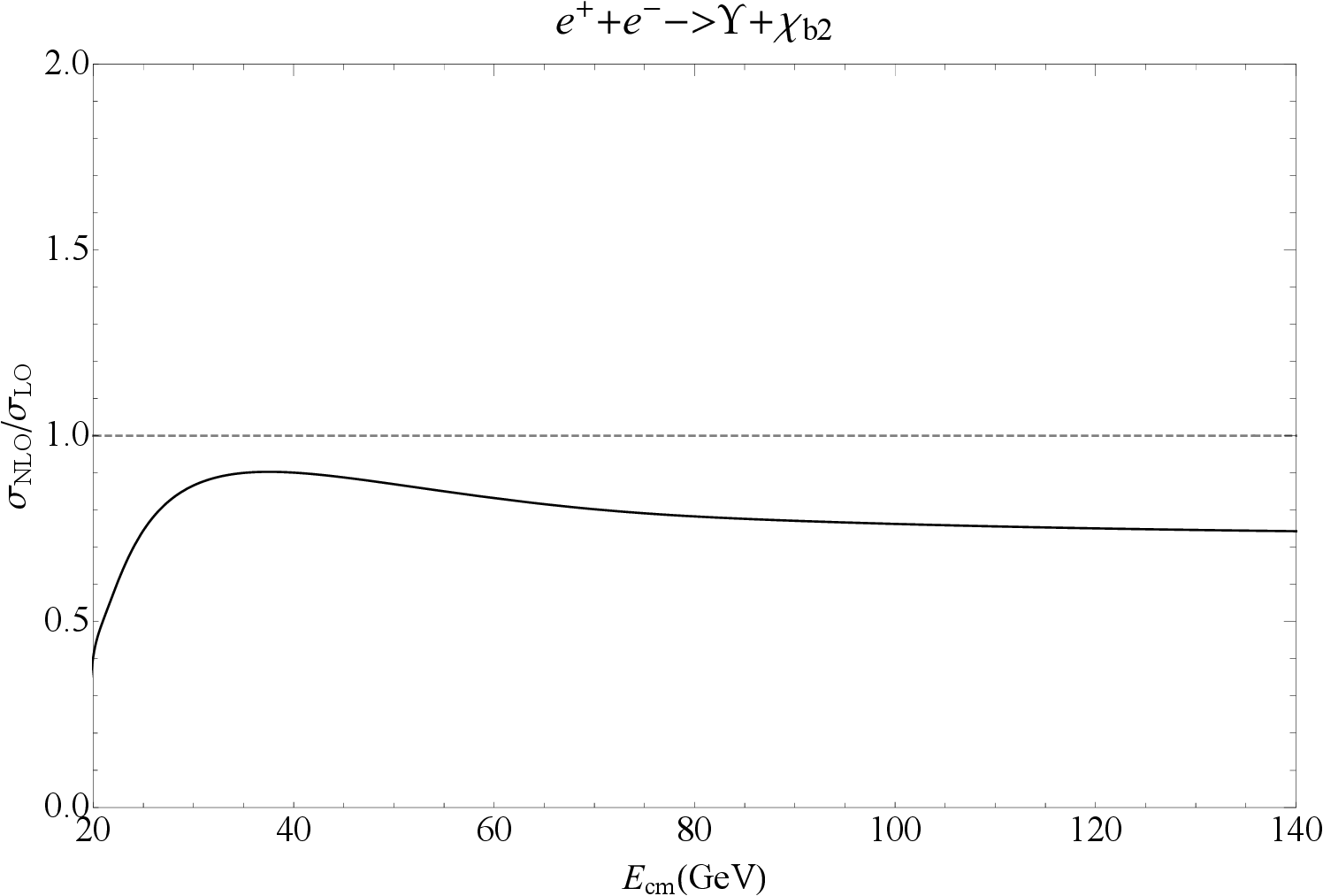}
			\includegraphics[width=0.333\textwidth]{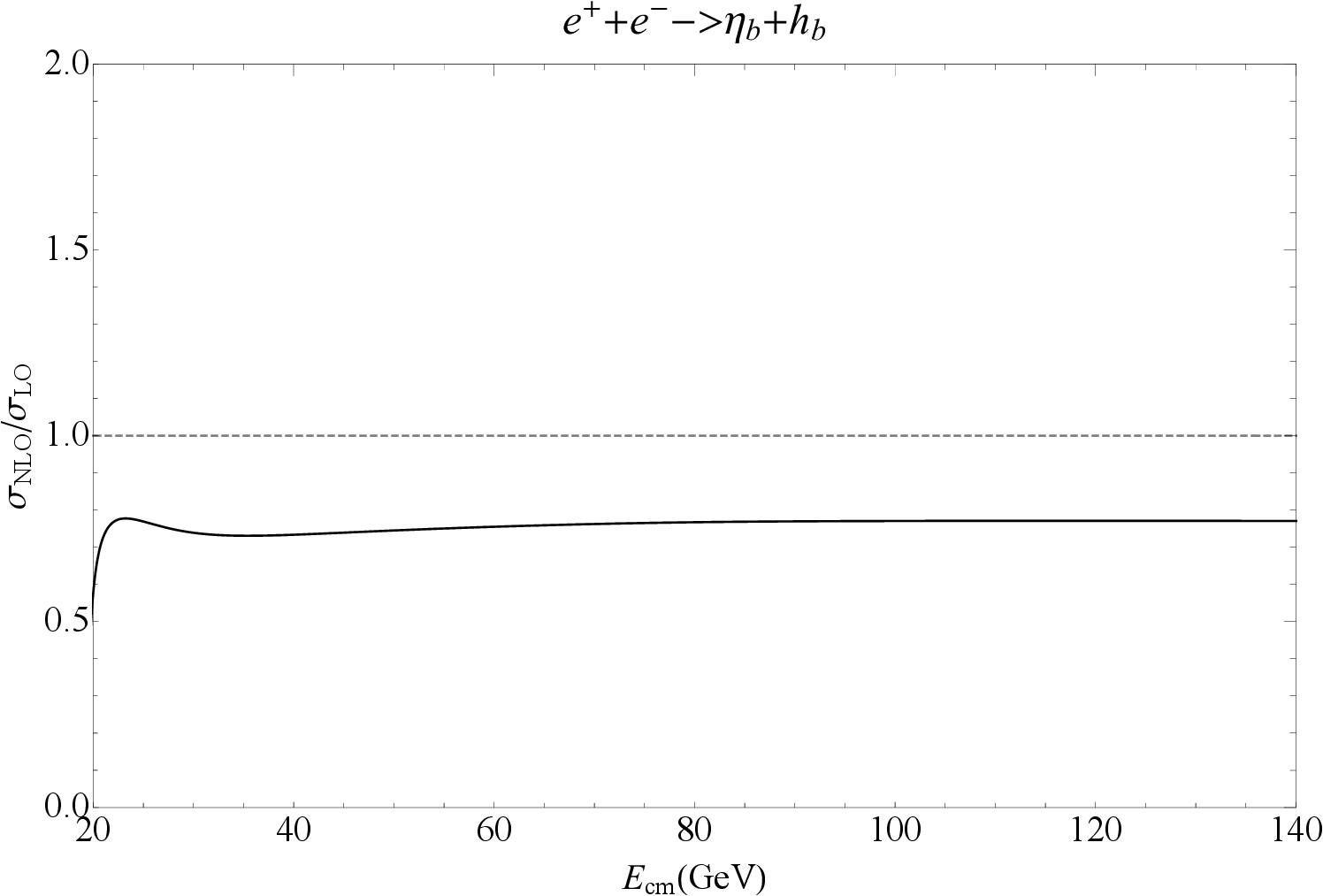}
			\includegraphics[width=0.333\textwidth]{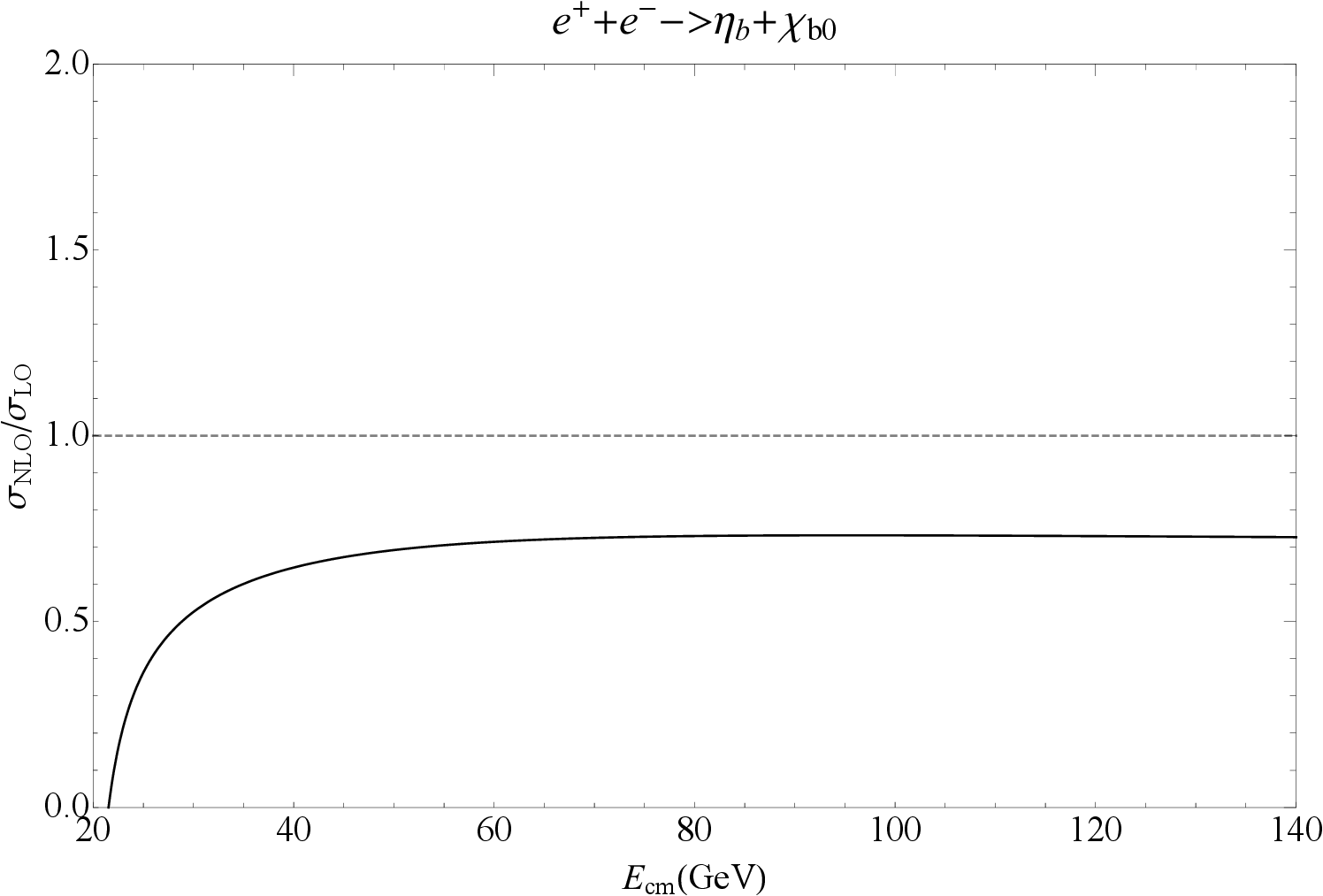}
		\end{tabular}
		\begin{tabular}{c c c }
			\includegraphics[width=0.333\textwidth]{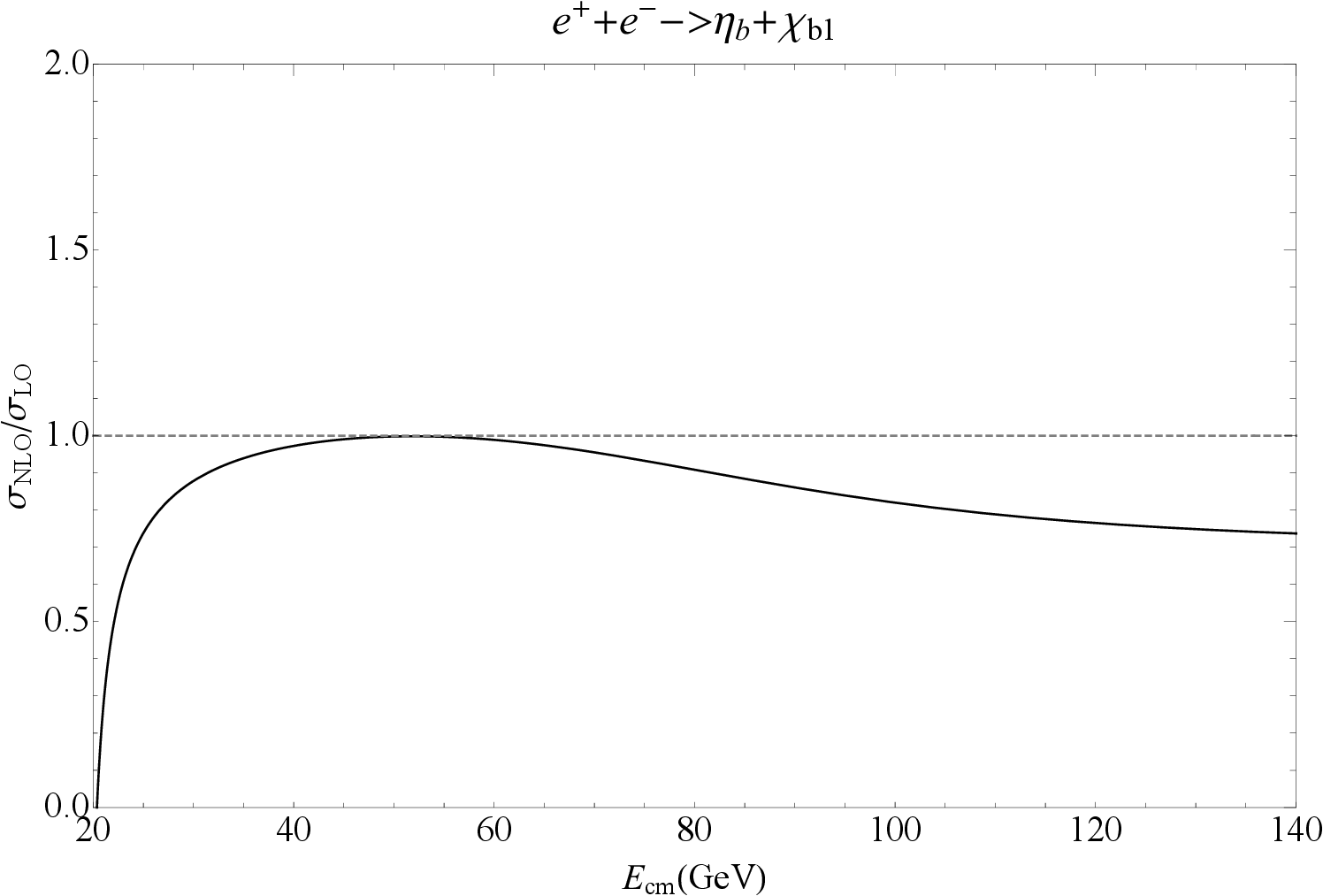}
			\includegraphics[width=0.333\textwidth]{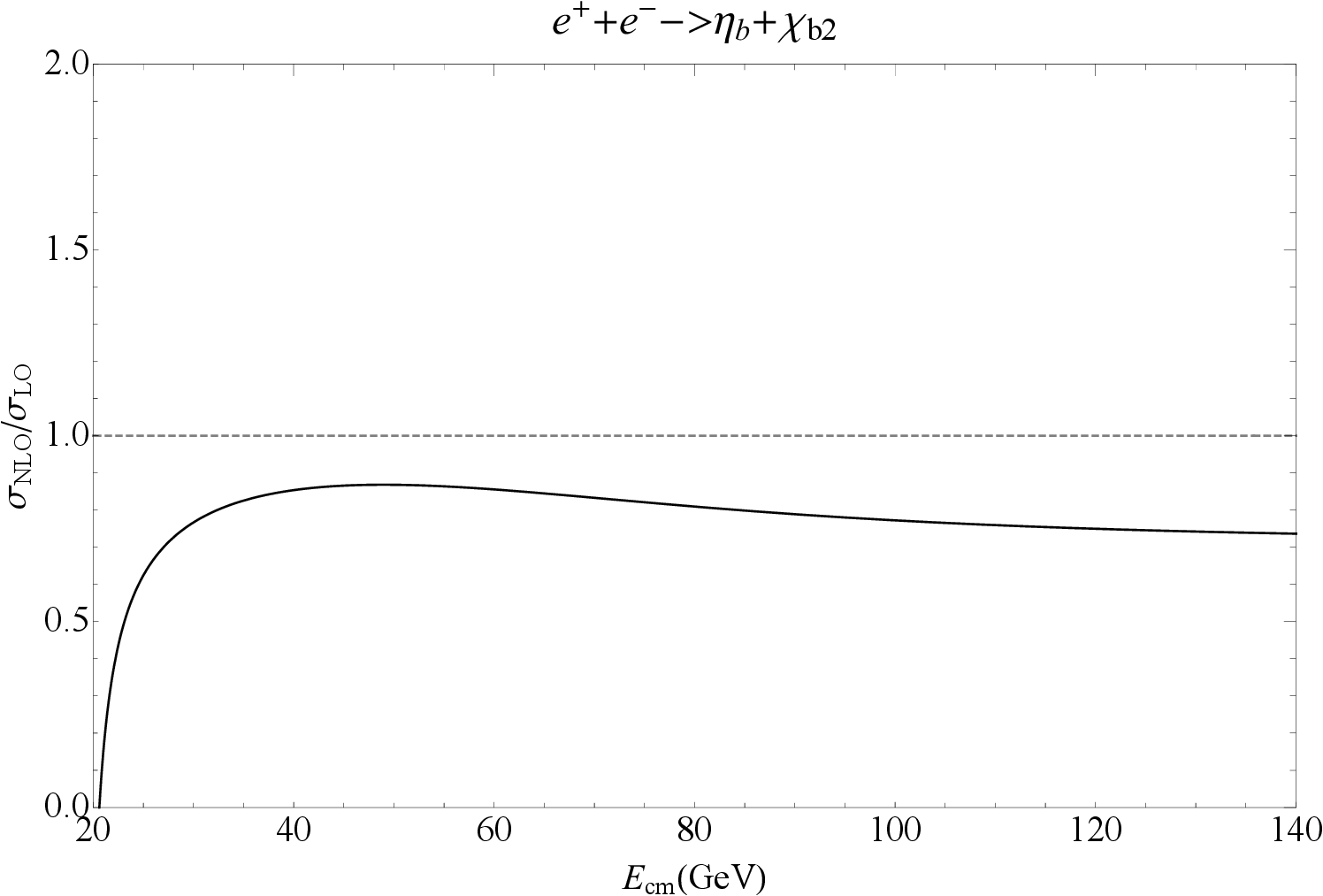}
			\includegraphics[width=0.333\textwidth]{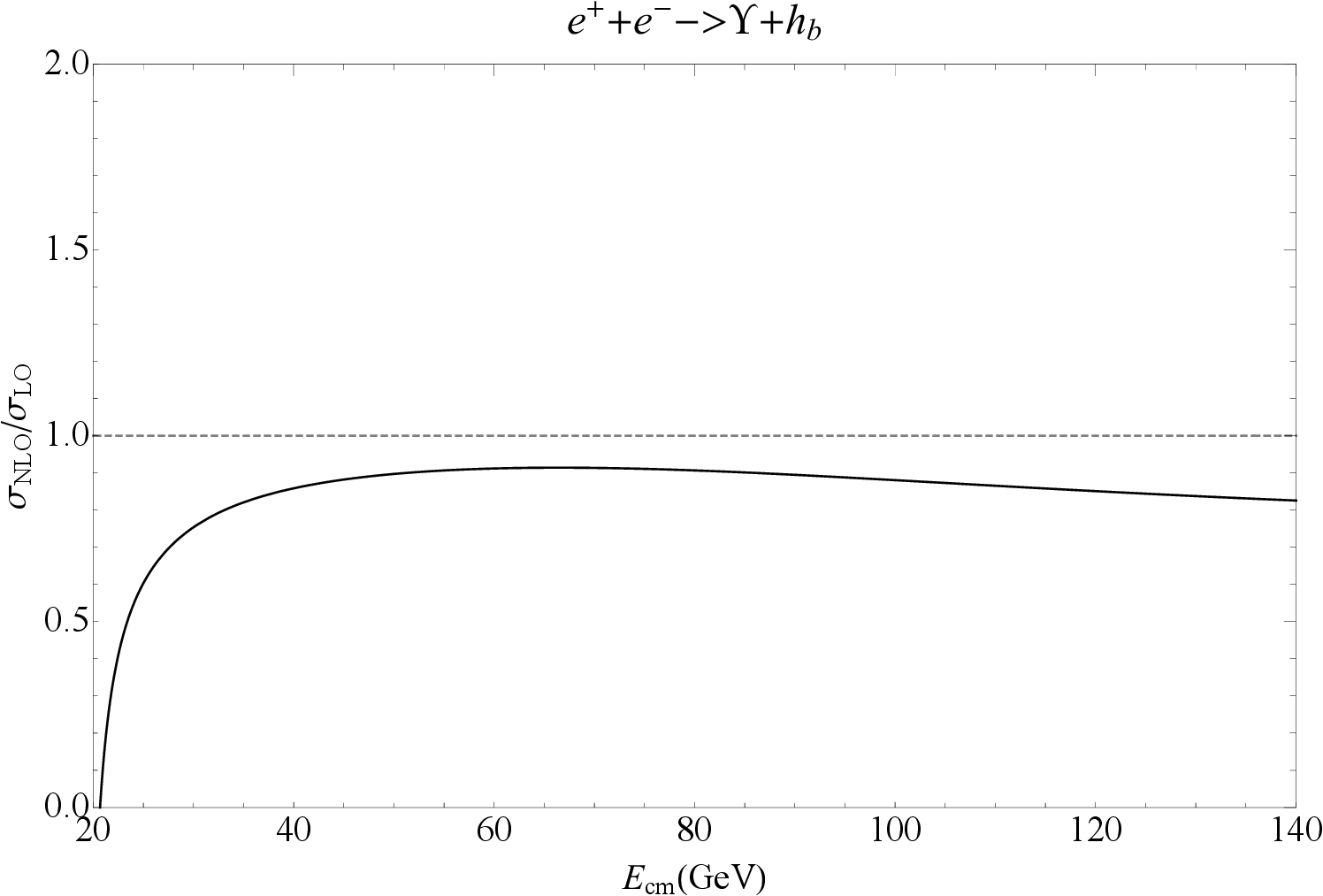}
		\end{tabular}
		\begin{tabular}{c c c}
			\includegraphics[width=0.333\textwidth]{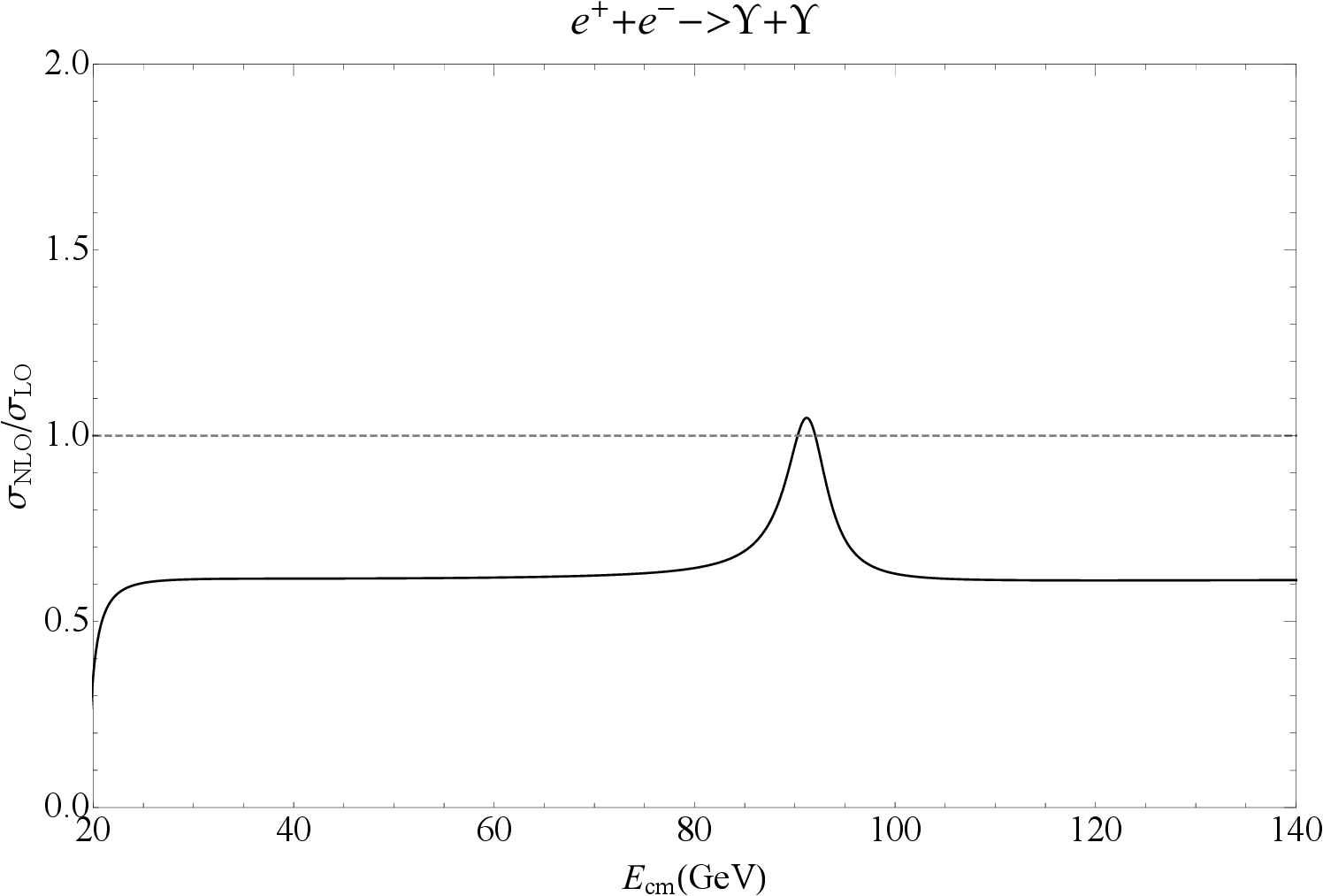}	
				\includegraphics[width=0.333\textwidth]{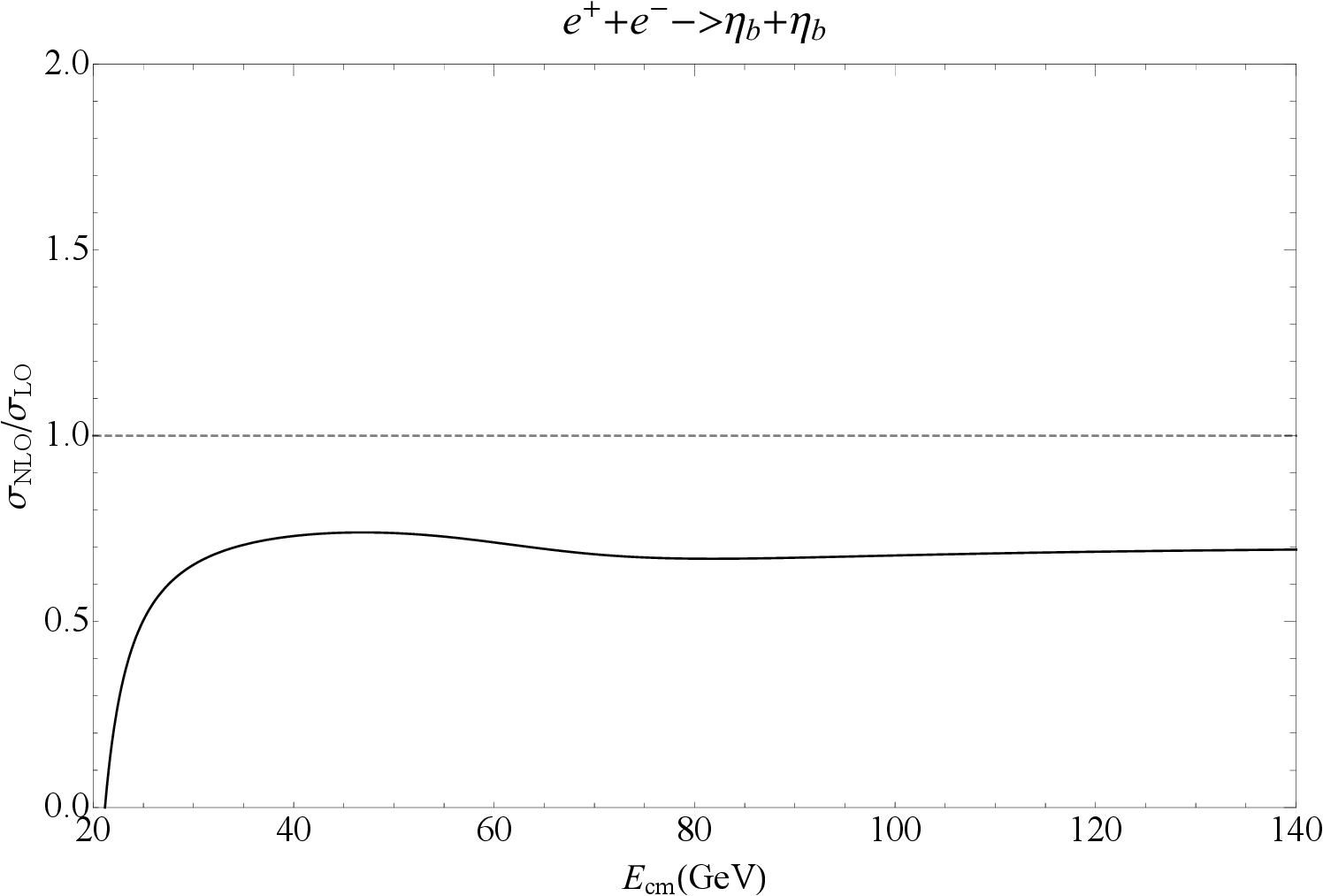}
		\end{tabular}
		\caption{ (Color online) The $K$ factors for next-to-leading order cross sections in $v^2$ as a function of c.m. energy for the production of double bottomonium. }
		\label{z0bbk}
	\end{figure*}
	\FloatBarrier
\end{widetext}

		\section{Appendix. B}
		\label{appdB}
	 Rather than presenting the analytical expression of the SDCs, we provide in Table \ref{Rratio} the ratios of relativistic correction SDCs (denoted as G) to LO SDCs (denoted as F) in the high c.m. energy limit ($m_Q^2<<s$). For specific states, the ratios are determined by the topologies of the Feynman diagrams. Consequently, each ratio is applicable to both charmonia and bottomonia in the channels with consistent states. The gluon fragmentation processes into $^3S_1^{[8]}$ are dominated for the $^3S_1^{[8]}$ channels associated with $^1S_0^{[8]}, ^3P_J^{[8]},  ^1P_1^{[8]}$ states. Their ratios can be verified against the Refs.\cite{Xu2012,YJLi}. And the photon fragmentation processes into $^3S_1^{[1]}$ are dominated for the CS channels of $J/\psi$ production associated with $\eta_c, \chi_{c0,1,2}, h_c$. Their ratios can be verified against the Refs. \cite{Xu:2014zra}. In the high c.m. energy limit, the $K$ factors for $NLO(v^2)$ can be obtained from the ratios.
In the energy region near $Z^0$ pole, the $K$ factors would be jointly determined by the ratios of the non-fragmentation and fragmentation processes for $J/\psi$ exclusive production.

	\begin{table}[ht!]
		\caption{Ratios of relativistic correction SDCs (denoted as G) to LO SDCs (denoted as F) in the high c.m. energy limit ($m_Q^2<<s$). In each cell, we define, $R_i\equiv G[n_i]/F[n_1+n_2]$, $c=1,8$ for CS or CO states and (nf/f) as the non-fragmentation/fragmentation processes corresponding to (a,b,e,f)/(c,d,g,h) diagrams in Fig.\ref{feynmandia}. The first three rows  correspond to positive total C-parity of final states and are applicable to $\gamma^*Z^0$-propagated processes, while the second group of three rows correspond to negative C-parity and are applicable only to $Z^0$-propagated processes. The last rows correspond to the t-channel processes.
		}
		\begin{tabular}{|cll|cll|cll|}
			\hline
(nf)${}^3S_1^{[1,8]}+{}^1S_0^{[1,8]}$&$R_1= \frac{3}{2}$&$R_2=\frac{11}{6}$&
(nf)${}^3S_1^{[1,8]}+{}^3P_0^{[1,8]}$&$R_1= -\frac{1}{6}$&$R_2=-\frac{13}{10}$&
(nf)${}^3S_1^{[1,8]}+{}^3P_1^{[1,8]}$&$R_1= \frac{3}{2}$&$R_2=\frac{1}{2}$\\
			\hline
(nf)${}^3S_1^{[1,8]}+{}^3P_2^{[1,8]}$&$R_1= -\frac{1}{6}$&$R_2=-\frac{7}{10}$&
(nf)${}^1S_0^{[1,8]}+{}^1P_1^{[1,8]}$&$R_1= -\frac{5}{6}$&$R_2=-\frac{13}{10}$&
(f)${}^3S_1^{[1,8]}+{}^1S_0^{[1,8]}$&$R_1=-\frac{11}{6}$&$R_2= -\frac{5}{6}$\\
			\hline
(f)${}^3S_1^{[1,8]}+{}^3P_0^{[1,8]}$&$R_1=-\frac{11}{6}$&$R_2= -\frac{13}{10}$&
(f)${}^3S_1^{[1,8]}+{}^3P_1^{[1,8]}$&$R_1=-\frac{11}{6}$&$R_2= -\frac{11}{10}$&
(f)${}^3S_1^{[1,8]}+{}^3P_2^{[1,8]}$&$R_1=-\frac{11}{6}$&$R_2= -\frac{7}{10}$\\
			\hline
(f)${}^3S_1^{[8]}+{}^3P_J^{[8]}$&$R_1=-\frac{11}{6}$&$R_2= -\frac{31}{30}$&
(nf)${}^3P_J^{[8]}+{}^1P_1^{[8]}$&$R_1=-\frac{11}{10}$&$R_2=-\frac{13}{10}$&
&&\\
            \hline
            \hline
(nf)${}^3S_1^{[1,8]}+{}^1P_1^{[1,8]}$&$R_1=-\frac{1}{6}$&$R_2= -\frac{13}{10}$&
(nf)${}^1S_0^{[1]}+{}^3P_0^{[1]}$&$R_1=-\frac{5}{6}$&$R_2= -\frac{13}{10}$&
(nf)${}^1S_0^{[1]}+{}^3P_1^{[1]}$&$R_1=\frac{7}{6}$&$R_2= \frac{3}{10}$\\
			\hline
(nf)${}^1S_0^{[1]}+{}^3P_2^{[1]}$&$R_1=-\frac{5}{6}$&$R_2= -\frac{7}{10}$&
(nf)${}^1S_0^{[8]}+{}^3P_J^{[8]}$&$R_1=-\frac{5}{6}$&$R_2= -\frac{9}{10}$&
(f)${}^3S_1^{[1,8]}+{}^1P_1^{[1,8]}$&$R_1=-\frac{11}{6}$&$R_2=-\frac{13}{10}$\\
			\hline
(f)${}^3S_1^{[8]}+{}^3S_1^{[8]}$&\multicolumn{2}{c |}{$R_1=R_2= -1$}&
(nf)${}^3P_J^{[8]}+{}^3P_J^{[8]}$&\multicolumn{2}{c |}{$R_1=R_2= -1$}&
(nf)${}^1P_1^{[8]}+{}^1P_1^{[8]}$&\multicolumn{2}{c |}{$R_1=R_2=\frac{6}{5}$}\\
			\hline
            \hline
${}^3S_1^{[1]}+{}^3S_1^{[1]}$&\multicolumn{2}{c |}{$R_1=R_2= -\frac{11}{6}$}&
${}^1S_0^{[1]}+{}^1S_0^{[1]}$&\multicolumn{2}{c |}{$R_1=R_2= -\frac{1}{2}$}&
&&\\
			\hline
		\end{tabular}
		\label{Rratio}
	\end{table}
\FloatBarrier

\section{Acknowledgements:} The authors would like to thank Dr. Guang-Yu Wang for providing the code of NLO radiative corrections for $J/\psi$ pair production. This work was supported by the National Natural Science Foundation of China (No 11705078).

\hspace{2cm}
	

\begin{thebibliography}{87}%
\makeatletter
\providecommand \@ifxundefined [1]{%
 \@ifx{#1\undefined}
}%
\providecommand \@ifnum [1]{%
 \ifnum #1\expandafter \@firstoftwo
 \else \expandafter \@secondoftwo
 \fi
}%
\providecommand \@ifx [1]{%
 \ifx #1\expandafter \@firstoftwo
 \else \expandafter \@secondoftwo
 \fi
}%
\providecommand \natexlab [1]{#1}%
\providecommand \enquote  [1]{``#1''}%
\providecommand \bibnamefont  [1]{#1}%
\providecommand \bibfnamefont [1]{#1}%
\providecommand \citenamefont [1]{#1}%
\providecommand \href@noop [0]{\@secondoftwo}%
\providecommand \href [0]{\begingroup \@sanitize@url \@href}%
\providecommand \@href[1]{\@@startlink{#1}\@@href}%
\providecommand \@@href[1]{\endgroup#1\@@endlink}%
\providecommand \@sanitize@url [0]{\catcode `\\12\catcode `\$12\catcode
  `\&12\catcode `\#12\catcode `\^12\catcode `\_12\catcode `\%12\relax}%
\providecommand \@@startlink[1]{}%
\providecommand \@@endlink[0]{}%
\providecommand \url  [0]{\begingroup\@sanitize@url \@url }%
\providecommand \@url [1]{\endgroup\@href {#1}{\urlprefix }}%
\providecommand \urlprefix  [0]{URL }%
\providecommand \Eprint [0]{\href }%
\providecommand \doibase [0]{https://doi.org/}%
\providecommand \selectlanguage [0]{\@gobble}%
\providecommand \bibinfo  [0]{\@secondoftwo}%
\providecommand \bibfield  [0]{\@secondoftwo}%
\providecommand \translation [1]{[#1]}%
\providecommand \BibitemOpen [0]{}%
\providecommand \bibitemStop [0]{}%
\providecommand \bibitemNoStop [0]{.\EOS\space}%
\providecommand \EOS [0]{\spacefactor3000\relax}%
\providecommand \BibitemShut  [1]{\csname bibitem#1\endcsname}%
\let\auto@bib@innerbib\@empty
\bibitem [{\citenamefont {Bodwin}\ \emph {et~al.}(1995)\citenamefont {Bodwin},
  \citenamefont {Braaten},\ and\ \citenamefont {Lepage}}]{NRQCD}%
  \BibitemOpen
  \bibfield  {author} {\bibinfo {author} {\bibfnamefont {G.~T.}\ \bibnamefont
  {Bodwin}}, \bibinfo {author} {\bibfnamefont {E.}~\bibnamefont {Braaten}},\
  and\ \bibinfo {author} {\bibfnamefont {G.~P.}\ \bibnamefont {Lepage}},\
  }\bibfield  {title} {\bibinfo {title} {Rigorous qcd analysis of inclusive
  annihilation and production of heavy quarkonium},\ }\href@noop {} {\bibfield
  {journal} {\bibinfo  {journal} {Physical Review D}\ }\textbf {\bibinfo
  {volume} {51}},\ \bibinfo {pages} {1125} (\bibinfo {year}
  {1995})}\BibitemShut {NoStop}%
\bibitem [{\citenamefont {Bodwin}\ \emph
  {et~al.}(1992{\natexlab{a}})\citenamefont {Bodwin}, \citenamefont {Braaten},
  \citenamefont {Yuan},\ and\ \citenamefont {Lepage}}]{PwaveIR}%
  \BibitemOpen
  \bibfield  {author} {\bibinfo {author} {\bibfnamefont {G.~T.}\ \bibnamefont
  {Bodwin}}, \bibinfo {author} {\bibfnamefont {E.}~\bibnamefont {Braaten}},
  \bibinfo {author} {\bibfnamefont {T.~C.}\ \bibnamefont {Yuan}},\ and\
  \bibinfo {author} {\bibfnamefont {G.~P.}\ \bibnamefont {Lepage}},\ }\bibfield
   {title} {\bibinfo {title} {P-wave charmonium production in b-meson decays},\
  }\href@noop {} {\bibfield  {journal} {\bibinfo  {journal} {Physical Review
  D}\ }\textbf {\bibinfo {volume} {46}},\ \bibinfo {pages} {R3703} (\bibinfo
  {year} {1992}{\natexlab{a}})}\BibitemShut {NoStop}%
\bibitem [{\citenamefont {Abe}\ \emph {et~al.}(1997{\natexlab{a}})\citenamefont
  {Abe}, \citenamefont {Akimoto}, \citenamefont {Akopian}, \citenamefont
  {Albrow}, \citenamefont {Amendolia}, \citenamefont {Amidei}, \citenamefont
  {Antos}, \citenamefont {Aota}, \citenamefont {Apollinari}, \citenamefont
  {Asakawa} \emph {et~al.}}]{psianomalye1}%
  \BibitemOpen
  \bibfield  {author} {\bibinfo {author} {\bibfnamefont {F.}~\bibnamefont
  {Abe}}, \bibinfo {author} {\bibfnamefont {H.}~\bibnamefont {Akimoto}},
  \bibinfo {author} {\bibfnamefont {A.}~\bibnamefont {Akopian}}, \bibinfo
  {author} {\bibfnamefont {M.}~\bibnamefont {Albrow}}, \bibinfo {author}
  {\bibfnamefont {S.}~\bibnamefont {Amendolia}}, \bibinfo {author}
  {\bibfnamefont {D.}~\bibnamefont {Amidei}}, \bibinfo {author} {\bibfnamefont
  {J.}~\bibnamefont {Antos}}, \bibinfo {author} {\bibfnamefont
  {S.}~\bibnamefont {Aota}}, \bibinfo {author} {\bibfnamefont {G.}~\bibnamefont
  {Apollinari}}, \bibinfo {author} {\bibfnamefont {T.}~\bibnamefont {Asakawa}},
  \emph {et~al.},\ }\bibfield  {title} {\bibinfo {title} {J/$\psi$ and $\psi$
  (2 s) production in p p{\={}} collisions at $\sqrt{s}$= 1.8 tev},\
  }\href@noop {} {\bibfield  {journal} {\bibinfo  {journal} {Physical review
  letters}\ }\textbf {\bibinfo {volume} {79}},\ \bibinfo {pages} {572}
  (\bibinfo {year} {1997}{\natexlab{a}})}\BibitemShut {NoStop}%
\bibitem [{\citenamefont {Abe}\ \emph {et~al.}(1997{\natexlab{b}})\citenamefont
  {Abe}, \citenamefont {Akimoto}, \citenamefont {Akopian}, \citenamefont
  {Albrow}, \citenamefont {Amendolia}, \citenamefont {Amidei}, \citenamefont
  {Antos}, \citenamefont {Aota}, \citenamefont {Apollinari}, \citenamefont
  {Asakawa} \emph {et~al.}}]{psianomalye2}%
  \BibitemOpen
  \bibfield  {author} {\bibinfo {author} {\bibfnamefont {F.}~\bibnamefont
  {Abe}}, \bibinfo {author} {\bibfnamefont {H.}~\bibnamefont {Akimoto}},
  \bibinfo {author} {\bibfnamefont {A.}~\bibnamefont {Akopian}}, \bibinfo
  {author} {\bibfnamefont {M.}~\bibnamefont {Albrow}}, \bibinfo {author}
  {\bibfnamefont {S.}~\bibnamefont {Amendolia}}, \bibinfo {author}
  {\bibfnamefont {D.}~\bibnamefont {Amidei}}, \bibinfo {author} {\bibfnamefont
  {J.}~\bibnamefont {Antos}}, \bibinfo {author} {\bibfnamefont
  {S.}~\bibnamefont {Aota}}, \bibinfo {author} {\bibfnamefont {G.}~\bibnamefont
  {Apollinari}}, \bibinfo {author} {\bibfnamefont {T.}~\bibnamefont {Asakawa}},
  \emph {et~al.},\ }\bibfield  {title} {\bibinfo {title} {Production of
  j/$\psi$ mesons from $\chi$ c meson decays in p p{\={}} collisions at
  $\sqrt{s}$= 1.8 tev},\ }\href@noop {} {\bibfield  {journal} {\bibinfo
  {journal} {Physical review letters}\ }\textbf {\bibinfo {volume} {79}},\
  \bibinfo {pages} {578} (\bibinfo {year} {1997}{\natexlab{b}})}\BibitemShut
  {NoStop}%
\bibitem [{\citenamefont {Braaten}\ and\ \citenamefont
  {Fleming}(1995)}]{psianomalyt}%
  \BibitemOpen
  \bibfield  {author} {\bibinfo {author} {\bibfnamefont {E.}~\bibnamefont
  {Braaten}}\ and\ \bibinfo {author} {\bibfnamefont {S.}~\bibnamefont
  {Fleming}},\ }\bibfield  {title} {\bibinfo {title} {Color-octet fragmentation
  and the $\psi^\prime$ surplus at the fermilab tevatron},\ }\href@noop {}
  {\bibfield  {journal} {\bibinfo  {journal} {Physical Review Letters}\
  }\textbf {\bibinfo {volume} {74}},\ \bibinfo {pages} {3327} (\bibinfo {year}
  {1995})}\BibitemShut {NoStop}%
\bibitem [{\citenamefont {Butenschoen}\ and\ \citenamefont
  {Kniehl}(2011)}]{universalityt1}%
  \BibitemOpen
  \bibfield  {author} {\bibinfo {author} {\bibfnamefont {M.}~\bibnamefont
  {Butenschoen}}\ and\ \bibinfo {author} {\bibfnamefont {B.~A.}\ \bibnamefont
  {Kniehl}},\ }\bibfield  {title} {\bibinfo {title} {World data of j/$\psi$
  production consolidate nonrelativistic qcd factorization<? format?> at
  next-to-leading order},\ }\href@noop {} {\bibfield  {journal} {\bibinfo
  {journal} {Physical Review D}\ }\textbf {\bibinfo {volume} {84}},\ \bibinfo
  {pages} {051501} (\bibinfo {year} {2011})}\BibitemShut {NoStop}%
\bibitem [{\citenamefont {Chao}\ \emph {et~al.}(2012)\citenamefont {Chao},
  \citenamefont {Ma}, \citenamefont {Shao}, \citenamefont {Wang},\ and\
  \citenamefont {Zhang}}]{universalityt2}%
  \BibitemOpen
  \bibfield  {author} {\bibinfo {author} {\bibfnamefont {K.-T.}\ \bibnamefont
  {Chao}}, \bibinfo {author} {\bibfnamefont {Y.-Q.}\ \bibnamefont {Ma}},
  \bibinfo {author} {\bibfnamefont {H.-S.}\ \bibnamefont {Shao}}, \bibinfo
  {author} {\bibfnamefont {K.}~\bibnamefont {Wang}},\ and\ \bibinfo {author}
  {\bibfnamefont {Y.-J.}\ \bibnamefont {Zhang}},\ }\bibfield  {title} {\bibinfo
  {title} {J/$\psi$ polarization at hadron colliders in nonrelativistic qcd},\
  }\href@noop {} {\bibfield  {journal} {\bibinfo  {journal} {Physical review
  letters}\ }\textbf {\bibinfo {volume} {108}},\ \bibinfo {pages} {242004}
  (\bibinfo {year} {2012})}\BibitemShut {NoStop}%
\bibitem [{\citenamefont {Gong}\ \emph {et~al.}(2013)\citenamefont {Gong},
  \citenamefont {Wan}, \citenamefont {Wang},\ and\ \citenamefont
  {Zhang}}]{universalityt3}%
  \BibitemOpen
  \bibfield  {author} {\bibinfo {author} {\bibfnamefont {B.}~\bibnamefont
  {Gong}}, \bibinfo {author} {\bibfnamefont {L.-P.}\ \bibnamefont {Wan}},
  \bibinfo {author} {\bibfnamefont {J.-X.}\ \bibnamefont {Wang}},\ and\
  \bibinfo {author} {\bibfnamefont {H.-F.}\ \bibnamefont {Zhang}},\ }\bibfield
  {title} {\bibinfo {title} {Polarization for prompt j<? format?>/<? format?>
  $\psi$ and $\psi$ (2 s) production at the tevatron and lhc},\ }\href@noop {}
  {\bibfield  {journal} {\bibinfo  {journal} {Physical review letters}\
  }\textbf {\bibinfo {volume} {110}},\ \bibinfo {pages} {042002} (\bibinfo
  {year} {2013})}\BibitemShut {NoStop}%
\bibitem [{\citenamefont {Bodwin}\ \emph
  {et~al.}(2014{\natexlab{a}})\citenamefont {Bodwin}, \citenamefont {Chung},
  \citenamefont {Kim},\ and\ \citenamefont {Lee}}]{universalityt4}%
  \BibitemOpen
  \bibfield  {author} {\bibinfo {author} {\bibfnamefont {G.~T.}\ \bibnamefont
  {Bodwin}}, \bibinfo {author} {\bibfnamefont {H.~S.}\ \bibnamefont {Chung}},
  \bibinfo {author} {\bibfnamefont {U.-R.}\ \bibnamefont {Kim}},\ and\ \bibinfo
  {author} {\bibfnamefont {J.}~\bibnamefont {Lee}},\ }\bibfield  {title}
  {\bibinfo {title} {Fragmentation contributions to j/$\psi$ production at the
  tevatron and the lhc},\ }\href@noop {} {\bibfield  {journal} {\bibinfo
  {journal} {Physical review letters}\ }\textbf {\bibinfo {volume} {113}},\
  \bibinfo {pages} {022001} (\bibinfo {year} {2014}{\natexlab{a}})}\BibitemShut
  {NoStop}%
\bibitem [{\citenamefont {Abe}\ \emph {et~al.}(2004)\citenamefont {Abe},
  \citenamefont {Abe}, \citenamefont {Aihara}, \citenamefont {Asano},
  \citenamefont {Aulchenko}, \citenamefont {Aushev}, \citenamefont
  {Bahinipati}, \citenamefont {Bakich}, \citenamefont {Ban}, \citenamefont
  {Bedny} \emph {et~al.}}]{Bfactorye1}%
  \BibitemOpen
  \bibfield  {author} {\bibinfo {author} {\bibfnamefont {K.}~\bibnamefont
  {Abe}}, \bibinfo {author} {\bibfnamefont {K.}~\bibnamefont {Abe}}, \bibinfo
  {author} {\bibfnamefont {H.}~\bibnamefont {Aihara}}, \bibinfo {author}
  {\bibfnamefont {Y.}~\bibnamefont {Asano}}, \bibinfo {author} {\bibfnamefont
  {V.}~\bibnamefont {Aulchenko}}, \bibinfo {author} {\bibfnamefont
  {T.}~\bibnamefont {Aushev}}, \bibinfo {author} {\bibfnamefont
  {S.}~\bibnamefont {Bahinipati}}, \bibinfo {author} {\bibfnamefont
  {A.}~\bibnamefont {Bakich}}, \bibinfo {author} {\bibfnamefont
  {Y.}~\bibnamefont {Ban}}, \bibinfo {author} {\bibfnamefont {I.}~\bibnamefont
  {Bedny}}, \emph {et~al.},\ }\bibfield  {title} {\bibinfo {title} {Study of
  double charmonium production in e+ e-annihilation at s= 10.6 g e v},\
  }\href@noop {} {\bibfield  {journal} {\bibinfo  {journal} {Physical Review
  D}\ }\textbf {\bibinfo {volume} {70}},\ \bibinfo {pages} {071102} (\bibinfo
  {year} {2004})}\BibitemShut {NoStop}%
\bibitem [{\citenamefont {Aubert}\ \emph {et~al.}(2005)\citenamefont {Aubert},
  \citenamefont {Barate}, \citenamefont {Boutigny}, \citenamefont {Couderc},
  \citenamefont {Karyotakis}, \citenamefont {Lees}, \citenamefont {Poireau},
  \citenamefont {Tisserand}, \citenamefont {Zghiche}, \citenamefont {Grauges}
  \emph {et~al.}}]{Bfactorye2}%
  \BibitemOpen
  \bibfield  {author} {\bibinfo {author} {\bibfnamefont {B.}~\bibnamefont
  {Aubert}}, \bibinfo {author} {\bibfnamefont {R.}~\bibnamefont {Barate}},
  \bibinfo {author} {\bibfnamefont {D.}~\bibnamefont {Boutigny}}, \bibinfo
  {author} {\bibfnamefont {F.}~\bibnamefont {Couderc}}, \bibinfo {author}
  {\bibfnamefont {Y.}~\bibnamefont {Karyotakis}}, \bibinfo {author}
  {\bibfnamefont {J.}~\bibnamefont {Lees}}, \bibinfo {author} {\bibfnamefont
  {V.}~\bibnamefont {Poireau}}, \bibinfo {author} {\bibfnamefont
  {V.}~\bibnamefont {Tisserand}}, \bibinfo {author} {\bibfnamefont
  {A.}~\bibnamefont {Zghiche}}, \bibinfo {author} {\bibfnamefont
  {E.}~\bibnamefont {Grauges}}, \emph {et~al.},\ }\bibfield  {title} {\bibinfo
  {title} {Measurement of double charmonium production in e+ e-annihilations at
  s= 10.6 gev},\ }\href@noop {} {\bibfield  {journal} {\bibinfo  {journal}
  {Physical Review D}\ }\textbf {\bibinfo {volume} {72}},\ \bibinfo {pages}
  {031101} (\bibinfo {year} {2005})}\BibitemShut {NoStop}%
\bibitem [{\citenamefont {Liu}\ \emph {et~al.}(2003)\citenamefont {Liu},
  \citenamefont {He},\ and\ \citenamefont {Chao}}]{KYLiu1}%
  \BibitemOpen
  \bibfield  {author} {\bibinfo {author} {\bibfnamefont {K.-Y.}\ \bibnamefont
  {Liu}}, \bibinfo {author} {\bibfnamefont {Z.-G.}\ \bibnamefont {He}},\ and\
  \bibinfo {author} {\bibfnamefont {K.-T.}\ \bibnamefont {Chao}},\ }\bibfield
  {title} {\bibinfo {title} {Problems of double-charm production in e+ e-
  annihilation at s= 10.6 gev},\ }\href@noop {} {\bibfield  {journal} {\bibinfo
   {journal} {Physics Letters B}\ }\textbf {\bibinfo {volume} {557}},\ \bibinfo
  {pages} {45} (\bibinfo {year} {2003})}\BibitemShut {NoStop}%
\bibitem [{\citenamefont {Braaten}\ and\ \citenamefont
  {Lee}(2003)}]{RCbraaten}%
  \BibitemOpen
  \bibfield  {author} {\bibinfo {author} {\bibfnamefont {E.}~\bibnamefont
  {Braaten}}\ and\ \bibinfo {author} {\bibfnamefont {J.}~\bibnamefont {Lee}},\
  }\bibfield  {title} {\bibinfo {title} {Exclusive double-charmonium production
  from e+ e- annihilation into a virtual photon},\ }\href@noop {} {\bibfield
  {journal} {\bibinfo  {journal} {Physical Review D}\ }\textbf {\bibinfo
  {volume} {67}},\ \bibinfo {pages} {054007} (\bibinfo {year}
  {2003})}\BibitemShut {NoStop}%
\bibitem [{\citenamefont {Hagiwara}\ \emph {et~al.}(2003)\citenamefont
  {Hagiwara}, \citenamefont {Kou},\ and\ \citenamefont {Qiao}}]{LOHag}%
  \BibitemOpen
  \bibfield  {author} {\bibinfo {author} {\bibfnamefont {K.}~\bibnamefont
  {Hagiwara}}, \bibinfo {author} {\bibfnamefont {E.}~\bibnamefont {Kou}},\ and\
  \bibinfo {author} {\bibfnamefont {C.-F.}\ \bibnamefont {Qiao}},\ }\bibfield
  {title} {\bibinfo {title} {Exclusive j/$\psi$ productions at e+ e-
  colliders},\ }\href@noop {} {\bibfield  {journal} {\bibinfo  {journal}
  {Physics Letters B}\ }\textbf {\bibinfo {volume} {570}},\ \bibinfo {pages}
  {39} (\bibinfo {year} {2003})}\BibitemShut {NoStop}%
\bibitem [{\citenamefont {Zhang}\ \emph {et~al.}(2006)\citenamefont {Zhang},
  \citenamefont {Gao},\ and\ \citenamefont {Chao}}]{Bfactorynlo1}%
  \BibitemOpen
  \bibfield  {author} {\bibinfo {author} {\bibfnamefont {Y.-J.}\ \bibnamefont
  {Zhang}}, \bibinfo {author} {\bibfnamefont {Y.-J.}\ \bibnamefont {Gao}},\
  and\ \bibinfo {author} {\bibfnamefont {K.-T.}\ \bibnamefont {Chao}},\
  }\bibfield  {title} {\bibinfo {title} {Next-to-leading-order qcd correction
  to e+ e-→ j/$\psi$+ $\eta$ c at s= 10.6 gev},\ }\href@noop {} {\bibfield
  {journal} {\bibinfo  {journal} {Physical review letters}\ }\textbf {\bibinfo
  {volume} {96}},\ \bibinfo {pages} {092001} (\bibinfo {year}
  {2006})}\BibitemShut {NoStop}%
\bibitem [{\citenamefont {Gong}\ and\ \citenamefont
  {Wang}(2008)}]{Bfactorynlo2}%
  \BibitemOpen
  \bibfield  {author} {\bibinfo {author} {\bibfnamefont {B.}~\bibnamefont
  {Gong}}\ and\ \bibinfo {author} {\bibfnamefont {J.-X.}\ \bibnamefont
  {Wang}},\ }\bibfield  {title} {\bibinfo {title} {Qcd corrections to j/$\psi$
  plus $\eta$ c production in e+ e-annihilation at s= 10.6 gev},\ }\href@noop
  {} {\bibfield  {journal} {\bibinfo  {journal} {Physical Review D}\ }\textbf
  {\bibinfo {volume} {77}},\ \bibinfo {pages} {054028} (\bibinfo {year}
  {2008})}\BibitemShut {NoStop}%
\bibitem [{\citenamefont {He}\ \emph {et~al.}(2007)\citenamefont {He},
  \citenamefont {Fan},\ and\ \citenamefont {Chao}}]{RC1}%
  \BibitemOpen
  \bibfield  {author} {\bibinfo {author} {\bibfnamefont {Z.-G.}\ \bibnamefont
  {He}}, \bibinfo {author} {\bibfnamefont {Y.}~\bibnamefont {Fan}},\ and\
  \bibinfo {author} {\bibfnamefont {K.-T.}\ \bibnamefont {Chao}},\ }\bibfield
  {title} {\bibinfo {title} {Relativistic corrections to j/$\psi$ exclusive and
  inclusive double charm production at b factories},\ }\href@noop {} {\bibfield
   {journal} {\bibinfo  {journal} {Physical Review D}\ }\textbf {\bibinfo
  {volume} {75}},\ \bibinfo {pages} {074011} (\bibinfo {year}
  {2007})}\BibitemShut {NoStop}%
\bibitem [{\citenamefont {Jia}(2010)}]{RC2}%
  \BibitemOpen
  \bibfield  {author} {\bibinfo {author} {\bibfnamefont {Y.}~\bibnamefont
  {Jia}},\ }\bibfield  {title} {\bibinfo {title} {Color-singlet relativistic
  correction to inclusive j/$\psi$ production associated<? format?> with light
  hadrons at b factories},\ }\href@noop {} {\bibfield  {journal} {\bibinfo
  {journal} {Physical Review D}\ }\textbf {\bibinfo {volume} {82}},\ \bibinfo
  {pages} {034017} (\bibinfo {year} {2010})}\BibitemShut {NoStop}%
\bibitem [{\citenamefont {He}\ \emph {et~al.}(2010)\citenamefont {He},
  \citenamefont {Fan},\ and\ \citenamefont {Chao}}]{RC3}%
  \BibitemOpen
  \bibfield  {author} {\bibinfo {author} {\bibfnamefont {Z.-G.}\ \bibnamefont
  {He}}, \bibinfo {author} {\bibfnamefont {Y.}~\bibnamefont {Fan}},\ and\
  \bibinfo {author} {\bibfnamefont {K.-T.}\ \bibnamefont {Chao}},\ }\bibfield
  {title} {\bibinfo {title} {Relativistic correction to e+ e-→ j/$\psi$+ gg
  at b factories and constraint<? format?> on color-octet matrix elements},\
  }\href@noop {} {\bibfield  {journal} {\bibinfo  {journal} {Physical Review
  D}\ }\textbf {\bibinfo {volume} {81}},\ \bibinfo {pages} {054036} (\bibinfo
  {year} {2010})}\BibitemShut {NoStop}%
\bibitem [{\citenamefont {Bodwin}\ \emph {et~al.}(2008)\citenamefont {Bodwin},
  \citenamefont {Lee},\ and\ \citenamefont {Yu}}]{RCbodwin}%
  \BibitemOpen
  \bibfield  {author} {\bibinfo {author} {\bibfnamefont {G.~T.}\ \bibnamefont
  {Bodwin}}, \bibinfo {author} {\bibfnamefont {J.}~\bibnamefont {Lee}},\ and\
  \bibinfo {author} {\bibfnamefont {C.}~\bibnamefont {Yu}},\ }\bibfield
  {title} {\bibinfo {title} {Resummation of relativistic corrections to e+
  e-→ j/$\psi$+ $\eta$ c},\ }\href@noop {} {\bibfield  {journal} {\bibinfo
  {journal} {Physical Review D}\ }\textbf {\bibinfo {volume} {77}},\ \bibinfo
  {pages} {094018} (\bibinfo {year} {2008})}\BibitemShut {NoStop}%
\bibitem [{\citenamefont {Dong}\ \emph {et~al.}(2012)\citenamefont {Dong},
  \citenamefont {Feng},\ and\ \citenamefont {Jia}}]{Bfactorynnlo}%
  \BibitemOpen
  \bibfield  {author} {\bibinfo {author} {\bibfnamefont {H.-R.}\ \bibnamefont
  {Dong}}, \bibinfo {author} {\bibfnamefont {F.}~\bibnamefont {Feng}},\ and\
  \bibinfo {author} {\bibfnamefont {Y.}~\bibnamefont {Jia}},\ }\bibfield
  {title} {\bibinfo {title} {{$O(\alpha_s v^2)$ correction to $e^+e^-\to
  J/\psi+\eta_c$ at $B$ factories}},\ }\href
  {https://doi.org/10.1103/PhysRevD.85.114018} {\bibfield  {journal} {\bibinfo
  {journal} {Phys. Rev. D}\ }\textbf {\bibinfo {volume} {85}},\ \bibinfo
  {pages} {114018} (\bibinfo {year} {2012})},\ \Eprint
  {https://arxiv.org/abs/1204.4128} {arXiv:1204.4128 [hep-ph]} \BibitemShut
  {NoStop}%
\bibitem [{\citenamefont {Huang}\ \emph {et~al.}(2023)\citenamefont {Huang},
  \citenamefont {Gong},\ and\ \citenamefont {Wang}}]{huangxd}%
  \BibitemOpen
  \bibfield  {author} {\bibinfo {author} {\bibfnamefont {X.-D.}\ \bibnamefont
  {Huang}}, \bibinfo {author} {\bibfnamefont {B.}~\bibnamefont {Gong}},\ and\
  \bibinfo {author} {\bibfnamefont {J.-X.}\ \bibnamefont {Wang}},\ }\bibfield
  {title} {\bibinfo {title} {{Next-to-next-to-leading-order QCD corrections to
  J/\ensuremath{\psi} plus \ensuremath{\eta}$_{c}$ production at the B
  factories}},\ }\href {https://doi.org/10.1007/JHEP02(2023)049} {\bibfield
  {journal} {\bibinfo  {journal} {JHEP}\ }\textbf {\bibinfo {volume} {02}},\
  \bibinfo {pages} {049}},\ \Eprint {https://arxiv.org/abs/2212.03631}
  {arXiv:2212.03631 [hep-ph]} \BibitemShut {NoStop}%
\bibitem [{\citenamefont {Brambilla}\ \emph {et~al.}(2011)\citenamefont
  {Brambilla} \emph {et~al.}}]{review1}%
  \BibitemOpen
  \bibfield  {author} {\bibinfo {author} {\bibfnamefont {N.}~\bibnamefont
  {Brambilla}} \emph {et~al.},\ }\bibfield  {title} {\bibinfo {title} {{Heavy
  Quarkonium: Progress, Puzzles, and Opportunities}},\ }\href
  {https://doi.org/10.1140/epjc/s10052-010-1534-9} {\bibfield  {journal}
  {\bibinfo  {journal} {Eur. Phys. J. C}\ }\textbf {\bibinfo {volume} {71}},\
  \bibinfo {pages} {1534} (\bibinfo {year} {2011})},\ \Eprint
  {https://arxiv.org/abs/1010.5827} {arXiv:1010.5827 [hep-ph]} \BibitemShut
  {NoStop}%
\bibitem [{\citenamefont {Andronic}\ \emph {et~al.}(2016)\citenamefont
  {Andronic} \emph {et~al.}}]{review2}%
  \BibitemOpen
  \bibfield  {author} {\bibinfo {author} {\bibfnamefont {A.}~\bibnamefont
  {Andronic}} \emph {et~al.},\ }\bibfield  {title} {\bibinfo {title}
  {{Heavy-flavour and quarkonium production in the LHC era: from
  proton\textendash{}proton to heavy-ion collisions}},\ }\href
  {https://doi.org/10.1140/epjc/s10052-015-3819-5} {\bibfield  {journal}
  {\bibinfo  {journal} {Eur. Phys. J. C}\ }\textbf {\bibinfo {volume} {76}},\
  \bibinfo {pages} {107} (\bibinfo {year} {2016})},\ \Eprint
  {https://arxiv.org/abs/1506.03981} {arXiv:1506.03981 [nucl-ex]} \BibitemShut
  {NoStop}%
\bibitem [{\citenamefont {Chung}(2018)}]{review3}%
  \BibitemOpen
  \bibfield  {author} {\bibinfo {author} {\bibfnamefont {H.~S.}\ \bibnamefont
  {Chung}},\ }\bibfield  {title} {\bibinfo {title} {{Review of quarkonium
  production: status and prospects}},\ }\href
  {https://doi.org/10.22323/1.336.0007} {\bibfield  {journal} {\bibinfo
  {journal} {PoS}\ }\textbf {\bibinfo {volume} {Confinement2018}},\ \bibinfo
  {pages} {007} (\bibinfo {year} {2018})},\ \Eprint
  {https://arxiv.org/abs/1811.12098} {arXiv:1811.12098 [hep-ph]} \BibitemShut
  {NoStop}%
\bibitem [{\citenamefont {Chen}\ \emph {et~al.}(2022)\citenamefont {Chen},
  \citenamefont {Ma},\ and\ \citenamefont {Zhang}}]{review4}%
  \BibitemOpen
  \bibfield  {author} {\bibinfo {author} {\bibfnamefont {A.-P.}\ \bibnamefont
  {Chen}}, \bibinfo {author} {\bibfnamefont {Y.-Q.}\ \bibnamefont {Ma}},\ and\
  \bibinfo {author} {\bibfnamefont {H.}~\bibnamefont {Zhang}},\ }\bibfield
  {title} {\bibinfo {title} {{A Short Theoretical Review of Charmonium
  Production}},\ }\href {https://doi.org/10.1155/2022/7475923} {\bibfield
  {journal} {\bibinfo  {journal} {Adv. High Energy Phys.}\ }\textbf {\bibinfo
  {volume} {2022}},\ \bibinfo {pages} {7475923} (\bibinfo {year} {2022})},\
  \Eprint {https://arxiv.org/abs/2109.04028} {arXiv:2109.04028 [hep-ph]}
  \BibitemShut {NoStop}%
\bibitem [{\citenamefont {Chapon}\ \emph {et~al.}(2022)\citenamefont {Chapon}
  \emph {et~al.}}]{review5}%
  \BibitemOpen
  \bibfield  {author} {\bibinfo {author} {\bibfnamefont {E.}~\bibnamefont
  {Chapon}} \emph {et~al.},\ }\bibfield  {title} {\bibinfo {title} {{Prospects
  for quarkonium studies at the high-luminosity LHC}},\ }\href
  {https://doi.org/10.1016/j.ppnp.2021.103906} {\bibfield  {journal} {\bibinfo
  {journal} {Prog. Part. Nucl. Phys.}\ }\textbf {\bibinfo {volume} {122}},\
  \bibinfo {pages} {103906} (\bibinfo {year} {2022})},\ \Eprint
  {https://arxiv.org/abs/2012.14161} {arXiv:2012.14161 [hep-ph]} \BibitemShut
  {NoStop}%
\bibitem [{\citenamefont {Dong}\ \emph {et~al.}(2018)\citenamefont {Dong} \emph
  {et~al.}}]{CEPC}%
  \BibitemOpen
  \bibfield  {author} {\bibinfo {author} {\bibfnamefont {M.}~\bibnamefont
  {Dong}} \emph {et~al.} (\bibinfo {collaboration} {CEPC Study Group}),\
  }\bibfield  {title} {\bibinfo {title} {{CEPC Conceptual Design Report: Volume
  2 - Physics \& Detector}},\ }\href@noop {} {\  (\bibinfo {year} {2018})},\
  \Eprint {https://arxiv.org/abs/1811.10545} {arXiv:1811.10545 [hep-ex]}
  \BibitemShut {NoStop}%
\bibitem [{\citenamefont {Agapov}\ \emph {et~al.}(2022)\citenamefont {Agapov}
  \emph {et~al.}}]{FCC}%
  \BibitemOpen
  \bibfield  {author} {\bibinfo {author} {\bibfnamefont {I.}~\bibnamefont
  {Agapov}} \emph {et~al.},\ }\bibfield  {title} {\bibinfo {title} {{Future
  Circular Lepton Collider FCC-ee: Overview and Status}},\ }in\ \href@noop {}
  {\emph {\bibinfo {booktitle} {{Snowmass 2021}}}}\ (\bibinfo {year} {2022})\
  \Eprint {https://arxiv.org/abs/2203.08310} {arXiv:2203.08310
  [physics.acc-ph]} \BibitemShut {NoStop}%
\bibitem [{\citenamefont {Koratzinos}(2016)}]{FCC1}%
  \BibitemOpen
  \bibfield  {author} {\bibinfo {author} {\bibfnamefont {M.}~\bibnamefont
  {Koratzinos}} (\bibinfo {collaboration} {FCC-ee study}),\ }\bibfield  {title}
  {\bibinfo {title} {{FCC-ee accelerator parameters, performance and
  limitations}},\ }\href {https://doi.org/10.1016/j.nuclphysbps.2015.09.380}
  {\bibfield  {journal} {\bibinfo  {journal} {Nucl. Part. Phys. Proc.}\
  }\textbf {\bibinfo {volume} {273-275}},\ \bibinfo {pages} {2326} (\bibinfo
  {year} {2016})},\ \Eprint {https://arxiv.org/abs/1411.2819} {arXiv:1411.2819
  [physics.acc-ph]} \BibitemShut {NoStop}%
\bibitem [{\citenamefont {Aarons}\ \emph {et~al.}(2007)\citenamefont {Aarons}
  \emph {et~al.}}]{ILC1}%
  \BibitemOpen
  \bibfield  {author} {\bibinfo {author} {\bibfnamefont {G.}~\bibnamefont
  {Aarons}} \emph {et~al.} (\bibinfo {collaboration} {ILC}),\ }\bibfield
  {title} {\bibinfo {title} {{International Linear Collider Reference Design
  Report Volume 2: Physics at the ILC}},\ }\href@noop {} {\  (\bibinfo {year}
  {2007})},\ \Eprint {https://arxiv.org/abs/0709.1893} {arXiv:0709.1893
  [hep-ph]} \BibitemShut {NoStop}%
\bibitem [{\citenamefont {Erler}\ \emph
  {et~al.}(2000{\natexlab{a}})\citenamefont {Erler}, \citenamefont
  {Heinemeyer}, \citenamefont {Hollik}, \citenamefont {Weiglein},\ and\
  \citenamefont {Zerwas}}]{ILC2}%
  \BibitemOpen
  \bibfield  {author} {\bibinfo {author} {\bibfnamefont {J.}~\bibnamefont
  {Erler}}, \bibinfo {author} {\bibfnamefont {S.}~\bibnamefont {Heinemeyer}},
  \bibinfo {author} {\bibfnamefont {W.}~\bibnamefont {Hollik}}, \bibinfo
  {author} {\bibfnamefont {G.}~\bibnamefont {Weiglein}},\ and\ \bibinfo
  {author} {\bibfnamefont {P.~M.}\ \bibnamefont {Zerwas}},\ }\bibfield  {title}
  {\bibinfo {title} {{Physics impact of GigaZ}},\ }\href
  {https://doi.org/10.1016/S0370-2693(00)00749-8} {\bibfield  {journal}
  {\bibinfo  {journal} {Phys. Lett. B}\ }\textbf {\bibinfo {volume} {486}},\
  \bibinfo {pages} {125} (\bibinfo {year} {2000}{\natexlab{a}})},\ \Eprint
  {https://arxiv.org/abs/hep-ph/0005024} {arXiv:hep-ph/0005024} \BibitemShut
  {NoStop}%
\bibitem [{\citenamefont {Aguilar-Saavedra}\ \emph {et~al.}(2001)\citenamefont
  {Aguilar-Saavedra} \emph {et~al.}}]{ILC3}%
  \BibitemOpen
  \bibfield  {author} {\bibinfo {author} {\bibfnamefont {J.~A.}\ \bibnamefont
  {Aguilar-Saavedra}} \emph {et~al.} (\bibinfo {collaboration} {ECFA/DESY LC
  Physics Working Group}),\ }\bibfield  {title} {\bibinfo {title} {{TESLA: The
  Superconducting electron positron linear collider with an integrated x-ray
  laser laboratory. Technical design report. Part 3. Physics at an e+ e- linear
  collider}},\ }\href@noop {} {\  (\bibinfo {year} {2001})},\ \Eprint
  {https://arxiv.org/abs/hep-ph/0106315} {arXiv:hep-ph/0106315} \BibitemShut
  {NoStop}%
\bibitem [{\citenamefont {Erler}\ \emph
  {et~al.}(2000{\natexlab{b}})\citenamefont {Erler}, \citenamefont
  {Heinemeyer}, \citenamefont {Hollik}, \citenamefont {Weiglein},\ and\
  \citenamefont {Zerwas}}]{ILC4}%
  \BibitemOpen
  \bibfield  {author} {\bibinfo {author} {\bibfnamefont {J.}~\bibnamefont
  {Erler}}, \bibinfo {author} {\bibfnamefont {S.}~\bibnamefont {Heinemeyer}},
  \bibinfo {author} {\bibfnamefont {W.}~\bibnamefont {Hollik}}, \bibinfo
  {author} {\bibfnamefont {G.}~\bibnamefont {Weiglein}},\ and\ \bibinfo
  {author} {\bibfnamefont {P.~M.}\ \bibnamefont {Zerwas}},\ }\bibfield  {title}
  {\bibinfo {title} {{Physics impact of GigaZ}},\ }\href
  {https://doi.org/10.1016/S0370-2693(00)00749-8} {\bibfield  {journal}
  {\bibinfo  {journal} {Phys. Lett. B}\ }\textbf {\bibinfo {volume} {486}},\
  \bibinfo {pages} {125} (\bibinfo {year} {2000}{\natexlab{b}})},\ \Eprint
  {https://arxiv.org/abs/hep-ph/0005024} {arXiv:hep-ph/0005024} \BibitemShut
  {NoStop}%
\bibitem [{\citenamefont {Ma}\ and\ \citenamefont {Zhang}(2010)}]{Zfact}%
  \BibitemOpen
  \bibfield  {author} {\bibinfo {author} {\bibfnamefont {J.-P.}\ \bibnamefont
  {Ma}}\ and\ \bibinfo {author} {\bibfnamefont {Z.-X.}\ \bibnamefont {Zhang}},\
  }\bibfield  {title} {\bibinfo {title} {Preface},\ }\href@noop {} {\bibfield
  {journal} {\bibinfo  {journal} {Sci. Chin. Phys. Mech. Astro.}\ }\textbf
  {\bibinfo {volume} {53}},\ \bibinfo {pages} {1947} (\bibinfo {year}
  {2010})}\BibitemShut {NoStop}%
\bibitem [{\citenamefont {Chen}\ \emph {et~al.}(2013)\citenamefont {Chen},
  \citenamefont {Wu}, \citenamefont {Sun}, \citenamefont {Wang},\ and\
  \citenamefont {Shen}}]{LOccchengu}%
  \BibitemOpen
  \bibfield  {author} {\bibinfo {author} {\bibfnamefont {G.}~\bibnamefont
  {Chen}}, \bibinfo {author} {\bibfnamefont {X.-G.}\ \bibnamefont {Wu}},
  \bibinfo {author} {\bibfnamefont {Z.}~\bibnamefont {Sun}}, \bibinfo {author}
  {\bibfnamefont {S.-Q.}\ \bibnamefont {Wang}},\ and\ \bibinfo {author}
  {\bibfnamefont {J.-M.}\ \bibnamefont {Shen}},\ }\bibfield  {title} {\bibinfo
  {title} {{Exclusive charmonium production from $e^+ e^-$ annihilation round
  the $Z^0$ peak}},\ }\href {https://doi.org/10.1103/PhysRevD.88.074021}
  {\bibfield  {journal} {\bibinfo  {journal} {Phys. Rev. D}\ }\textbf {\bibinfo
  {volume} {88}},\ \bibinfo {pages} {074021} (\bibinfo {year} {2013})},\
  \Eprint {https://arxiv.org/abs/1308.5375} {arXiv:1308.5375 [hep-ph]}
  \BibitemShut {NoStop}%
\bibitem [{\citenamefont {Likhoded}\ and\ \citenamefont
  {Luchinsky}(2018)}]{LOccLikhode}%
  \BibitemOpen
  \bibfield  {author} {\bibinfo {author} {\bibfnamefont {A.~K.}\ \bibnamefont
  {Likhoded}}\ and\ \bibinfo {author} {\bibfnamefont {A.~V.}\ \bibnamefont
  {Luchinsky}},\ }\bibfield  {title} {\bibinfo {title} {{Double Charmonia
  Production in Exclusive $Z$ Boson Decays}},\ }\href
  {https://doi.org/10.1142/S0217732318500785} {\bibfield  {journal} {\bibinfo
  {journal} {Mod. Phys. Lett. A}\ }\textbf {\bibinfo {volume} {33}},\ \bibinfo
  {pages} {1850078} (\bibinfo {year} {2018})},\ \Eprint
  {https://arxiv.org/abs/1712.03108} {arXiv:1712.03108 [hep-ph]} \BibitemShut
  {NoStop}%
\bibitem [{\citenamefont {Berezhnoy}\ \emph {et~al.}(2021)\citenamefont
  {Berezhnoy}, \citenamefont {Belov}, \citenamefont {Poslavsky},\ and\
  \citenamefont {Likhoded}}]{Berezhnoy}%
  \BibitemOpen
  \bibfield  {author} {\bibinfo {author} {\bibfnamefont {A.~V.}\ \bibnamefont
  {Berezhnoy}}, \bibinfo {author} {\bibfnamefont {I.~N.}\ \bibnamefont
  {Belov}}, \bibinfo {author} {\bibfnamefont {S.~V.}\ \bibnamefont
  {Poslavsky}},\ and\ \bibinfo {author} {\bibfnamefont {A.~K.}\ \bibnamefont
  {Likhoded}},\ }\bibfield  {title} {\bibinfo {title} {{One-loop corrections to
  the processes e+e-\textrightarrow{}\ensuremath{\gamma},
  Z\textrightarrow{}J/\ensuremath{\psi}\,\ensuremath{\eta}c and
  e+e-\textrightarrow{}Z\textrightarrow{}J/\ensuremath{\psi}\,J/\ensuremath{\psi}}},\
  }\href {https://doi.org/10.1103/PhysRevD.104.034029} {\bibfield  {journal}
  {\bibinfo  {journal} {Phys. Rev. D}\ }\textbf {\bibinfo {volume} {104}},\
  \bibinfo {pages} {034029} (\bibinfo {year} {2021})},\ \Eprint
  {https://arxiv.org/abs/2101.01477} {arXiv:2101.01477 [hep-ph]} \BibitemShut
  {NoStop}%
\bibitem [{\citenamefont {Luo}\ \emph {et~al.}(2022)\citenamefont {Luo},
  \citenamefont {Fu}, \citenamefont {Tian},\ and\ \citenamefont
  {Li}}]{LuoxuanNLO}%
  \BibitemOpen
  \bibfield  {author} {\bibinfo {author} {\bibfnamefont {X.}~\bibnamefont
  {Luo}}, \bibinfo {author} {\bibfnamefont {H.-B.}\ \bibnamefont {Fu}},
  \bibinfo {author} {\bibfnamefont {H.-J.}\ \bibnamefont {Tian}},\ and\
  \bibinfo {author} {\bibfnamefont {C.}~\bibnamefont {Li}},\ }\bibfield
  {title} {\bibinfo {title} {{Next-to-leading-order QCD correction to the
  exclusive double charmonium production via $Z$ decays}},\ }\href@noop {} {\
  (\bibinfo {year} {2022})},\ \Eprint {https://arxiv.org/abs/2209.08802}
  {arXiv:2209.08802 [hep-ph]} \BibitemShut {NoStop}%
\bibitem [{\citenamefont {Belov}\ \emph {et~al.}(2023)\citenamefont {Belov},
  \citenamefont {Berezhnoy},\ and\ \citenamefont {Leshchenko}}]{Belov}%
  \BibitemOpen
  \bibfield  {author} {\bibinfo {author} {\bibfnamefont {I.~N.}\ \bibnamefont
  {Belov}}, \bibinfo {author} {\bibfnamefont {A.~V.}\ \bibnamefont
  {Berezhnoy}},\ and\ \bibinfo {author} {\bibfnamefont {E.~A.}\ \bibnamefont
  {Leshchenko}},\ }\bibfield  {title} {\bibinfo {title} {{Associated Quarkonia
  Production in a Single Boson ${e^{+}e^{-}}$ Annihilation}},\ }\href
  {https://doi.org/10.1134/S1063778823060273} {\bibfield  {journal} {\bibinfo
  {journal} {Phys. Atom. Nucl.}\ }\textbf {\bibinfo {volume} {86}},\ \bibinfo
  {pages} {1474} (\bibinfo {year} {2023})},\ \Eprint
  {https://arxiv.org/abs/2303.03362} {arXiv:2303.03362 [hep-ph]} \BibitemShut
  {NoStop}%
\bibitem [{\citenamefont {Berezhnoy}\ \emph {et~al.}(2017)\citenamefont
  {Berezhnoy}, \citenamefont {Likhoded}, \citenamefont {Onishchenko},\ and\
  \citenamefont {Poslavsky}}]{Berezhn-nlo-doubleBc}%
  \BibitemOpen
  \bibfield  {author} {\bibinfo {author} {\bibfnamefont {A.~V.}\ \bibnamefont
  {Berezhnoy}}, \bibinfo {author} {\bibfnamefont {A.~K.}\ \bibnamefont
  {Likhoded}}, \bibinfo {author} {\bibfnamefont {A.~I.}\ \bibnamefont
  {Onishchenko}},\ and\ \bibinfo {author} {\bibfnamefont {S.~V.}\ \bibnamefont
  {Poslavsky}},\ }\bibfield  {title} {\bibinfo {title} {{Next-to-leading order
  QCD corrections to paired Bc production in e+e\ensuremath{-} annihilation}},\
  }\href {https://doi.org/10.1016/j.nuclphysb.2016.12.013} {\bibfield
  {journal} {\bibinfo  {journal} {Nucl. Phys. B}\ }\textbf {\bibinfo {volume}
  {915}},\ \bibinfo {pages} {224} (\bibinfo {year} {2017})},\ \Eprint
  {https://arxiv.org/abs/1610.00354} {arXiv:1610.00354 [hep-ph]} \BibitemShut
  {NoStop}%
\bibitem [{\citenamefont {Belov}\ \emph {et~al.}(2021)\citenamefont {Belov},
  \citenamefont {Berezhnoy},\ and\ \citenamefont
  {Leshchenko}}]{Belov-lo-charm+bottom}%
  \BibitemOpen
  \bibfield  {author} {\bibinfo {author} {\bibfnamefont {I.~N.}\ \bibnamefont
  {Belov}}, \bibinfo {author} {\bibfnamefont {A.}~\bibnamefont {Berezhnoy}},\
  and\ \bibinfo {author} {\bibfnamefont {E.}~\bibnamefont {Leshchenko}},\
  }\bibfield  {title} {\bibinfo {title} {{Associated Charmonium-Bottomonium
  Production in a Single Boson e+e\ensuremath{-} Annihilation}},\ }\href
  {https://doi.org/10.3390/sym13071262} {\bibfield  {journal} {\bibinfo
  {journal} {Symmetry}\ }\textbf {\bibinfo {volume} {13}},\ \bibinfo {pages}
  {1262} (\bibinfo {year} {2021})},\ \Eprint {https://arxiv.org/abs/2105.06174}
  {arXiv:2105.06174 [hep-ph]} \BibitemShut {NoStop}%
\bibitem [{\citenamefont {Liao}\ \emph
  {et~al.}(2023{\natexlab{a}})\citenamefont {Liao}, \citenamefont {Jiang},\
  and\ \citenamefont {Zhao}}]{Liao_2023}%
  \BibitemOpen
  \bibfield  {author} {\bibinfo {author} {\bibfnamefont {Q.-L.}\ \bibnamefont
  {Liao}}, \bibinfo {author} {\bibfnamefont {J.}~\bibnamefont {Jiang}},\ and\
  \bibinfo {author} {\bibfnamefont {Y.-H.}\ \bibnamefont {Zhao}},\ }\bibfield
  {title} {\bibinfo {title} {Production of double heavy quarkonia at super z
  factory},\ }\bibfield  {journal} {\bibinfo  {journal} {The European Physical
  Journal C}\ }\textbf {\bibinfo {volume} {83}},\ \href
  {https://doi.org/10.1140/epjc/s10052-023-11174-x}
  {10.1140/epjc/s10052-023-11174-x} (\bibinfo {year}
  {2023}{\natexlab{a}})\BibitemShut {NoStop}%
\bibitem [{\citenamefont {Liao}\ and\ \citenamefont {Jiang}(2024)}]{liaoqili2}%
  \BibitemOpen
  \bibfield  {author} {\bibinfo {author} {\bibfnamefont {Q.-L.}\ \bibnamefont
  {Liao}}\ and\ \bibinfo {author} {\bibfnamefont {J.}~\bibnamefont {Jiang}},\
  }\bibfield  {title} {\bibinfo {title} {{Production of higher excited
  quarkonium pair at the super Z factory}},\ }\href
  {https://doi.org/10.1088/1674-1137/ad3c2e} {\bibfield  {journal} {\bibinfo
  {journal} {Chin. Phys. C}\ }\textbf {\bibinfo {volume} {48}},\ \bibinfo
  {pages} {073102} (\bibinfo {year} {2024})}\BibitemShut {NoStop}%
\bibitem [{\citenamefont {Liao}\ \emph
  {et~al.}(2023{\natexlab{b}})\citenamefont {Liao}, \citenamefont {Jiang},\
  and\ \citenamefont {Zhao}}]{liaoqili}%
  \BibitemOpen
  \bibfield  {author} {\bibinfo {author} {\bibfnamefont {Q.-L.}\ \bibnamefont
  {Liao}}, \bibinfo {author} {\bibfnamefont {J.}~\bibnamefont {Jiang}},\ and\
  \bibinfo {author} {\bibfnamefont {Y.-H.}\ \bibnamefont {Zhao}},\ }\bibfield
  {title} {\bibinfo {title} {{Production of double P-wave heavy quarkonia at a
  super Z factory}},\ }\href {https://doi.org/10.1140/epjc/s10052-024-13292-6}
  {\bibfield  {journal} {\bibinfo  {journal} {Eur. Phys. J. C}\ }\textbf
  {\bibinfo {volume} {83}},\ \bibinfo {pages} {22} (\bibinfo {year}
  {2023}{\natexlab{b}})},\ \Eprint {https://arxiv.org/abs/2206.06123}
  {arXiv:2206.06123 [hep-ph]} \BibitemShut {NoStop}%
\bibitem [{\citenamefont {Sun}\ \emph {et~al.}(2013)\citenamefont {Sun},
  \citenamefont {Wu}, \citenamefont {Chen}, \citenamefont {Jiang},\ and\
  \citenamefont {Yang}}]{sunzhan}%
  \BibitemOpen
  \bibfield  {author} {\bibinfo {author} {\bibfnamefont {Z.}~\bibnamefont
  {Sun}}, \bibinfo {author} {\bibfnamefont {X.-G.}\ \bibnamefont {Wu}},
  \bibinfo {author} {\bibfnamefont {G.}~\bibnamefont {Chen}}, \bibinfo {author}
  {\bibfnamefont {J.}~\bibnamefont {Jiang}},\ and\ \bibinfo {author}
  {\bibfnamefont {Z.}~\bibnamefont {Yang}},\ }\bibfield  {title} {\bibinfo
  {title} {{Heavy quarkonium production through the semi-exclusive $e^+e^-$
  annihilation channels round the $Z^0$ peak}},\ }\href
  {https://doi.org/10.1103/PhysRevD.87.114008} {\bibfield  {journal} {\bibinfo
  {journal} {Phys. Rev. D}\ }\textbf {\bibinfo {volume} {87}},\ \bibinfo
  {pages} {114008} (\bibinfo {year} {2013})},\ \Eprint
  {https://arxiv.org/abs/1302.4282} {arXiv:1302.4282 [hep-ph]} \BibitemShut
  {NoStop}%
\bibitem [{\citenamefont {Hahn}(2001)}]{FA}%
  \BibitemOpen
  \bibfield  {author} {\bibinfo {author} {\bibfnamefont {T.}~\bibnamefont
  {Hahn}},\ }\bibfield  {title} {\bibinfo {title} {{Generating Feynman diagrams
  and amplitudes with FeynArts 3}},\ }\href
  {https://doi.org/10.1016/S0010-4655(01)00290-9} {\bibfield  {journal}
  {\bibinfo  {journal} {Comput. Phys. Commun.}\ }\textbf {\bibinfo {volume}
  {140}},\ \bibinfo {pages} {418} (\bibinfo {year} {2001})},\ \Eprint
  {https://arxiv.org/abs/hep-ph/0012260} {arXiv:hep-ph/0012260} \BibitemShut
  {NoStop}%
\bibitem [{\citenamefont {Mertig}\ \emph {et~al.}(1991)\citenamefont {Mertig},
  \citenamefont {Bohm},\ and\ \citenamefont {Denner}}]{FC}%
  \BibitemOpen
  \bibfield  {author} {\bibinfo {author} {\bibfnamefont {R.}~\bibnamefont
  {Mertig}}, \bibinfo {author} {\bibfnamefont {M.}~\bibnamefont {Bohm}},\ and\
  \bibinfo {author} {\bibfnamefont {A.}~\bibnamefont {Denner}},\ }\bibfield
  {title} {\bibinfo {title} {{FEYN CALC: Computer algebraic calculation of
  Feynman amplitudes}},\ }\href {https://doi.org/10.1016/0010-4655(91)90130-D}
  {\bibfield  {journal} {\bibinfo  {journal} {Comput. Phys. Commun.}\ }\textbf
  {\bibinfo {volume} {64}},\ \bibinfo {pages} {345} (\bibinfo {year}
  {1991})}\BibitemShut {NoStop}%
\bibitem [{\citenamefont {Li}\ \emph {et~al.}(2013)\citenamefont {Li},
  \citenamefont {Xu}, \citenamefont {Liu},\ and\ \citenamefont {Zhang}}]{YJLi}%
  \BibitemOpen
  \bibfield  {author} {\bibinfo {author} {\bibfnamefont {Y.-J.}\ \bibnamefont
  {Li}}, \bibinfo {author} {\bibfnamefont {G.-Z.}\ \bibnamefont {Xu}}, \bibinfo
  {author} {\bibfnamefont {K.-Y.}\ \bibnamefont {Liu}},\ and\ \bibinfo {author}
  {\bibfnamefont {Y.-J.}\ \bibnamefont {Zhang}},\ }\bibfield  {title} {\bibinfo
  {title} {{Relativistic Correction to J/psi and Upsilon Pair Production}},\
  }\href {https://doi.org/10.1007/JHEP07(2013)051} {\bibfield  {journal}
  {\bibinfo  {journal} {JHEP}\ }\textbf {\bibinfo {volume} {07}},\ \bibinfo
  {pages} {051}},\ \Eprint {https://arxiv.org/abs/1303.1383} {arXiv:1303.1383
  [hep-ph]} \BibitemShut {NoStop}%
\bibitem [{\citenamefont {Li}\ \emph {et~al.}(2014)\citenamefont {Li},
  \citenamefont {Xu}, \citenamefont {Liu},\ and\ \citenamefont
  {Zhang}}]{Li:2013nna}%
  \BibitemOpen
  \bibfield  {author} {\bibinfo {author} {\bibfnamefont {Y.-J.}\ \bibnamefont
  {Li}}, \bibinfo {author} {\bibfnamefont {G.-Z.}\ \bibnamefont {Xu}}, \bibinfo
  {author} {\bibfnamefont {K.-Y.}\ \bibnamefont {Liu}},\ and\ \bibinfo {author}
  {\bibfnamefont {Y.-J.}\ \bibnamefont {Zhang}},\ }\bibfield  {title} {\bibinfo
  {title} {{Search for $C=+$ charmonium and XYZ states in $e^+e^-\to \gamma+ H$
  at BESIII}},\ }\href {https://doi.org/10.1007/JHEP01(2014)022} {\bibfield
  {journal} {\bibinfo  {journal} {JHEP}\ }\textbf {\bibinfo {volume} {01}},\
  \bibinfo {pages} {022}},\ \Eprint {https://arxiv.org/abs/1310.0374}
  {arXiv:1310.0374 [hep-ph]} \BibitemShut {NoStop}%
\bibitem [{\citenamefont {Tanabashi}\ \emph {et~al.}(2018)\citenamefont
  {Tanabashi} \emph {et~al.}}]{PDG}%
  \BibitemOpen
  \bibfield  {author} {\bibinfo {author} {\bibfnamefont {M.}~\bibnamefont
  {Tanabashi}} \emph {et~al.} (\bibinfo {collaboration} {Particle Data
  Group}),\ }\bibfield  {title} {\bibinfo {title} {{Review of Particle
  Physics}},\ }\href {https://doi.org/10.1103/PhysRevD.98.030001} {\bibfield
  {journal} {\bibinfo  {journal} {Phys. Rev. D}\ }\textbf {\bibinfo {volume}
  {98}},\ \bibinfo {pages} {030001} (\bibinfo {year} {2018})}\BibitemShut
  {NoStop}%
\bibitem [{\citenamefont {Yu}\ \emph {et~al.}(2017)\citenamefont {Yu},
  \citenamefont {Cai}, \citenamefont {Li},\ and\ \citenamefont {Wang}}]{LDMEs}%
  \BibitemOpen
  \bibfield  {author} {\bibinfo {author} {\bibfnamefont {G.-M.}\ \bibnamefont
  {Yu}}, \bibinfo {author} {\bibfnamefont {Y.-B.}\ \bibnamefont {Cai}},
  \bibinfo {author} {\bibfnamefont {Y.-D.}\ \bibnamefont {Li}},\ and\ \bibinfo
  {author} {\bibfnamefont {J.-S.}\ \bibnamefont {Wang}},\ }\bibfield  {title}
  {\bibinfo {title} {{Heavy quarkonium photoproduction in ultrarelativistic
  heavy ion collisions}},\ }\href {https://doi.org/10.1103/PhysRevC.95.014905}
  {\bibfield  {journal} {\bibinfo  {journal} {Phys. Rev. C}\ }\textbf {\bibinfo
  {volume} {95}},\ \bibinfo {pages} {014905} (\bibinfo {year} {2017})},\
  \bibinfo {note} {[Addendum: Phys.Rev.C 95, 069901 (2017)]},\ \Eprint
  {https://arxiv.org/abs/1703.03194} {arXiv:1703.03194 [hep-ph]} \BibitemShut
  {NoStop}%
\bibitem [{\citenamefont {Bodwin}\ and\ \citenamefont {Lee}(2004)}]{GKrela1}%
  \BibitemOpen
  \bibfield  {author} {\bibinfo {author} {\bibfnamefont {G.~T.}\ \bibnamefont
  {Bodwin}}\ and\ \bibinfo {author} {\bibfnamefont {J.}~\bibnamefont {Lee}},\
  }\bibfield  {title} {\bibinfo {title} {{Relativistic corrections to gluon
  fragmentation into spin triplet S wave quarkonium}},\ }\href
  {https://doi.org/10.1103/PhysRevD.69.054003} {\bibfield  {journal} {\bibinfo
  {journal} {Phys. Rev. D}\ }\textbf {\bibinfo {volume} {69}},\ \bibinfo
  {pages} {054003} (\bibinfo {year} {2004})},\ \Eprint
  {https://arxiv.org/abs/hep-ph/0308016} {arXiv:hep-ph/0308016} \BibitemShut
  {NoStop}%
\bibitem [{\citenamefont {Gremm}\ and\ \citenamefont
  {Kapustin}(1997)}]{GKrela2}%
  \BibitemOpen
  \bibfield  {author} {\bibinfo {author} {\bibfnamefont {M.}~\bibnamefont
  {Gremm}}\ and\ \bibinfo {author} {\bibfnamefont {A.}~\bibnamefont
  {Kapustin}},\ }\bibfield  {title} {\bibinfo {title} {{Annihilation of S wave
  quarkonia and the measurement of alpha-s}},\ }\href
  {https://doi.org/10.1016/S0370-2693(97)00744-2} {\bibfield  {journal}
  {\bibinfo  {journal} {Phys. Lett. B}\ }\textbf {\bibinfo {volume} {407}},\
  \bibinfo {pages} {323} (\bibinfo {year} {1997})},\ \Eprint
  {https://arxiv.org/abs/hep-ph/9701353} {arXiv:hep-ph/9701353} \BibitemShut
  {NoStop}%
\bibitem [{\citenamefont {Cho}\ and\ \citenamefont
  {Leibovich}(1996{\natexlab{a}})}]{LDMEs1cc}%
  \BibitemOpen
  \bibfield  {author} {\bibinfo {author} {\bibfnamefont {P.~L.}\ \bibnamefont
  {Cho}}\ and\ \bibinfo {author} {\bibfnamefont {A.~K.}\ \bibnamefont
  {Leibovich}},\ }\bibfield  {title} {\bibinfo {title} {{Color octet quarkonia
  production}},\ }\href {https://doi.org/10.1103/PhysRevD.53.150} {\bibfield
  {journal} {\bibinfo  {journal} {Phys. Rev. D}\ }\textbf {\bibinfo {volume}
  {53}},\ \bibinfo {pages} {150} (\bibinfo {year} {1996}{\natexlab{a}})},\
  \Eprint {https://arxiv.org/abs/hep-ph/9505329} {arXiv:hep-ph/9505329}
  \BibitemShut {NoStop}%
\bibitem [{\citenamefont {Cho}\ and\ \citenamefont
  {Leibovich}(1996{\natexlab{b}})}]{LDMEs2}%
  \BibitemOpen
  \bibfield  {author} {\bibinfo {author} {\bibfnamefont {P.~L.}\ \bibnamefont
  {Cho}}\ and\ \bibinfo {author} {\bibfnamefont {A.~K.}\ \bibnamefont
  {Leibovich}},\ }\bibfield  {title} {\bibinfo {title} {{Color octet quarkonia
  production. 2.}},\ }\href {https://doi.org/10.1103/PhysRevD.53.6203}
  {\bibfield  {journal} {\bibinfo  {journal} {Phys. Rev. D}\ }\textbf {\bibinfo
  {volume} {53}},\ \bibinfo {pages} {6203} (\bibinfo {year}
  {1996}{\natexlab{b}})},\ \Eprint {https://arxiv.org/abs/hep-ph/9511315}
  {arXiv:hep-ph/9511315} \BibitemShut {NoStop}%
\bibitem [{\citenamefont {Butenschoen}\ \emph {et~al.}(2015)\citenamefont
  {Butenschoen}, \citenamefont {He},\ and\ \citenamefont {Kniehl}}]{LDMEs3}%
  \BibitemOpen
  \bibfield  {author} {\bibinfo {author} {\bibfnamefont {M.}~\bibnamefont
  {Butenschoen}}, \bibinfo {author} {\bibfnamefont {Z.-G.}\ \bibnamefont
  {He}},\ and\ \bibinfo {author} {\bibfnamefont {B.~A.}\ \bibnamefont
  {Kniehl}},\ }\bibfield  {title} {\bibinfo {title} {{$\eta_c$ production at
  the LHC challenges nonrelativistic-QCD factorization}},\ }\href
  {https://doi.org/10.1103/PhysRevLett.114.092004} {\bibfield  {journal}
  {\bibinfo  {journal} {Phys. Rev. Lett.}\ }\textbf {\bibinfo {volume} {114}},\
  \bibinfo {pages} {092004} (\bibinfo {year} {2015})},\ \Eprint
  {https://arxiv.org/abs/1411.5287} {arXiv:1411.5287 [hep-ph]} \BibitemShut
  {NoStop}%
\bibitem [{\citenamefont {Han}\ \emph {et~al.}(2015)\citenamefont {Han},
  \citenamefont {Ma}, \citenamefont {Meng}, \citenamefont {Shao},\ and\
  \citenamefont {Chao}}]{LDMEs4}%
  \BibitemOpen
  \bibfield  {author} {\bibinfo {author} {\bibfnamefont {H.}~\bibnamefont
  {Han}}, \bibinfo {author} {\bibfnamefont {Y.-Q.}\ \bibnamefont {Ma}},
  \bibinfo {author} {\bibfnamefont {C.}~\bibnamefont {Meng}}, \bibinfo {author}
  {\bibfnamefont {H.-S.}\ \bibnamefont {Shao}},\ and\ \bibinfo {author}
  {\bibfnamefont {K.-T.}\ \bibnamefont {Chao}},\ }\bibfield  {title} {\bibinfo
  {title} {{$\eta_c$ production at LHC and indications on the understanding of
  $J/\psi$ production}},\ }\href
  {https://doi.org/10.1103/PhysRevLett.114.092005} {\bibfield  {journal}
  {\bibinfo  {journal} {Phys. Rev. Lett.}\ }\textbf {\bibinfo {volume} {114}},\
  \bibinfo {pages} {092005} (\bibinfo {year} {2015})},\ \Eprint
  {https://arxiv.org/abs/1411.7350} {arXiv:1411.7350 [hep-ph]} \BibitemShut
  {NoStop}%
\bibitem [{\citenamefont {Zhang}\ \emph {et~al.}(2015)\citenamefont {Zhang},
  \citenamefont {Sun}, \citenamefont {Sang},\ and\ \citenamefont
  {Li}}]{LDMEs5}%
  \BibitemOpen
  \bibfield  {author} {\bibinfo {author} {\bibfnamefont {H.-F.}\ \bibnamefont
  {Zhang}}, \bibinfo {author} {\bibfnamefont {Z.}~\bibnamefont {Sun}}, \bibinfo
  {author} {\bibfnamefont {W.-L.}\ \bibnamefont {Sang}},\ and\ \bibinfo
  {author} {\bibfnamefont {R.}~\bibnamefont {Li}},\ }\bibfield  {title}
  {\bibinfo {title} {{Impact of $\eta_c$ hadroproduction data on charmonium
  production and polarization within NRQCD framework}},\ }\href
  {https://doi.org/10.1103/PhysRevLett.114.092006} {\bibfield  {journal}
  {\bibinfo  {journal} {Phys. Rev. Lett.}\ }\textbf {\bibinfo {volume} {114}},\
  \bibinfo {pages} {092006} (\bibinfo {year} {2015})},\ \Eprint
  {https://arxiv.org/abs/1412.0508} {arXiv:1412.0508 [hep-ph]} \BibitemShut
  {NoStop}%
\bibitem [{\citenamefont {He}\ and\ \citenamefont {Kniehl}(2015)}]{LDMEs6}%
  \BibitemOpen
  \bibfield  {author} {\bibinfo {author} {\bibfnamefont {Z.-G.}\ \bibnamefont
  {He}}\ and\ \bibinfo {author} {\bibfnamefont {B.~A.}\ \bibnamefont
  {Kniehl}},\ }\bibfield  {title} {\bibinfo {title} {{Complete
  Nonrelativistic-QCD Prediction for Prompt Double J/\ensuremath{\psi}
  Hadroproduction}},\ }\href {https://doi.org/10.1103/PhysRevLett.115.022002}
  {\bibfield  {journal} {\bibinfo  {journal} {Phys. Rev. Lett.}\ }\textbf
  {\bibinfo {volume} {115}},\ \bibinfo {pages} {022002} (\bibinfo {year}
  {2015})},\ \Eprint {https://arxiv.org/abs/1609.02786} {arXiv:1609.02786
  [hep-ph]} \BibitemShut {NoStop}%
\bibitem [{\citenamefont {Wang}\ and\ \citenamefont {Zhang}(2015)}]{LDMEs7}%
  \BibitemOpen
  \bibfield  {author} {\bibinfo {author} {\bibfnamefont {J.-X.}\ \bibnamefont
  {Wang}}\ and\ \bibinfo {author} {\bibfnamefont {H.-F.}\ \bibnamefont
  {Zhang}},\ }\bibfield  {title} {\bibinfo {title} {{$h_c$ production at hadron
  colliders}},\ }\href {https://doi.org/10.1088/0954-3899/42/2/025004}
  {\bibfield  {journal} {\bibinfo  {journal} {J. Phys. G}\ }\textbf {\bibinfo
  {volume} {42}},\ \bibinfo {pages} {025004} (\bibinfo {year} {2015})},\
  \Eprint {https://arxiv.org/abs/1403.5944} {arXiv:1403.5944 [hep-ph]}
  \BibitemShut {NoStop}%
\bibitem [{\citenamefont {Braaten}\ \emph {et~al.}(2001)\citenamefont
  {Braaten}, \citenamefont {Fleming},\ and\ \citenamefont
  {Leibovich}}]{LDMEs8}%
  \BibitemOpen
  \bibfield  {author} {\bibinfo {author} {\bibfnamefont {E.}~\bibnamefont
  {Braaten}}, \bibinfo {author} {\bibfnamefont {S.}~\bibnamefont {Fleming}},\
  and\ \bibinfo {author} {\bibfnamefont {A.~K.}\ \bibnamefont {Leibovich}},\
  }\bibfield  {title} {\bibinfo {title} {{NRQCD analysis of bottomonium
  production at the Tevatron}},\ }\href
  {https://doi.org/10.1103/PhysRevD.63.094006} {\bibfield  {journal} {\bibinfo
  {journal} {Phys. Rev. D}\ }\textbf {\bibinfo {volume} {63}},\ \bibinfo
  {pages} {094006} (\bibinfo {year} {2001})},\ \Eprint
  {https://arxiv.org/abs/hep-ph/0008091} {arXiv:hep-ph/0008091} \BibitemShut
  {NoStop}%
\bibitem [{\citenamefont {Sharma}\ and\ \citenamefont
  {Vitev}(2013)}]{LDMEs9cc}%
  \BibitemOpen
  \bibfield  {author} {\bibinfo {author} {\bibfnamefont {R.}~\bibnamefont
  {Sharma}}\ and\ \bibinfo {author} {\bibfnamefont {I.}~\bibnamefont {Vitev}},\
  }\bibfield  {title} {\bibinfo {title} {{High transverse momentum quarkonium
  production and dissociation in heavy ion collisions}},\ }\href
  {https://doi.org/10.1103/PhysRevC.87.044905} {\bibfield  {journal} {\bibinfo
  {journal} {Phys. Rev. C}\ }\textbf {\bibinfo {volume} {87}},\ \bibinfo
  {pages} {044905} (\bibinfo {year} {2013})},\ \Eprint
  {https://arxiv.org/abs/1203.0329} {arXiv:1203.0329 [hep-ph]} \BibitemShut
  {NoStop}%
\bibitem [{\citenamefont {Domenech}\ and\ \citenamefont
  {Sanchis-Lozano}(2000)}]{LDMEs10bb}%
  \BibitemOpen
  \bibfield  {author} {\bibinfo {author} {\bibfnamefont {J.~L.}\ \bibnamefont
  {Domenech}}\ and\ \bibinfo {author} {\bibfnamefont {M.~A.}\ \bibnamefont
  {Sanchis-Lozano}},\ }\bibfield  {title} {\bibinfo {title} {{Bottomonium
  production at the Tevatron and the LHC}},\ }\href
  {https://doi.org/10.1016/S0370-2693(00)00119-2} {\bibfield  {journal}
  {\bibinfo  {journal} {Phys. Lett. B}\ }\textbf {\bibinfo {volume} {476}},\
  \bibinfo {pages} {65} (\bibinfo {year} {2000})},\ \Eprint
  {https://arxiv.org/abs/hep-ph/9911332} {arXiv:hep-ph/9911332} \BibitemShut
  {NoStop}%
\bibitem [{\citenamefont {Eichten}\ and\ \citenamefont
  {Quigg}(1995)}]{LDMEs11}%
  \BibitemOpen
  \bibfield  {author} {\bibinfo {author} {\bibfnamefont {E.~J.}\ \bibnamefont
  {Eichten}}\ and\ \bibinfo {author} {\bibfnamefont {C.}~\bibnamefont
  {Quigg}},\ }\bibfield  {title} {\bibinfo {title} {{Quarkonium wave functions
  at the origin}},\ }\href {https://doi.org/10.1103/PhysRevD.52.1726}
  {\bibfield  {journal} {\bibinfo  {journal} {Phys. Rev. D}\ }\textbf {\bibinfo
  {volume} {52}},\ \bibinfo {pages} {1726} (\bibinfo {year} {1995})},\ \Eprint
  {https://arxiv.org/abs/hep-ph/9503356} {arXiv:hep-ph/9503356} \BibitemShut
  {NoStop}%
\bibitem [{\citenamefont {Buchmuller}\ and\ \citenamefont {Tye}(1981)}]{LDBT}%
  \BibitemOpen
  \bibfield  {author} {\bibinfo {author} {\bibfnamefont {W.}~\bibnamefont
  {Buchmuller}}\ and\ \bibinfo {author} {\bibfnamefont {S.~H.~H.}\ \bibnamefont
  {Tye}},\ }\bibfield  {title} {\bibinfo {title} {{Quarkonia and Quantum
  Chromodynamics}},\ }\href {https://doi.org/10.1103/PhysRevD.24.132}
  {\bibfield  {journal} {\bibinfo  {journal} {Phys. Rev. D}\ }\textbf {\bibinfo
  {volume} {24}},\ \bibinfo {pages} {132} (\bibinfo {year} {1981})}\BibitemShut
  {NoStop}%
\bibitem [{\citenamefont {Igi}\ and\ \citenamefont {Ono}(1986)}]{LDIO}%
  \BibitemOpen
  \bibfield  {author} {\bibinfo {author} {\bibfnamefont {K.}~\bibnamefont
  {Igi}}\ and\ \bibinfo {author} {\bibfnamefont {S.}~\bibnamefont {Ono}},\
  }\bibfield  {title} {\bibinfo {title} {{Heavy Quarkonium Systems and the
  {QCD} Scale Parameter $\Lambda$ Ms}},\ }\href
  {https://doi.org/10.1103/PhysRevD.33.3349} {\bibfield  {journal} {\bibinfo
  {journal} {Phys. Rev. D}\ }\textbf {\bibinfo {volume} {33}},\ \bibinfo
  {pages} {3349} (\bibinfo {year} {1986})}\BibitemShut {NoStop}%
\bibitem [{\citenamefont {Liao}\ and\ \citenamefont {Xie}(2014)}]{LDLiao}%
  \BibitemOpen
  \bibfield  {author} {\bibinfo {author} {\bibfnamefont {Q.-L.}\ \bibnamefont
  {Liao}}\ and\ \bibinfo {author} {\bibfnamefont {G.-Y.}\ \bibnamefont {Xie}},\
  }\bibfield  {title} {\bibinfo {title} {{Heavy quarkonium wave functions at
  the origin and excited heavy quarkonium production via top quark decays at
  the LHC}},\ }\href {https://doi.org/10.1103/PhysRevD.90.054007} {\bibfield
  {journal} {\bibinfo  {journal} {Phys. Rev. D}\ }\textbf {\bibinfo {volume}
  {90}},\ \bibinfo {pages} {054007} (\bibinfo {year} {2014})},\ \Eprint
  {https://arxiv.org/abs/1408.5563} {arXiv:1408.5563 [hep-ph]} \BibitemShut
  {NoStop}%
\bibitem [{\citenamefont {Chen}\ and\ \citenamefont {Kuang}(1992)}]{LDCK}%
  \BibitemOpen
  \bibfield  {author} {\bibinfo {author} {\bibfnamefont {Y.-Q.}\ \bibnamefont
  {Chen}}\ and\ \bibinfo {author} {\bibfnamefont {Y.-P.}\ \bibnamefont
  {Kuang}},\ }\bibfield  {title} {\bibinfo {title} {{Improved QCD motivated
  heavy quark potentials with explicit Lambda(ms) dependence}},\ }\href
  {https://doi.org/10.1103/PhysRevD.47.350} {\bibfield  {journal} {\bibinfo
  {journal} {Phys. Rev. D}\ }\textbf {\bibinfo {volume} {46}},\ \bibinfo
  {pages} {1165} (\bibinfo {year} {1992})},\ \bibinfo {note} {[Erratum:
  Phys.Rev.D 47, 350 (1993)]}\BibitemShut {NoStop}%
\bibitem [{\citenamefont {Zhang}\ \emph
  {et~al.}(2010{\natexlab{a}})\citenamefont {Zhang}, \citenamefont {Li},\ and\
  \citenamefont {Liu}}]{Zhang:2010uia}%
  \BibitemOpen
  \bibfield  {author} {\bibinfo {author} {\bibfnamefont {Y.-J.}\ \bibnamefont
  {Zhang}}, \bibinfo {author} {\bibfnamefont {B.-Q.}\ \bibnamefont {Li}},\ and\
  \bibinfo {author} {\bibfnamefont {K.-Y.}\ \bibnamefont {Liu}},\ }\bibfield
  {title} {\bibinfo {title} {{$J / \psi$ electromagnetic production associated
  with light hadrons at $B$ factories}},\ }\href@noop {} {\  (\bibinfo {year}
  {2010}{\natexlab{a}})},\ \Eprint {https://arxiv.org/abs/1003.5566}
  {arXiv:1003.5566 [hep-ph]} \BibitemShut {NoStop}%
\bibitem [{\citenamefont {Trunin}(2024)}]{rcresum}%
  \BibitemOpen
  \bibfield  {author} {\bibinfo {author} {\bibfnamefont {A.}~\bibnamefont
  {Trunin}},\ }\bibfield  {title} {\bibinfo {title} {{Resummation of
  relativistic corrections to heavy quarkonium $+\,\gamma$ production at $Z$
  factory}},\ }\href@noop {} {\  (\bibinfo {year} {2024})},\ \Eprint
  {https://arxiv.org/abs/2406.05729} {arXiv:2406.05729 [hep-ph]} \BibitemShut
  {NoStop}%
\bibitem [{\citenamefont {Bhatnagar}\ and\ \citenamefont
  {Negash}(2025)}]{Bhatnagar:2024ykb}%
  \BibitemOpen
  \bibfield  {author} {\bibinfo {author} {\bibfnamefont {S.}~\bibnamefont
  {Bhatnagar}}\ and\ \bibinfo {author} {\bibfnamefont {H.}~\bibnamefont
  {Negash}},\ }\bibfield  {title} {\bibinfo {title}
  {{(J/\ensuremath{\psi},J/\ensuremath{\psi}), and
  (\ensuremath{\eta}c,\ensuremath{\eta}c) production through two intermediate
  photons in electron-positron annihilation at B-factories}},\ }\href
  {https://doi.org/10.1016/j.nuclphysa.2024.122969} {\bibfield  {journal}
  {\bibinfo  {journal} {Nucl. Phys. A}\ }\textbf {\bibinfo {volume} {1053}},\
  \bibinfo {pages} {122969} (\bibinfo {year} {2025})},\ \Eprint
  {https://arxiv.org/abs/2406.07508} {arXiv:2406.07508 [hep-ph]} \BibitemShut
  {NoStop}%
\bibitem [{\citenamefont {Zheng}\ \emph {et~al.}(2021)\citenamefont {Zheng},
  \citenamefont {Chang}, \citenamefont {Wu}, \citenamefont {Huang},\ and\
  \citenamefont {Wang}}]{zhxc}%
  \BibitemOpen
  \bibfield  {author} {\bibinfo {author} {\bibfnamefont {X.-C.}\ \bibnamefont
  {Zheng}}, \bibinfo {author} {\bibfnamefont {C.-H.}\ \bibnamefont {Chang}},
  \bibinfo {author} {\bibfnamefont {X.-G.}\ \bibnamefont {Wu}}, \bibinfo
  {author} {\bibfnamefont {X.-D.}\ \bibnamefont {Huang}},\ and\ \bibinfo
  {author} {\bibfnamefont {G.-Y.}\ \bibnamefont {Wang}},\ }\bibfield  {title}
  {\bibinfo {title} {{Inclusive production of heavy quarkonium
  \ensuremath{\eta}Q via Z boson decays within the framework of nonrelativistic
  QCD}},\ }\href {https://doi.org/10.1103/PhysRevD.104.054044} {\bibfield
  {journal} {\bibinfo  {journal} {Phys. Rev. D}\ }\textbf {\bibinfo {volume}
  {104}},\ \bibinfo {pages} {054044} (\bibinfo {year} {2021})},\ \Eprint
  {https://arxiv.org/abs/2104.03808} {arXiv:2104.03808 [hep-ph]} \BibitemShut
  {NoStop}%
\bibitem [{\citenamefont {Liu}\ \emph {et~al.}(2008)\citenamefont {Liu},
  \citenamefont {He},\ and\ \citenamefont {Chao}}]{KYLiu2}%
  \BibitemOpen
  \bibfield  {author} {\bibinfo {author} {\bibfnamefont {K.-Y.}\ \bibnamefont
  {Liu}}, \bibinfo {author} {\bibfnamefont {Z.-G.}\ \bibnamefont {He}},\ and\
  \bibinfo {author} {\bibfnamefont {K.-T.}\ \bibnamefont {Chao}},\ }\bibfield
  {title} {\bibinfo {title} {{Search for excited charmonium states in e+ e-
  annihilation at s**(1/2) = 10.6-GeV}},\ }\href
  {https://doi.org/10.1103/PhysRevD.77.014002} {\bibfield  {journal} {\bibinfo
  {journal} {Phys. Rev. D}\ }\textbf {\bibinfo {volume} {77}},\ \bibinfo
  {pages} {014002} (\bibinfo {year} {2008})},\ \Eprint
  {https://arxiv.org/abs/hep-ph/0408141} {arXiv:hep-ph/0408141} \BibitemShut
  {NoStop}%
\bibitem [{\citenamefont {Beneke}\ and\ \citenamefont
  {Kr\"amer}(1997{\natexlab{a}})}]{jpsiCOLDMEs1}%
  \BibitemOpen
  \bibfield  {author} {\bibinfo {author} {\bibfnamefont {M.}~\bibnamefont
  {Beneke}}\ and\ \bibinfo {author} {\bibfnamefont {M.}~\bibnamefont
  {Kr\"amer}},\ }\bibfield  {title} {\bibinfo {title} {{Direct $J/\psi$ and
  $\psi^\prime$ polarization and cross-sections at the Tevatron}},\ }\href
  {https://doi.org/10.1103/PhysRevD.55.R5269} {\bibfield  {journal} {\bibinfo
  {journal} {Phys. Rev. D}\ }\textbf {\bibinfo {volume} {55}},\ \bibinfo
  {pages} {5269} (\bibinfo {year} {1997}{\natexlab{a}})},\ \Eprint
  {https://arxiv.org/abs/hep-ph/9611218} {arXiv:hep-ph/9611218} \BibitemShut
  {NoStop}%
\bibitem [{\citenamefont {Braaten}\ \emph {et~al.}(2000)\citenamefont
  {Braaten}, \citenamefont {Kniehl},\ and\ \citenamefont {Lee}}]{jpsiCOLDMEs2}%
  \BibitemOpen
  \bibfield  {author} {\bibinfo {author} {\bibfnamefont {E.}~\bibnamefont
  {Braaten}}, \bibinfo {author} {\bibfnamefont {B.~A.}\ \bibnamefont
  {Kniehl}},\ and\ \bibinfo {author} {\bibfnamefont {J.}~\bibnamefont {Lee}},\
  }\bibfield  {title} {\bibinfo {title} {{Polarization of prompt $J/\psi$ at
  the Tevatron}},\ }\href {https://doi.org/10.1103/PhysRevD.62.094005}
  {\bibfield  {journal} {\bibinfo  {journal} {Phys. Rev. D}\ }\textbf {\bibinfo
  {volume} {62}},\ \bibinfo {pages} {094005} (\bibinfo {year} {2000})},\
  \Eprint {https://arxiv.org/abs/hep-ph/9911436} {arXiv:hep-ph/9911436}
  \BibitemShut {NoStop}%
\bibitem [{\citenamefont {Bodwin}\ \emph
  {et~al.}(2014{\natexlab{b}})\citenamefont {Bodwin}, \citenamefont {Chung},
  \citenamefont {Kim},\ and\ \citenamefont {Lee}}]{etacCOLDMEs2}%
  \BibitemOpen
  \bibfield  {author} {\bibinfo {author} {\bibfnamefont {G.~T.}\ \bibnamefont
  {Bodwin}}, \bibinfo {author} {\bibfnamefont {H.~S.}\ \bibnamefont {Chung}},
  \bibinfo {author} {\bibfnamefont {U.-R.}\ \bibnamefont {Kim}},\ and\ \bibinfo
  {author} {\bibfnamefont {J.}~\bibnamefont {Lee}},\ }\bibfield  {title}
  {\bibinfo {title} {{Fragmentation contributions to $J/\psi$ production at the
  Tevatron and the LHC}},\ }\href
  {https://doi.org/10.1103/PhysRevLett.113.022001} {\bibfield  {journal}
  {\bibinfo  {journal} {Phys. Rev. Lett.}\ }\textbf {\bibinfo {volume} {113}},\
  \bibinfo {pages} {022001} (\bibinfo {year} {2014}{\natexlab{b}})},\ \Eprint
  {https://arxiv.org/abs/1403.3612} {arXiv:1403.3612 [hep-ph]} \BibitemShut
  {NoStop}%
\bibitem [{\citenamefont {Fleming}\ and\ \citenamefont
  {Mehen}(1998)}]{hcCOLDMEs1}%
  \BibitemOpen
  \bibfield  {author} {\bibinfo {author} {\bibfnamefont {S.}~\bibnamefont
  {Fleming}}\ and\ \bibinfo {author} {\bibfnamefont {T.}~\bibnamefont
  {Mehen}},\ }\bibfield  {title} {\bibinfo {title} {{Photoproduction of
  h(c)}},\ }\href {https://doi.org/10.1103/PhysRevD.58.037503} {\bibfield
  {journal} {\bibinfo  {journal} {Phys. Rev. D}\ }\textbf {\bibinfo {volume}
  {58}},\ \bibinfo {pages} {037503} (\bibinfo {year} {1998})},\ \Eprint
  {https://arxiv.org/abs/hep-ph/9801328} {arXiv:hep-ph/9801328} \BibitemShut
  {NoStop}%
\bibitem [{\citenamefont {Braaten}\ \emph {et~al.}(1996)\citenamefont
  {Braaten}, \citenamefont {Fleming},\ and\ \citenamefont {Yuan}}]{hcCOLDMEs2}%
  \BibitemOpen
  \bibfield  {author} {\bibinfo {author} {\bibfnamefont {E.}~\bibnamefont
  {Braaten}}, \bibinfo {author} {\bibfnamefont {S.}~\bibnamefont {Fleming}},\
  and\ \bibinfo {author} {\bibfnamefont {T.~C.}\ \bibnamefont {Yuan}},\
  }\bibfield  {title} {\bibinfo {title} {{Production of heavy quarkonium in
  high-energy colliders}},\ }\href
  {https://doi.org/10.1146/annurev.nucl.46.1.197} {\bibfield  {journal}
  {\bibinfo  {journal} {Ann. Rev. Nucl. Part. Sci.}\ }\textbf {\bibinfo
  {volume} {46}},\ \bibinfo {pages} {197} (\bibinfo {year} {1996})},\ \Eprint
  {https://arxiv.org/abs/hep-ph/9602374} {arXiv:hep-ph/9602374} \BibitemShut
  {NoStop}%
\bibitem [{\citenamefont {Beneke}\ \emph {et~al.}(1999)\citenamefont {Beneke},
  \citenamefont {Maltoni},\ and\ \citenamefont {Rothstein}}]{hcCOLDMEs4}%
  \BibitemOpen
  \bibfield  {author} {\bibinfo {author} {\bibfnamefont {M.}~\bibnamefont
  {Beneke}}, \bibinfo {author} {\bibfnamefont {F.}~\bibnamefont {Maltoni}},\
  and\ \bibinfo {author} {\bibfnamefont {I.~Z.}\ \bibnamefont {Rothstein}},\
  }\bibfield  {title} {\bibinfo {title} {{QCD analysis of inclusive B decay
  into charmonium}},\ }\href {https://doi.org/10.1103/PhysRevD.59.054003}
  {\bibfield  {journal} {\bibinfo  {journal} {Phys. Rev. D}\ }\textbf {\bibinfo
  {volume} {59}},\ \bibinfo {pages} {054003} (\bibinfo {year} {1999})},\
  \Eprint {https://arxiv.org/abs/hep-ph/9808360} {arXiv:hep-ph/9808360}
  \BibitemShut {NoStop}%
\bibitem [{\citenamefont {Jia}\ \emph {et~al.}(2012)\citenamefont {Jia},
  \citenamefont {Sang},\ and\ \citenamefont {Xu}}]{hcCOLDMEs5}%
  \BibitemOpen
  \bibfield  {author} {\bibinfo {author} {\bibfnamefont {Y.}~\bibnamefont
  {Jia}}, \bibinfo {author} {\bibfnamefont {W.-L.}\ \bibnamefont {Sang}},\ and\
  \bibinfo {author} {\bibfnamefont {J.}~\bibnamefont {Xu}},\ }\bibfield
  {title} {\bibinfo {title} {{Inclusive $h_c$ Production at $B$ Factories}},\
  }\href {https://doi.org/10.1103/PhysRevD.86.074023} {\bibfield  {journal}
  {\bibinfo  {journal} {Phys. Rev. D}\ }\textbf {\bibinfo {volume} {86}},\
  \bibinfo {pages} {074023} (\bibinfo {year} {2012})},\ \Eprint
  {https://arxiv.org/abs/1206.5785} {arXiv:1206.5785 [hep-ph]} \BibitemShut
  {NoStop}%
\bibitem [{\citenamefont {Beneke}\ and\ \citenamefont
  {Kr\"amer}(1997{\natexlab{b}})}]{hcCOLDMEs6}%
  \BibitemOpen
  \bibfield  {author} {\bibinfo {author} {\bibfnamefont {M.}~\bibnamefont
  {Beneke}}\ and\ \bibinfo {author} {\bibfnamefont {M.}~\bibnamefont
  {Kr\"amer}},\ }\bibfield  {title} {\bibinfo {title} {{Direct $J/\psi$ and
  $\psi^\prime$ polarization and cross-sections at the Tevatron}},\ }\href
  {https://doi.org/10.1103/PhysRevD.55.R5269} {\bibfield  {journal} {\bibinfo
  {journal} {Phys. Rev. D}\ }\textbf {\bibinfo {volume} {55}},\ \bibinfo
  {pages} {5269} (\bibinfo {year} {1997}{\natexlab{b}})},\ \Eprint
  {https://arxiv.org/abs/hep-ph/9611218} {arXiv:hep-ph/9611218} \BibitemShut
  {NoStop}%
\bibitem [{\citenamefont {Bodwin}\ \emph
  {et~al.}(1992{\natexlab{b}})\citenamefont {Bodwin}, \citenamefont {Braaten},
  \citenamefont {Yuan},\ and\ \citenamefont {Lepage}}]{chicjCOLDMEs1}%
  \BibitemOpen
  \bibfield  {author} {\bibinfo {author} {\bibfnamefont {G.~T.}\ \bibnamefont
  {Bodwin}}, \bibinfo {author} {\bibfnamefont {E.}~\bibnamefont {Braaten}},
  \bibinfo {author} {\bibfnamefont {T.~C.}\ \bibnamefont {Yuan}},\ and\
  \bibinfo {author} {\bibfnamefont {G.~P.}\ \bibnamefont {Lepage}},\ }\bibfield
   {title} {\bibinfo {title} {{P wave charmonium production in B meson
  decays}},\ }\href {https://doi.org/10.1103/PhysRevD.46.R3703} {\bibfield
  {journal} {\bibinfo  {journal} {Phys. Rev. D}\ }\textbf {\bibinfo {volume}
  {46}},\ \bibinfo {pages} {R3703} (\bibinfo {year} {1992}{\natexlab{b}})},\
  \Eprint {https://arxiv.org/abs/hep-ph/9208254} {arXiv:hep-ph/9208254}
  \BibitemShut {NoStop}%
\bibitem [{\citenamefont {Shao}\ \emph {et~al.}(2014)\citenamefont {Shao},
  \citenamefont {Ma}, \citenamefont {Wang},\ and\ \citenamefont
  {Chao}}]{chicjCOLDMEs3}%
  \BibitemOpen
  \bibfield  {author} {\bibinfo {author} {\bibfnamefont {H.-S.}\ \bibnamefont
  {Shao}}, \bibinfo {author} {\bibfnamefont {Y.-Q.}\ \bibnamefont {Ma}},
  \bibinfo {author} {\bibfnamefont {K.}~\bibnamefont {Wang}},\ and\ \bibinfo
  {author} {\bibfnamefont {K.-T.}\ \bibnamefont {Chao}},\ }\bibfield  {title}
  {\bibinfo {title} {{Polarizations of $\chi_{c1}$ and $\chi_{c2}$ in prompt
  production at the LHC}},\ }\href
  {https://doi.org/10.1103/PhysRevLett.112.182003} {\bibfield  {journal}
  {\bibinfo  {journal} {Phys. Rev. Lett.}\ }\textbf {\bibinfo {volume} {112}},\
  \bibinfo {pages} {182003} (\bibinfo {year} {2014})},\ \Eprint
  {https://arxiv.org/abs/1402.2913} {arXiv:1402.2913 [hep-ph]} \BibitemShut
  {NoStop}%
\bibitem [{\citenamefont {Zhang}\ \emph
  {et~al.}(2010{\natexlab{b}})\citenamefont {Zhang}, \citenamefont {Ma},
  \citenamefont {Wang},\ and\ \citenamefont {Chao}}]{COM-Bfac2}%
  \BibitemOpen
  \bibfield  {author} {\bibinfo {author} {\bibfnamefont {Y.-J.}\ \bibnamefont
  {Zhang}}, \bibinfo {author} {\bibfnamefont {Y.-Q.}\ \bibnamefont {Ma}},
  \bibinfo {author} {\bibfnamefont {K.}~\bibnamefont {Wang}},\ and\ \bibinfo
  {author} {\bibfnamefont {K.-T.}\ \bibnamefont {Chao}},\ }\bibfield  {title}
  {\bibinfo {title} {{QCD radiative correction to color-octet $J/\psi$
  inclusive production at B Factories}},\ }\href
  {https://doi.org/10.1103/PhysRevD.81.034015} {\bibfield  {journal} {\bibinfo
  {journal} {Phys. Rev. D}\ }\textbf {\bibinfo {volume} {81}},\ \bibinfo
  {pages} {034015} (\bibinfo {year} {2010}{\natexlab{b}})},\ \Eprint
  {https://arxiv.org/abs/0911.2166} {arXiv:0911.2166 [hep-ph]} \BibitemShut
  {NoStop}%
\bibitem [{\citenamefont {Xu}\ \emph {et~al.}(2012)\citenamefont {Xu},
  \citenamefont {Li}, \citenamefont {Liu},\ and\ \citenamefont
  {Zhang}}]{Xu2012}%
  \BibitemOpen
  \bibfield  {author} {\bibinfo {author} {\bibfnamefont {G.-Z.}\ \bibnamefont
  {Xu}}, \bibinfo {author} {\bibfnamefont {Y.-J.}\ \bibnamefont {Li}}, \bibinfo
  {author} {\bibfnamefont {K.-Y.}\ \bibnamefont {Liu}},\ and\ \bibinfo {author}
  {\bibfnamefont {Y.-J.}\ \bibnamefont {Zhang}},\ }\bibfield  {title} {\bibinfo
  {title} {{Relativistic Correction to Color Octet J/psi Production at Hadron
  Colliders}},\ }\href {https://doi.org/10.1103/PhysRevD.86.094017} {\bibfield
  {journal} {\bibinfo  {journal} {Phys. Rev. D}\ }\textbf {\bibinfo {volume}
  {86}},\ \bibinfo {pages} {094017} (\bibinfo {year} {2012})},\ \Eprint
  {https://arxiv.org/abs/1203.0207} {arXiv:1203.0207 [hep-ph]} \BibitemShut
  {NoStop}%
\bibitem [{\citenamefont {Xu}\ \emph {et~al.}(2014)\citenamefont {Xu},
  \citenamefont {Li}, \citenamefont {Liu},\ and\ \citenamefont
  {Zhang}}]{Xu:2014zra}%
  \BibitemOpen
  \bibfield  {author} {\bibinfo {author} {\bibfnamefont {G.-Z.}\ \bibnamefont
  {Xu}}, \bibinfo {author} {\bibfnamefont {Y.-J.}\ \bibnamefont {Li}}, \bibinfo
  {author} {\bibfnamefont {K.-Y.}\ \bibnamefont {Liu}},\ and\ \bibinfo {author}
  {\bibfnamefont {Y.-J.}\ \bibnamefont {Zhang}},\ }\bibfield  {title} {\bibinfo
  {title} {{$\alpha_sv^2$ corrections to $\eta_c$ and $\chi_{cJ}$ production
  recoiled with a photon at $e^+e^-$ colliders}},\ }\href
  {https://doi.org/10.1007/JHEP10(2014)071} {\bibfield  {journal} {\bibinfo
  {journal} {JHEP}\ }\textbf {\bibinfo {volume} {10}},\ \bibinfo {pages}
  {071}},\ \Eprint {https://arxiv.org/abs/1407.3783} {arXiv:1407.3783 [hep-ph]}
  \BibitemShut {NoStop}%
\end{thebibliography}

%

\end{document}